\documentclass[twocolumn,nofootinbib,preprintnumbers,superscriptaddress]{revtex4}

\usepackage[colorlinks=true,linkcolor=blue,urlcolor=magenta,citecolor=blue,pagebackref=true]{hyperref}

\usepackage{graphicx}
\usepackage{amsmath}
\usepackage{color}
\usepackage[dvipsnames]{xcolor}
\usepackage{subfigure}
\usepackage{longtable}
\usepackage{bm}
\usepackage{multirow}

\numberwithin{equation}{section}

\newcommand{\crit}{{1\rm z}}   

\begin{document}

\preprint{KANAZAWA-24-02,\ MPP-2024-18}

\title{Spectrum of global string networks and the axion dark matter mass}

\author{Ken'ichi Saikawa}
\affiliation{Institute for Theoretical Physics, 
			Kanazawa University,
			Kakuma-machi, Kanazawa, 
			Ishikawa 920-1192, 
			Japan}
   
\author{Javier Redondo}
\affiliation{Max-Planck-Institut f\"{u}r Physik (Werner-Heisenberg-Institut),  
			Boltzmannstr. 8,
			85748 Garching,
			Germany}
\affiliation{CAPA \& Departamento de F\'isica Te\'orica, 
			Universidad de Zaragoza, 
			C. Pedro Cerbuna 12,
			50009 Zaragoza,
			Spain}

\author{Alejandro Vaquero}
\affiliation{CAPA \& Departamento de F\'isica Te\'orica, 
			Universidad de Zaragoza, 
			C. Pedro Cerbuna 12,
			50009 Zaragoza,
			Spain}

\author{Mathieu Kaltschmidt}
\affiliation{CAPA \& Departamento de F\'isica Te\'orica, 
			Universidad de Zaragoza, 
			C. Pedro Cerbuna 12,
			50009 Zaragoza,
			Spain}

\date{\today}

\begin{abstract}
Cold dark matter axions produced in the post-inflationary Peccei-Quinn symmetry breaking scenario serve as clear targets for their experimental detection, 
since it is in principle possible to give a sharp prediction for their mass once we understand precisely how they are produced from the decay of global cosmic strings in the early Universe.
In this paper, we perform a dedicated analysis of the spectrum of axions radiated from strings based on large scale numerical simulations of the cosmological evolution of the Peccei-Quinn field on a static lattice.
Making full use of the massively parallel code and computing resources, we executed the simulations with up to $11264^3$ lattice sites, 
which allows us to improve our understanding of the dependence on the parameter controlling the string tension and thus give a more accurate extrapolation of the numerical results.
We found that there are several systematic effects that have been overlooked in previous works, such as the dependence on the initial conditions,
contaminations due to oscillations in the spectrum, and discretisation effects, some of which could explain the discrepancy in the literature.
We confirmed the trend that the spectral index of the axion emission spectrum increases with the string tension, 
but did not find a clear evidence of whether it continues to increase or saturates to a constant at larger values of the string tension due to the severe discretisation effects.
Taking this uncertainty into account and performing the extrapolation with a simple power law assumption on the spectrum, 
we find that the dark matter mass is predicted in the range of $m_a \approx 95\text{--}450\,\mu\mathrm{eV}$. 
\end{abstract}

\maketitle

\tableofcontents

\section{Introduction}
\label{sec:Introduction}

A network of global cosmic strings forms in the early Universe by the Kibble mechanism~\cite{Kibble:1976sj} when a global continuous symmetry is spontaneously broken~\cite{Vilenkin:1982ks}.  
Explicit breaking of the symmetry may eventually lead to the formation of domain walls, which triggers the destruction of the network~\cite{Vilenkin:1982ks}, a domain wall problem~\cite{Zeldovich:1974uw} or interesting late Universe phenomenology, e.g.~\cite{Pospelov:2012mt,Benabou:2023ghl}. In any case, the networks can have very spectacular consequences in cosmology. Global string networks can trigger structure formation~\cite{Vilenkin:1982ks}, explain the baryon asymmetry through leptogenesis~\cite{Ballesteros:2016xej,Bhattacharjee:1982zj}, lead to the formation of primordial black holes \cite{Hawking:1987bn,Fort:1993zb,Ferrer:2018uiu,Gelmini:2022nim,Gelmini:2023ngs}, and, perhaps most importantly, produce the cold dark matter (CDM)~\cite{Davis:1986xc,Arias:2012az}.

The axion appears as a pseudo Nambu-Goldstone boson of the breaking of the Peccei-Quinn (PQ) symmetry invoked to solve the strong CP problem~\cite{Peccei:1977hh,Peccei:1977ur,Weinberg:1977ma,Wilczek:1977pj}. The main parameter of the axion theory is the decay constant, which suppresses its couplings to matter and its mass as $\sim 1/f_a$. Laboratory experiments and astrophysics have excluded values $f_a\lesssim 10^8$ GeV (corresponding to $m_a<$ 0.1 eV) and a band around $m_a\in 10^{-12}\sim10^{-11}$ eV. See the recent Particle Data Group (PDG) review~\cite{Workman:2022ynf} and recent ones about axion models~\cite{DiLuzio:2020wdo} and experimental searches~\cite{Irastorza:2018dyq}. 
In the unexplored parameter space, the axion is an excellent candidate for the CDM of the Universe. 

The most remarkable aspect of axions as CDM is that despite their minute couplings to the Standard Model particles, there are \emph{feasible} experimental techniques to detect them in almost all remaining parameter space. See for instance,\footnote{Purely laboratory experiments as: solar axion searches with IAXO~\cite{IAXO:2019mpb}, long range forces with ARIADNE~\cite{ARIADNE:2017tdd}, or flavour violating decays~\cite{Goudzovski:2022vbt} can be very powerful just at the meV frontier.} oscillating EDMs with CASPER~\cite{JacksonKimball:2017elr}, DM radio~\cite{DMRadio:2022pkf}, cavity haloscopes such as ADMX~\cite{Stern:2016bbw} and CAPP~\cite{Semertzidis:2019gkj}, dielectric haloscopes as MADMAX~\cite{Beurthey:2020yuq}, plasma haloscopes as ALPHA~\cite{Lawson:2019brd}, and dish antennas as BRASS~\cite{Ringwald:2023plk} or BREAD~\cite{BREAD:2021tpx}. 
But most of these techniques heavily rely on coupling a resonator to the very coherent local dark matter axion field, which would today oscillate in a frequency band $\delta \omega/\omega \sim 10^{-6}$ with $\omega \sim m_a$. Since we do not know the axion mass, we need to tune our devices to different resonance frequencies until the axion signal is picked. In practice, this scanning process slows down current axion experimental search prospects to an octave (a factor $\sim 2$) in frequency a year. In this sense, it would be extremely advantageous if we could guide the experimental search with theoretical arguments to accelerate a discovery or simply obtain the most information in the shortest time. Another strong point of axion CDM is that it has very interesting peculiarities that distinguish it from other CDM candidates by purely cosmological or astrophysical probes like: the production of axion miniclusters~\cite{Hogan:1988mp,Kolb:1994fi} (and stars~\cite{Levkov:2018kau,Eggemeier:2019jsu,Dmitriev:2023ipv}), which could act as gravitational lenses~\cite{Kolb:1995bu,Fairbairn:2017dmf,Fairbairn:2017sil,Katz:2018zrn,Dai:2019lud,Ellis:2022grh},
some interplay with inflationary models~\cite{Graham:2018jyp,Takahashi:2018tdu},
or large isocurvature fluctuations in the cosmic microwave background (CMB)~\cite{Linde:1985yf,Seckel:1985tj,Planck:2018jri}. 

If the PQ symmetry is broken after inflation (often referred to as the \textit{post-inflationary scenario}), 
it is possible to predict the axion mass from the calculation of the CDM yield from the Big Bang, $\rho_{a,c}=\rho_{a,c}(m_a)$, and focus the experimental search. Moreover, a sizable fraction of the axion CDM is in axion miniclusters \cite{Eggemeier:2019khm,Pierobon:2023ozb} (but not all~\cite{Eggemeier:2022hqa}!). 
Here, the axion appears as an effective degree of freedom after a phase transition where the PQ symmetry becomes spontaneously broken. The vacuum expectation value of the axion field is uncorrelated at distances longer than the causal horizon so that the uncertainty in the initial conditions disappears after an average over relatively small scales ($L\gg $ mpc). The most important uncertainty, which has jeopardised attempts to pinpoint exactly the axion CDM mass, is the radiation of axions from the network of strings and walls. In this paper, we will perform extensive numerical simulations of global string networks to understand as much as possible the radiation of axions from strings with its main systematics and estimate the axion CDM mass with sensible uncertainties. 

\begin{figure*}[htbp]
\includegraphics[width=0.8\textwidth]{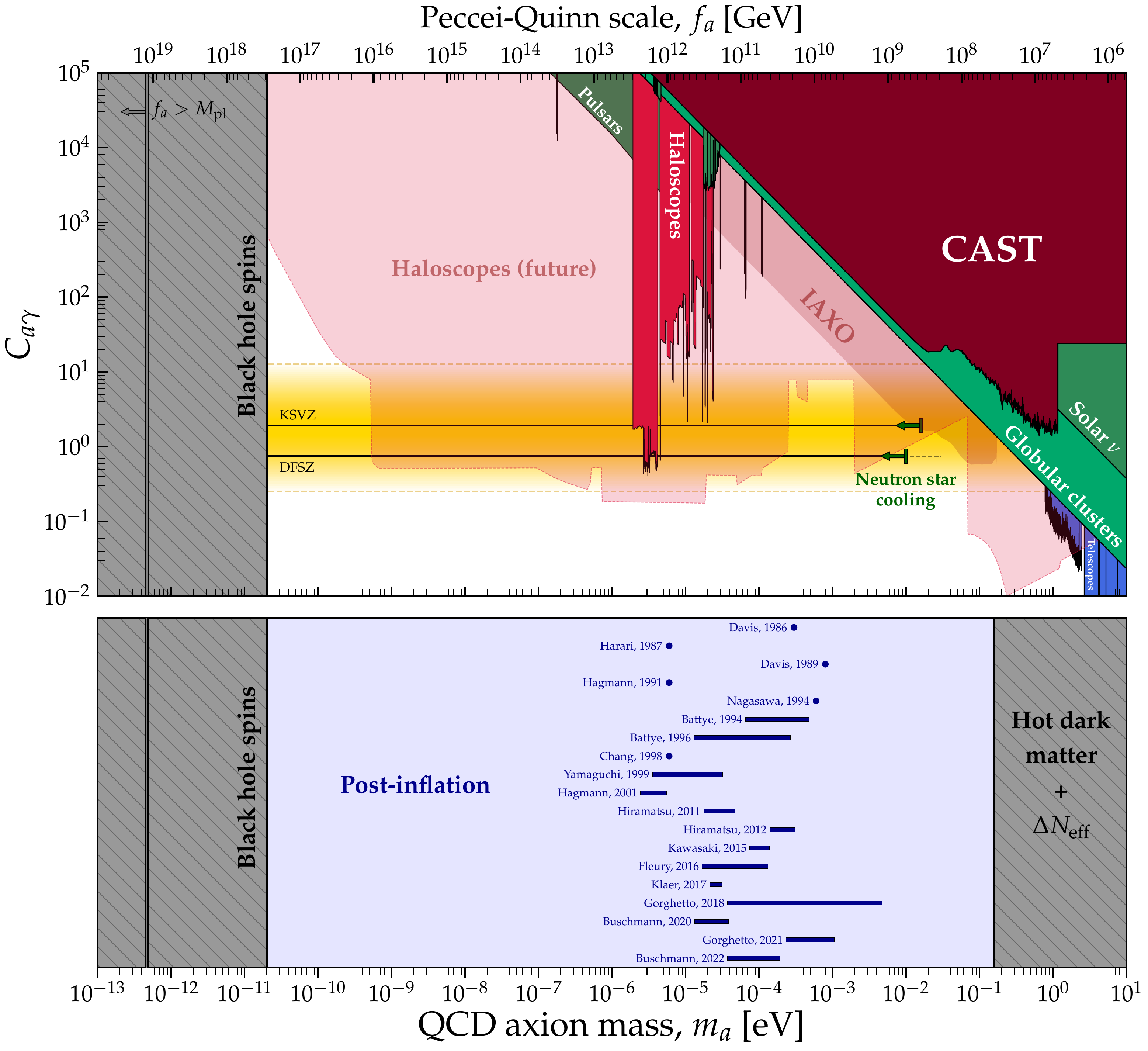}
\caption{Estimates on the axion dark matter mass in the post-inflationary radiation dominated Universe obtained in the previous works~\cite{Davis:1986xc,Harari:1987ht,Davis:1989nj,Hagmann:1990mj,Nagasawa:1994qu,Battye:1994au,Chang:1998tb,Yamaguchi:1998gx,Hagmann:2000ja,Hiramatsu:2010yu,Hiramatsu:2012gg,Kawasaki:2014sqa,Fleury:2015aca,Klaer:2017ond,Gorghetto:2018myk,Buschmann:2019icd,Gorghetto:2020qws,Buschmann:2021sdq},
confronted with constraints and current/future experimental sensitivities on the dimensionless axion-photon coupling $C_{a\gamma}$ (adapted from~\cite{AxionLimits}).}
\label{axionmass}
\end{figure*}

Many studies have already endeavoured the calculation of $\rho_{a,c}(m_a)$ and the axion CDM mass~\cite{Davis:1986xc,Harari:1987ht,Davis:1989nj,Hagmann:1990mj,Lyth:1991bb,Battye:1993jv,Nagasawa:1994qu,Battye:1994au,Battye:1994sp,Chang:1998tb,Yamaguchi:1998gx,Hagmann:2000ja,Hiramatsu:2010yu,Hiramatsu:2012gg,Kawasaki:2014sqa,Fleury:2015aca,Klaer:2017ond,Gorghetto:2018myk,Vaquero:2018tib,Buschmann:2019icd,Gorghetto:2020qws,Buschmann:2021sdq,OHare:2021zrq,Kim:2024wku}, but still the uncertainties span more than one order of magnitude, see Fig. \ref{axionmass}. The main source of uncertainty is the (differential) spectrum of axions radiated by the network, which is described by a function $\mathcal{F}$ of the wavenumber $k$ of radiated axions [see Eq.~\eqref{F_definition} for definition]. In the literature, it is often characterised by a simple power law function $\mathcal{F} \propto k^{-q}$ for the relevant momentum range, and in such a parameterisation the value of $q$ is crucial to determine the axion CDM yield.
In short, one can have many axions of small energy for $q>1$, few of high energy for $q<1$, or a scale invariant ($q=1$) for the intermediate case. When by the redshift, they become non-relativistic, 
all the axion quanta eventually have the same energy given by the axion mass,
and hence $\rho_{a,c}$ is a growing function of $q$. 
Considerations based on the Nambu-Goto effective theory coupled to a Kalb-Ramond field supported $q>1$ \cite{Davis:1986xc,Vilenkin:1986ku,Davis:1989nj,Battye:1993jv,Battye:1994au,Battye:1994sp} but a set of different arguments point to a scale invariant radiation spectrum ($q=1$)~\cite{Harari:1987ht,Hagmann:1990mj,Hagmann:2000ja}. 

Other important uncertainties are the precise location of the infrared cutoff of the spectrum and the total density of strings. 
The former can be used to estimate a typical momentum of radiated axions, and the latter defines
the total energy density of the radiation source.
It was argued that the density of strings for the network of global strings can be smaller than that of local strings~\cite{Yamaguchi:1998gx,Yamaguchi:1999yp,Moore:2001px}, 
but the estimation of its precise value remains somewhat controversial~\cite{Fleury:2015aca,Klaer:2017ond,Gorghetto:2018myk,Vaquero:2018tib,Buschmann:2019icd,Gorghetto:2020qws,Buschmann:2021sdq,OHare:2021zrq,Kawasaki:2018bzv,Hindmarsh:2019csc,Klaer:2019fxc,Hindmarsh:2021vih,Hindmarsh:2021zkt}.

In principle, numerical simulations of the dynamics of global strings can determine precisely these parameters. 
However, they suffer from a huge dynamical range issue. The smallest distances to resolve, the string cores, have typical sizes $f_a^{-1}$, but the typical distance between strings is associated with the causal horizon $H^{-1}$. The ratio at the latest times is $f_a/H\sim 10^{30}$, much beyond anything that we will be able to simulate even in a futuristic setup. We are thus doomed to simulate at unphysical values of the ratio and consider the extrapolation over a huge range. Fortunately, the string \emph{energies} depends only logarithmically on the ratio, 
\begin{align}
\ell \equiv \ln\left(\frac{m_r}{H}\right) \propto \ln\left(\frac{f_a}{H}\right), 
\end{align}
where $m_r$ is the mass of the heavier field degree of freedom [see below Eq.~\eqref{field_decomposition}],
and the extrapolation in $\ell$ appears much more feasible, if still very ambitious. 
Interestingly, it has been recently explained how using mixed local-global strings one can simulate at realistic values of the $\ell$ parameter ($\sim 70$) in an ``effective"\footnote{Here the effective tension parameter $\kappa$ is defined as the ratio of the full string tension normalised to the axionic contribution to the core $\kappa = \mu/(\pi f_a^2)$ (which is $\ell$ for pure global strings), and the trick is to make $f_a$ small, instead of increasing $\mu$. } manner~\cite{Klaer:2017qhr}. 
 
In this paper, we aim at describing the spectrum of relativistic Goldstone modes (axions) from \emph{pure} global string networks with unprecedented accuracy. To that effect, we perform numerical simulations of the simplest system presenting a spontaneously broken $U(1)$ symmetry, a complex scalar field with a ``Mexican hat" potential~\cite{Shellard:1987bv}. We perform simulations in up to 11264$^3$ lattices in the supercomputers of the Max Planck Computing and Data Facility (MPCDF, Garching) and Cybermedia Center (CMC, Osaka) to reach the most precise spectra up to date. Our results confirm the presence of an attractor~\cite{Gorghetto:2018myk} and a scaling solution but with logarithmic ($\ell$) corrections to the parameters, previously found by some studies~\cite{Fleury:2015aca,Klaer:2017ond,Gorghetto:2018myk,Vaquero:2018tib,Buschmann:2019icd,Gorghetto:2020qws,Buschmann:2021sdq,OHare:2021zrq,Kawasaki:2018bzv,Klaer:2019fxc} albeit challenged by some others~\cite{Hindmarsh:2019csc,Hindmarsh:2021vih}. 
Our study builds on the recent studies~\cite{Gorghetto:2018myk,Gorghetto:2020qws} by refining the definition of the attractor, and a more careful estimate of discretisation and systematic effects, made possible by conducting many simulations and utilising larger grids. 
Finally, we perform the extrapolation to physical values of $\ell$ by considering carefully the discretisation effects and compute the axion dark matter mass to be in the range 
\begin{equation}
m_a\in (95\text{--}450)  \,\mu\mathrm{eV}, 
\end{equation}
given the uncertainties. Unfortunately, we cannot assess whether $q=1$ or $q>1$ with the current simulation power. 

Recently, the use of adaptive-mesh-refinement (AMR) in simulations~\cite{Buschmann:2021sdq,Drew:2019mzc,Drew:2022iqz,Drew:2023ptp} has allowed to reach better dynamical range. 
The AMR simulation of the global string networks performed in Ref.~\cite{Buschmann:2021sdq} found $q=1$ while previous studies~\cite{Gorghetto:2018myk,Gorghetto:2020qws} showed $q<1$ with a growing trend with $\ell$ that points to $q>1$ once extrapolated. Thanks to our extensive simulation suite, we discuss why discretisation and systematic effects such as the role of initial conditions or the procedure to calculate $q$ from data can explain this difference. Unfortunately, more AMR simulations will be needed to fully exploit the extended dynamical range.  

The rest of this paper is organised as follows.
In Sec.~\ref{sec:basics}, we review theoretical basics for the calculation of the spectrum of axions produced by strings and introduce relevant quantities to characterise it.
After that, in Sec.~\ref{sec:attractor} we start by presenting our numerical results on the evolution of the string density and refine the notion of the attractor
to establish a baseline for the rest of the analysis and extrapolation.
We show that the attractor behaviour of the string networks can be described by a model inspired by Ref.~\cite{Klaer:2019fxc}, which allows us to interpret the attractor solution as a ``tracking" solution that lags behind the perfect scaling solution found in the system with a constant string tension.
The subsequent two sections are devoted to the discussion on systematic effects that can bias the results of the analysis:
One is the existence of the oscillations in the spectrum (Sec.~\ref{sec:oscillations}), which is an inevitable consequence of the dynamics of the axion field in the system.
The other is the effect of discretisation errors (Sec.~\ref{sec:disc_effect}), 
which distorts the spectrum and can lead to a misinterpretation of the value of the spectral index that is the most important quantity for the  estimation of the dark matter relic abundance.
In Sec.~\ref{sec:extrapolation}, we present further results on the quantities characterising the axion production efficiency, and give estimates for the axion relic abundance by extrapolating the results obtained from the simulations.
We summarise our results and conclude in Sec.~\ref{sec:conclusions}.
Some technical details and further results of the simulations are described in the appendices.
We describe details on our simulation method, including the scheme to solve the equation of motion and choice of parameters in Appendix~\ref{app:code}.
In Appendix~\ref{app:initial_conditons}, the method to prepare the initial conditions for the simulation is discussed.
In Appendix~\ref{app:masking}, we explain our method to mask the contribution of the data around the string core and discuss systematics associated with it.
We also describe technical details on the method to calculate the axion emission spectrum, energy density emission rate, and the spectral index in Appendix~\ref{app:calc_F_and_q}.
Appendix~\ref{app:other_systematics} is devoted to discussions on systematics not thoroughly mentioned in the main text.
Furthermore, in Appendix~\ref{app:evolution_analytical} we perform an analytical estimate on the evolution of the axion field, which helps to interpret some of the numerical results.
Finally, we summarise the differences with recent simulations by other groups in Appendix~\ref{app:comparisons}.

\section{String network and spectrum}
\label{sec:basics}

We simulate a complex scalar field $\phi$ with the Lagrangian, 
\begin{align}
\mathcal{L} = \frac{1}{2}|\partial_{\mu}\phi|^2 - V(\phi), \quad V(\phi) = \frac{\lambda}{4}\left(|\phi|^2 - f_a^2\right)^2,
\label{Lagrangian_PQ}
\end{align}
where $f_a$ defines the scale of symmetry breaking. 
Expanding the field around the potential minima $|\phi|=f_a$,
\begin{align}
\phi(x) = \left(f_a + r(x)\right)e^{i\theta(x)}, \label{field_decomposition}
\end{align}
we identify $a(x) = \theta(x) f_a$ as the massless Goldstone boson (the axion) and the radial field $r(x)$ as a particle with mass $m_r = \sqrt{2\lambda}f_a$ (saxion).\footnote{Note that the use of the term ``saxion" in this paper is different from the commonly used terminology that it refers to the scalar partner of the axion in supersymmetric models. Here we call the radial field ``saxion" just for convenience, even though the model does not necessarily have to be supersymmetric.}  

The field described by the Lagrangian~\eqref{Lagrangian_PQ} has vortex configurations, referred to as global strings, around which its phase wraps around $\theta\in [0,2\pi)$. They feature a core of size $m_r^{-1}$ where the saxion field transitions smoothly from the energy minimum $r=0$ to the unbroken PQ symmetry value $r=-f_a$ at the vortex center.  
The tension of the string (i.\, e. energy per unit length) is\footnote{The $c's$ are $\mathcal{O}(1)$ constants which depend on the geometry of the string. } $\mu \sim \pi f_a^2 (c_1+c_2\ell)$. The  dominant logarithmic piece comes from the axion gradient energy density $|\nabla a|^2/2$ outside the core. It is regularised by ultraviolet (UV) cutoffs and infrared (IR) scales given by $m_r^{-1}$ and the string radius or distance to closest strings, respectively~\cite{Dabholkar:1989ju}.

Throughout this paper, we consider only the evolution of the network during the radiation domination with constant degrees of freedom, most relevant for the axion dark matter case.
The evolution of the string network exhibits a pseudo-scaling regime in which the length of strings $l_s$ in a causal volume $\mathcal{V}= t^3$ is of the order of the horizon $t\sim H^{-1}$ itself. The latter becomes the typical string inter-distance and the IR cut-off. 
Thus, for cosmological simulations we define $\ell = \ln(m_r/H)$,
which increases with time because of the time evolution of the Hubble parameter $H(t)$.

Defining the ${\cal  O}(1)$ string density parameter as, 
\begin{align}
\xi = \frac{l_s}{\mathcal{V}}t^2, \label{xi_definition}
\end{align}
the energy density of strings can be defined as  
\begin{align}
\rho_{\rm str} = \xi\frac{\mu_{\rm th}}{t^2}, \label{rho_string_scaling}
\end{align}
where $\mu_{\rm th}$ is an network average string tension, 
\begin{align}
\mu_{\rm th} = \pi f_a^2 \ln\left(\frac{m_r\eta}{H\sqrt{\xi}}\right). \label{tension_theory}
\end{align}
The denominator $1/(H\sqrt{\xi})$ captures an expectation that the average distance between long strings 
is given by $\sim H^{-1}/\sqrt{\xi}$~\cite{Hagmann:2000ja}, and $\eta$ parametrises the non-trivial average shape.\footnote{ 
For straight parallel strings, $\eta=1/\sqrt{4\pi}$~\cite{Gorghetto:2018myk}.}
The string network must then release energy at a rate~\cite{Gorghetto:2018myk},\footnote{Dots represent derivatives with respect to $t$.}
\begin{align}
\label{Gamma_th}
\Gamma_{\rm th} &= \frac{\xi\mu_{\rm th}}{t^2}\left[2H - \frac{\dot{\xi}}{\xi} - \frac{\pi f_a^2}{\mu_{\rm th}}\left(H+\frac{\dot{\eta}}{\eta} - \frac{1}{2}\frac{\dot{\xi}}{\xi}\right)\right] \times f_L, \\
 \label{Gamma_a_large_log}
&\simeq  8\pi f_L\xi \ell f_a^2 H^3\quad \text{for}\ \ell \gg 1, 
\end{align}
to satisfy Eqs.~\eqref{rho_string_scaling} and~\eqref{tension_theory}. 
We introduce the extra factor $f_L$, to be determined by simulations, because of several reasons. First, only the ``long" strings ($l\gg H^{-1}$) are separated by the assumed $1/(H\sqrt{\xi})$ distances. The IR cutoff of small loops ($l\ll H^{-1}$) is their radius, and thus their $\ell$ is smaller. Since small loops only account for $\sim 20\%$ of the string length~\cite{Gorghetto:2018myk} and their $\ell$ has a lower IR scale, we neglect them in $\Gamma_{\rm th}$ and expect  
$\xi \to \xi_L \approx 0.8\xi$, where $\xi_L$ is the contribution of long strings. Furthermore, Eq.~\eqref{tension_theory} is a static result and the energy per unit length increases because of transverse motion, so we expect that string velocities contribute an extra multiplicative factor $\gamma=1/\sqrt{1-v^2}$, where $\gamma$ and $v$ are suitably averaged boost factors and transverse velocities, respectively.\footnote{In other words, if the string length $l_s$ in Eq.~\eqref{xi_definition} is measured in the comoving coordinate (we actually do so in our simulations), $\xi$ is suppressed by the Lorentz contraction, and we can convert it into the proper length by multiplying $\gamma$. Hence we can interpret $\gamma l_s\mu_{\rm th}$ as the mass energy in the rest frame of the string.}

The evolution of the massless axion field outside the string cores is given by 
$\psi_{\tau\tau} - \nabla^2 \psi = 0$,   
where $\psi =\tau\theta$, 
$\tau$ is the conformal time, $d\tau = dt/R$, $R$ is the scale factor of the Universe, 
and the subscript $\tau$ denotes the derivative with respect to $\tau$. 
Axion \emph{particle} solutions are plane waves, 
$\psi({\bm x}) = \int\frac{d^3{\bm k}}{(2\pi)^3}{\widetilde \psi}({\bm k})e^{i {\bm k}\cdot {\bm x}}$
(${\bm x}$ and ${\bm k}$ are comoving coordinates and wavenumbers),
with \emph{conserved}\footnote{In the absence of interactions, i.e. far from the cores, ${\cal N}$ is conserved in time but only after the wave has entered the horizon, i.e. $k\tau\gg 2\pi$.} occupation number, 
${\cal N}({\bm k}) = (|\partial_\tau \widetilde \psi({\bm k})|^2 + k^2 |\widetilde \psi({\bm k})|^2)/(2kV)$, 
where $k=|{\bm k}|$ and $V$ is the comoving volume. 
The energy density in massless axion waves is,
\begin{align}
\rho_a =(f_aH)^2 \int dk \frac{k^3}{2\pi^2}{\cal N}(k), 
\end{align}
and should increase with a rate $\Gamma_a = R^{-4}\frac{d}{dt}(R^4\rho_a)$ \emph{approximately} equal to the rate at which the string network releases energy according to Eq.~\eqref{Gamma_th}. We denote ${\cal N}(k)$ as the angle average version of ${\cal N}(\bm{k})$. The relation $\Gamma_a\approx \Gamma_{\rm th}$ is only \emph{approximate} because it neglects saxion radiation from the network and because we have neglected small loops, both of which are good approximations.\footnote{We will see in Sec.~\ref{sec:continuum_ext_Gamma} how saxion radiation becomes negligible at large $\ell$. The fact that small loops are negligible is supported by the analysis of Refs.~\cite{Gorghetto:2018myk,Gorghetto:2020qws,Gorghetto:2021fsn}, but Ref.~\cite{Gorghetto:2021fsn} mentions a caveat. If small loops at large $\ell$ would become extremely long lived, the estimate of the axion abundance would be significantly altered~\cite{Battye:1994au,Wantz:2009it}. 
However, there is no sign of this in simulations up to date.}

In order to calculate the axion CDM yield we need to study the distribution of axions radiated as a function of energy. We thus define the spectral production rate,
\begin{align}
\frac{\partial\Gamma_a(k)}{\partial k} = (f_aH)^2\frac{k^3}{2\pi^2}\frac{d\mathcal{N}(k)}{dt},
\end{align}
or its dimensionless variant,\footnote{Note that $\mathcal{F}(x,y)$ defined in Eq.~\eqref{F_definition} is different from $F(x,y)$
introduced in Ref.~\cite{Gorghetto:2018myk}. The latter is accompanied by the normalization condition $\int dx F(x,y) = 1$
and is related to $\mathcal{F}(x,y)$ as
\begin{align}
F(x,y) = \frac{f_a^2H^3}{\Gamma_a}\mathcal{F}(x,y).
\end{align}
In the numerical study, we prefer to use $\mathcal{F}(x,y)$ rather than $F(x,y)$, since $\Gamma_a$
is not reliably calculated at late times due to severe discretisation effects (see Sec.~\ref{sec:disc_effect}).}
\begin{align}
\mathcal{F}(x,y) &\equiv \frac{1}{(f_aH)^2}\frac{\partial \Gamma_a}{\partial(k/R)} \nonumber\\
&= \frac{1}{(f_aH)^2}\frac{1}{R^3}\frac{\partial}{\partial t}\left(R^4\frac{\partial\rho_a}{\partial k}\right),
\label{F_definition}
\end{align}
which characterises the spectrum through $x = k/(RH)$ and 
the time evolution through $y = m_r/H$.
Hereafter we call $\mathcal{F}(x,y)$ the \textit{instantaneous emission spectrum}.

Integrating in time we can calculate the axion number as, 
\begin{align}
\frac{n_a}{f_a^2H} &= \frac{1}{f_a^2H}\int dk\frac{1}{\omega}\frac{\partial\rho_a}{\partial k} \quad (\text{with}\ \omega = k/R) \nonumber\\
&= \int^{\tau}\frac{d\tau'}{\tau}\int\frac{dx'}{x'}\mathcal{F}(x',y'), \label{na_analytical_1}
\end{align}
and get an estimate of the variance of $\theta$, 
\begin{align}
\langle \theta^2\rangle=\frac{R^2}{f_a^2}\int\frac{dk}{k^2}\frac{\partial\rho_a}{\partial k} = \int^{\tau}\frac{\tau'd\tau'}{\tau^2}\int\frac{dx'}{x'^2}\mathcal{F}(x',y'). \label{amp_analytical_1}
\end{align}
Note that these quantities can be written as
\begin{align}
\frac{n_a}{f_a^2H} &= \int^{\tau}\frac{d\tau'}{\tau}\frac{\Gamma_a'}{f_a^2H'^3}\langle x^{-1}\rangle(y'), \label{na_analytical_2}\\
\langle \theta^2\rangle &= \int^{\tau}\frac{\tau'd\tau'}{\tau^2}\frac{\Gamma_a'}{f_a^2H'^3}\langle x^{-2}\rangle(y'), \label{amp_analytical_2}
\end{align}
where we define the mean inverse momenta,
\begin{align}
\langle x^{-1}\rangle(y) \equiv \frac{\int\frac{dx}{x}\mathcal{F}(x,y)}{\int dx\mathcal{F}(x,y)},\quad \langle x^{-2}\rangle(y) \equiv \frac{\int\frac{dx}{x^2}\mathcal{F}(x,y)}{\int dx\mathcal{F}(x,y)}.
\label{mean_inverse_momenta_definition}
\end{align}

The instantaneous emission spectrum has been studied numerically in Refs.~\cite{Gorghetto:2018myk,Gorghetto:2020qws,Buschmann:2021sdq} 
and seems to be well approximated by a power law $d\mathcal{N}/dt \sim 1/k^{3+q}$ (or $\mathcal{F} \propto 1/x^q$)
with IR and UV cutoffs at comoving momenta $k \sim HR \propto R^{-1}$ and $k \sim m_r R \propto R$, respectively.

\begin{figure}[htbp]
\includegraphics[width=0.4\textwidth]{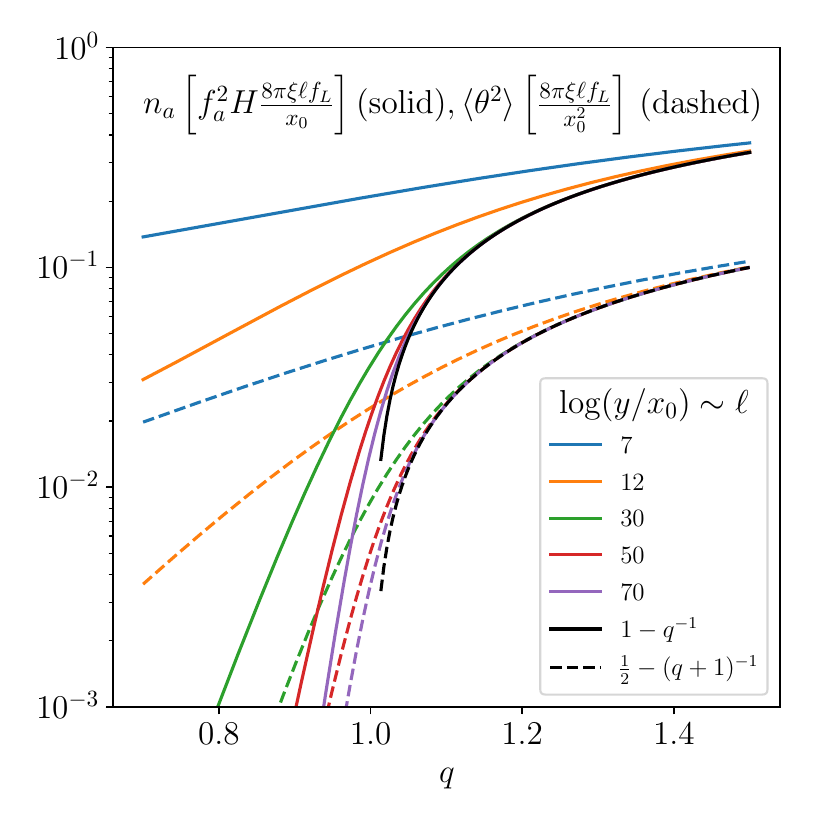}
\caption{Axion number and angle variance in the simplest radiation model \eqref{F_simple_power_law},
normalized by the multiplicative factor in the square bracket.}
\label{fig:naqmodel}
\end{figure}

Consider the following simple power law model,
\begin{align}
\mathcal{F} = 
\begin{cases}
\mathcal{F}_0x^{-q} & (x_0 < x< y),\\
0 & (\text{otherwise}),
\label{F_simple_power_law}
\end{cases}
\end{align}
where $\mathcal{F}_0$ and $x_0$ are independent of $x$ but may depend on $y$. 
We can evaluate Eq.~\eqref{mean_inverse_momenta_definition} analytically and calculate 
$n_a$ and $\langle\theta^2\rangle$ through Eqs.~\eqref{na_analytical_2} and~\eqref{amp_analytical_2}. 
In Fig.~\ref{fig:naqmodel} we show the results assuming $\Gamma_a=\Gamma_{\rm th}(\ell\gg 1)$ [Eq.~\eqref{Gamma_a_large_log}] with 
constant (or $\log$-varying) $\xi$, $f_L$, and $x_0$.
In this approximation, the non-trivial parameters are $q$ and the final value of $y/x_0$ while $\xi$, $x_0$, and $f_L$ enter only through a multiplicative factor. The most important parameter appears to be $q$. For $q>1$, the spectrum is dominated by the IR cut-off, so $n_a$ and $\langle\theta^2\rangle$ become the largest and also relatively insensitive to both $q$ and the UV cut-off parameter $y/x_0$. For $q<1$, the emission is dominated by the UV and thus, it strongly depends on $q$ and $y/x_0$, growing with the former and decreasing with the latter. In the scale-invariant case, $q=1$, the $y$-dependence of the shown curves $n_a,\langle\theta^2\rangle$ is a mild logarithmic suppression $1/\ell$, which cancels the multiplicative enhancement factor $\ell$.  
 
Note that the axion field becomes highly nonlinear ($\sqrt{\langle \theta^2\rangle} \gg \pi$) for $q>1$ and $\ell\gg 1$ (as long as $x_0$ is not too large). 
This non-linearity reduces the axion number around the epoch of the QCD phase transition~\cite{Gorghetto:2020qws}. 
We will take this into account in our estimates of the axion CDM relic abundance in Sec.~\ref{sec:axion_abundance}. 

The shape of $\cal F$ as well as the parameters $\xi$, $q$, $x_0$, and $f_L$ can be computed through numerical simulations, though only for a limited range of 
values of $\ell \lesssim 8\text{--}9$. Most of them turn out to have a small dependence on $\ell$ but full consensus on the $\ell$ dependence of these quantities has not been reached in the literature. There remains a serious discrepancy in the prediction for the axion dark matter mass
obtained from the extrapolation of the numerical results to $\ell \gg 1$.
In this paper, we study these dependencies with precision so that this extrapolation can be adequately improved.

\section{Attractor, radial field emission, and network density}
\label{sec:attractor}

The extrapolation of the parameters of the string network, $\xi$, $q$, $x_0$, and $f_L$ to the relevant values of $\ell$ depends to some extent on the way we initialise the network in our simulations. It is therefore crucial to note that the system exhibits an attractor behaviour~\cite{Gorghetto:2018myk} around which differences in initial conditions become less significant at late times. Our extrapolation would be based on the properties of the attractor solution, in the hope to be insensitive to properties of the transient behaviours.   
In this section, we characterise the properties of the attractor by initialising networks with different densities and studying the evolution of $\xi$ and the radiation spectra.  Dependence on other parameters for initial conditions is discussed in Appendix~\ref{app:initial_conditons}. 

Figure~\ref{fig:attr_xi} shows the evolution of the $\xi$ parameter for different initial string densities. 
Although $\xi$ seems to converge, the convergence seems rather slow, particularly at small values of $\xi$.
In the spirit of Ref.~\cite{Klaer:2019fxc}, we can \emph{model} the evolution of the string network density $\xi$ as, 
\begin{equation}
\label{eq:attractive_xi_evol}
\frac{d\xi}{dt} = \frac{C}{t}\left(\xi_c(\ell(t))-\xi(t)\right), 
\end{equation}
where $\xi_c(\ell)$ is the equilibrium density of the network of conformal strings~\cite{Klaer:2019fxc} at a given value of $\ell$ and $t/C$ is the characteristic time scale of the restoration of the network. 
The right-hand side is expected to vanish in the limit $\xi\to\xi_c$, and for a small difference it may be modeled by a function linear in $\xi_c-\xi$.
Assuming a constant $C$ yields a poor fit to the evolution. The main reasons are: 1) that $\xi=0$ must be a fixed point (string length is hardly created if there are no strings to begin with) and 2) that over-dense networks should have a relaxation scale parametrically smaller because the distance between strings is smaller. 
We model these effects with $C(x)=x/(1+\sqrt{x}/c_0)$, where $x=\xi/\xi_c$, to have $C(x\to 0)\to \xi$ in the small $\xi\ll\xi_c$ limit (which corresponds to a frozen network, $\xi\propto 1/t \propto 1/R^2$) and $C(x\gg1)\sim c_0\sqrt{\xi/\xi_c}$, i.e. a $1/\sqrt{\xi}$-suppressed restoration time scale corresponding to a $1/\sqrt{\xi}$-suppressed interdistance. We interpolate the $\xi_c$ data from Ref.~\cite{Klaer:2019fxc} and find that $c_0$ in the range $c_0\sim1.5^{+0.8}_{-0.4}$ gives reasonable fits to the data.

The density parameter $\xi(t)$ of physical strings shown in Fig.~\ref{fig:attr_xi} is approaching the perfect scaling solution $\xi_c(\ell)$ but it is chasing a moving target
because $\xi_c(\ell)$ increases in time. Moreover, the values of $C$ are relatively small, particularly at small $\xi_{\ell=3}$.  This explains that the lowest curves show almost no convergence. Rather than reaching the perfect-scaling solution, the evolutions approach a ``tracking" solution that lags behind. 
In the case where $\xi_c=\xi_c(\ell)$ is a linear function of $\ell$ and $C$ is constant, the tracking solution is $\xi_t(\ell)=\xi_c(\ell)-(d\xi_c/d\ell)/C$~\cite{Klaer:2019fxc}. The delay clearly increases with the speed at which $\xi_c$ moves and with the smallness of the ``restoring force" $C$. Perturbations around this solution decrease exponentially as  
$\Delta\xi(\ell)=\Delta\xi(\ell_0)\exp(-C(\ell-\ell_0))$. 
We use the simple linear tracking formula to iteratively find the tracking solution of the system with $C=C(x)$, which we identify with the initial condition $\xi_{\ell=3}=0.3(1)$. Naturally, we identify this to the true attractor solution of the system of physical strings and show it in Fig.~\ref{fig:attr_xi} as a dashed black line. 
Deviations from $\xi_{\ell=3}=0.3$ will decrease, although not with a simple exponential form due to the fact that $C$ depends on the trajectory through the ratio $\xi/\xi_c$ and possibly on $\ell$. 
From the $C$ dependence we expect that over-dense networks approach the attractor faster than under-dense ones, and this trend is clearly visible from Fig.~\ref{fig:attr_xi}. We have to be thus very careful with under-dense networks. 

\begin{figure}[tbp]
\includegraphics[width=0.48\textwidth]{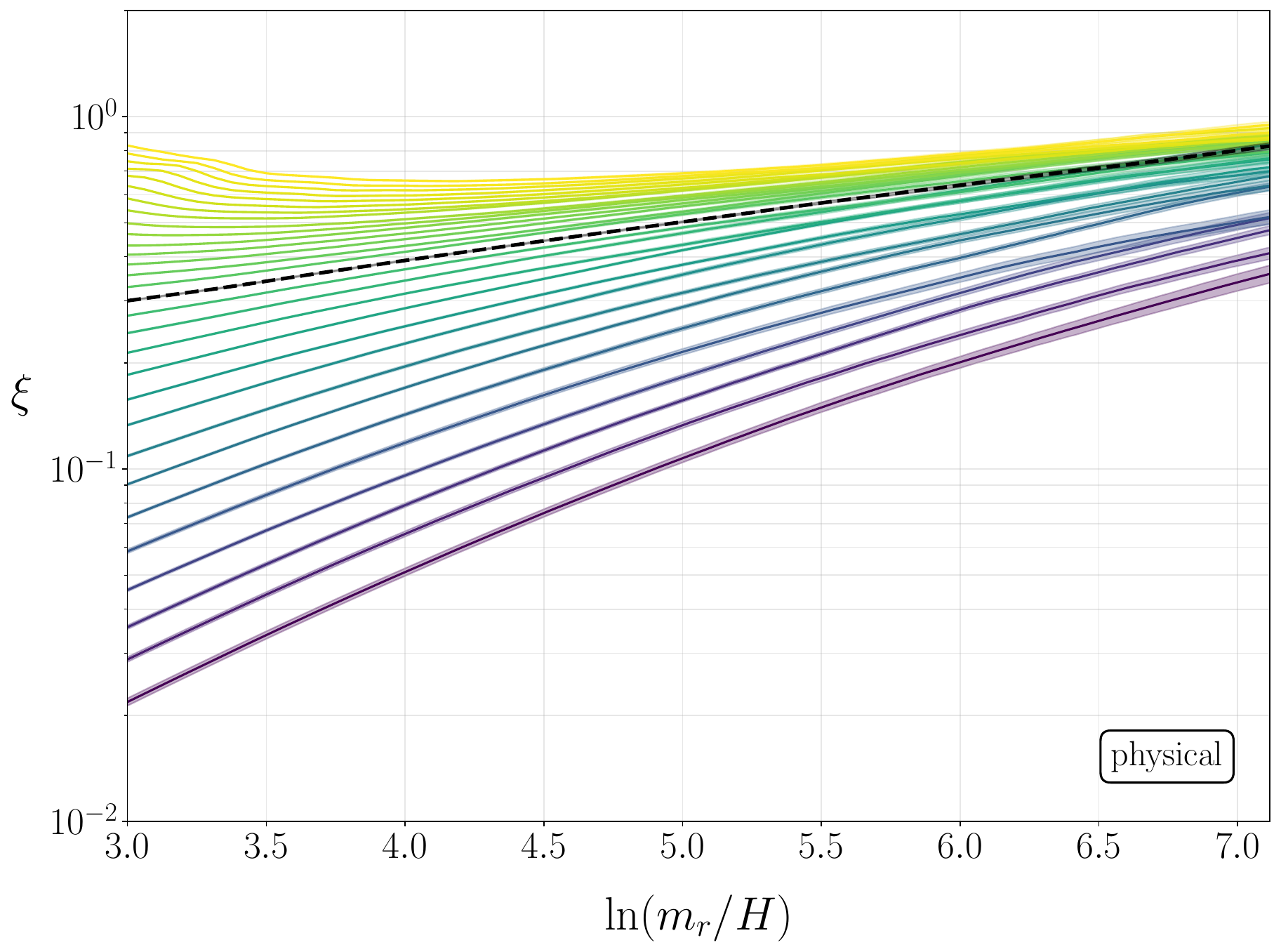}
\caption{Evolution of the string density parameter $\xi$ for different initial string densities 
(simulations with $2048^3$ boxes).
The coloured bands represent statistical uncertainties.
The black dashed line corresponding to $\xi_{\ell =3} \simeq 0.3$ is identified as the attractor/tracking solution.}
\label{fig:attr_xi}
\end{figure}

\begin{figure}[tbp]
\includegraphics[width=0.48\textwidth]{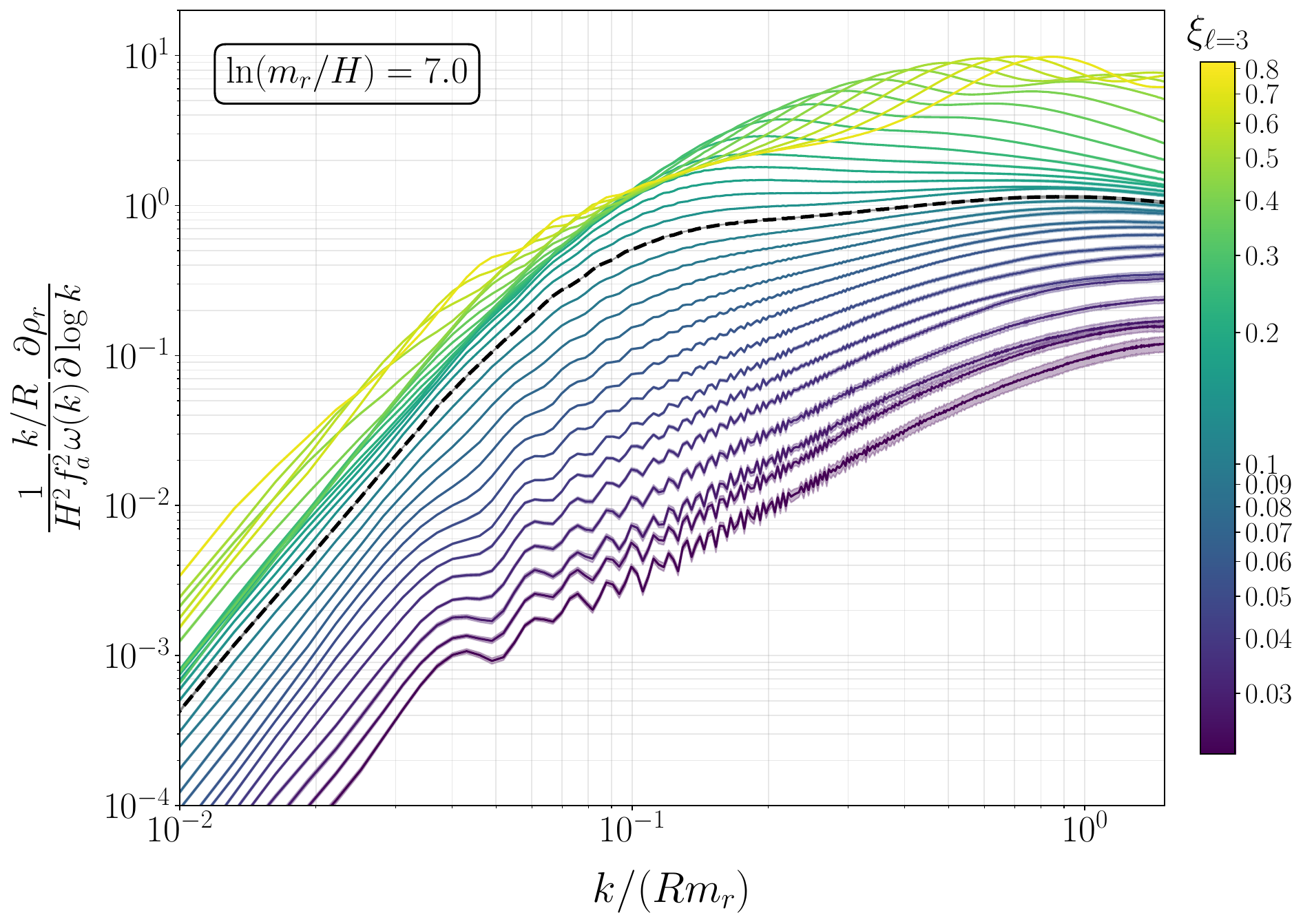}
\caption{
Proxy for the saxion occupation number in networks with different initial string densities at time $\ln(m_r/H) = 7$. Colours as in Fig.~\ref{fig:attr_xi}. 
The black dashed line corresponds to the attractor/tracking solution with $\xi_{\ell =3} = 0.3$.}
\label{fig:attr_ES_invariant}
\end{figure}

We also studied the attractor using the energy spectrum of axions and saxions based on 
twice of their kinetic energy densities, $\rho_a =  2\rho_{a,\mathrm{kin}} = \langle \dot{a}^2\rangle$
and $\rho_r = 2\rho_{r,\mathrm{kin}} = \langle \dot{r}^2\rangle$, see Appendices~\ref{app:masking} and~\ref{app:calc_F_and_q}. 
They both tend to evolve into a uniform shape at late times, but we find the saxion spectrum more helpful to define the attractor. 
In Fig.~\ref{fig:attr_ES_invariant} we plot a proxy for the saxion occupation number, 
\begin{align}
\frac{1}{f_a^2H^2}\frac{k/R}{\omega(k)}\frac{\partial\rho_r}{\partial\log k}, 
\end{align}
at a late time $\ell=7$. Here $k$ is the comoving wavenumber (momentum) and $\omega(k) = \sqrt{m_r^2 + k^2/R^2}$. 
The most relevant feature happens at momenta around the saxion mass, $k/R\sim m_r$, where most saxions are produced. 
We see how spectra with different initial conditions approach a common flat value $\sim 1$. 

\begin{figure}[htbp]
$\begin{array}{c}
\subfigure{
\includegraphics[width=0.48\textwidth]{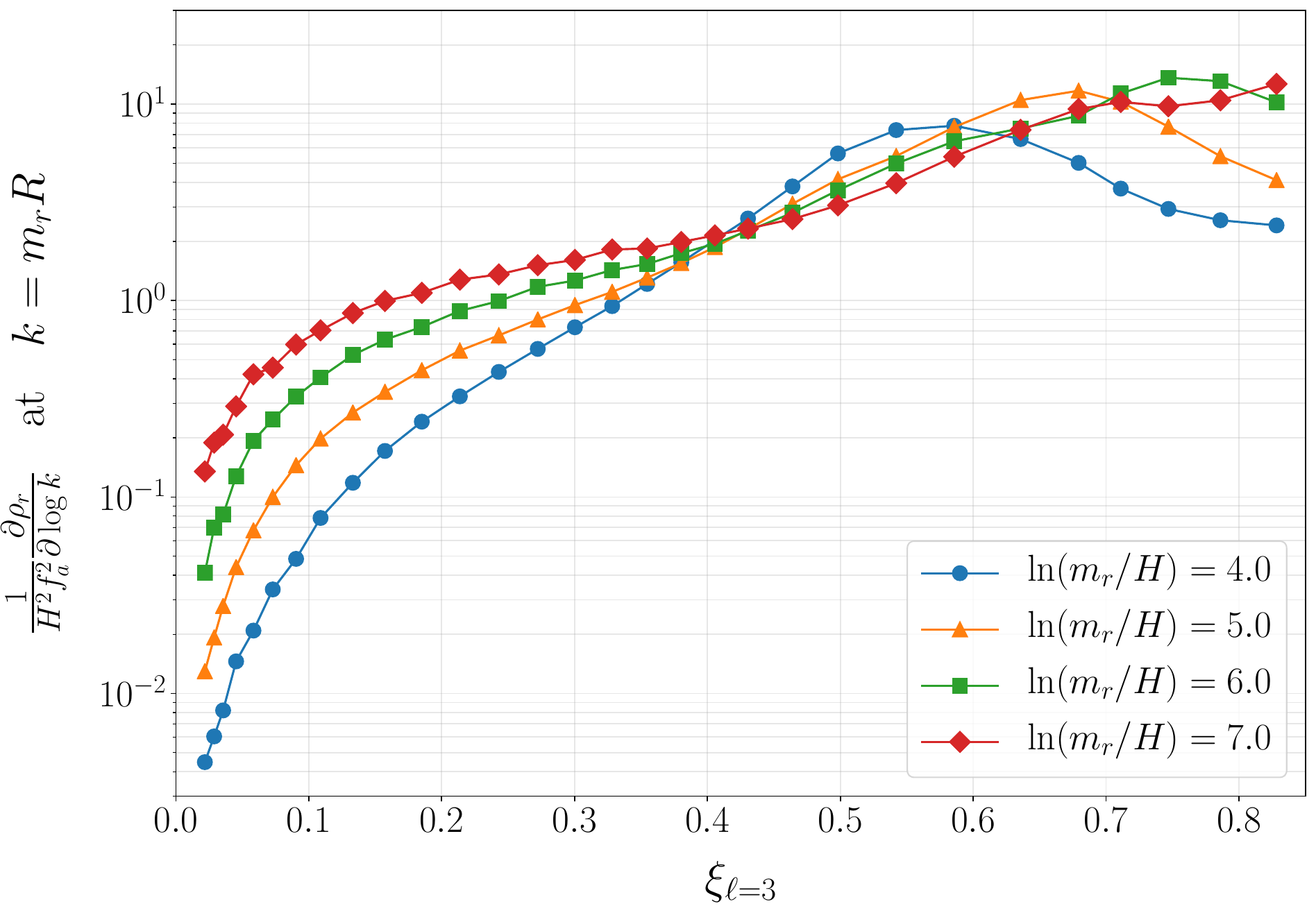}}
\\
\subfigure{
\includegraphics[width=0.48\textwidth]{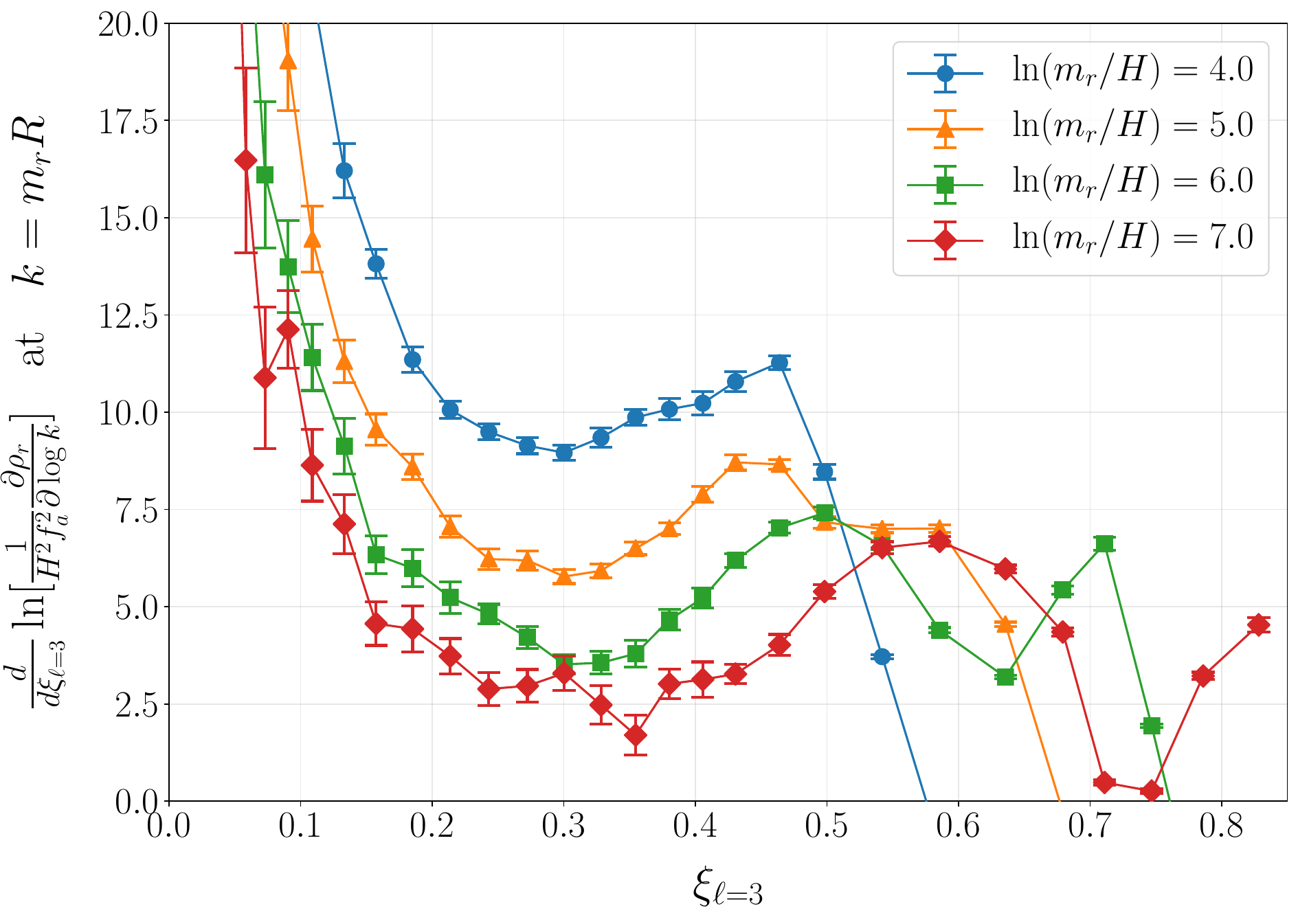}}
\end{array}$
\caption{Amplitude of the saxion energy spectrum at $k=m_rR$ (top panel) and its derivative with respect to the initial string density $\xi_{\ell=3}$ (bottom panel)
plotted as functions of $\xi_{\ell=3}$.
Errors plotted are explained in footnote~\ref{fn:ES_xi_derivative}. 
}
\label{fig:attr_ES_rmass}
\end{figure}

To examine this feature in detail, we study the amplitude of the saxion spectrum at $k=m_rR$ as a function of $\xi_{\ell=3}$ in Fig.~\ref{fig:attr_ES_rmass} (top panel). 
The slope of the plotted line becomes flatter with time, implying that the amplitude at $k=m_rR$ becomes less sensitive to the initial conditions, i.e. a convergence to the attractor, and takes a similar value at late times. The values with the smallest derivative must be associated to the attractor. 
Indeed, in Fig.~\ref{fig:attr_ES_rmass} (bottom panel) we observe that the logarithmic derivative has a minimum at $\xi_{\ell=3}\simeq 0.3$, if we neglect the data at large $\ell$, where the data exhibits larger statistical fluctuations.\footnote{\label{fn:ES_xi_derivative}We computed the derivative with respect to $\xi_{\ell=3}$ via finite differences
and estimated uncertainties in this procedure by using the jackknife method.
Namely, we took the finite difference by using the average of 29 simulations (all the data except one) and iterated it for 30 different ways of the resampling.
After that we estimated the derivative and its error by using the average and variance of 30 results of the resampling, respectively.}
The value of the attractor $\xi_{\ell=3}=0.3$ agrees very well with the previous analysis of the evolution of $\xi$. Moreover, it does not rely on the artificial modelling of the evolution of $\xi$. Both results therefore complement each other supporting the attractive behaviour of the string network and the estimate $\xi_{\rm attractor}({\ell=3})=0.3$. Therefore, we use $\xi_{\ell=3}\simeq 0.3$ as a fiducial initial condition for simulations with larger box sizes.

\begin{figure}[htbp]
\includegraphics[width=0.48\textwidth]{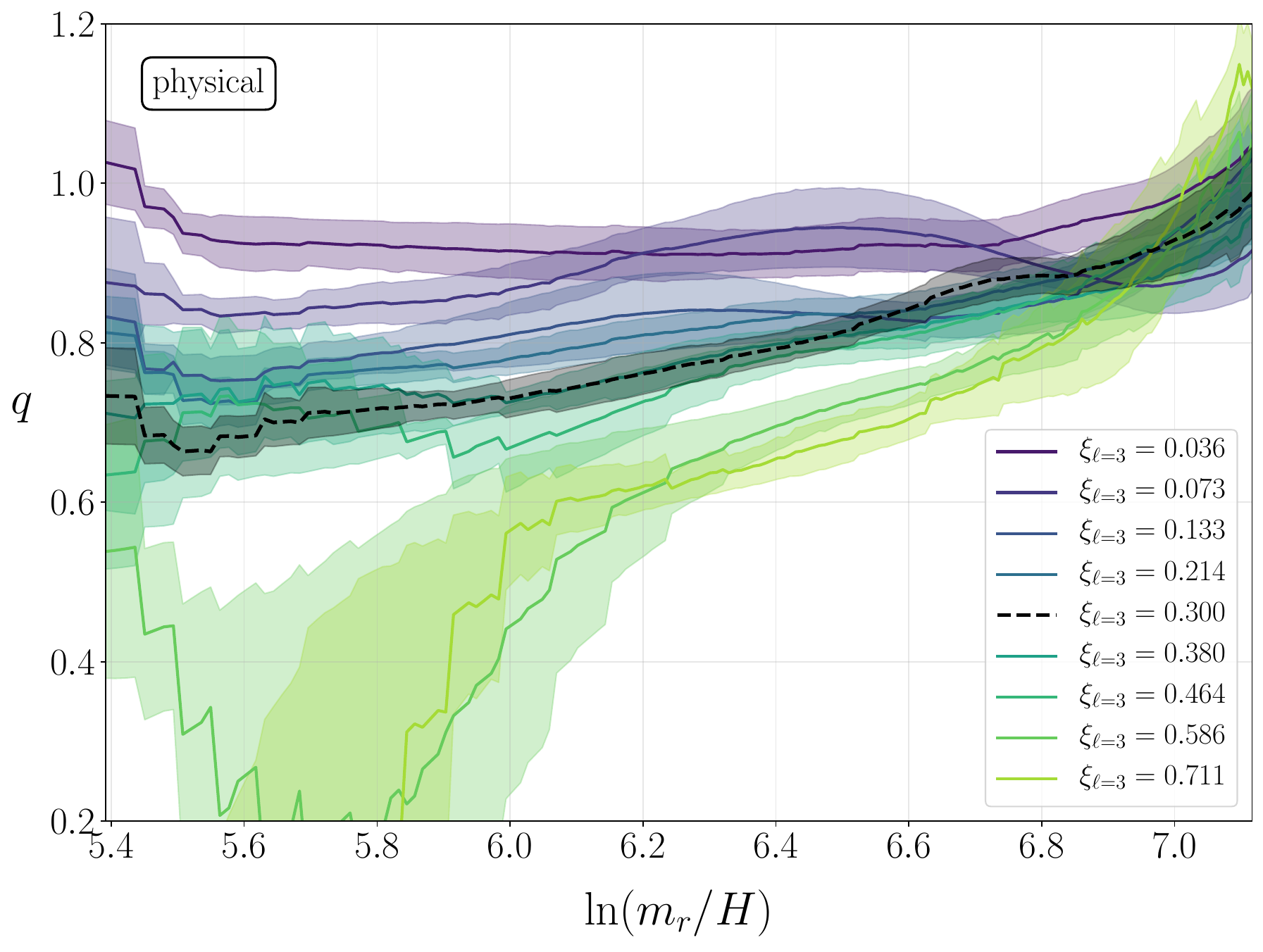}
\caption{Evolution of the spectral index $q$ of the instantaneous emission spectrum for various different values of the initial string density $\xi_{\ell=3}$.
The black dashed line corresponds to the attractor with $\xi_{\ell =3} \simeq 0.3$.
The coloured bands represent the error induced by changing the parameter $\sigma_{\rm filter}$ for the filtering procedure to calculate $\mathcal{F}$ [Eq.~\eqref{filter_for_F}] in addition to statistical uncertainties.
The results are obtained from simulations of physical strings with $2048^3$ lattice sites.}
\label{fig:attr_q}
\end{figure}

From the data of the evolution of the axion spectra, we also computed the instantaneous emission spectra defined by Eq.~\eqref{F_definition}
and estimated their slope $q$ (see Appendix~\ref{app:calc_F_and_q} for technical details of the computation).
Figure~\ref{fig:attr_q} shows the comparison of the evolution of $q$ between different initial string densities.
We see that the value of $q$ becomes large (small) for under-dense (over-dense) initial conditions. This trend can be understood as follows. 
For the under-dense case, the interactions between strings are less efficient, and there is little structure at small scales.
As a consequence, axions are dominantly produced at IR scales corresponding to the size of long strings, which makes the spectrum more red-tilted.
On the other hand, for the over-dense case, there are a lot of small scale fluctuations that produce axions with higher momenta, leading to a smaller values of $q$.
As we see in Fig.~\ref{fig:attr_q}, such a difference due to the initial string density becomes less important at late times, and $q$ also exhibits the attractor behaviour.
However, the convergence is slow again, and one should keep in mind that a small difference in the initial string density could change the value of $q$ measured at small $\ell$
and affect the extrapolation to large $\ell$.
Moreover, we also see some oscillatory features in the plots shown in Fig.~\ref{fig:attr_q}. Those features can be regarded as another systematics biasing the estimate of $q$
and will be elaborated in Sec.~\ref{sec:oscillations}.

\begin{figure}[htbp]
\includegraphics[width=0.48\textwidth]{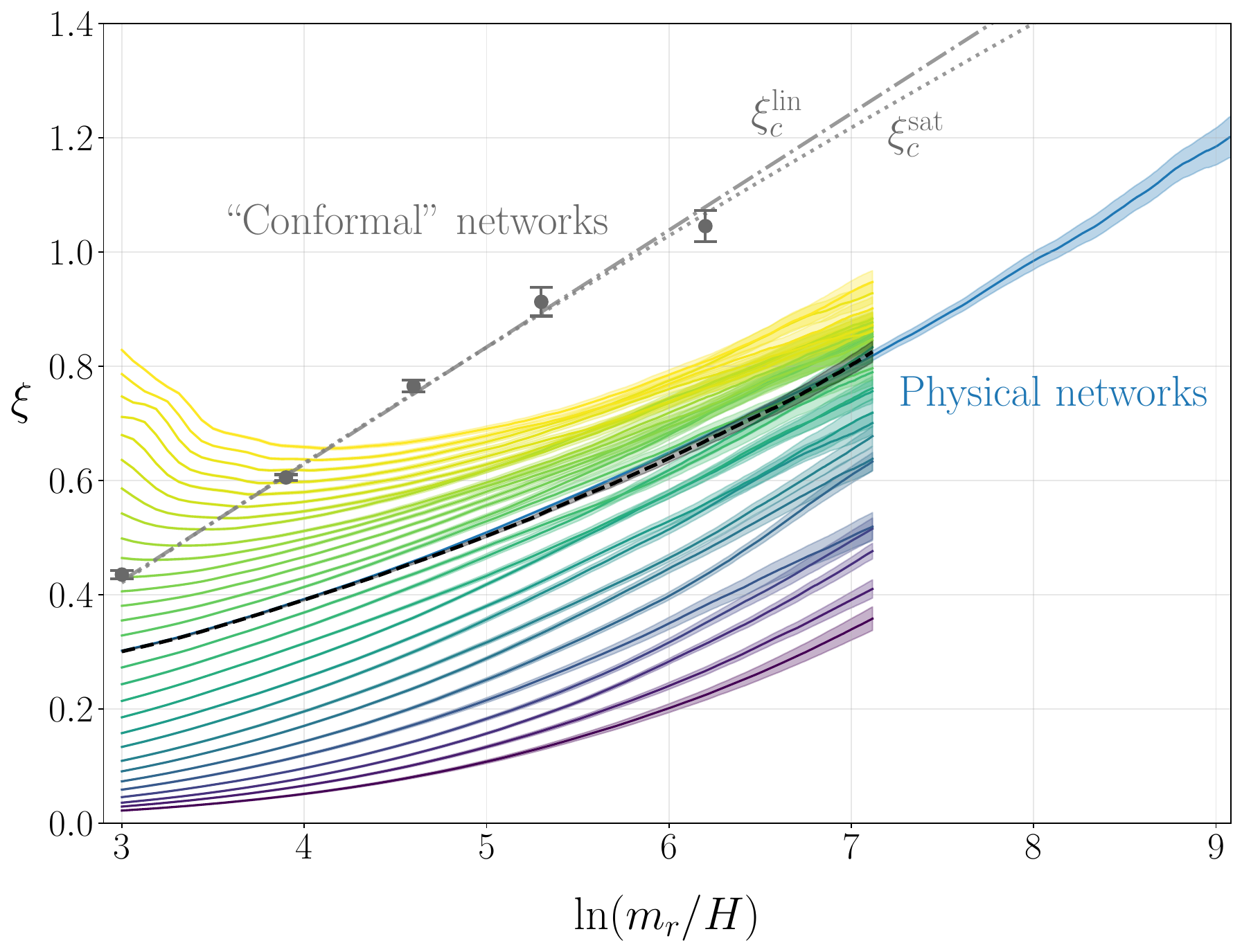}
\caption{Evolution of the string density parameter $\xi$ for physical strings, obtained from simulations with $11264^3$ lattice sites started from the initial string density closest to the attractor (blue line)
and that obtained from $2048^3$ simulations started from 30 different values of the initial string density (colours as in Fig.~\ref{fig:attr_xi}).
The coloured bands represent statistical uncertainties.
The black dashed line represents our best estimate of the attractor or tracking solution.
The data of conformal string simulations from Ref.~\cite{Klaer:2019fxc} are also shown as gray dots with error bars.
The gray dash-dotted and dotted lines represent the fits to the linear function~\eqref{eq:xi_c_fit_lin} and saturating function~\eqref{eq:xi_c_fit_sat}, respectively.}
\label{fig:attr_xi_physical}
\end{figure}

After identifying the initial conditions of the attractor, we used it for simulations with larger number of lattice sites to
investigate the $\ell$ dependence of the quantities such as $\xi$ and $q$. 
In particular, we performed simulations with $11264^3$ lattice sites that can reach $\ell \approx 9$.
Figure~\ref{fig:attr_xi_physical} shows the evolution of the $\xi$ parameter computed at simulations with the largest box, 
together with the results from smaller simulations with different initial string densities.
The result indicates that $\xi$ continues to grow with $\ell$ at the attractor, 
supporting the findings of Refs.~\cite{Fleury:2015aca,Klaer:2017ond,Gorghetto:2018myk,Vaquero:2018tib,Buschmann:2019icd,Gorghetto:2020qws,Buschmann:2021sdq,OHare:2021zrq,Kawasaki:2018bzv,Klaer:2019fxc}.
They also support the interpretation as a ``tracking" solution~\cite{Klaer:2019fxc} chasing the $\ell$-increasing conformal perfect-scaling solution $\xi_c=\xi_c(\ell)$ (also shown in Fig.~\ref{fig:attr_xi_physical}). 

We have several options for the extrapolation. On the one hand, we can take the tracking interpretation~\cite{Klaer:2019fxc} seriously, extrapolate $\xi_c$ and $C$ and solve the differential equation until reaching physically meaningful values. On the other hand, we can take an agnostic point of view, perform global fits of the $\xi$-data with sensible fitting functions and extrapolate within them as \cite{Gorghetto:2018myk,Gorghetto:2020qws,Hindmarsh:2021zkt}. At the end both extrapolations are as good as the functional guesses. Since the second approach has already been explored and uses less insight we will focus on the first. 

We take two models for $\xi_c=\xi_c(\ell)$ that give satisfactory fits to the conformal data~\cite{Klaer:2019fxc}: 
\begin{align}
\xi_c^{\rm lin}&=-0.19(3) +0.205(7)\ell, \label{eq:xi_c_fit_lin}\\
\xi_c^{\rm sat}&=\frac{-0.25(15)+0.23(6) \ell}{1+0.02(4)\ell}, \label{eq:xi_c_fit_sat}
\end{align}
a linear function capturing the main visible trend, and one that eventually saturates,\footnote{The errors of this fit are strongly correlated.} which reflects a slightly visible flattening of the growth at large $\ell$, see Fig.~\ref{fig:attr_xi_physical}. 
Using the $C=C(x)$ function previously described fitted to each model, we evolve the attractor solution and find 
\begin{align}
\xi^{\rm lin}(\ell = 70)&\sim 13.8(5),\\ 
\label{eq:extrapolation_sat}
\xi^{\rm sat}(\ell=70)&\sim 7(3).
\end{align}
 We will take these values as brackets from our uncertainty in the extrapolation.  

The interpretation of the logarithmic growth of the string density parameter $\xi$ has been challenged by Refs.~\cite{Hindmarsh:2019csc,Hindmarsh:2021vih,Hindmarsh:2021zkt}, where it is claimed that the growth in $\xi$ can be interpreted as the slow approach to the ``scaling" solution with a constant $\xi\sim 1.2(2)$. Our  $11264^3$ box data reaches this density without signs of levelling off and strongly disfavours an early saturation. However, as evident from Eq.~\eqref{eq:extrapolation_sat}, a ``late" saturation at $\xi\gtrsim 4$ cannot be excluded. 

\begin{figure*}[tbp]
$\begin{array}{cc}
\subfigure{
\includegraphics[width=0.48\textwidth]{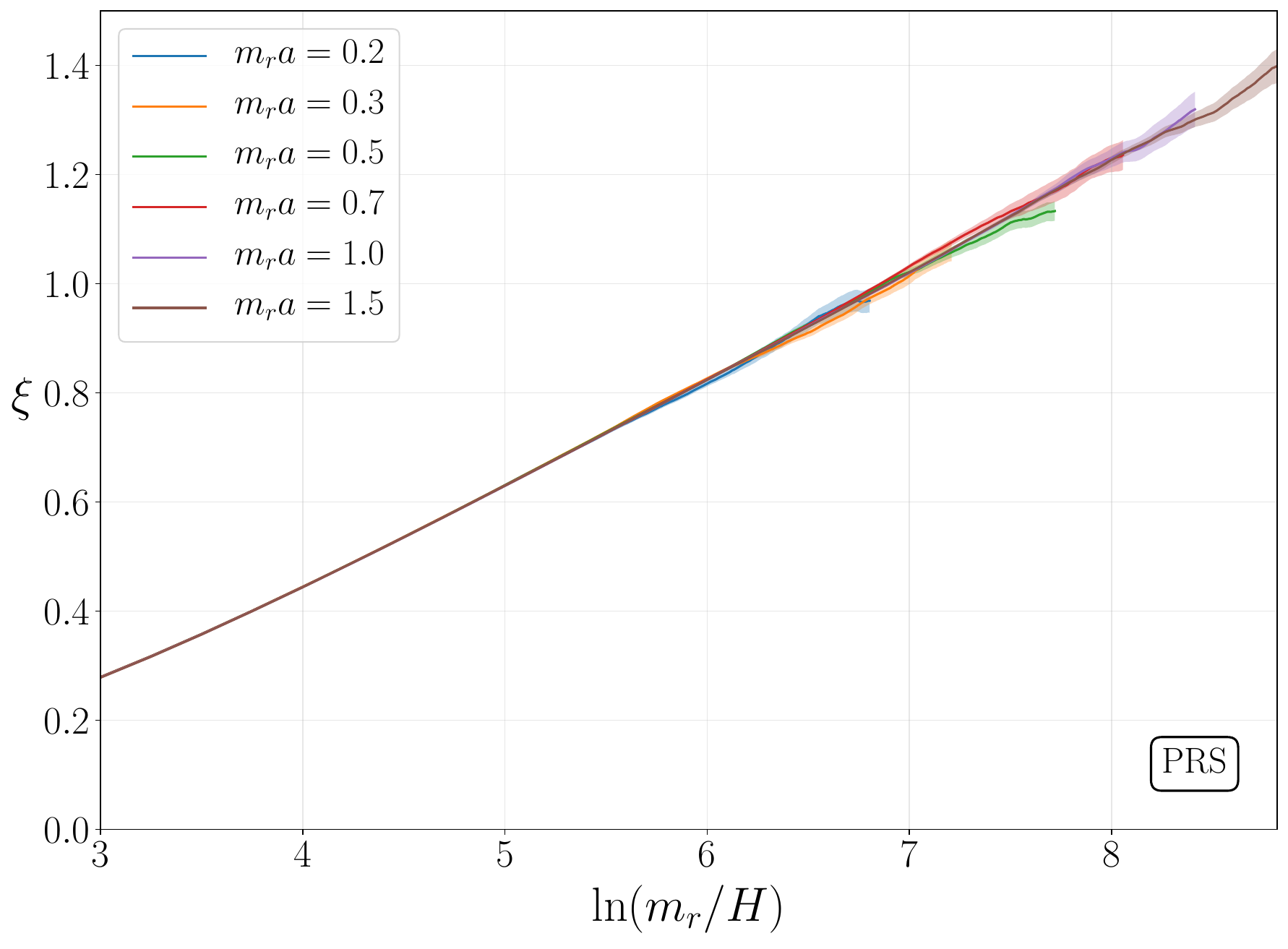}}
\hspace{1mm}
\subfigure{
\includegraphics[width=0.48\textwidth]{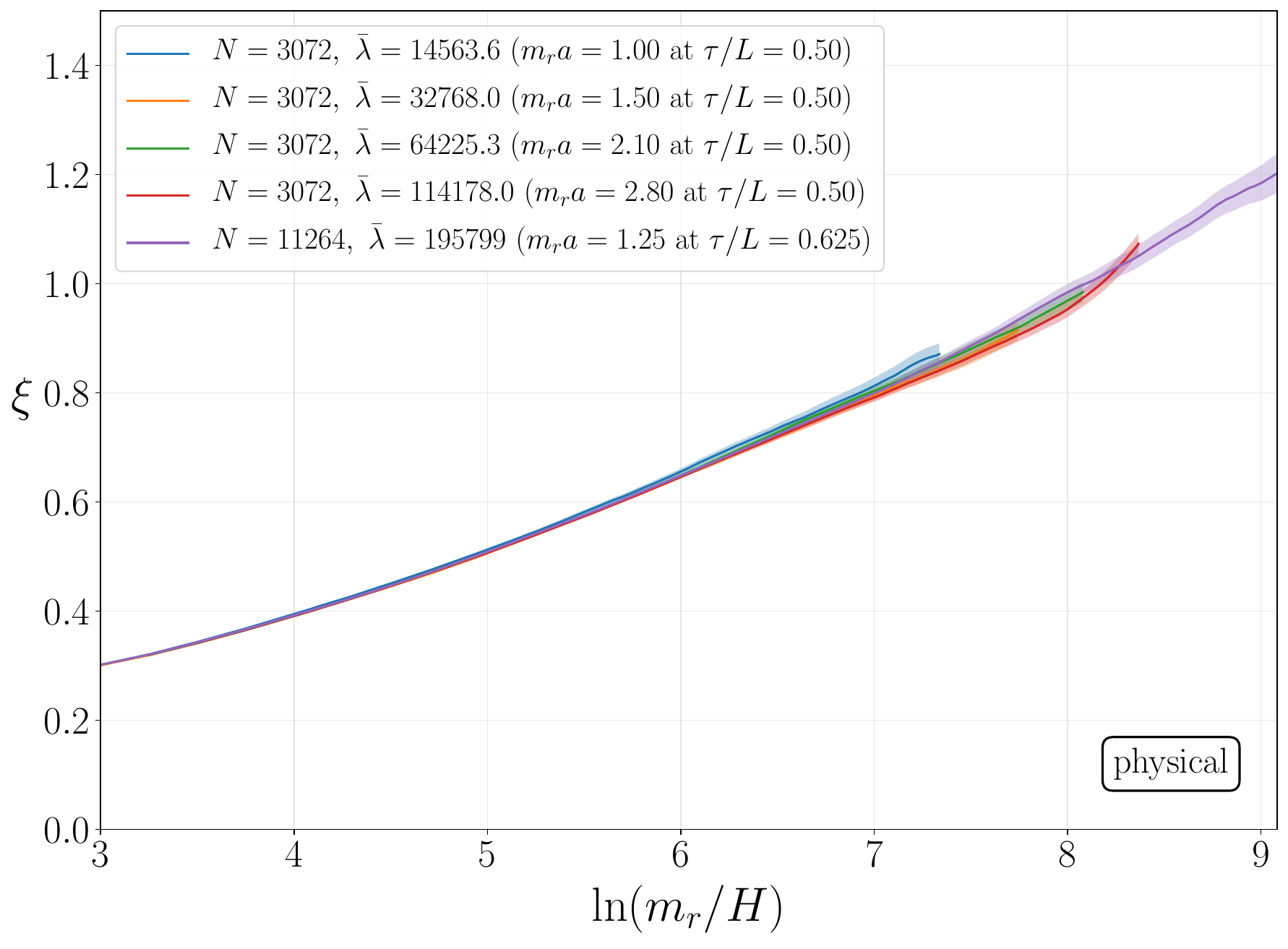}}
\end{array}$
\caption{
Evolution of the string density for different values of the string core width for PRS strings (constant $m_ra$) with $8192^3$ lattice sites (left) and physical strings ($m_ra\propto R$) with one set of $11264^3$ simulations and four sets of $3072^3$ simulations. The coupling parameter $\bar{\lambda}$ is defined by Eq.~\eqref{lambda_definition}. The coloured bands represent statistical uncertainties.
}
\label{fig:attr_xi_PRS_vs_physical}
\end{figure*}

In Refs. \cite{Fleury:2015aca,Gorghetto:2018myk,Vaquero:2018tib,Klaer:2019fxc,Gorghetto:2020qws}, 
simulations were also performed with the Press-Ryden-Spergel (PRS) trick,\footnote{This method is inspired by the work by Press, Ryden and Spergel~\cite{Press:1989yh} in which a similar trick was used for the simulation of domain wall networks.} 
where the parameter corresponding to the saxion mass
is replaced by $m_r \to (R_1/R)m_r$ (with $R_1$ being the scale factor at some reference time) by hand,
such that the ratio between the string core radius $m_r^{-1}$ and 
the physical lattice spacing $a = RL/N$ remains constant.\footnote{Sometimes this is also called the \emph{fat string} trick, 
since in this scheme the string core radius grows with time ($m_r^{-1} \propto R$).}
Although such an artificial growth in the string core is unphysical, this method has an advantage that 
the system converges into the attractor faster than 
the physical case. This is evident when we write the evolution of $\xi$ as a function of $\ell$ rather than time, as Eq.~\eqref{eq:attractive_xi_evol} becomes $d\xi/d\ell=(C/f)(\xi_c(\ell)-\xi)$ with a factor $f=d\ell/d\log t$, which is 1 in the physical case and 1/2 for PRS strings ($m_r/H\propto t/R\propto \sqrt{t}$ in this case). 

In the left panel of Fig.~\ref{fig:attr_xi_PRS_vs_physical}, we show the results on the evolution of the string density parameter $\xi$
in our simulations with PRS strings for several different values of $m_ra$. 
The initial conditions were taken to have similar $\xi$ to the attractor solution for PRS string simulations of Ref.~\cite{Gorghetto:2018myk}. 
The evolution of $\xi$ that is increasing as expected from the growth of $\xi_c(\ell)$ and, indeed, 
the simulations agree very well with the numerical solution of Eq.~(\ref{eq:attractive_xi_evol}) with the same linear $\xi_c$ and $C$-function that fitted the evolution of physical strings after we rescale $C\to2C$ according to the prediction. 
Indeed we expected and found that the PRS attractor solution should be approximately twice as close to $\xi_c$ than the physical case. 
This gives greater credibility to our results and interpretation. 
 
We take the opportunity to study discretisation errors in $\xi$ induced by a poor resolution of the string core (see also Sec.~\ref{sec:disc_effect}).  
The resolution is parametrised by the value of $m_ra$, which measures how many lattice points fit in the core. 
Figure~\ref{fig:attr_xi_PRS_vs_physical} shows that these errors do not significantly affect the evolution of $\xi$ in PRS and physical strings close to the attractors (see Table~\ref{tab:simu_parameters} in Appendix~\ref{app:code} for the summary of our simulation parameters).  
The situation will be very different when we study the spectra of radiated axions and saxions in Sec.~\ref{sec:disc_effect}.

\section{Oscillations in the spectrum}
\label{sec:oscillations}

The instantaneous emission spectrum $\mathcal{F}(x,y)$ defined by Eq.~\eqref{F_definition} is calculated by taking the time derivative of $R^4\partial\rho_a/\partial k$
for each comoving momentum $k$. In order to adequately evaluate the time derivative, we need to keep track of the evolution of each mode in the simulation.
However, it turns out that these modes typically oscillate with time, which gives rise to contamination of $\mathcal{F}(x,y)$, and hence of $q$.
In this section, we elaborate on the nature of oscillations in the spectrum and their impact on the estimation of the instantaneous spectrum $\mathcal{F}(x,y)$ and spectral index $q$.

\begin{figure}[tbp]
$\begin{array}{c}
\subfigure{
\includegraphics[width=0.48\textwidth]{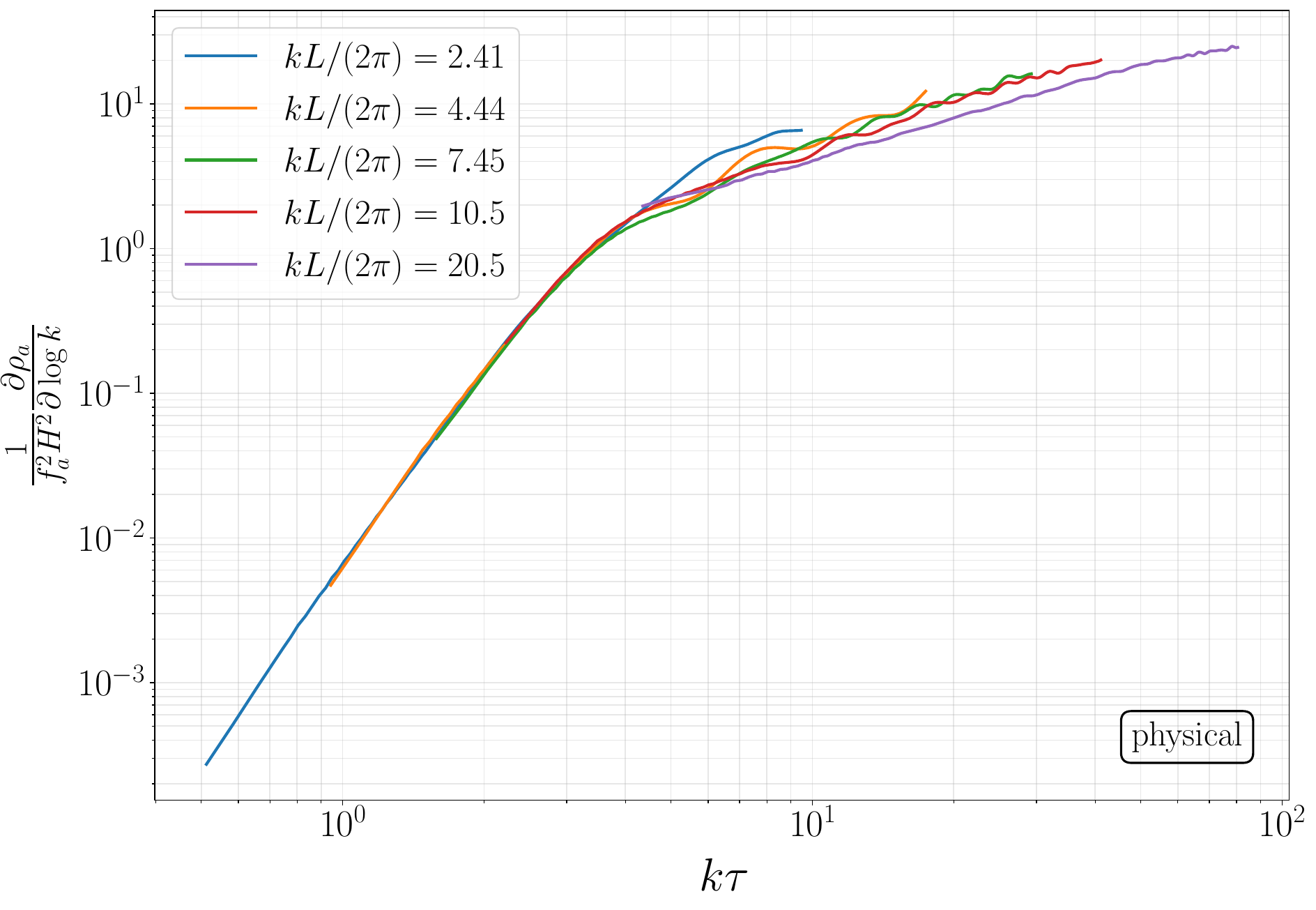}}
\\
\subfigure{
\includegraphics[width=0.48\textwidth]{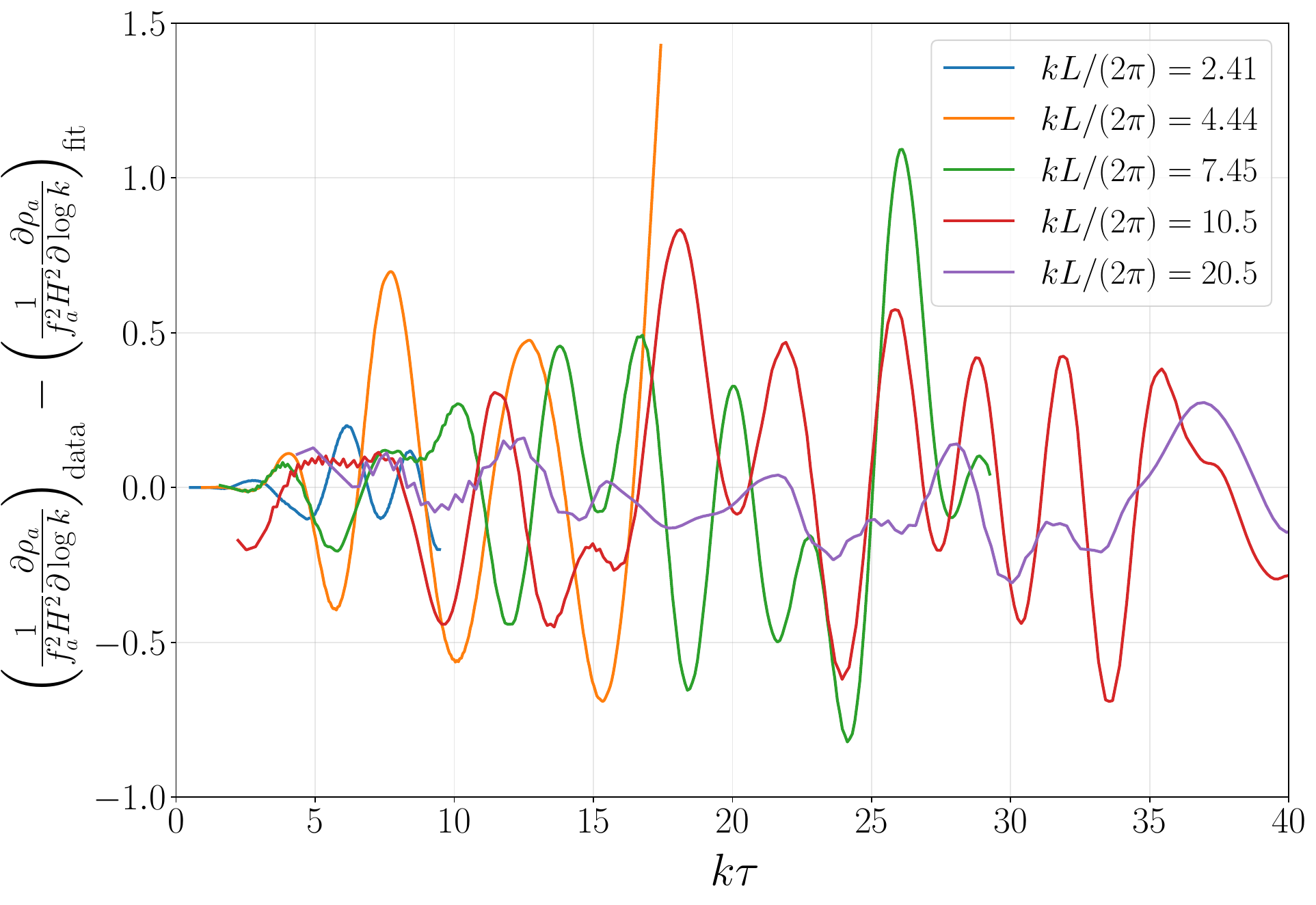}}
\end{array}$
\caption{Top panel: Time evolution of the energy density of one Fourier mode of the axion field for various different comoving wavenumbers. 
These modes cross the horizon $k\tau \sim \mathcal{O}(1)$ during the simulation and show a change in the rate of growth around that time.
After that they oscillate with a frequency $\sim 2k$.
Bottom panel: Time evolution of the residue obtained by subtracting the fit result given by Eq.~\eqref{mode_fit_function} 
from the data of one mode (shown in the top panel) for various different comoving wavenumbers.
These plots are obtained from one simulation (not the average) of physical strings with $11264^3$ lattice sites.}
\label{fig:osc_mode_evol_lowk}
\end{figure}

In the top panel of Fig.~\ref{fig:osc_mode_evol_lowk}, we show the time evolution of a few modes of the comoving box. 
The evolution of different modes is apparently the same outside the horizon ($k\tau \lesssim \pi$),\footnote{It is possible to derive these features analytically by considering the evolution of the Fourier components of the free axion field across the horizon crossing. See Appendix~\ref{app:evolution_analytical}.}
but they get a different kick at horizon crossing ($k\tau \gtrsim \pi$) and have slightly different subsequent evolution.
In particular, we find oscillatory features in the mode evolution after the horizon crossing. 

Oscillations are expected on physical grounds. 
The equation of motion of the \emph{free axion field} admits solutions $\tilde{\psi}({\bm k}) \propto e^{\pm ik\tau}$ for the Fourier component ${\bm k}$. 
We can call these misalignment axions, inhomogeneities in the axion field originated from the finite correlation length at the PQ phase transition \emph{unrelated to strings}. 
Since we estimate $\rho_a$ from the kinetic energy alone $\propto a_{\tau}^2$, the solution of the free field equation naturally predicts that $\frac{d^3\rho_a}{d {\bm k}^3}$ oscillates coherently with a frequency $2k$ (in conformal time). 
When we bin modes with the same frequency $k$ and random phases, the oscillations should decrease in amplitude as $(\text{number of modes in the bin})^{-1}$ and $\frac{\partial \rho_a}{\partial \log k}$ converges to the average power. Thus, if there is some coherence between modes or simply when we bin a \emph{finite number} of modes we shall have oscillations. 
In the expanding Universe, all the modes \emph{start in phase} as $\tilde{\psi}({\bm k})\propto \sin(k\tau)$~\cite{Kolb:1993hw} (see Appendix~\ref{app:evolution_analytical} for details). 
However, axions in our simulations are not free. 
The string network slowly decays/radiates somewhat incoherently into axion waves making the $\tilde{\psi}({\bm k})$ amplitudes grow and phases drift. 
Moreover, string motion also contributes to $a_\tau$. 
Thus, we expect that misalignment axions enter the horizon oscillating more or less in phase, and the coherence decreases with time as the network pumps the axion field incoherently. 
Moreover, the amplitude of oscillations should be larger for smaller values of $k$, for which we sum less modes in each bin. 
These are the trends obtained from simulations and shown in Fig.~\ref{fig:osc_mode_evol_lowk}. 

It is clear that the oscillations can affect the extraction of the time derivative of the mode's average power, thus implying a systematic error. 
To deal with the oscillatory component, we extract the overall trend in the evolution by fitting the following function to the data of time evolution of $\frac{1}{f_a^2H^2}\frac{\partial\rho_a}{\partial\log k}$ for each $k$,
\begin{align}
\left(\frac{1}{f_a^2H^2}\frac{\partial\rho_a}{\partial\log k}\right)_{\rm fit} = \frac{e^{a_0}x^{a_1}}{1+(a_2x)^{a_3+a_4\ln x +a_5(\ln x)^2}},
\label{mode_fit_function}
\end{align}
where $x=k\tau$, $a_i$ ($i=0,1,2,3,4,5$) are constants to be determined by the fit, and
terms of $a_4$ and $a_5$ are added to model the evolution at subhorizon.\footnote{Although we have included these terms to allow the possibility that
the evolution at subhorizon is not given by an exact power law in $x$, the detailed form of these terms should not be important, 
if we properly add the residue when we calculate $\mathcal{F}(x,y)$.
For instance, we confirmed that the results on $q$ do not change much when we use other polynomials of $\ln x$ instead of $a_4\ln x +a_5(\ln x)^2$.}
The bottom panel of Fig.~\ref{fig:osc_mode_evol_lowk} shows the evolution of the residue obtained by subtracting the fit from the data.
We see that each mode actually oscillates with a frequency approximately given by $\sim 2k$, as one period of the oscillation roughly corresponds to the change in $k\tau$ by $\pi$.\footnote{
In Fig.~\ref{fig:osc_mode_evol_lowk}, we also observe tiny oscillations with a frequency much faster than $\sim 2k$. They can be interpreted as the oscillation in the radial direction of the 
PQ field, which can be associated with an imperfectness in the initial condition.
This issue is elaborated in Appendix~\ref{app:initial_conditons}.}
As expected, the oscillation is not a pure sinusoidal function,  but we note that amplitudes and phases do not change significantly. 
Unfortunately, the fact that it is not straightforward to separate the misalignment component from string-induced ones turns out to be a hurdle 
for the calculation of the axion production rate, as we will see below.

\begin{figure}[tbp]
\includegraphics[width=0.48\textwidth]{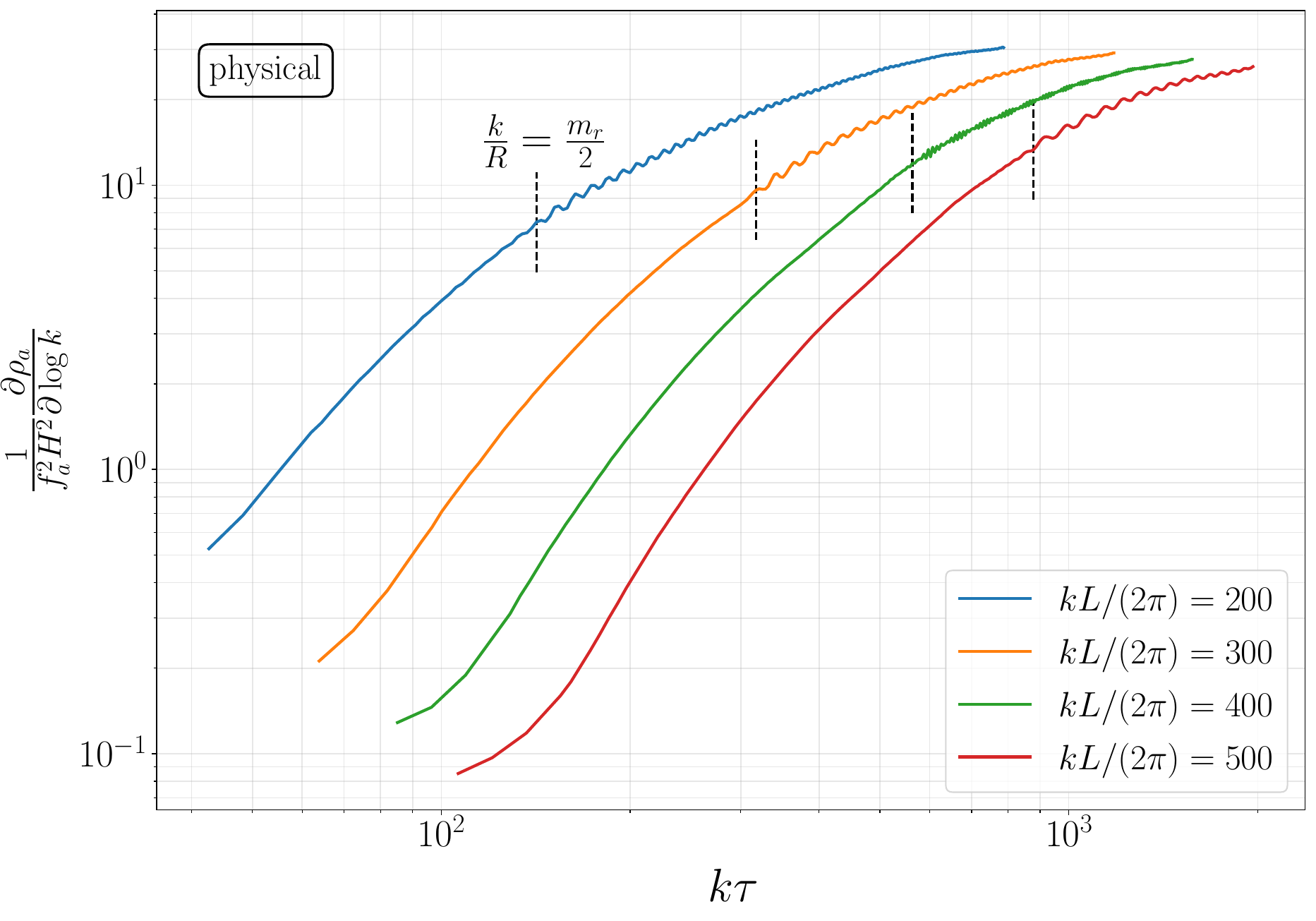}
\caption{Time evolution of the energy density of one Fourier mode of the axion field for higher comoving wavenumbers,
obtained from one simulation (not the average) of physical strings with $11264^3$ lattice sites.
Times at which the physical momenta of the modes become equal to the half of the saxion mass 
are marked by dashed lines.}
\label{fig:osc_mode_evol_highk}
\end{figure}

In addition to the oscillations in the low frequency modes, we also observe oscillations when the physical wavenumber of a mode crosses the value 
$k/R=m_r/2$. A parametric resonance exchanges energy between axions and saxions coherently~\cite{Gorghetto:2020qws}, 
and thus creates $2k$ oscillations, see Fig.~\ref{fig:osc_mode_evol_highk}.\footnote{The frequency of oscillations
shown in Fig.~\ref{fig:osc_mode_evol_highk} does not look like $2k$.
This is just because the frequency of the measurements of axion spectra in our simulations is not high enough to resolve fast oscillations of these high-$k$ modes.
In that case, the oscillations show up as a spurious feature with a lower frequency, which can be predicted from the genuine frequency ($2k$) and the frequency of measurements (see Appendix~\ref{app:method_F}).}
Note that this effect is less problematic in simulations with the PRS method, since in that case $m_rR$ is constant, and the resonant production of axions is relevant only for a certain mode whose comoving momentum is given by $k=m_rR/2$.

As the $2k$-oscillations start at a determined time, such as the horizon crossing or saxion mass crossing, it is in principle not possible to eliminate them by performing simulations many times and averaging the results. 
For instance, see Figs.~\ref{fig:ext_F_evolution},~\ref{fig:init_mode_evol}, and~\ref{fig:evol_EA}, which explicitly show that the oscillations remain even after averaging over many realisations. 
Some tricks could help to average out the oscillations. For instance, one can shift the misalignment phase by using different initial string densities, one could change the phase of the saxion field at the beginning of the simulations, or one can simply bin more modes together.  
For this paper, we applied a filter that suppresses high-frequency components in the time series data of the residue of the mode evolution. 
To minimise the systematic error due to our choice of the fit function when we compute ${\cal F}$, we computed the time derivative of the filtered residue and added it to the time derivative of the fit (see Appendix~\ref{app:method_F} for technical details).

\begin{figure}[tbp]
\includegraphics[width=0.48\textwidth]{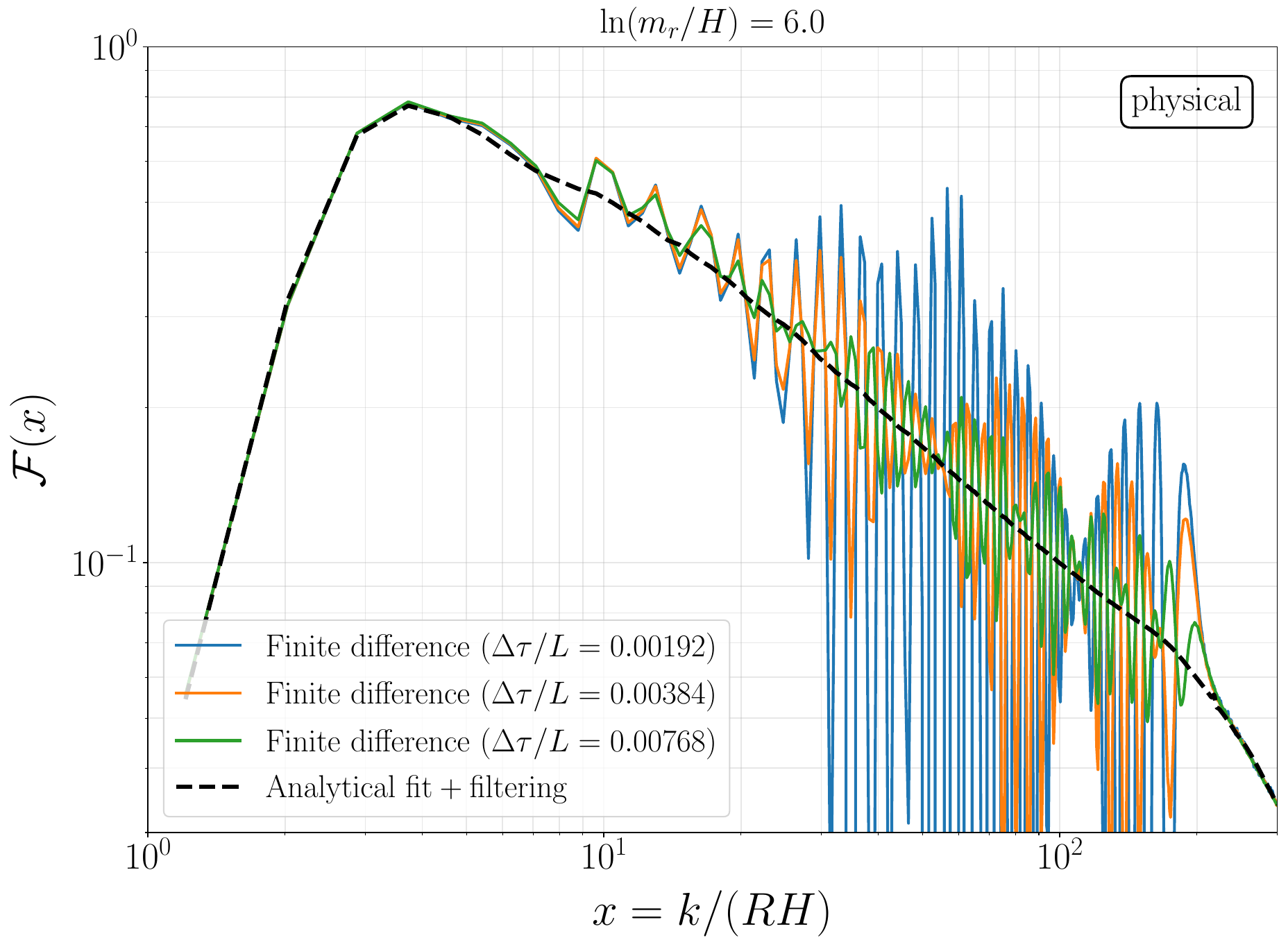}
\caption{Instantaneous emission spectrum at $\ln(m_r/H) = 6$, obtained by using the finite difference method
for three different choices of the interval $\Delta\tau$ (coloured lines) 
and by the sum of the analytical fit and time derivative of the filtered residue (black dashed line).
Each line represents the average of 20 simulations of physical strings with $11264^3$ lattice sites.}
\label{fig:osc_F_diff_vs_fit}
\end{figure}

In Fig.~\ref{fig:osc_F_diff_vs_fit}, we compare the instantaneous spectrum ${\cal F}$ obtained by our procedure to a calculation that evaluates the time derivative with finite differences. 
To ease the discussion, we model 
\begin{align}
\label{eq:amodel}
\frac{1}{(f_a H)^2} \frac{\partial \rho_a}{\partial \log k} = x f_0 +f_1\cos(2k\tau+\alpha), 
\end{align}
where $f_{0,1}\sim {\cal O}(1)$ and $\alpha$ are \emph{smooth} functions of $k$, see Fig.~\ref{fig:osc_mode_evol_lowk}. 
The instantaneous spectrum calculated with finite differences with an interval $\Delta\tau$ is thus 
\begin{align}
\mathcal{F}_{\rm finite\mathchar`-diff.} &= f_0-2f_1\sin(2k\tau+\alpha){\rm sinc}(2k\Delta \tau)\nonumber\\
&\quad+{\cal O}(f_0',f_1',\alpha')\text{-terms}, \label{F_finite_diff_analytical}
\end{align}
where the prime represents the derivative with respect to $\tau$.
The finite difference version of $\cal F$ features the same $2k$ (here $2x$) oscillations than $\partial \rho_a/\partial k$. 
The envelope ${\rm sinc}(2x\Delta\tau/\tau)$ can also be clearly seen in Fig.~\ref{fig:osc_F_diff_vs_fit}. 
It acts as a low-pass filter suppressing the oscillations above a critical value 
\begin{align}
x_{\crit} = \frac{\pi}{2} \frac{\tau}{\Delta\tau}, 
\end{align} 
(chosen arbitrarily to be the first zero of the sync function), which depends on the time-interval. 
For the values of $\Delta\tau$ shown in Fig.~\ref{fig:osc_F_diff_vs_fit} in increasing order, we find $x_\crit\simeq 110,55,28$.  
It is clearly visible that all the envelopes become zero at $110$, as follows from the fact that the intervals are multiples of 2.  
Below the smallest $x_\crit\sim 25$, all the finite difference curves agree very well, as all the different filter values are $\sim 1$ here.  
The relative amplitude $2f_1/f_0$ can be estimated from the curve of the finest discretisation and reaches more than a factor of $\sim 2$ (!). 

Increasing the time interval in the finite difference estimate of the derivative is a very convenient way of calculating ${\cal F}$, 
as one decreases the oscillating features automatically and one needs less evaluations of the spectrum during the simulation (which are very time-consuming). 
This seems to be well suited for our problem because the main growing trend in $(f_a H)^{-2}\partial \rho_a/\partial \log k$ is a \emph{linear} growth $\propto x$ and calculating the derivative with finite differences does not bias the result, even with relatively long time-intervals. 
However, the finite difference estimate has its limitations. 
First, it does not remove the oscillations below $x_\crit$, and above $x_\crit$, only by a factor $\sim x_\crit/x$. 
Second, the finite difference method picks up the derivative of the non-oscillating part $x f_0$ exactly as long as $f_0$ does not depend on $x$. 
If there is a small, for instance logarithmic, extra dependence, finite differences will produce a bias in the estimator which grows with $\Delta\tau$. 
For instance, when $f_0\propto\ln (x/x_c)$ with $x_c$ some constant, 
\begin{align}
\frac{{\cal F}_{\rm finite\mathchar`-diff.}}{{\cal F}}-1 \sim \frac{-\frac{1}{6} \left(\Delta\tau/\tau\right)^2}{1+\ln (x/x_c)}. 
\end{align}
Thus, we would need to find a compromise between the minimisation of this bias with a small $\Delta \tau$ and the oscillation amplitude with a large $\Delta\tau$. 
Our fitting procedure nicely overcomes these limitations. 

The oscillations in the spectrum will affect the determination of $q$. The estimation of $q$ is based on fitting a power law $\mathcal{F} \propto x^{-q}$ to the ${\cal F}$ data within the range $c_{\rm IR}HR < k < c_{\rm UV}m_rR$ 
($c_{\rm IR}<x<c_{\rm UV} e^\ell$) with some fixed values of coefficients $c_{\rm IR}$ and $c_{\rm UV}$.
Since the fit model does not include oscillations, it will produce a bias in the estimation of $q$. 
We expect this bias to oscillate in time, i.e. with $\ell$, but not necessarily around the true value. 
As long as oscillations average over the data, the bias should be small. 
However, since we cut the fitting range at high and low values of $k$, we can have only partially cancelled oscillations there. 
These uncanceled oscillations can contribute significantly for instance if the data range is small (which happens at low $\ell$, early times, when the UV and IR cutoffs are closer), 
if the weights in the fit favour the extremes of the fit range, etc.~\cite{Buschmann:2021sdq}.
In such a case, one expects that the bias will depend on $c_{\rm IR}$ and $c_{\rm UV}$ as well.\footnote{In what follows
we discuss how the bias due to the oscillations in the spectrum 
appears according to the choices of $c_{\rm IR}$ and $c_{\rm UV}$.
Even if we minimise the bias by removing oscillations from $\mathcal{F}$ (e.g. by using the filtering method),
there remains another systematics associated with $c_{\rm IR}$ and $c_{\rm UV}$ due to the fact that $\mathcal{F}(x)$ 
does not exhibit a perfect power law behaviour around the momentum ranges close to the IR and UV cutoffs. 
The errors associated with the latter effect are discussed in Appendix~\ref{app:syst_cutoffs}.}

The effect of oscillations will be maximum when the phases are less random. 
In particular, since the $k$-modes are discrete in our finite-volume simulation, $2k\tau=2x=4\pi n \tau/L$ with $n$ an integer. 
Thus, when 
$\tau = L/2$, $2x$ becomes a multiple of $2\pi$ and independent of the mode,\footnote{Same happens for multiples of $\tau=L/2$ but in practice, we seldom exceed $\tau=L/2$ in our simulations to avoid unphysical effects from the periodic boundary conditions and never reach $\tau=L$.} leaving only the smoother dependence with $\alpha$.
 
We note that when the time intervals used for calculating the derivative are logarithmically distributed, 
for instance the popular $\Delta \ell=0.25$ used in Refs.~\cite{Gorghetto:2018myk,Gorghetto:2020qws,Buschmann:2021sdq}, 
the critical value of the filter is constant in time, $x_\crit \simeq (\pi/2)(4/\Delta\ell)$, 
which is $x_\crit\simeq 25.1$ for $\Delta \ell = 0.25$. 
Then, by choosing $c_{\rm IR}=x_\crit$, we can ensure that 
the amplitude of the oscillations is minimum at the lower end of the fit and is always suppressed in the fit region. 
Coincidentally, the work of Ref.~\cite{Buschmann:2021sdq} 
chose approximately $c_{\rm IR}\simeq 2x_\crit$ for their results without a discussion of the oscillations presented here. 
The main results of Refs.~\cite{Gorghetto:2018myk,Gorghetto:2020qws} were obtained by using $c_{\rm IR} =30$,
which is slightly larger than $x_{\crit}$, and hence the effect of oscillations were not minimised.

\begin{figure}[htbp]
\includegraphics[width=0.48\textwidth]{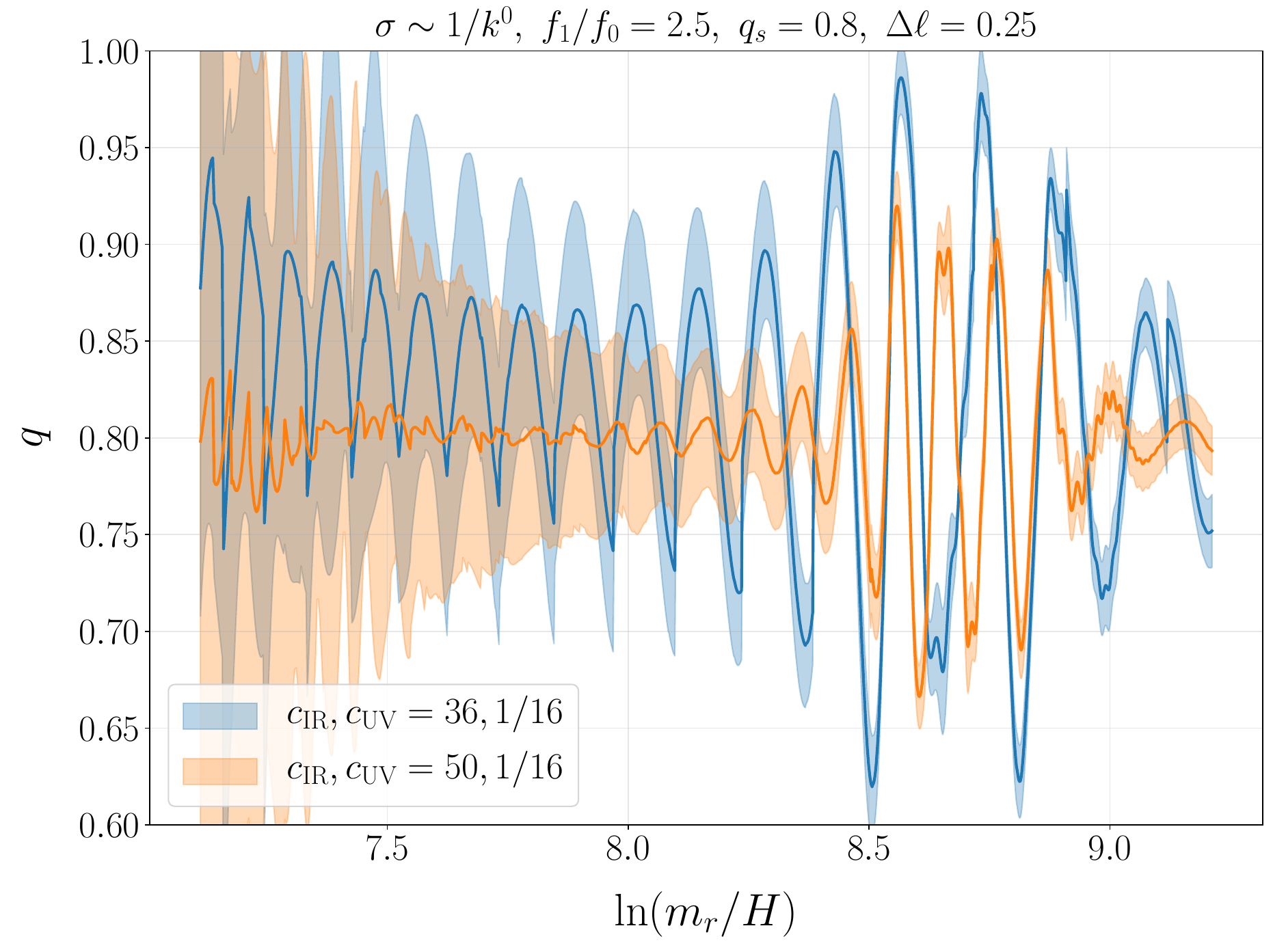}
\caption{Value of $q$ fitted from a synthetic $1/k^{q_s}$ spectrum with 
constant $q_s=0.8$ and oscillations as in Eq.~\eqref{eq:amodel} for two choices of $c_{\rm IR}$.
The fits are performed by assuming flat errors ($\sigma \sim 1/k$).}
\label{fig:q-synthetic}
\end{figure}

We can evaluate directly how the oscillation affect the extraction of $q$ by considering
the simple toy model given by Eq.~\eqref{eq:amodel}. 
Here we build a synthetic spectrum with $f_0\sim 1/k^{q_s}$ and introduce oscillations with $f_1/f_0=2.5$. 
Based on this synthetic spectrum, we build the time derivative with logarithmically spaced $\Delta \ell=0.25$ 
and fit the result to a simple power law $\propto 1/k^q$ without oscillations. 
The $m_r/H$ ratio and mode discretisation are the same as we have in our $11264^3$ simulations.  
In Fig.~\ref{fig:q-synthetic}, we show that the value of $q$ extracted in this way contains itself oscillations of order $\sigma_q\sim 0.1$ with slowly shifting frequency. 
We also see a sizable bias at small $\ell$ when $c_{\rm IR}$ is not a multiple of $x_\crit$ (case $c_{\rm IR}=36$ here).  
Note that the errors are larger too, even though the fitting range for $c_{\rm IR}=50$ is smaller. 
In the $\ell\sim8\text{--}9$ interval, the period is approximately $0.1$. 
As mentioned before, we expect the phases between the oscillations of the different Fourier modes to be less random at $\tau=L/2$. 
This leads to the increase of amplitude at $\ell=8.5\text{--}9$ where this condition is met in our simulations. This can explain the offset of the $\ell=8.5$ point in the results of Ref.~\cite{Buschmann:2021sdq}. 

\begin{figure}[htbp]
\includegraphics[width=0.48\textwidth]{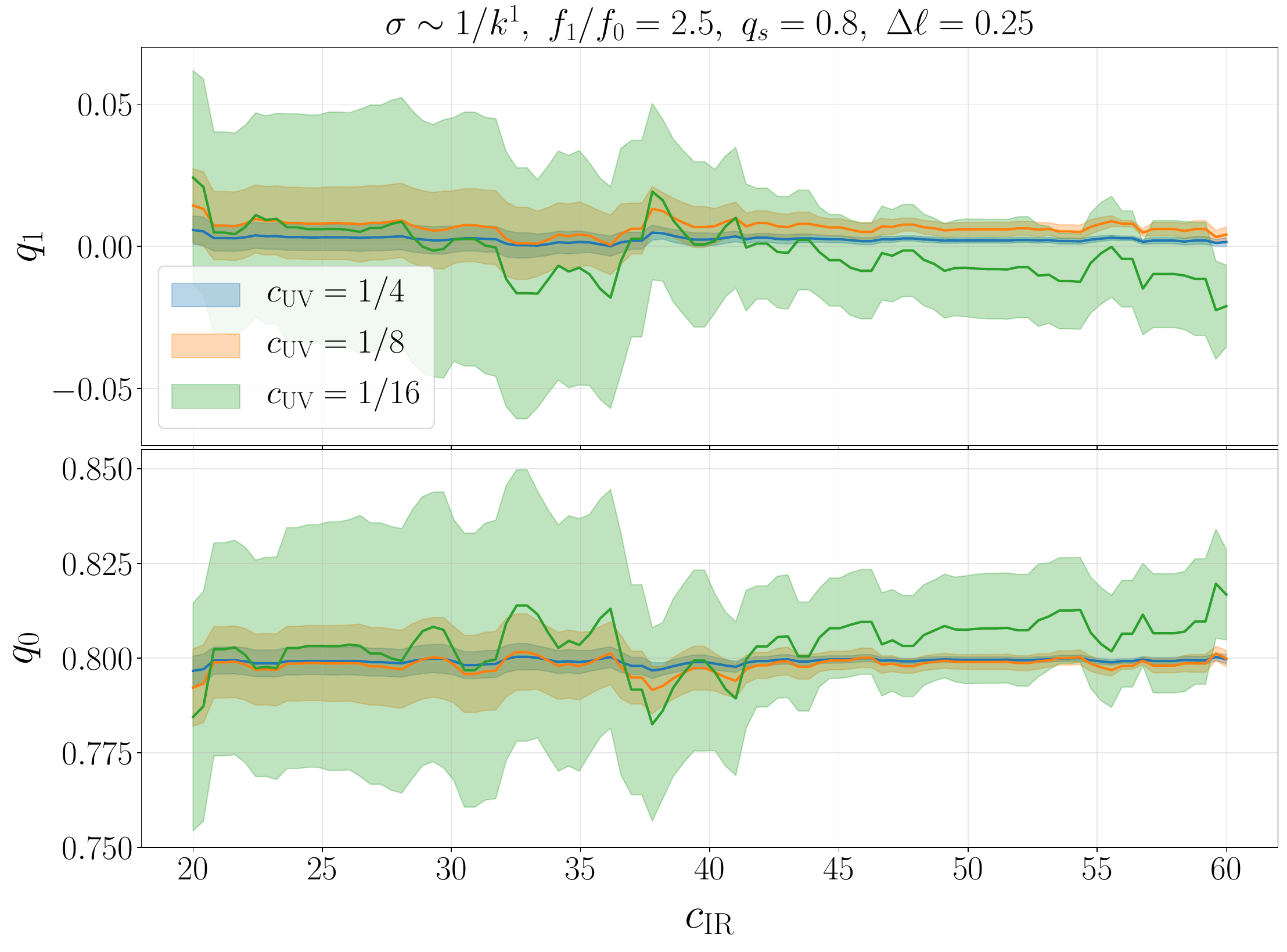}
\caption{Fitted values of $q=q_0+q_1(\ell-8)$ produced by mode oscillations in the example discussed in the text. 
Here the errors in the spectra are assumed to be proportional to $\sim 1/k$ in the fits.
The mock data has an $\ell$-independent $q_s=0.8$ but a positive bias $q_0>q_s$ and a negative slope $q_1<0$ are created by oscillations in the fit. 
}
\label{fig:q-synthetic-fit}
\end{figure}

In reality one would calculate the spectrum at $\Delta\ell= 0.25$ intervals, such that only a few points of the lines shown in Fig.~\ref{fig:q-synthetic}
would be sampled. Oscillations contribute then as a systematic error, which would reveal itself with less sparse data.  
If we try to fit a linear trend to, for instance a function $q=q_0+q_1(\ell-8)$, the effects of oscillations in the extraction of $q_0$ and $q_1$ will be somehow suppressed. 
When we fit our mock data, we find good fits but small systematic biases, which decrease with the fitting range. 
Assuming flat errors for the modes in the fit, this bias is of order 0.02 for the typical values of current simulations, while if errors 
in the spectra are taken to be proportional to it ($\sim 1/k^{q_s}$) the bias becomes negative at small $c_{\rm UV}\sim 1/16$. 
An example of the latter is Fig.~\ref{fig:q-synthetic-fit}, where we show the results for $q_0$ and $q_1$ as functions of $c_{\rm IR}$ and $c_{\rm UV}$ 
for a synthetic spectrum with a spectral index $q_s=0.8$ constant in time. 
We observe a positive bias of the order 0.01 in $q_0$ accompanying a negative slope bias. 
We note that these are results from synthetic spectra using Fourier modes equal to the model, without the further dispersion expected in simulations, 
due to statistical fluctuations caused by a finite number of modes added per bin, for instance. 
Adding statistical fluctuations will amplify the oscillation effects. Averaging over simulations and extrapolating to the infinite-volume limit will decrease the effects discussed here. 

\begin{figure}[htbp]
\includegraphics[width=0.48\textwidth]{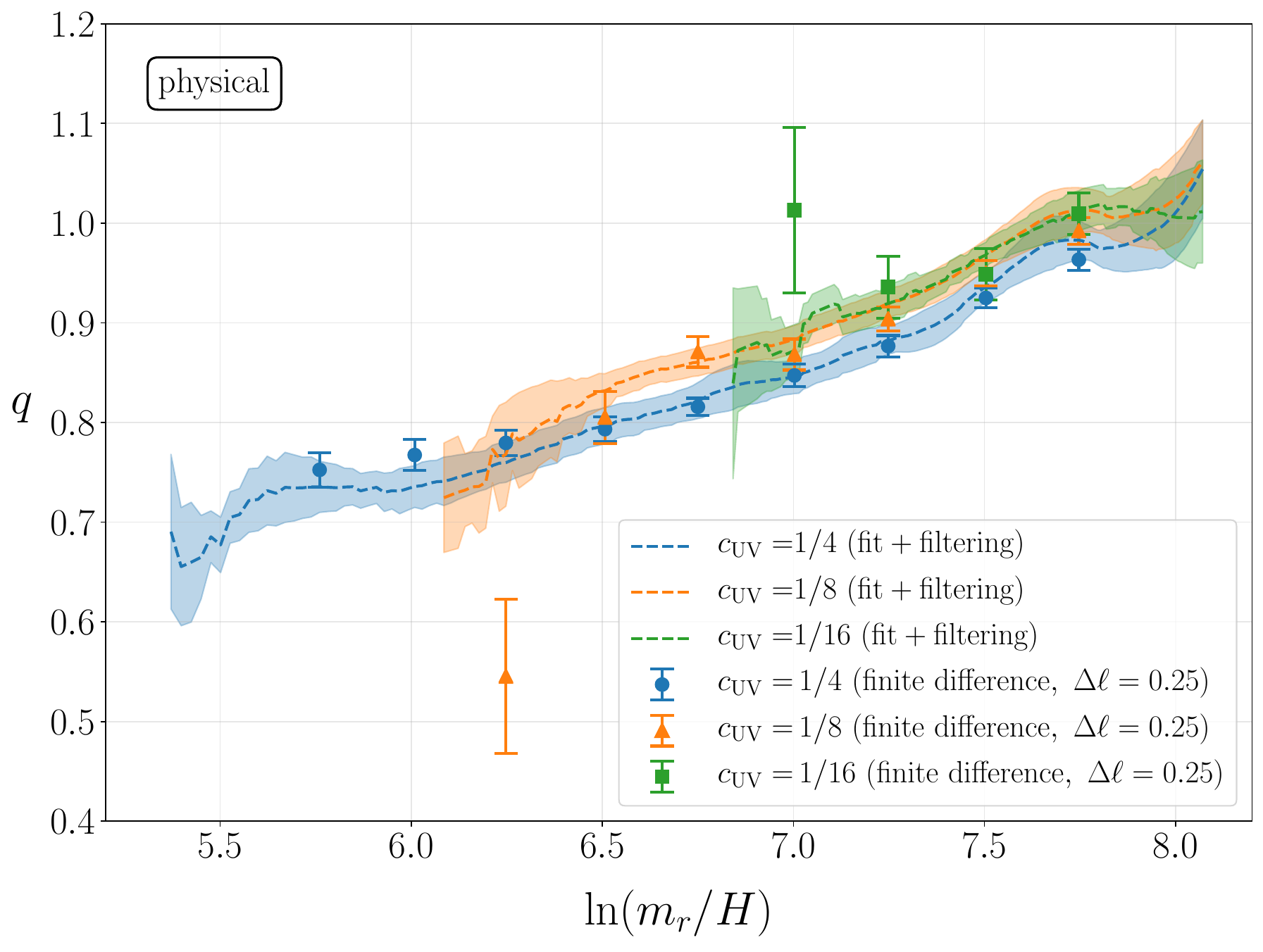}
\caption{Evolution of the spectral index $q$ for three different choices of $c_{\rm UV}$, obtained from simulations of physical strings with $4096^2$ lattice sites 
and the parameter $\bar{\lambda}=25890.8$.
Markers with error bars represent the results of the finite difference method, and dashed lines represent those obtained by the filtering method.
The coloured bands represent the error induced by changing the parameter $\sigma_{\rm filter}$ for the filtering procedure to calculate $\mathcal{F}$ [Eq.~\eqref{filter_for_F}] in addition to statistical uncertainties.}
\label{fig:osc_q_cUV}
\end{figure}

We can look at our data to search for effects of the oscillations through the time discretisation of the derivative. 
In Fig.~\ref{fig:osc_q_cUV}, we  compare $q(\ell)$ extracted from the $4096^3$ data 
with finite differences and our filtering method for three different values of $c_{\rm UV}$ using $c_{\rm IR}=50$. 
The finite difference results follow the filter results within errors except when the fitting range is small, where results diverge significantly. 
For smaller values of $c_{\rm UV}$, the relative importance of the IR part of the spectrum becomes more pronounced in the fit,
and the bias due to the oscillation in the IR modes contributes to the larger fluctuations in $q$.
In the figure we also see that the amplitude of the oscillation of $q$ is larger at early times, see also Fig.~\ref{fig:q-synthetic}. 
This is because the range for the comoving wavenumber $c_{\rm IR}HR < k < c_{\rm UV}m_rR$ is narrow at early times, and hence 
the value of $q$ is more likely to be affected by a local feature in $\mathcal{F}(x)$.
Furthermore, the range for the fit becomes even narrower for smaller values of $c_{\rm UV}$, 
and for that case the measurement of $q$ becomes possible
only at late times when the number of $k$-bins satisfying the condition $c_{\rm IR}HR < k < c_{\rm UV}m_rR$ becomes sufficiently large.\footnote{When we perform the fit,
we re-bin the data of $\mathcal{F}(x)$ such that the number of bins in the interval remains the same at all times (see Appendix~\ref{app:method_q}).}

In Ref.~\cite{Buschmann:2021sdq}, it was reported that $q$ does not exhibit a logarithmic growth for $c_{\rm UV} = 1/16$,
which is taken as a fiducial value in that work, while it shows a growing trend for $c_{\rm UV}=1/4$, which was adopted in the analysis of Ref.~\cite{Gorghetto:2020qws}.
Contrary to those results, our simulation results shown in Fig.~\ref{fig:osc_q_cUV} exhibit a trend that the value of $q$ increases with $\ell$
irrespective of the value of $c_{\rm UV}$ for both the filter and the finite difference case.
The bias of shifting the values of $q$ and its slope 
for a smaller value of $c_{\rm UV}$ caused by the oscillations could contribute to the difference,
but the systematics due to these effects is not so dramatic, typically $\sigma_q\sim\mathcal{O}(0.01)$ as shown in Fig.~\ref{fig:q-synthetic-fit}.
A caveat is that only one simulation was performed in Ref.~\cite{Buschmann:2021sdq}, 
and we cannot exclude the possibility that the effects were amplified by statistical fluctuations.

\section{Discretisation effects}
\label{sec:disc_effect}
\setcounter{equation}{0}

In addition to the contamination from oscillations in the spectrum, there is a bias due to the discretisation of the lattice,
which turns out to be the most serious issue for the measurement of $q$.
This section is devoted to the problem of discretisation effects observed in our simulations.
The effects can be categorized into two types: the discretisation of Laplacian and the choice of the lattice spacing compared to the string core radius.
The latter effect is particularly hard to control.
In the following, we first describe the effects associated to the Laplacian and to the lattice spacing in Sec.~\ref{sec:lap} and Sec.~\ref{sec:msa}, respectively.
After discussing how these effects show up in the numerical results, we perform analytical fits to get rid of them in Sec.~\ref{sec:continuum_ext}.
We will see that 
the corrected values of $q$ at large $\ell$ become smaller than what is measured from the numerical results.

\subsection{Laplacian}
\label{sec:lap}

\begin{figure*}[htbp]
\includegraphics[width=0.85\textwidth]{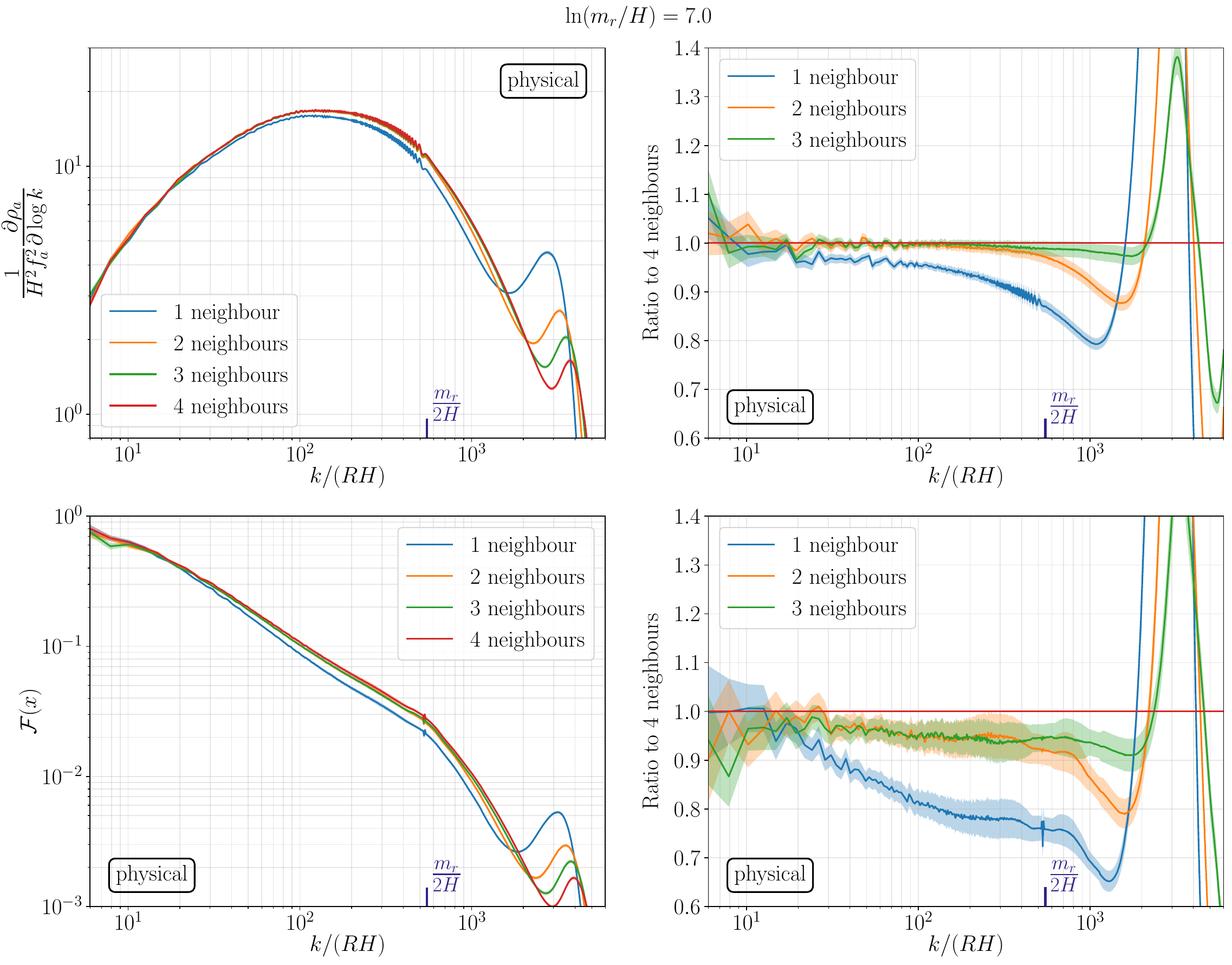}
\caption{The axion energy density spectrum (top left) and instantaneous emission spectrum (bottom left) at $\ln(m_r/H)=7$
obtained from simulations of physical strings with $4096^3$ lattice sites for 4 different choices on the neighbouring points $N_g$ used in the discretisation of the Laplacian.
The ratio of the energy density spectrum (instantaneous emission spectrum) for each choice of $N_g$ to that for $N_g=4$ is also shown in the top right (bottom right) panel.
The coloured bands represent statistical uncertainties, and the momentum corresponding to $k/R = m_r/2$ is marked with dark blue ticks.}
\label{fig:disc_lap_spectrum}
\end{figure*}

In our code, discretisation of the Laplacian in the equation of motion for the PQ field is implemented 
with $N_g$ neighbouring points, where $N_g$ can take either 1, 2, 3, or 4 (see Appendix~\ref{app:code}).
For a larger value of $N_g$, the field associated to a lattice point can react to incoming waves in advance,
and hence there is an improvement in the propagation speed of perturbations.
Such an improvement is expected to provide a more accurate result for the spectrum.
To investigate the systematics associated with the discretisation scheme of the Laplacian, 
we performed simulations of physical strings with $4096^3$ lattice sites for four different values of $N_g$. 
In Fig.~\ref{fig:disc_lap_spectrum}, we show the axion spectrum and $\mathcal{F}(x)$ for each choice of $N_g$.
We see that the spectrum at intermediate momenta is underestimated for smaller values of $N_g$.
Furthermore, there is a peak-like feature at a very high momentum, and the height of the peak is pronounced for smaller $N_g$.
These features can be regarded as unphysical effects caused by the discretisation error in the Laplacian,
which can be alleviated by choosing a larger value of $N_g$.

\begin{figure}[htbp]
\includegraphics[width=0.48\textwidth]{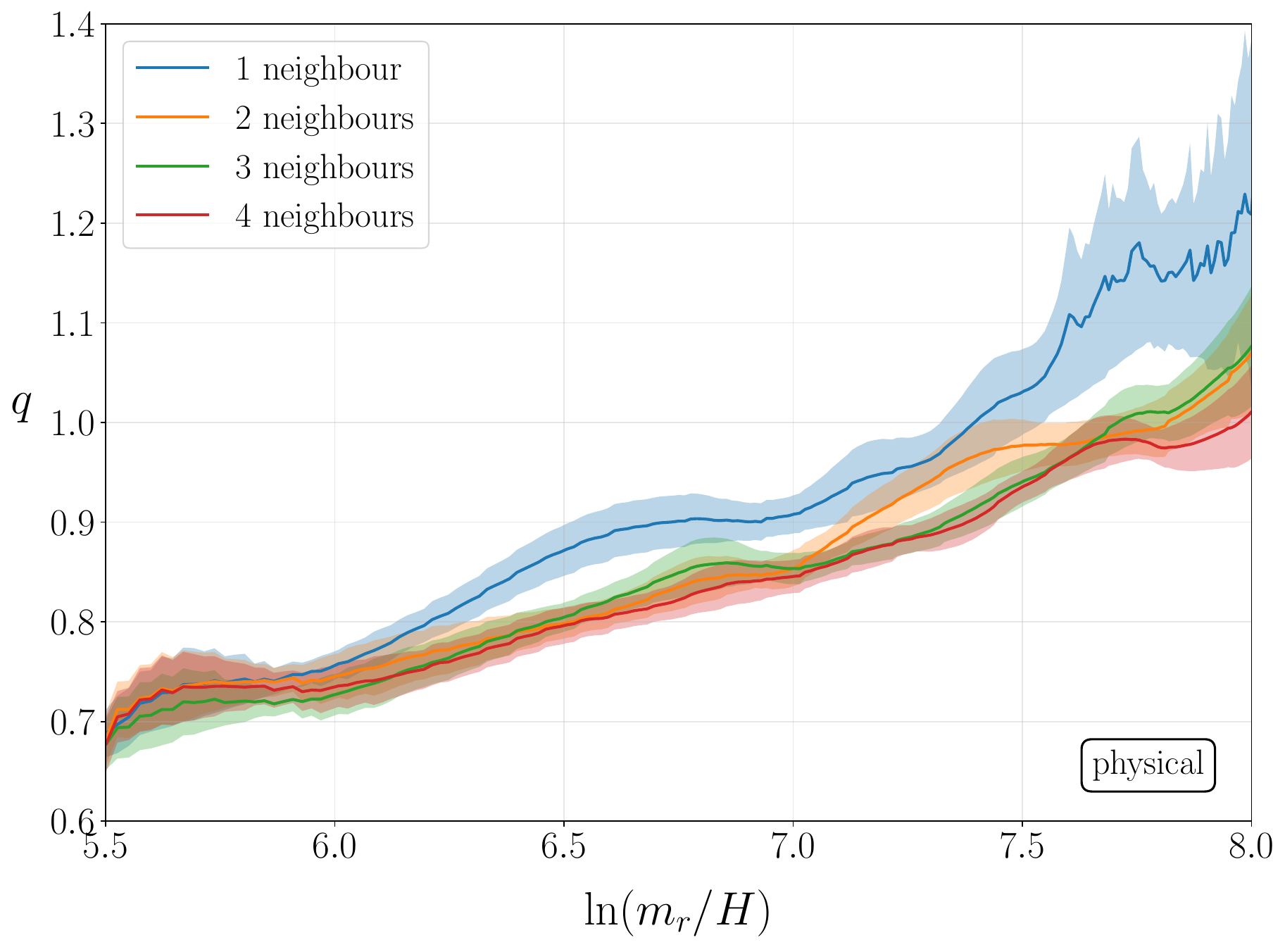}
\caption{Evolution of the spectral index $q$ of the instantaneous emission spectrum
obtained from simulations of physical strings with $4096^3$ lattice sites for 4 different choices on the neighbouring points $N_g$ used in the discretisation of the Laplacian.
The coloured bands represent the error induced by changing the parameter $\sigma_{\rm filter}$ for the filtering procedure to calculate $\mathcal{F}$ [Eq.~\eqref{filter_for_F}] in addition to statistical uncertainties.}
\label{fig:disc_lap_q}
\end{figure}

The discretisation effects distort the spectrum such that the slope at the intermediate momenta
becomes steeper for a smaller $N_g$, as shown in Fig.~\ref{fig:disc_lap_spectrum}.
This implies that the value of $q$ can be overestimated for smaller $N_g$.
Such a bias is clearly shown in Fig.~\ref{fig:disc_lap_q}, where we plot the evolution of $q$ for different choices of $N_g$.
From this figure we can also see that $N_g \ge 3$ is sufficient to suppress the effect.

The effect of the Laplacian could partially explain the difference in $q$ in the literature.
We observe that the values of $q$ obtained for our fiducial choice $N_g=4$ is slightly smaller than 
Ref.~\cite{Gorghetto:2020qws}, which used the Laplacian with $N_g=2$ (see Appendix~\ref{app:comparisons} for an explicit comparison).
Furthermore, $N_g=1$ was adopted in Ref.~\cite{Buschmann:2021sdq}, where systematically larger values of $q$ were reported.
These facts are consistent with the trend shown in Fig.~\ref{fig:disc_lap_q}.

\subsection{Lattice spacing}
\label{sec:msa}

\begin{figure*}[htbp]
\includegraphics[width=0.85\textwidth]{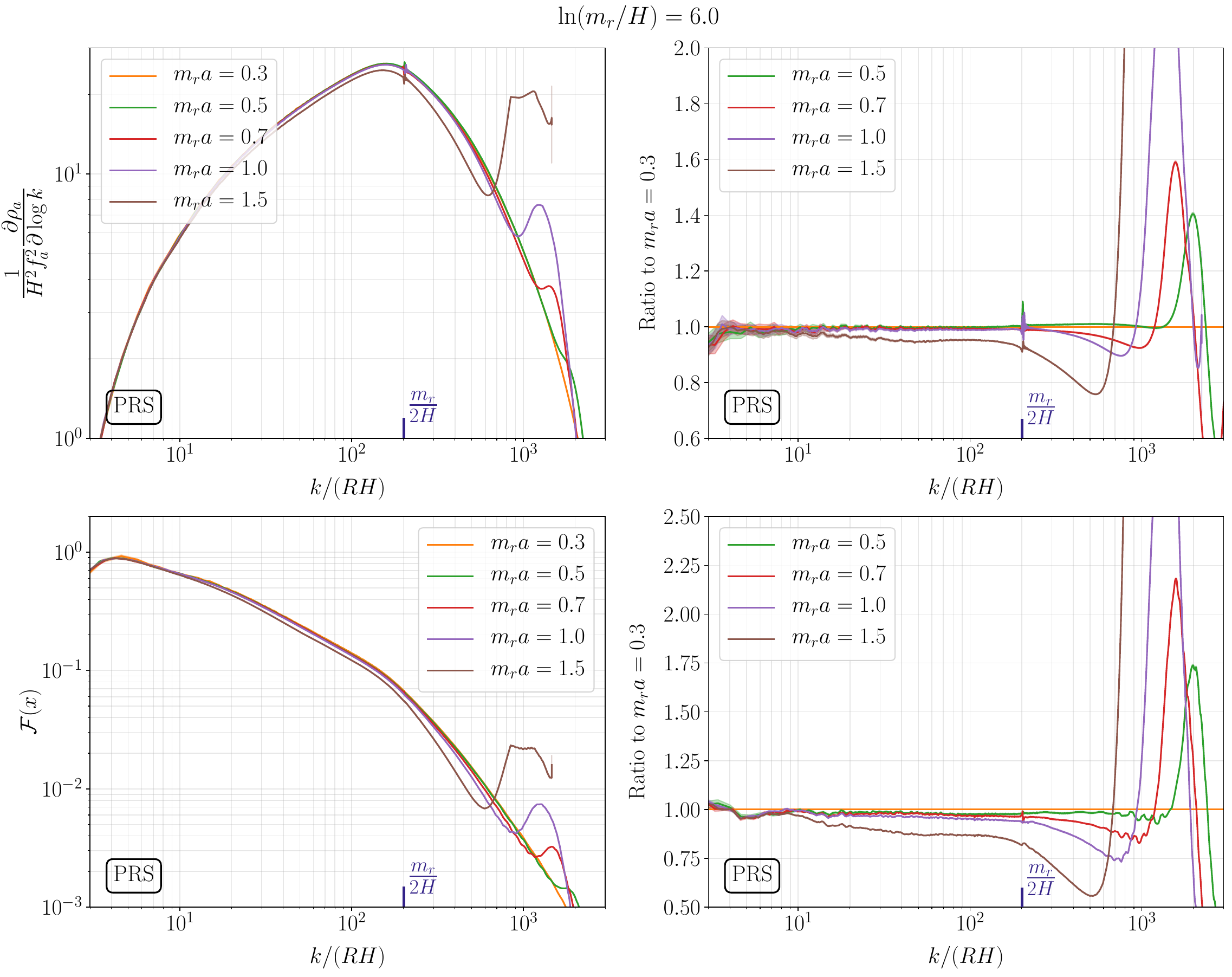}
\caption{The axion energy density spectrum (top left) and instantaneous emission spectrum (bottom left) at $\ln(m_r/H)=6$
obtained from simulations of PRS strings with $8192^3$ lattice sites for 5 different choices of $m_ra$.
The ratio of the energy density spectrum (instantaneous emission spectrum) for each choice of $m_ra$ to that for $m_ra=0.3$ is also shown in the top right (bottom right) panel.
The coloured bands represent statistical uncertainties.
The momentum corresponding to $k/R = m_r/2$ is marked with dark blue ticks.
Here we do not show the results from simulations with $m_r a=0.2$, since they are consistent with those with $m_r a= 0.3$ but have larger statistical uncertainties.}
\label{fig:disc_mra_spectrum_PRS}
\end{figure*}

The effect of the lattice spacing compared to the string core radius can be parameterised by the quantity $m_r a$,
where $a = RL/N$ is the physical size of the lattice spacing.
The simulation with PRS strings is useful to monitor this effect, since in that case
$m_r \propto R^{-1}$ and $m_ra$ remains constant.
In Fig.~\ref{fig:disc_mra_spectrum_PRS}, we compare the axion spectrum and $\mathcal{F}$ 
obtained from PRS-type simulations among different values of $m_ra$.
We can see features similar to Fig.~\ref{fig:disc_lap_spectrum}:
The amplitude is underestimated at intermediate momenta, and there is a peak at very high momenta.
The effect is pronounced for larger values of $m_ra$, where the resolution of the string core becomes worse.

\begin{figure}[htbp]
$\begin{array}{c}
\subfigure{
\includegraphics[width=0.48\textwidth]{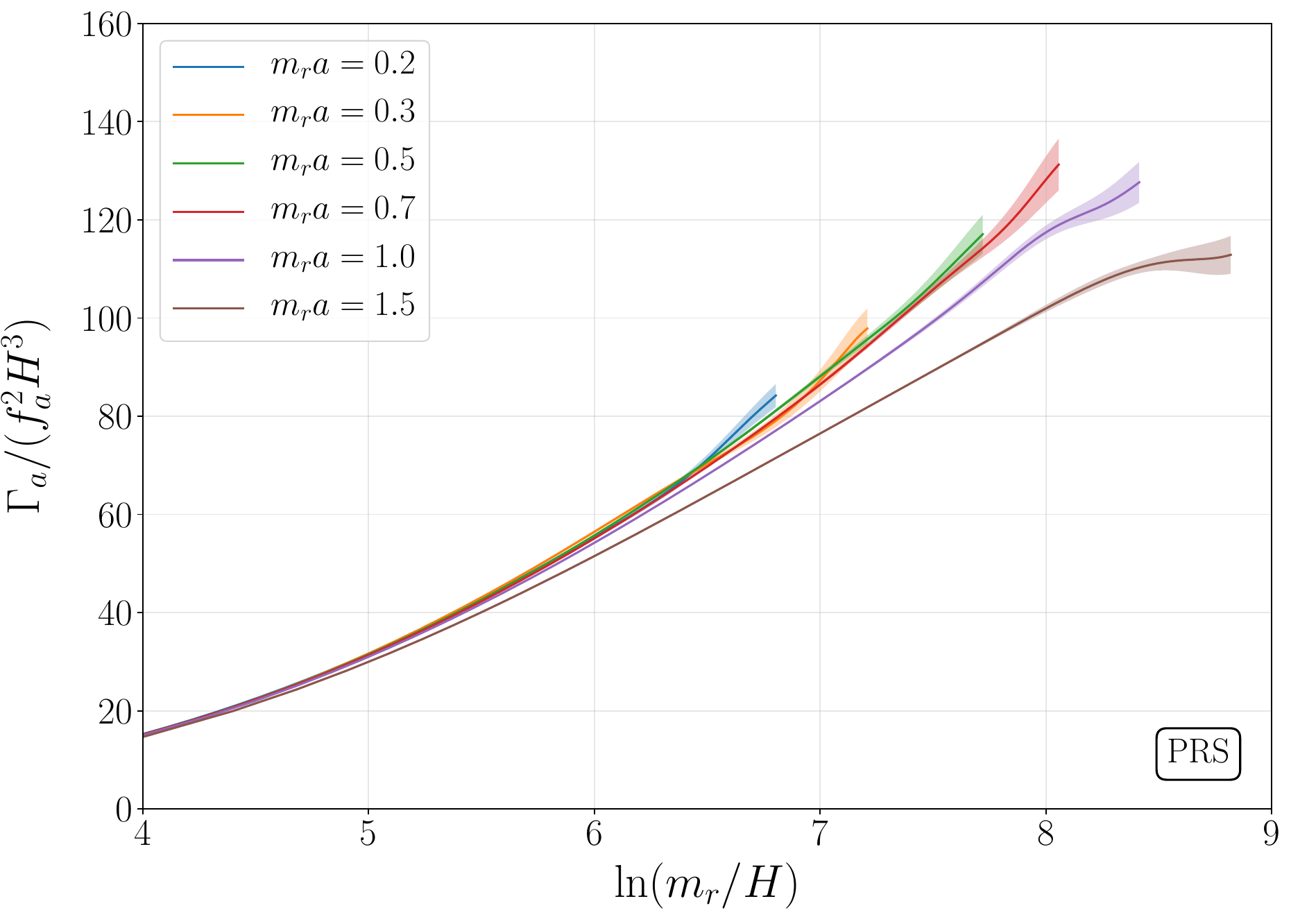}}
\\
\subfigure{
\includegraphics[width=0.48\textwidth]{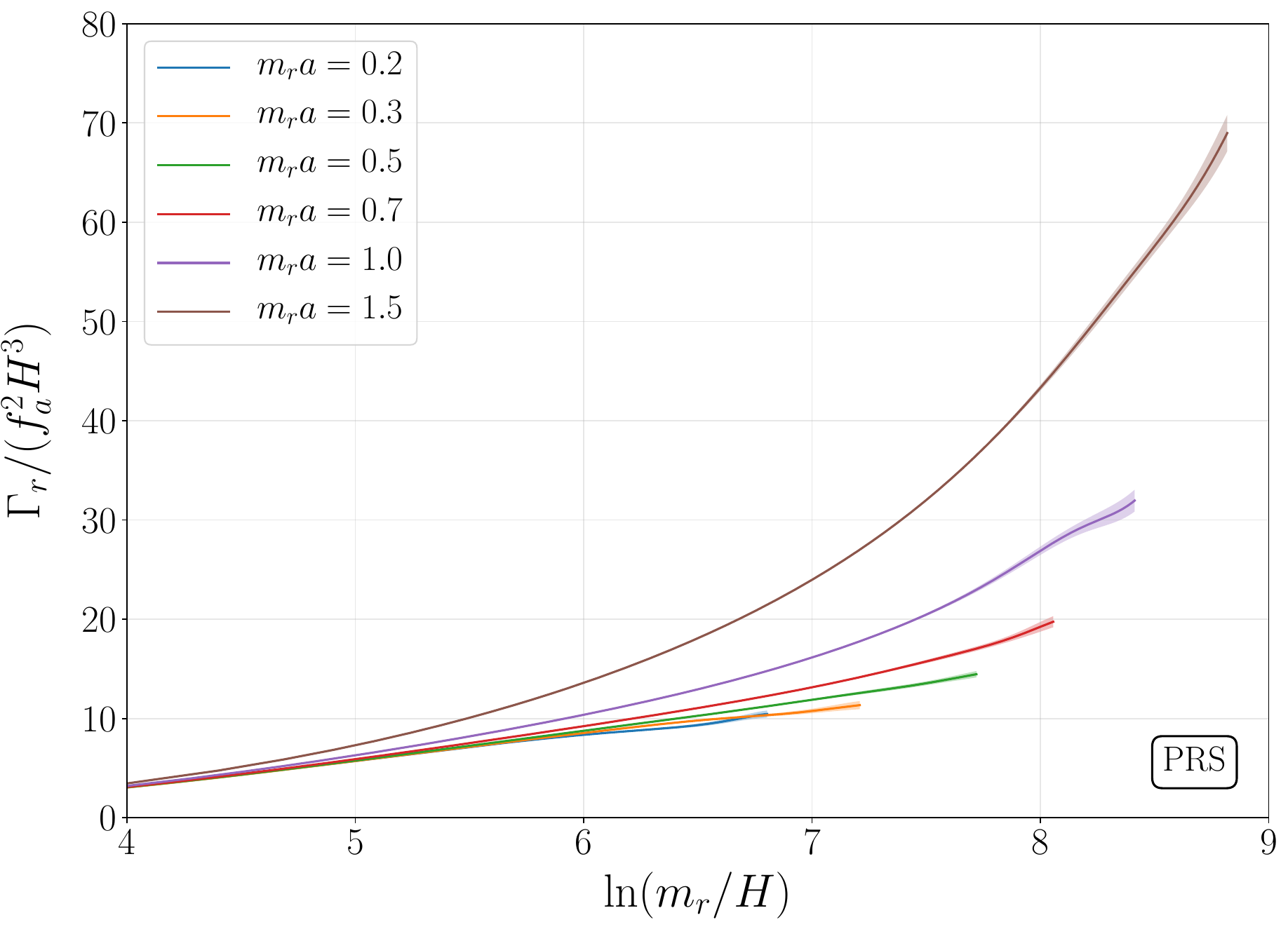}}
\end{array}$
\caption{Evolution of the energy density emission rate of axions (top panel) and that of saxions (bottom panel) for PRS strings with different values of $m_ra$.
The coloured bands represent statistical uncertainties.}
\label{fig:disc_Gamma_PRS}
\end{figure}

The discretisation effect is also manifested in the energy density emission rates for axions and saxions,
\begin{align}
\Gamma_a = R^{-4}\frac{d}{dt}(R^4\rho_a), \quad \Gamma_r = R^{-\langle z\rangle}\frac{d}{dt}(R^{\langle z\rangle}\rho_r), \label{Gamma_definition}
\end{align}
where $\langle z\rangle$ is a suitably averaged redshift exponent [see Eq.~\eqref{saxion_mean_z}].
We computed $\Gamma_a$ and $\Gamma_r$ for both physical and PRS simulations (see Appendix~\ref{app:method_Gamma} for more technical details).
Figure~\ref{fig:disc_Gamma_PRS} shows the plots of $\Gamma_a$ and $\Gamma_r$ obtained from PRS-type simulations.
We see that the discretisation effects make the axion emission less efficient, and lead to the production of large amounts of saxions.
Note that the effect is more pronounced for larger values of $\ell$. Namely, the effect does not only depend on $m_ra$ but also on $\ell$.
The fact that the effect is less manifest at small $\ell$ may explain why the systematics associated with this effect have been overlooked previously.

For simulations with physical strings, the effect becomes more intricate than the PRS case, since $m_ra$ increases with time ($\propto R$)
so that the resolution gets worse at late times.
When considered in conjunction with the possible $\ell$-dependence observed in the PRS case, 
we expect that the discretisation effects rapidly come into play at the latest times in the physical case.

\begin{figure}[htbp]
$\begin{array}{c}
\subfigure{
\includegraphics[width=0.48\textwidth]{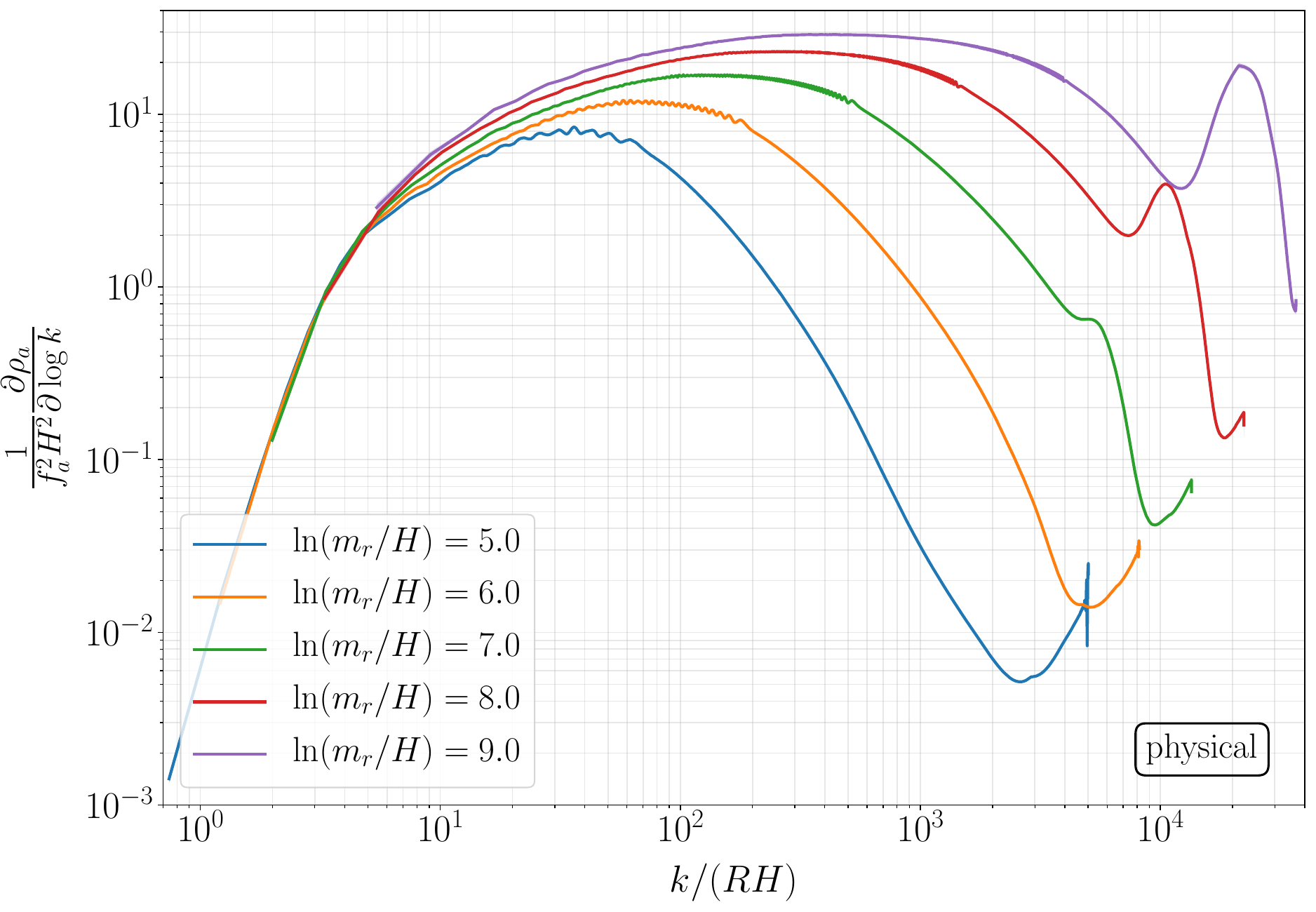}}
\\
\subfigure{
\includegraphics[width=0.48\textwidth]{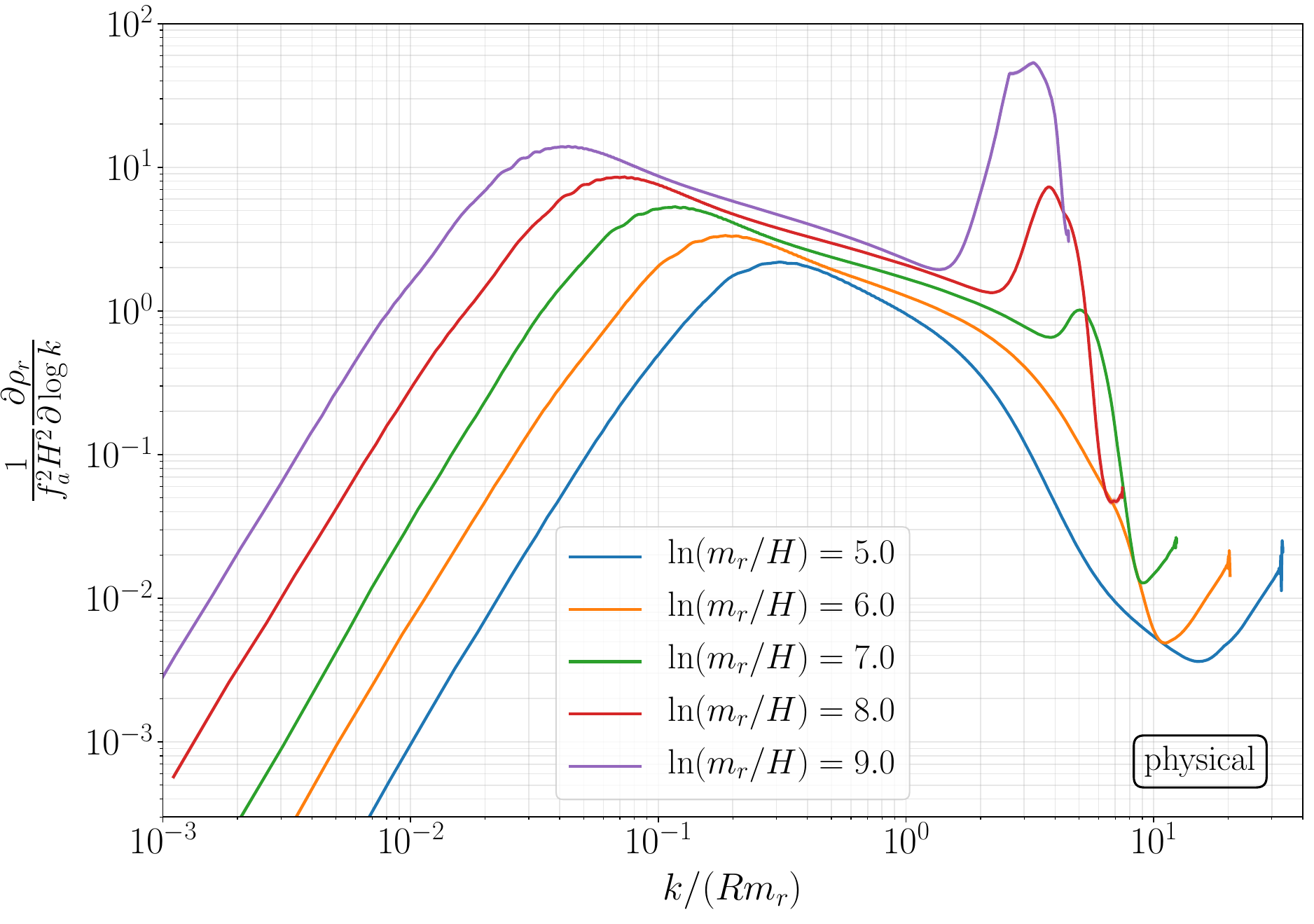}}
\end{array}$
\caption{Energy density spectrum of axions (top panel) and that of saxions (bottom panel) 
obtained from simulations of physical strings with $11264^3$ lattice sites.
Different lines correspond to the spectra measured at different values of $\ln(m_r/H)$.}
\label{fig:disc_spectrum_physical}
\end{figure}

Figure~\ref{fig:disc_spectrum_physical} shows the evolution of the spectrum of axions (top panel) and saxions (bottom panel) 
in the simulations of physical strings with $11264^3$ lattice sites.
Overall, the axion spectrum exhibits the expected behaviour: There are cutoffs at $k/R \sim H$ and $k/R \sim m_r$, 
which means that the location of the IR cutoff remains constant while the UV cutoff moves right when the spectrum is plotted in terms of $k/(RH)$.
The growth in the amplitude of the axion spectrum can be attributed to the logarithmic growth in the production rate,
which originates from the increase in the string tension $\mu_{\rm th}$ and density $\xi$ [see Eq.~\eqref{Gamma_th}].
However, we also see a peak-like feature at very high momenta, which starts to appear at $\ell \approx 7$ and grows rapidly.
A similar feature is seen in the plot of the saxion spectrum.
These features can be regarded as a signal that the system is affected by the unphysical discretisation effects at late times of the simulations.
In that regime, instead of radiating axions and saxions with proper momenta ($k/R \lesssim m_r$),
strings dissipate their energy by producing a huge amount of high frequency modes.

\begin{figure}[htbp]
\includegraphics[width=0.48\textwidth]{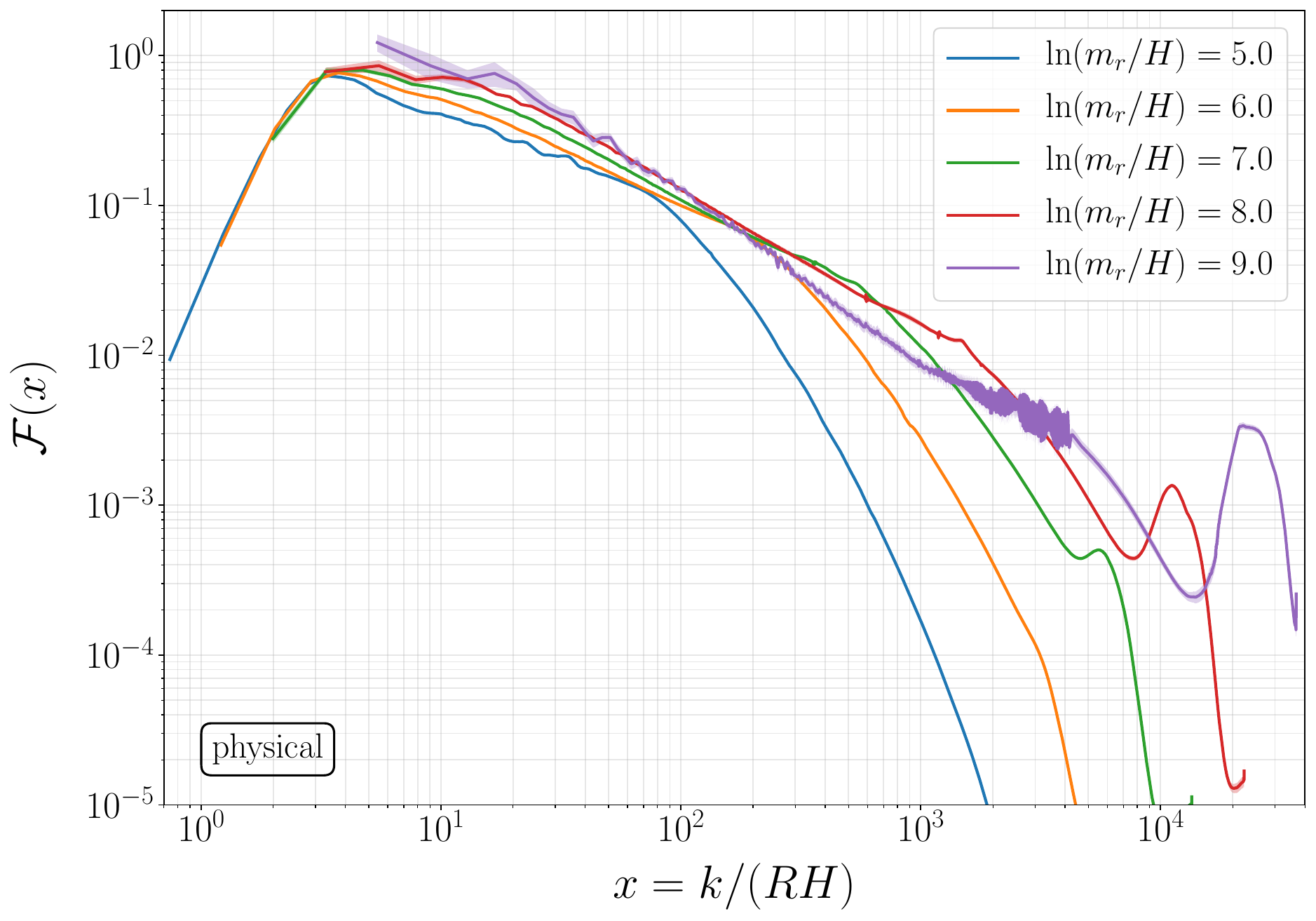}
\caption{Instantaneous emission spectrum of axions obtained from simulations of physical strings with $11264^3$ lattice sites.
Different lines correspond to the emission spectra measured at different values of $\ln(m_r/H)$, and the coloured bands represent statistical uncertainties.
For the calculation of $\mathcal{F}(x,y)$ shown in this figure, we used the filtering method introduced in Sec.~\ref{sec:oscillations} and described in detail in Appendix~\ref{app:method_F},
rather than the finite difference method.}
\label{fig:disc_F_physical}
\end{figure}

Figure~\ref{fig:disc_F_physical} shows the evolution of the instantaneous emission spectrum in the simulations of physical strings with $11264^3$ lattice sites.
Comparing the instantaneous emission spectrum at $\ell = 9$ with those at earlier times, we see that $\mathcal{F}$ is considerably distorted at the latest times,
as there is a suppression in the amplitude at intermediate momenta and a peak-like feature appears at very high momenta.
The distortion of $\mathcal{F}$ can bias $q$ towards larger values. We have seen a similar effect in Fig.~\ref{fig:disc_lap_q} for the case of the discretisation of the Laplacian,
and we expect that the effect of the Laplacian can be suppressed by taking a sufficiently larger value of $N_g$.
On the other hand, the effect we see in Fig.~\ref{fig:disc_F_physical} is attributed to the large value of $m_ra$, which is not possible to suppress
unless we perform simulations with higher resolutions that realise a smaller value of $m_ra$ at the same value of $\ell$.

\begin{figure}[htbp]
$\begin{array}{c}
\subfigure{
\includegraphics[width=0.48\textwidth]{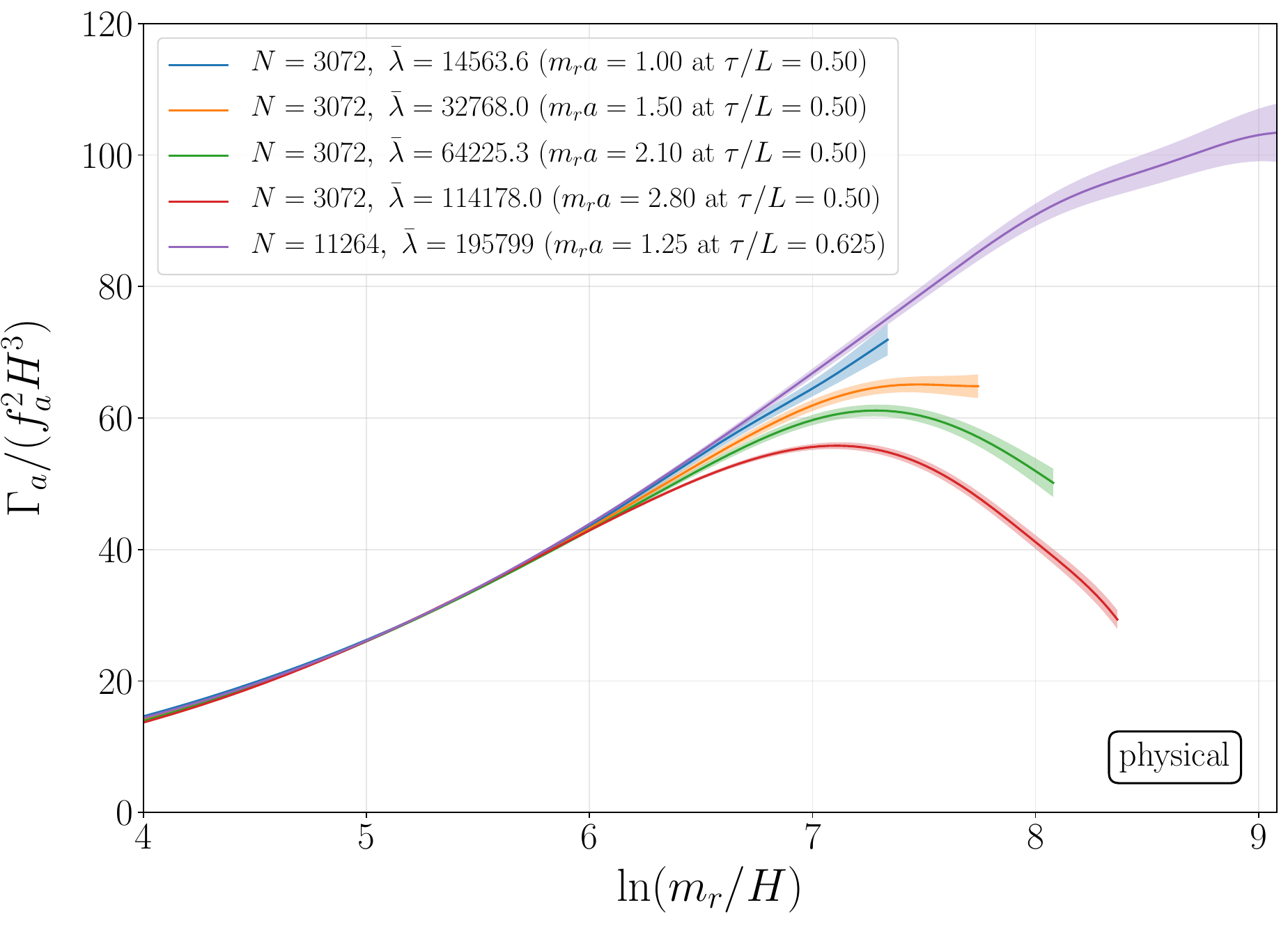}}
\\
\subfigure{
\includegraphics[width=0.48\textwidth]{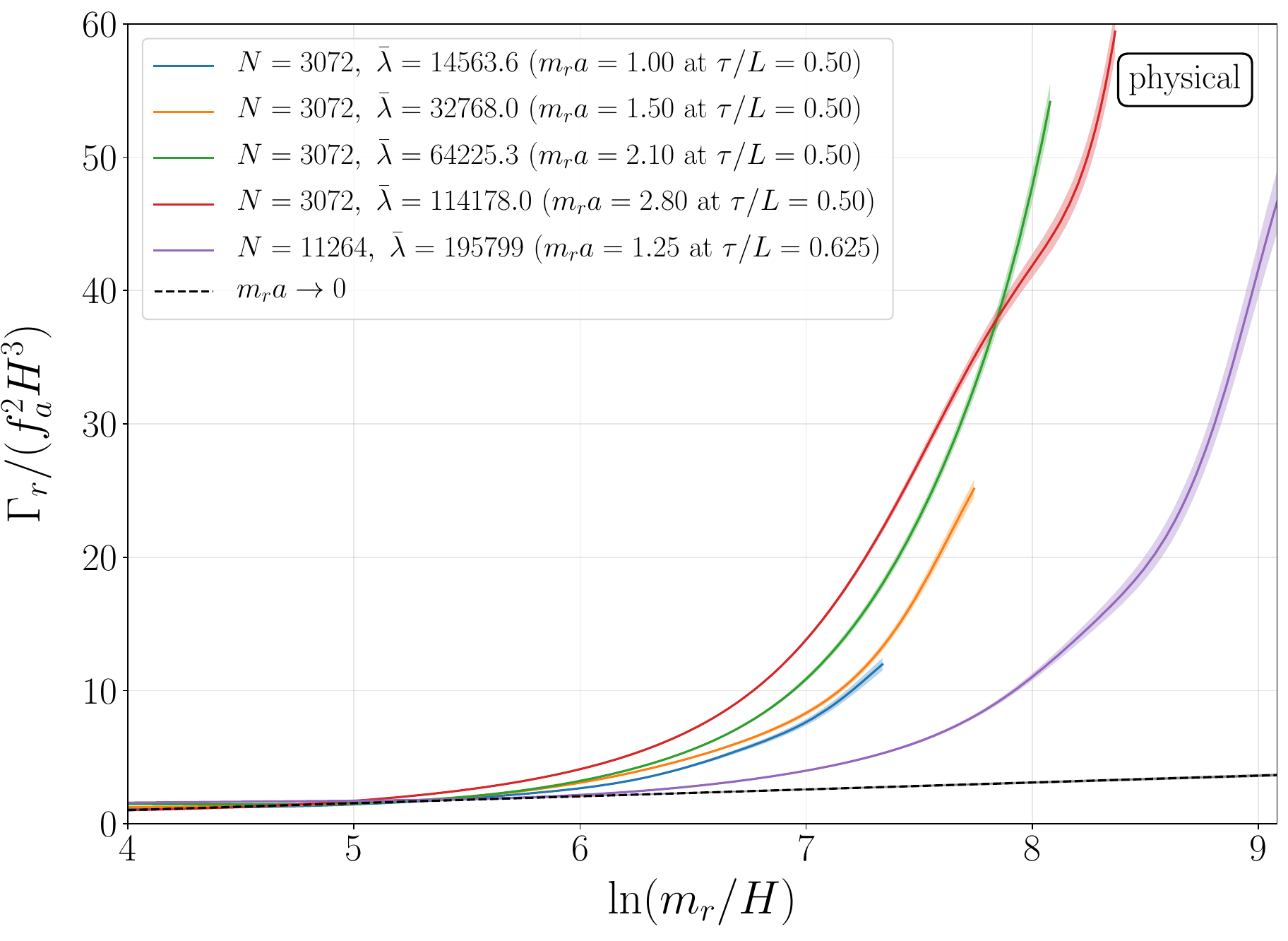}}
\end{array}$
\caption{Evolution of the energy density emission rate of axions (top panel) and that of saxions (bottom panel) 
for physical strings with different values of the self coupling parameter $\bar{\lambda}$ defined by Eq.~\eqref{lambda_definition}.
The results are obtained from one set of $11264^3$ simulations and four sets of $3072^3$ simulations of physical strings, 
where $m_r a$ evolves with time and its value at the end of the simulation is also shown in the legend for each choice of $\bar{\lambda}$.
The coloured bands represent statistical uncertainties.
In the bottom panel, the continuum extrapolation of the saxion energy density emission rate is also shown by the dashed line.}
\label{fig:disc_Gamma_physical}
\end{figure}

In order to investigate the effect of $m_ra$ on the instantaneous spectrum and $q$ for simulations with physical strings, we performed additional simulations
with $3072^3$ lattice sites for three different choices of the self coupling of the PQ field that lead to different values of $m_ra$ at the end of the simulation
(see Fig.~\ref{fig:code_mra} for the value of $m_ra$ as a function of $\ell$ for each choice of the simulation parameters).
Figure~\ref{fig:disc_Gamma_physical} shows the evolution of the energy density emission rates of axions and saxions obtained from
physical string simulations including those with $3072^3$ and $11264^3$ lattice sites.
Similarly to the case of PRS strings shown in Fig.~\ref{fig:disc_Gamma_PRS}, we see that $\Gamma_a$ ceases to increase and $\Gamma_r$ blows up 
at large $\ell$, and that the effect is more pronounced for simulations that lead to the larger value of $m_ra$ at late times.
Compared to the PRS case, the effect shows up rapidly at large logs, due to the exponential dependence on $\ell$,
i.e. $m_ra \propto R \propto \exp(\frac{1}{2}\ell)$.

\begin{figure}[htbp]
$\begin{array}{c}
\subfigure{
\includegraphics[width=0.48\textwidth]{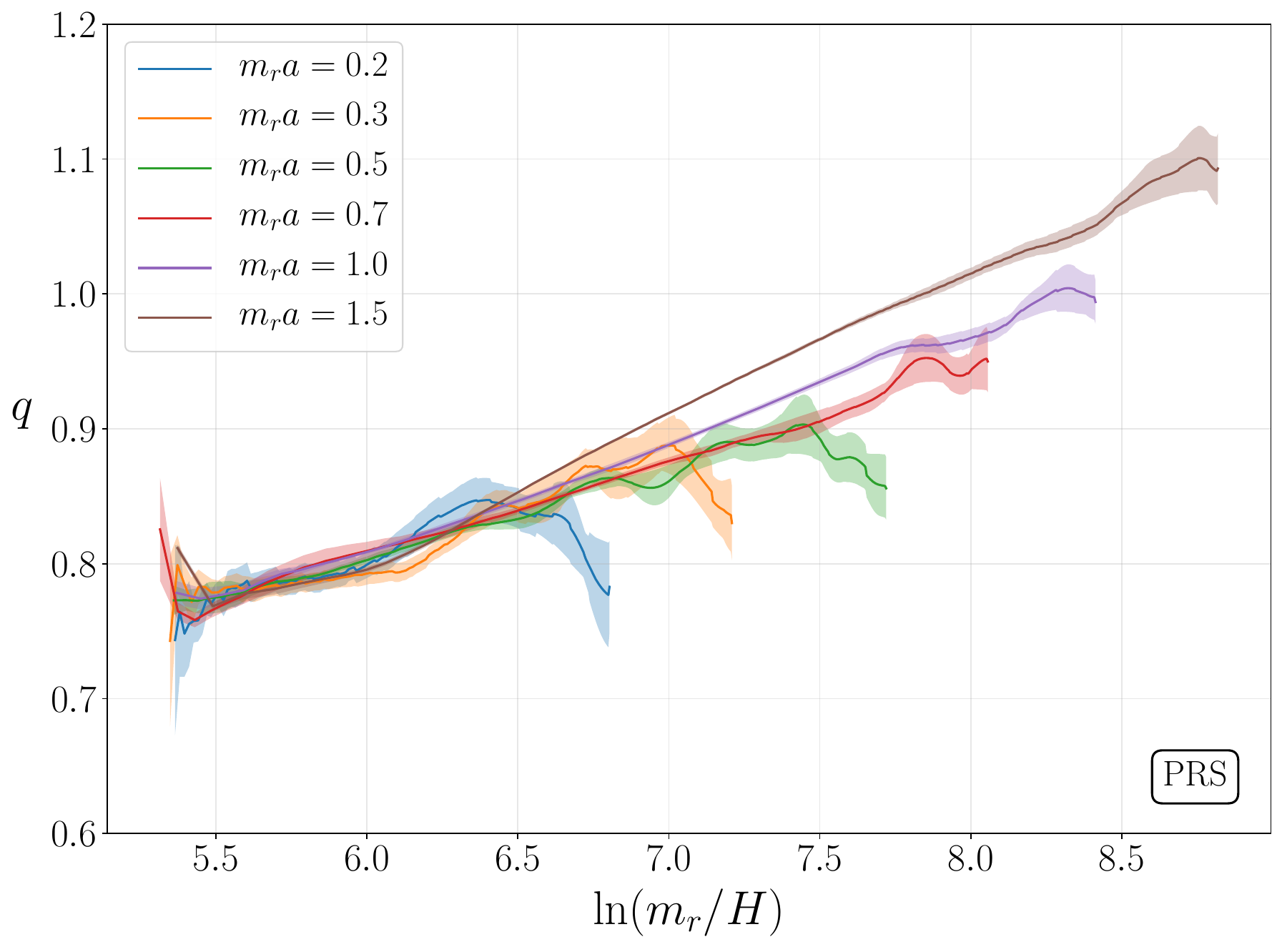}}
\\
\subfigure{
\includegraphics[width=0.48\textwidth]{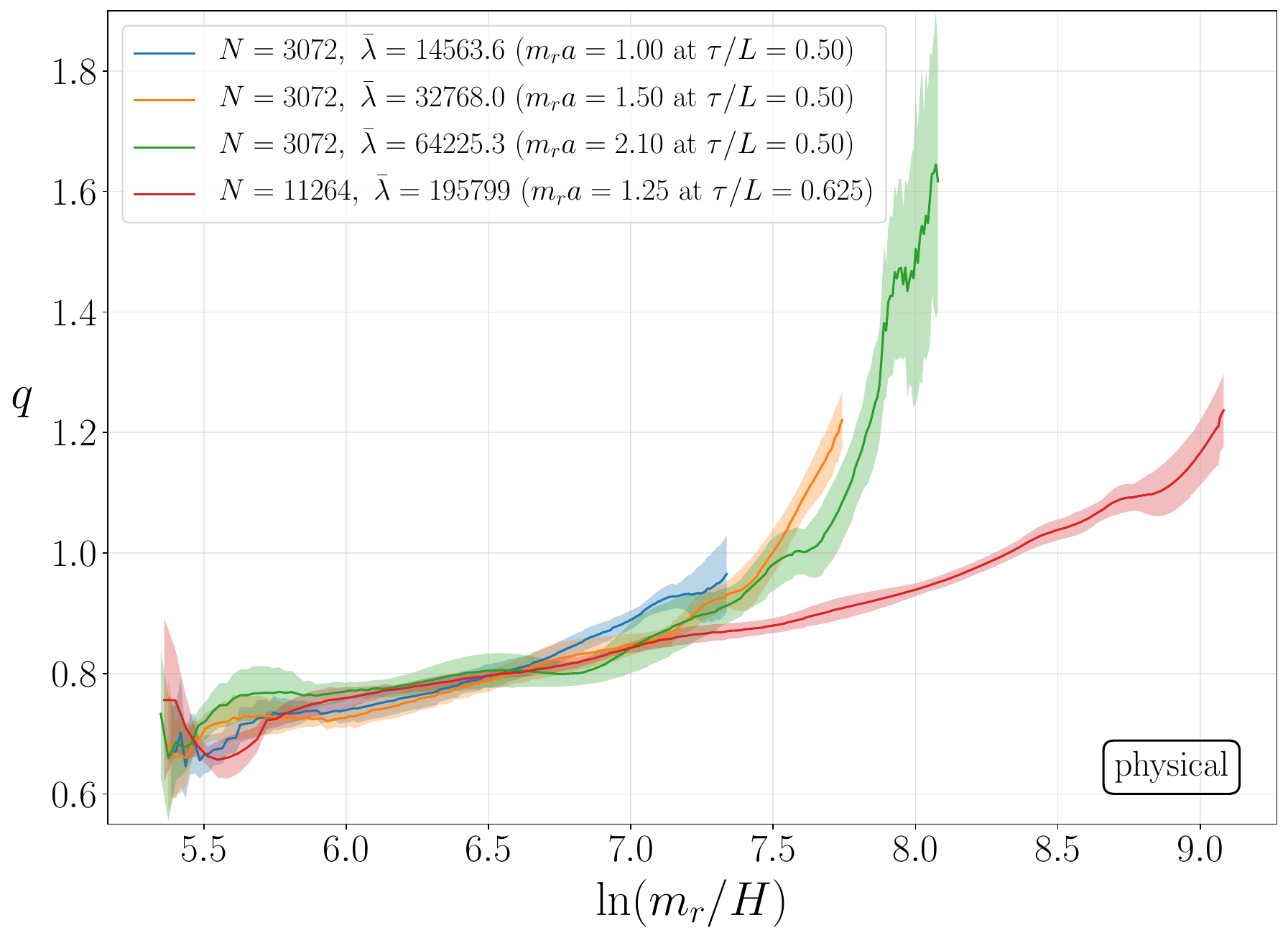}}
\end{array}$
\caption{Evolution of the spectral index $q$ of the instantaneous emission spectrum for different choices of the coupling parameter of the PQ field that controls the string core width.
Top panel shows the results of simulations of PRS strings with $8192^3$ lattice sites.
Bottom panel shows the results of one set of $11264^3$ simulations and three sets of $3072^3$ simulations of physical strings.
The coloured bands represent the error induced by changing the parameter $\sigma_{\rm filter}$ for the filtering procedure to calculate $\mathcal{F}$ [Eq.~\eqref{filter_for_F}] in addition to statistical uncertainties.}
\label{fig:disc_mra_q}
\end{figure}

Figure~\ref{fig:disc_mra_q} shows the evolution of $q$ for simulations with PRS (top panel) and physical strings (bottom panel).
In both cases, we see that the value of $q$ converges at small $\ell$, but takes larger values at larger $\ell$ and $m_ra$.\footnote{For the PRS case, 
the convergence at small $\ell$ is less clear since the data with small $m_ra$ contains large oscillations at late times of the simulations. 
As discussed in Sec.~\ref{sec:oscillations}, the largest oscillation is expected to occur at around $\tau = L/2$, 
which corresponds to $\ell\simeq 6.7,7.1,7.6,8.0,8.3,8.7$ for $m_ra = 0.2,0.3,0.5,0.7,1.0,1.5$.
It turns out that these oscillations cannot be eliminated even if we use the filtering method rather than the finite difference method to compute $\mathcal{F}$.
Furthermore, for smaller $m_ra$, these oscillatory features could be amplified due to the effect of the finite volume (see Appendix~\ref{app:syst_finite_V}).
The feature due to the maximum oscillation at $\tau \simeq L/2$ is also seen in the results of physical strings.
For instance, the plot of $N=11264$ data shown in the bottom panel of Fig.~\ref{fig:disc_mra_q}
exhibits the oscillatory feature at $\ell = 8.5\text{--}9$, in accord with what was shown in Fig.~\ref{fig:q-synthetic}.}
We also observe that the effect increases more rapidly for the physical case compared to the PRS case.

The rapid increase in the discretisation effect can be understood by considering the fact that the horizon-sized string loops
acquire a large velocity when they collapse.
To illustrate this, let us consider the motion of a circular loop in the Nambu-Goto effective theory~\cite{Vilenkin:2000jqa}, where the string is described by
spacetime coordinates $X^{\mu}(\tau,\sigma)$, and $\tau$ and $\sigma$ are parameters of the world sheet swept by the string.\footnote{For simplicity, here we ignore the effect of back-reaction 
on the loop due to the radiation of axions. It was confirmed that including the effect of the axion radiation does not significantly alter the motion of the loop described by the Nambu-Goto solution
even if $\ell$ is as small as $\ell \approx 5$~\cite{Gorghetto:2018myk}.}
Here we choose the time-like parameter as the conformal time $\tau$.
After imposing the gauge condition ${\bm X}_{\tau}\cdot{\bm X}'=0$, where the prime denotes the derivative with respect to $\sigma$,
the equations of motion for the Nambu-Goto string in a flat Friedmann-Robertson-Walker (FRW) background ~\cite{Vilenkin:2000jqa,Turok:1984db} read,
\begin{align}
\bm{X}_{\tau\tau} + 2\frac{R_{\tau}}{R}(1-\bm{X}_{\tau}^2)\bm{X}_{\tau} &= \epsilon^{-1}(\epsilon^{-1}\bm{X}')', \label{NG_EOM_1}\\
\epsilon_{\tau} &= -2\frac{R_{\tau}}{R}\epsilon\bm{X}_{\tau}^2, \label{NG_EOM_2}
\end{align}
where $\epsilon = [\bm{X}'^2/(1-\bm{X}_{\tau}^2)]^{1/2}$.
For a circular loop represented by $X^{\mu}(\tau,\sigma) = (\tau,l(\tau)\cos\sigma,l(\tau)\sin\sigma,0)$ with $-\pi\le \sigma < \pi$,
Eq.~\eqref{NG_EOM_1} reduces to
\begin{align}
l_{\tau\tau} + 2\frac{R_{\tau}}{R}(1-l_{\tau}^2)l_{\tau} + \frac{1-l_{\tau}^2}{l} = 0,
\end{align}
and $\epsilon = [l^2/(1-l_{\tau}^2)]^{1/2}$.
The second term in the left-hand side of the above equation represents the effect of damping due to the cosmic expansion.
When $l$ is larger than the horizon size, the velocity of the string is suppressed due to the damping term.
In fact, by ignoring the terms with $l_{\tau\tau}$ and with higher orders in $l_{\tau}$, we see from the above equation that the string velocity is of order
$l_{\tau} \sim R/(R_{\tau} l) \sim \tau/l$, which is very small for $\tau/l \ll 1$.

The effect of the Hubble damping becomes irrelevant when the loop size becomes smaller than the horizon size, and after that it starts to collapse.
The motion of the string loop in this regime may be approximated by dropping the terms proportional to $R_{\tau}/R$ in Eqs.~\eqref{NG_EOM_1} and~\eqref{NG_EOM_2},
\begin{align}
l_{\tau\tau} + \frac{1-l_{\tau}^2}{l} \approx 0,\quad \epsilon_{\tau} \approx 0,
\end{align}
whose solution is given by $l \approx l_0\cos(\tau/\epsilon)$ with $\epsilon \approx \text{constant}$, and $l_0$ is the radius of the loop when it starts to collapse.
Using the fact that $\epsilon$ remains constant, we can estimate the loop velocity and the corresponding Lorentz factor $\gamma$ at the time when the loop size becomes
of the order of the string core radius, $l \approx m_r^{-1}$: 
\begin{align}
\gamma = \left[1-(l_{\tau}^2)_{l\approx m_r^{-1}}\right]^{-1/2} \approx l_0m_r,
\end{align} 
which implies $\gamma \approx e^{\ell}$ for $l_0 \sim H^{-1}$.

Based on the above result, we expect that the collapsing strings can have an exponentially large Lorentz factor in the comoving coordinates at large $\ell$.
This implies that the size of the string core is suppressed by the Lorentz contraction, and the $m_ra$ parameter is effectively modified as
\begin{align}
m_ra \to \gamma m_r a \approx e^{\ell}m_r a. \label{mra_boosted} 
\end{align}
Therefore, the discretisation effect occurring at large $m_ra$ can be further pronounced at large $\ell$.
Note that, for simulations of physical strings $m_ra$ also grows as $e^{\ell/2}$, and hence the effect arises more dramatically than for the case of PRS strings.

\subsection{Continuum extrapolation}
\label{sec:continuum_ext}
\subsubsection{Energy density emission rate}
\label{sec:continuum_ext_Gamma}

\begin{figure}[htbp]
$\begin{array}{c}
\subfigure{
\includegraphics[width=0.48\textwidth]{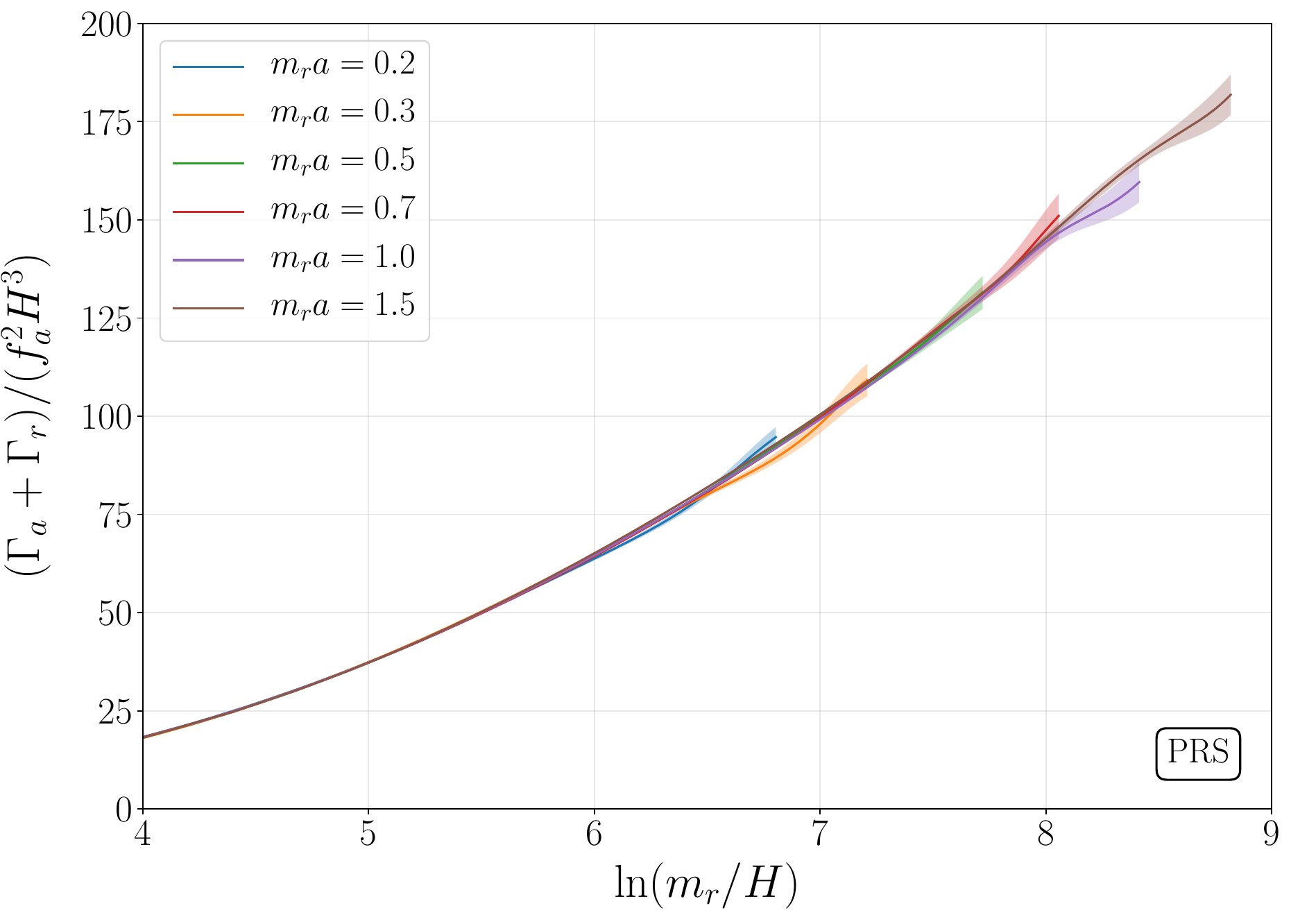}}
\\
\subfigure{
\includegraphics[width=0.48\textwidth]{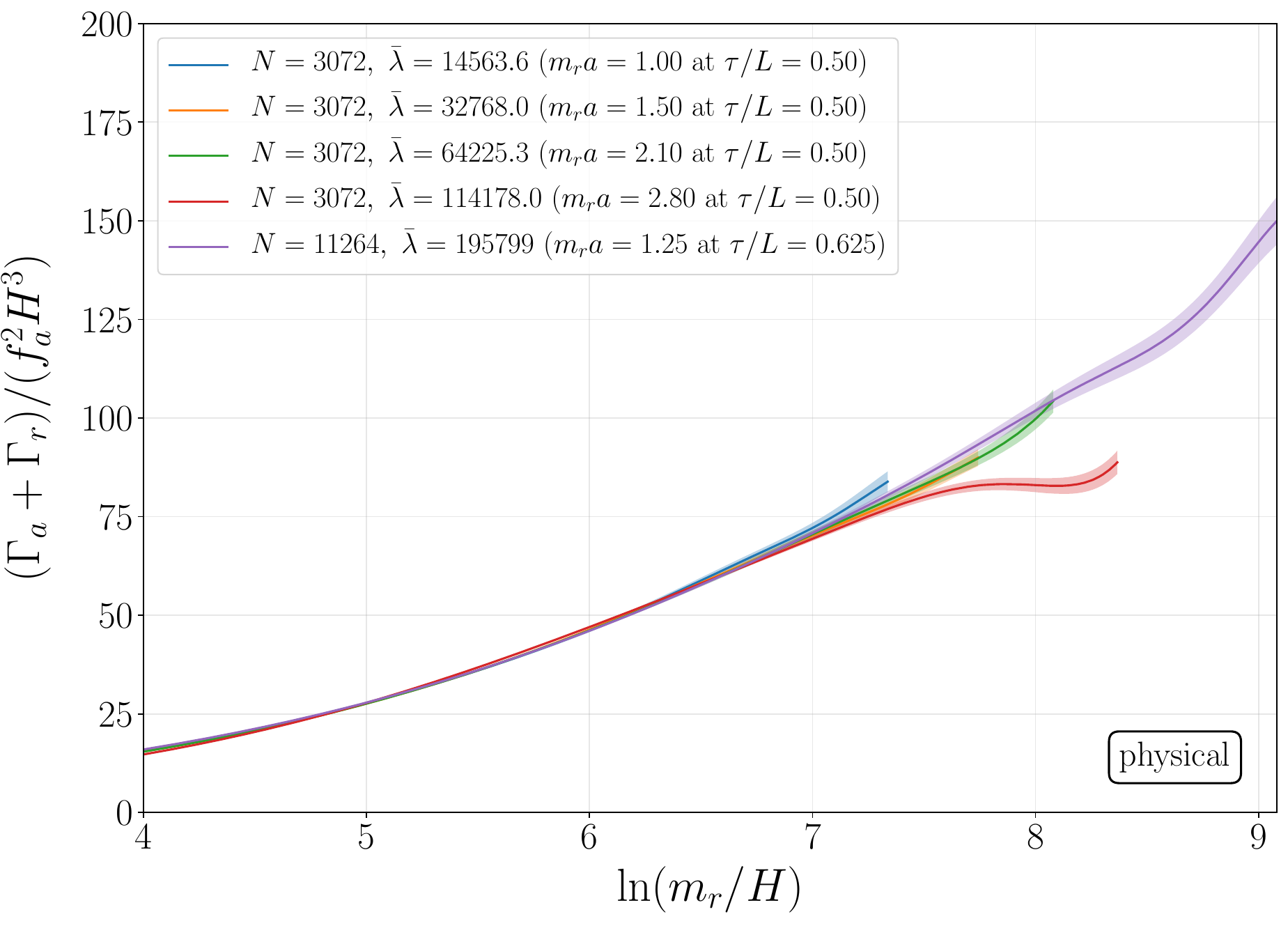}}
\end{array}$
\caption{Evolution of the sum of the energy density emission rate of axions and that of saxions for different choices of the coupling parameter of the PQ field that controls the string core width.
Top panel shows the results of simulations of PRS strings with $8192^3$ lattice sites.
Bottom panel shows the results of one set of $11264^3$ simulations and four sets of $3072^3$ simulations of physical strings.
The coloured bands represent statistical uncertainties.}
\label{fig:disc_Gamma_tot}
\end{figure}

There is a huge unphysical effect that distorts the instantaneous emission spectrum at large $\ell$ and large $m_ra$.
We have seen that it also changes the net emission rates for axions ($\Gamma_a$) and saxions ($\Gamma_r$), see Fig.~\ref{fig:disc_Gamma_physical}.
However, we find that the total emission rate $\Gamma_a+\Gamma_r$ does not depend much on $m_ra$.
This is true for both the PRS case and physical case, as shown in Fig.~\ref{fig:disc_Gamma_tot}.\footnote{The exceptional case is 
$N=3072$ and $\bar{\lambda}=114178$, where $\Gamma_a+\Gamma_r$ deviates from other lines at $\ell \gtrsim 7.5$.
In that case the discretisation effect turns out to be so large that the convergence in $\Gamma_a+\Gamma_r$ is not guaranteed.
However, it shows a good agreement with others for $\ell \lesssim 7$ (corresponding to $m_ra \lesssim 1.5$), and we use the data in that region
to estimate $(\Gamma_r)_{m_r a\to 0}$ appearing in Eq~\eqref{bar_Gamma_a}.}
Comparing Figs.~\ref{fig:disc_Gamma_PRS} and~\ref{fig:disc_Gamma_physical} to Fig.~\ref{fig:disc_Gamma_tot}, we see that
the diminishing trend of $\Gamma_a$ and increasing trend of $\Gamma_r$ compensate each other,
such that the total emission rate becomes almost insensitive to the discretisation effect.
The fact that the total emission rate is almost independent of $m_ra$ is consistent with 
the behaviour of the string density parameter $\xi$.
We have seen that $\xi$ is also less sensitive to $m_ra$ (see Fig.~\ref{fig:attr_xi_PRS_vs_physical}),
which is naturally expected as the total emission rate is tied to the string density through Eq.~\eqref{Gamma_th}.

Assuming that the discretisation effect is less harmful for the total emission rate $\Gamma_a+\Gamma_r$,
we may indirectly estimate the axion emission rate in the continuum limit as
\begin{align}
\bar{\Gamma}_a = (\Gamma_a+\Gamma_r)_{\rm data} - (\Gamma_r)_{m_r a\to 0},
\label{bar_Gamma_a}
\end{align}
where $(\Gamma_a+\Gamma_r)_{\rm data}$ represents the data obtained from simulations (shown in Fig.~\ref{fig:disc_Gamma_tot}),
and $(\Gamma_r)_{m_r a\to 0}$ is the continuum extrapolation of $\Gamma_r$.
We expect that this indirect method to estimate $\Gamma_a$ is better under control than taking the continuum extrapolation of $\Gamma_a$ directly,
since the error in the continuum extrapolation of $\Gamma_r$ in Eq.~\eqref{bar_Gamma_a} should be negligible at large $\ell$,
where the value of $(\Gamma_r)_{m_r a\to 0}$ itself becomes much smaller than $(\Gamma_a+\Gamma_r)_{\rm data}$, as shown below.

We perform the continuum extrapolation of $\Gamma_r$ by using the data from four sets of $3072^3$ simulations with different values of $\bar{\lambda}$
and one set of $11264^3$ simulations, shown in the bottom panel of Fig.~\ref{fig:disc_Gamma_physical}.
At each value of $\ell$ within the interval $4\le\ell\le7$, we fit the five data points to a quadratic function of $m_ra$ and perform the extrapolation to $m_ra=0$.
We observe that the result of the extrapolation still exhibits a trend of slightly increasing $\Gamma_r/(f_a^2H^3)$ with $\ell$, and we model it by a simple function
in the form $\Gamma_r/(f_a^2H^3) = c_0 + c_1\ell$ with $\mathcal{O}(1)$ coefficients $c_0$ and $c_1$.
The estimate of $\Gamma_r$ obtained based on this procedure is shown in the bottom panel of Fig.~\ref{fig:disc_Gamma_physical} as a dashed line.
Within the range of $4 \le \ell \le 9$, the value of $\Gamma_r/(f_a^2H^3)$ slightly increases from $\Gamma_r/(f_a^2H^3) \approx 1$ (at $\ell = 4$) to $\Gamma_r/(f_a^2H^3) \approx 3.6$ (at $\ell = 9$).
In the following, we use this estimate based on the linear function up to $\ell \sim 9$, where the $m_ra \to 0$ extrapolation cannot be performed adequately.

\begin{figure}[htbp]
\includegraphics[width=0.48\textwidth]{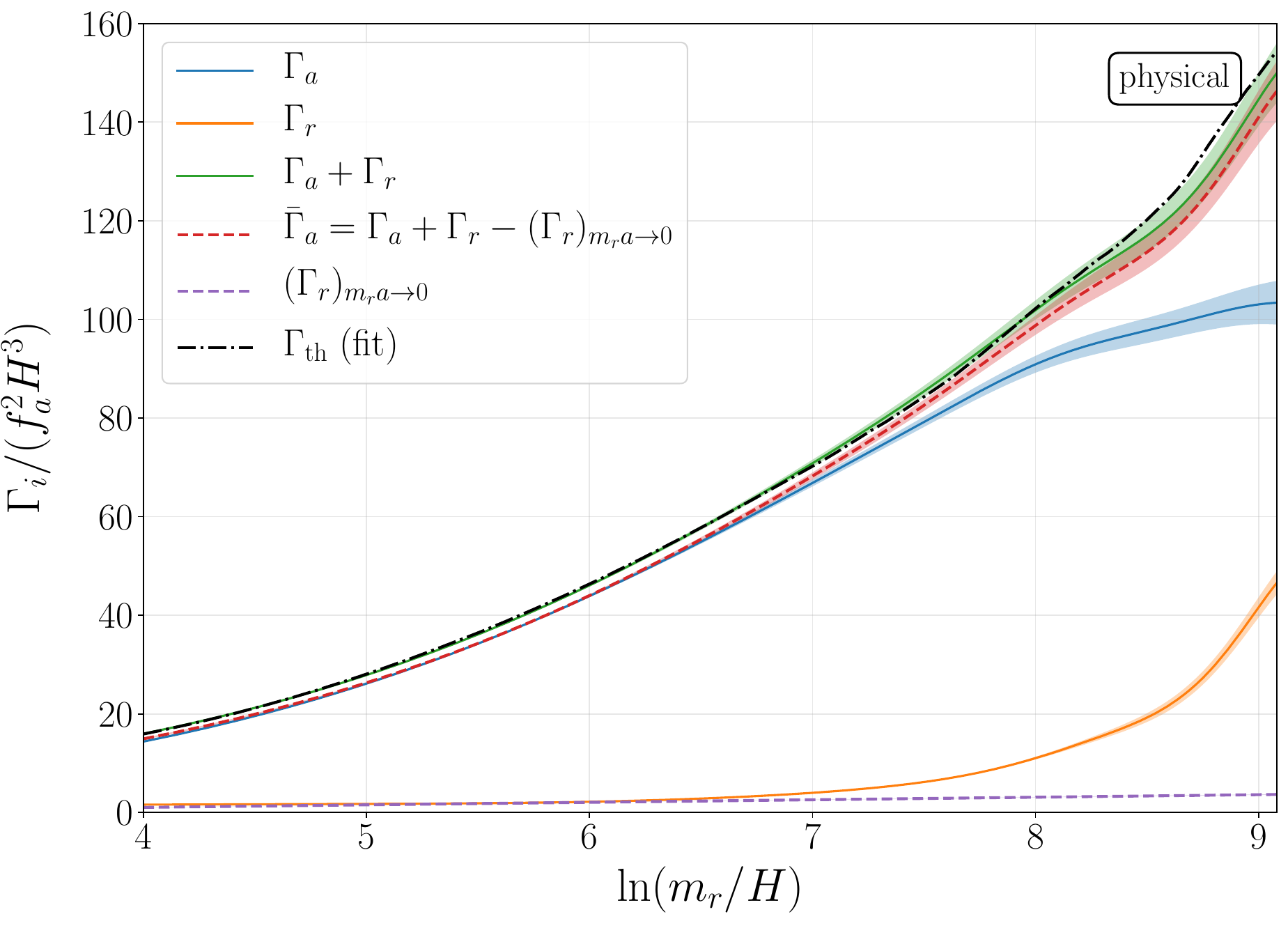}
\caption{Evolution of the energy density emission rate of axions (blue line), saxions (orange line), and the sum of them (green line) obtained from simulations of physical strings with $11264^3$ lattice sites.
The coloured bands represent statistical uncertainties.
The axion energy density emission rate in the continuum limit defined by Eq.~\eqref{bar_Gamma_a} and the continuum extrapolation of $\Gamma_r/(f_a^2H^3)$ are also shown as red dashed line and purple dashed line, respectively.
The dash-dotted line represents the theoretical expectation given by Eq.~\eqref{Gamma_th}
with parameters [Eqs.~\eqref{fL_fit_result} and~\eqref{eta_fit_result}] determined by the fit.}
\label{fig:disc_Gamma_physical_comparison}
\end{figure}

Figure~\ref{fig:disc_Gamma_physical_comparison} shows the evolution of axion, saxion, and total emission rates together with 
the continuum extrapolation of $\Gamma_r$ and indirect estimate of $\Gamma_a$ [Eq.~\eqref{bar_Gamma_a}],
obtained for the simulations with $11264^3$ lattice sites.
The result of the extrapolation of $\Gamma_r$ implies that $\bar{\Gamma}_a$ continues to grow, and the change in $\Gamma_r$ is small compared to the growth in $\bar{\Gamma}_a$.
Therefore we expect that the energy stored in the string network is dominantly transferred into axions at large $\ell$.
By looking at the difference between the plots of $\Gamma_a$ and $\bar{\Gamma}_a$, we can also see that
about 40\% of the axion radiation power is lost into unphysical high-frequency saxions at around the end of simulations due to the discretisation effects.

\subsubsection{Spectral index}
\label{sec:continuum_ext_q}

We have seen that the discretisation effects due to finite values of $1/(m_ra)$ distort the spectrum and can lead to the overestimation of $q$.
Here we try to model this effect in terms of the following function,
\begin{align}
q = q_{\rm model}(\ell) + q_{\rm disc}(\ell,m_ra), 
\label{q_model_plus_disc}
\end{align}
where $q_{\rm model}(\ell)$ represents the genuine behaviour given by some function of $\ell$,
and $q_{\rm disc}(\ell,m_ra)$ represents the bias due to the discretisation effects that depends not only on $\ell$ but also on $m_ra$.
We expect that $q_{\rm disc}(\ell,m_ra)$ monotonically increases with $\ell$ and $m_ra$, and vanishes in the limit $m_ra \to 0$.
The argument of collapsing Nambu-Goto string loops that led to Eq.~\eqref{mra_boosted} inspires us to think of the form $q_{\rm disc} \sim \alpha e^{\beta\ell}$,
where $\alpha$ and $\beta$ are coefficients that do not depend on $\ell$. We found that such a $\ell$-dependence nicely reproduces the behaviour of $q$
for the PRS strings for each $m_ra$ (shown in the top panel of Fig.~\ref{fig:disc_mra_q}), 
and also observed a trend that the coefficient $\alpha$ increases while $\beta$ decreases for larger values of $m_ra$.
Taking into account such a trend, we introduce the following function to model the effects for physical strings,
\begin{align}
q_{\rm disc}(\ell,m_ra) = d_0(m_ra)^{d_1}\exp\left[\frac{d_2\ell}{1+(m_ra)^{d_3}}\right], \label{q_model_disc}
\end{align}
where $d_0$, $d_1$, $d_2$, and $d_3$ are constant parameters.

With the ansatz~\eqref{q_model_disc} for the discretisation effects, we performed the global fits of the data of $q$ measured in the simulations of physical strings
to the function given by Eq.~\eqref{q_model_plus_disc}.
We assume that the model part $q_{\rm model}$ should be given by a simple function of $\ell$, and consider the following possibilities:
\begin{align}
\renewcommand{\arraystretch}{1.5}
\begin{array}{lll}
\text{Model\ A}: & q= q_0+q_1\ell  & (q_1>0), \\
\text{Model\ B}: & q= q_0+q_1\ell^2 & (q_1>0), \\
\text{Model\ C}: & q= q_0+q_1/\ell & (q_1<0), \\
\text{Model\ D}: & q= q_0+q_1/\ell^2 & (q_1<0). 
\label{q_models}
\end{array}
\renewcommand{\arraystretch}{1}
\end{align}
In the fits, we use three sets of $3072^3$ simulations with different values of $\bar{\lambda}$ and one set of $11264^3$ simulations, 
shown in the bottom panel of Fig.~\ref{fig:disc_mra_q}.\footnote{In this analysis, we do not include the data from $3072^3$ simulations with $\bar{\lambda} = 114178$ ($m_ra = 2.8$ at $\tau/L$ = 0.5),
since in that case the spectrum is too distorted at late times to be accurately modeled by a simple function of $q(\ell,m_ra)$.}
For each data set, we sampled the value of $q$ with an interval of $\Delta\ell=0.25$ starting from $\ell = 5.5$, and the contribution of each data is weighted by the error interval
including the statistical errors and systematics due to the filtering procedure to calculate $\mathcal{F}(x,y)$ (see Appendix~\ref{app:method_F}).
In order to make sure that the fit parameters end up within a reasonable interval, we performed the fits in three steps:
First we fitted the model part $q_{\rm model}$ only by using data with $m_ra < 0.7$, and
then fitted the full data to the function given by Eq.~\eqref{q_model_plus_disc} with the parameters in $q_{\rm model}$ fixed to values obtained in the first step.
After that, we fitted full data again to the function given by Eq.~\eqref{q_model_plus_disc} by allowing all the parameters in $q_{\rm model}$ and $q_{\rm disc}$ to change.
In this third step, we specified the boundaries for the parameters $d_0,d_1,d_2,d_3$ to ranges corresponding to one sigma
intervals obtained in the second step.

\begin{table}
\centering
\caption{
Best fit parameters and $\chi^2/\mathrm{d.o.f.}$ for the models characterising the $\ell$-dependence of $q$.
}
\label{tab:fits_q_results}
\begin{tabular}{l c c c}
\hline \hline
Model & $\chi^2/{\rm d.o.f.}$ & $q_0$ & $q_1$ \\
\hline 
(A)~$q_0+q_1\ell$ & 0.89 & 0.19(3) & 0.093(5) \\
(B)~$q_0+q_1\ell^2$ & 1.02 & 0.50(2) & 0.0069(4) \\
(C)~$q_0+q_1/\ell$ & 0.82 & 1.39(4) & $-$3.9(3) \\
(D)~$q_0+q_1/\ell^2$ & 0.81 & 1.09(2) & $-$12.4(8) \\
\hline
\end{tabular}
\end{table}

\begin{figure}[htbp]
\includegraphics[width=0.48\textwidth]{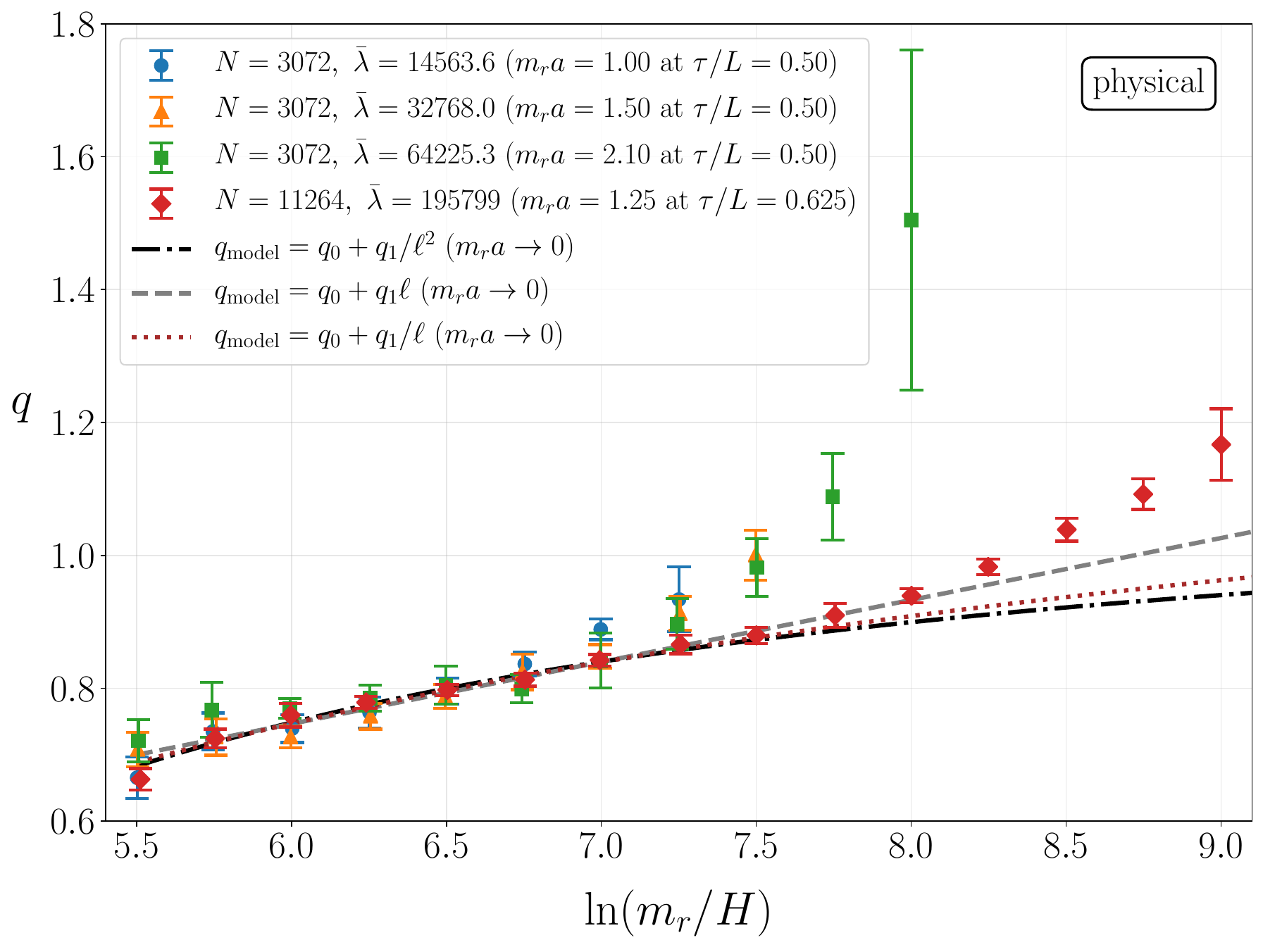}
\caption{Comparison of the data on the evolution of the spectral index $q$ of the instantaneous emission spectrum obtained from simulations of physical strings (markers with error bars)
and their continuum extrapolations given by Model A (gray dashed line), Model C (brown dotted line), 
and Model D (black dash-dotted line) in Eq.~\eqref{q_models} with coefficients given by Table~\ref{tab:fits_q_results},
which show good fits to the data when the additional term [Eq.~\eqref{q_model_disc}] characterising the discretisation effect is included.
The error bars include systematics induced by changing the parameter $\sigma_{\rm filter}$ for the filtering procedure to calculate $\mathcal{F}$ [Eq.~\eqref{filter_for_F}] in addition to statistical uncertainties.
}
\label{fig:disc_q_physical_continuum}
\end{figure}

The result of the fit for each model, including the value of $\chi^2/\mathrm{d.o.f.}$ and those for the best fit parameters, is summarised in table~\ref{tab:fits_q_results}.
Values of the parameters $d_0$, $d_1$, $d_2$, and $d_3$ in $q_{\rm disc}$ also differ according to the model, 
which are not shown in the table as their precise values become irrelevant after taking the continuum limit ($m_ra \to 0$). 
We find that the simple two-parameter models enumerated in Eq.~\eqref{q_models} yield almost equally good fits.\footnote{We also considered a set of three-parameter models
such as $q_0+q_1\ell+q_2\ell^2$, $q_0+q_1/\ell+q_2/\ell^2$, and $q_0+q_1\ell+q_2/\ell$,
and found that they can also fit the data very well.
However, in those cases the values of the coefficients $q_0$,$q_1$,$q_2$ become much more uncertain.
This result signals that any of these coefficients could be potentially important,
and explains why different two-parameter models (model A, C and D) are almost equally preferred.}
Model C and D are slightly more preferred than A and B, but the latter (especially model A) also shows a good fit.
The evolution of $q$ predicted by 
three preferred models
and the data set used for the fits are shown in Fig.~\ref{fig:disc_q_physical_continuum}.
The data is consistent with $q_{\rm model}$ at $\ell\lesssim 7$, 
which implies that the discretisation error is not important at that range.
On the other hand, the discrepancy between different models as well as differences between the models and data
stand out at large $\ell$, which indicates that the discretisation error is quite serious in that range.

The fact that 
different models are almost equally preferred
by the data implies that the value of $q$ at large $\ell$ remains quite uncertain.
We can basically consider two possibilities: one is that $q$ grows linearly with $\ell$ (model A, which was also claimed in Ref.~\cite{Gorghetto:2020qws}), 
and the other is that $q$ continues to grow but asymptotes to a constant value 
with the $\ell$ dependence suppressed as $1/\ell$ (model C) or $1/\ell^2$ (model D).
Although differences between these 
models are marginal at small $\ell$ simulated so far,
it leads to a big impact on the calculation of the axion dark matter abundance when extrapolated to large $\ell$.
Implications for the estimation of the axion relic abundance and dark matter mass will be discussed in Sec.~\ref{sec:axion_abundance}.

\section{Results: Extrapolation}
\label{sec:extrapolation}

In this section, we estimate the axion relic abundance obtained by extrapolating numerical results to large $\ell$.
In order to perform such an extrapolation, we need several quantities that characterise the axion emission spectrum.
Among them, we have investigated the systematics on the string density parameter $\xi$ in Sec.~\ref{sec:attractor},
and the spectral index $q$ of the instantaneous axion
emission spectrum in Secs.~\ref{sec:oscillations} and~\ref{sec:disc_effect}.
In the following two subsections, we present further results on the total energy density emission rate (Sec.~\ref{sec:Gamma_th}) and IR peak of the instantaneous emission spectrum (Sec.~\ref{sec:IRpeak}).
All these results are combined to calculate the axion dark matter abundance and its uncertainty in Sec.~\ref{sec:axion_abundance}.

\subsection{Energy density emission rate}
\label{sec:Gamma_th}

As shown in Fig.~\ref{fig:disc_Gamma_tot}, the total energy density emission rate $(\Gamma_a+\Gamma_r)/(f_a^2H^3)$ increases with $\ell$.
This trend is expected to be captured by the analytical expression shown in Eq.~\eqref{Gamma_th}, which can be used for the extrapolation.
We utilise the orthogonal distance regression (ODR) algorithm~\cite{1148166} to fit the data and determine the values of the parameters $f_L$ and $\eta$ appearing in Eq.~\eqref{Gamma_th}.
In this procedure, we treat Eq.~\eqref{Gamma_th} as a model function $f(\bm{x};\bm{\beta})$ that depends on three explanatory variables $\bm{x} = (\xi,d\xi/d\ell,\ell)$ and two parameters $\bm{\beta} = (f_L,\eta)$,
and we use the data of $(\Gamma_a+\Gamma_r)/(f_a^2H^3)$, $\xi$, and $d\xi/d\ell$ obtained from the simulations with 11264$^3$ lattice sites.\footnote{Here we treat $\eta$ as a constant, since its time-dependence is expected 
to be subdominant, suppressed by $1/\ell$ [see Eq.~\eqref{Gamma_th}].}
The data is sampled with an interval of $\Delta\ell = 0.25$, while the value of $d\xi/d\ell$ is evaluated by the simple finite difference between the nearest neighbour (much closer than $\Delta\ell = 0.25$) at each sample point.
The data of the response variable, $(\Gamma_a+\Gamma_r)/(f_a^2H^3)$, are summed with weights given by the reciprocal of their statistical uncertainties.
For the data of $\xi$ and $d\xi/d\ell$, we evaluate their covariance matrix at each sample point, 
and use it as a weight matrix in ODR in order to take account of the possible correlation between them.
On the other hand, the data of $\ell$ for each sample point is just introduced as an auxiliary variable, and we set its weight values to 1 with no correlation with $\xi$ and $d\xi/d\ell$.

As a result of the fitting procedure described above, the parameters are determined as
\begin{align}
f_L &= 0.83(1), \label{fL_fit_result} \\
\eta &= 0.27(2). \label{eta_fit_result}
\end{align}
With these values of parameters, the behaviour of $(\Gamma_a+\Gamma_r)/(f_a^2H^3)$ is well reproduced by Eq.~\eqref{Gamma_th}, as shown in Fig.~\ref{fig:disc_Gamma_physical_comparison}.
It is notable that the value of $f_L$ agrees with our expectation of $\sim 80\,\%$ reduction due to the abundance of long strings discussed in Sec.~\ref{sec:basics}.
Furthermore, the value of $\eta$ matches the theoretical estimate $\eta = 1/\sqrt{4\pi} \simeq 0.28$ remarkably well.
The above results for the system of physical strings are also compatible with $f_L \approx 0.84$ and $\eta \approx 0.27$ found in Ref.~\cite{Gorghetto:2021fsn} for PRS strings.

\subsection{IR peak}
\label{sec:IRpeak}

The IR cutoff $x_0$ of the instantaneous emission spectrum has some impact on the estimation of the axion number radiated from strings (see Fig.~\ref{fig:naqmodel}).
A non-trivial question is whether the value of $x_0$ depends on $\ell$.
One might expect that the location of the IR cutoff should be proportional to $\sqrt{\xi}$~\cite{Hagmann:2000ja,Kawasaki:2018bzv,Gorghetto:2020qws,Dine:2020pds,Buschmann:2021sdq}, 
since the increase in the string density leads to the shortening of the typical distance of neighbouring strings.
Given the logarithmic growth of $\xi$, this expectation implies that the IR cutoff would shift to higher momenta at large $\ell$.
The actual spectrum does not have a sharp cutoff but has a peak-like feature in the IR, and we can have an idea on how $x_0$ evolves by looking at the behaviour of the IR peak of $\mathcal{F}(x)$.

The existence of the IR peak in the instantaneous emission spectrum $\mathcal{F}(x)$ implies that, 
when we keep track of the time evolution of $\mathcal{F}$ for a fixed value of the comoving momentum $k$, 
there is a point at which $\mathcal{F}$ becomes maximum: 
Before that time (outside the horizon, $k \lesssim RH$) it monotonically increases, and after that time (inside the horizon, $k\gtrsim RH$) it decreases and oscillates.
Such a time of turnaround $\tau_{\rm t.a.}$ can be different according to the comoving wavenumber, but we expect that the product $x_{\rm t.a.} = k\tau_{\rm t.a.}$
should take a similar value, as it should happen around the time of the horizon crossing, where $k^{-1}$ becomes comparable to $\tau = 1/(RH)$.
Nevertheless, the exact value of $x_{\rm t.a.}$ could be slightly different according to the value of the string density parameter $\xi$ or $\ln(m_r/H)$
at the time of turnaround, from which we can infer the relation between $\xi$ and the location of the IR peak.
The identification of the peak location in terms of the turnaround of the mode evolution is advantageous in the numerical study,
since it is hard to identify the IR peak directly from the spectrum ($k$-dependence) of $\mathcal{F}(x)$ due to the fact that the number of modes in the IR are very few, 
which prevents us from extracting the feature of the IR peak precisely.

\begin{figure}[htbp]
$\begin{array}{c}
\subfigure{
\includegraphics[width=0.48\textwidth]{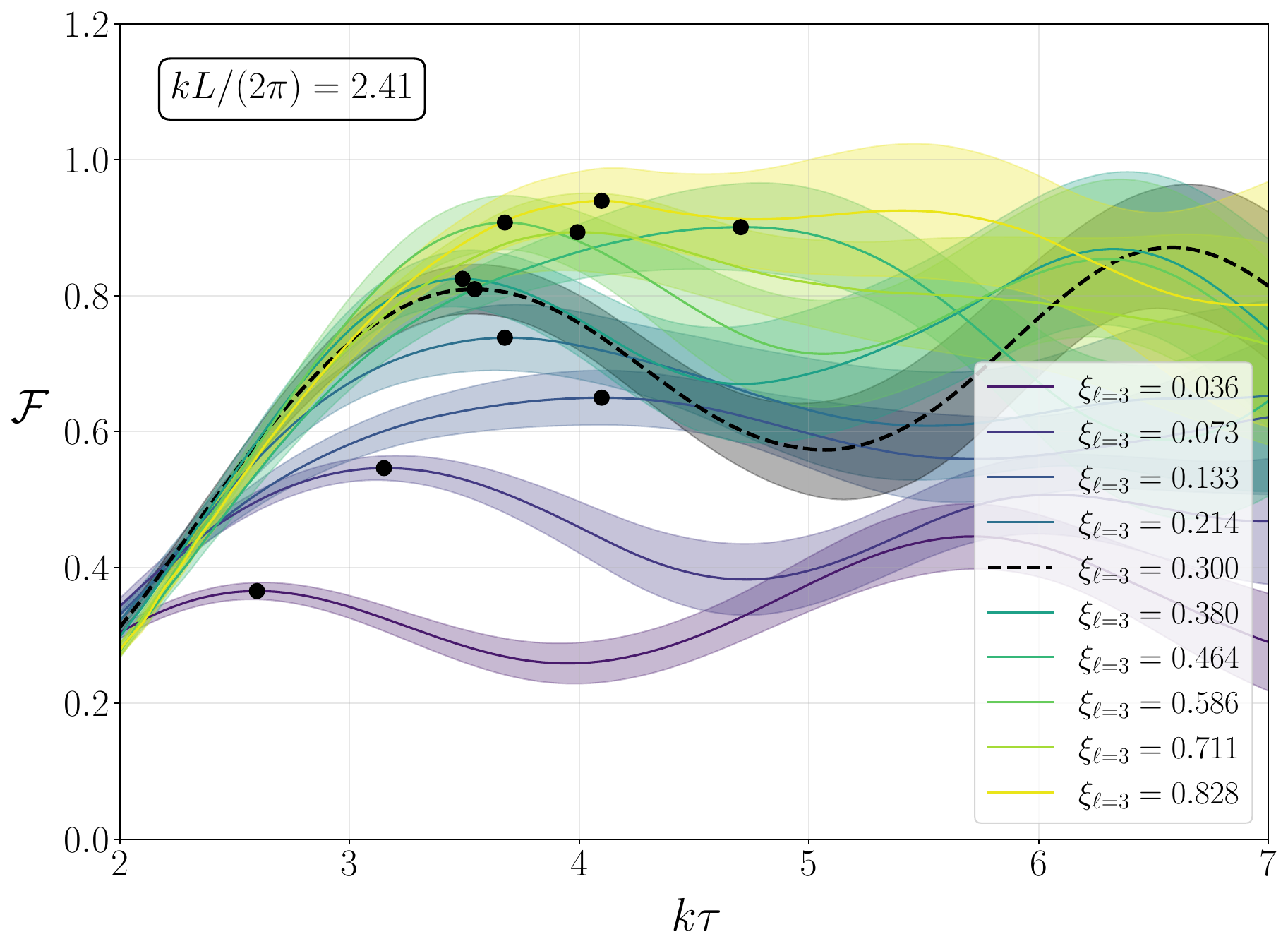}}
\\
\subfigure{
\includegraphics[width=0.48\textwidth]{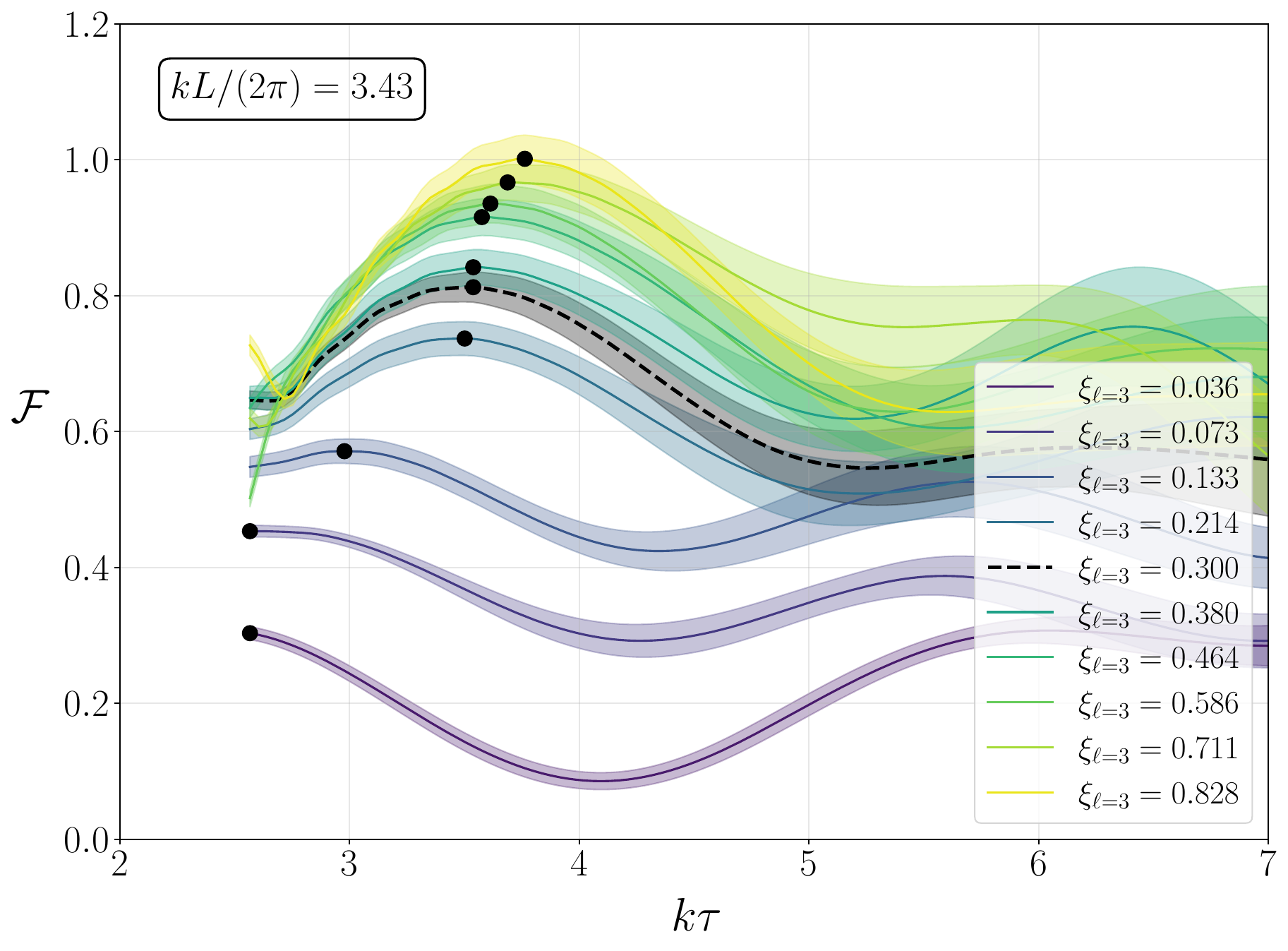}}
\end{array}$
\caption{Time evolution of the spectral energy density production rate $\mathcal{F}$ of one Fourier mode specified by $kL/(2\pi) = 2.41$ (top panel) and $kL/(2\pi) = 3.43$ (bottom panel) 
for various different values of the initial strings density $\xi_{\ell=3}$.
The coloured bands represent statistical uncertainties, and the black dashed line corresponds to the attractor with $\xi_{\ell=3} = 0.3$. 
The locations of first turnaround are marked by black dots.
The results are obtained from simulations of physical strings with $2048^3$ lattice sites.}
\label{fig:ext_F_evolution}
\end{figure}

Figure~\ref{fig:ext_F_evolution} shows the time evolution\footnote{Here we change two independent variables of $\mathcal{F}$ from
$(x,y)$ to $(k,\tau)$, and study the dependence on $\tau$ for fixed $k$. In this context, the change in $x= k\tau$ should be regarded as the time evolution rather than the momentum dependence.}
of $\mathcal{F}$ for two fixed values of comoving momenta, $kL/(2\pi)=2.41$ (top panel) and $kL/(2\pi)=3.43$ (bottom panel).\footnote{Note that for the computation of 
$\mathcal{F}$ used in this subsection, we do not use the filtering method described in Appendix~\ref{app:method_F}, which was used in Sec.~\ref{sec:oscillations} to obtain the smooth spectrum of $\mathcal{F}$.
Such a method is beneficial for removing the $2k$-oscillation that acts as a contamination in characterising the emission spectrum at the intermediate momenta,
but here we are interested in the turnaround behaviour of a mode with a very low momentum, which is a part of the $2k$-oscillation.
Instead of applying such a filtering procedure, here we compute the time derivative in $\mathcal{F}$ by taking the finite difference and applying another Gaussian filter to
remove fluctuations that show up in a period much shorter than the oscillation frequency $\sim 2k$.}
In this figure we compare the results from simulations with different values for the initial string density (the same data sets as Sec.~\ref{sec:attractor}).
We see that the turnaround happens at $k\tau \sim 3.5$ for the attractor (simulations with the initial string density $\xi_{\ell=3}=0.3$),
while there is a trend that it happens earlier (later) for under-dense (over-dense) cases.
This trend agrees with our intuition that the location of the IR peak shifts towards larger values of $x=k\tau$ for dense string networks. 

Note that in this analysis we can only use the data with two comoving momenta, 
the second-lowest mode [$kL/(2\pi)=2.41$] and the third-lowest mode [$kL/(2\pi)=3.43$].
This limitation comes from the fact that we have to monitor the evolution of the modes that cross the horizon within the simulated time scale in order to observe the turnaround.
For instance, the lowest mode [$kL/(2\pi)\sim 1$] is not suitable, since it does not enter well inside the horizon before the end of the simulations.
Furthermore, the modes with higher momenta $kL/(2\pi)\gtrsim 4$ are also inappropriate for the identification of the turnaround, since 
they are already inside the horizon at the beginning of the simulations in most cases.
This constraint also affects the identification of the turnaround for the third-lowest mode [$kL/(2\pi)=3.43$] shown in the bottom panel of Fig.~\ref{fig:ext_F_evolution}.
In that figure, we see that in some of the simulations with under-dense strings the mode is already inside the horizon at the beginning of the simulations,
and hence it is not possible to identify the time of the turnaround accurately for such data sets.

It should also be noted that the turnaround behaviour can be observed even in the absence of strings ($\xi \to 0$).
In that case the mode evolution can be described as a simple misalignment oscillation of the free axion field, 
and the location of the turnaround corresponds to the time at which $\mathcal{F}$ reaches the first oscillation maximum.
It is possible to calculate the evolution due to such misalignment oscillations analytically, 
and we find $x\simeq 1.87$ at turnaround (see Appendix~\ref{app:evolution_analytical}).

\begin{figure}[htbp]
$\begin{array}{c}
\subfigure{
\includegraphics[width=0.48\textwidth]{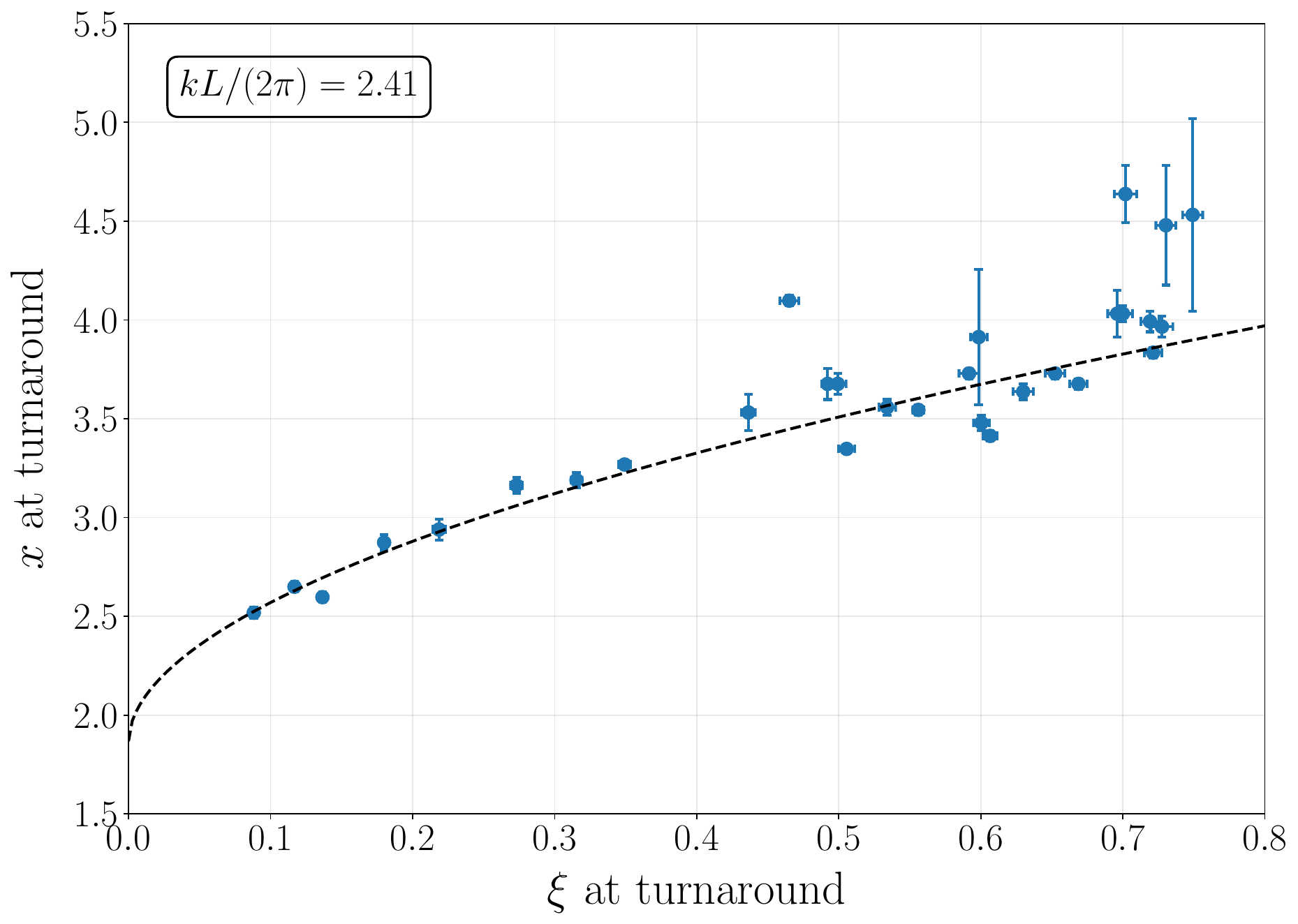}}
\\
\subfigure{
\includegraphics[width=0.48\textwidth]{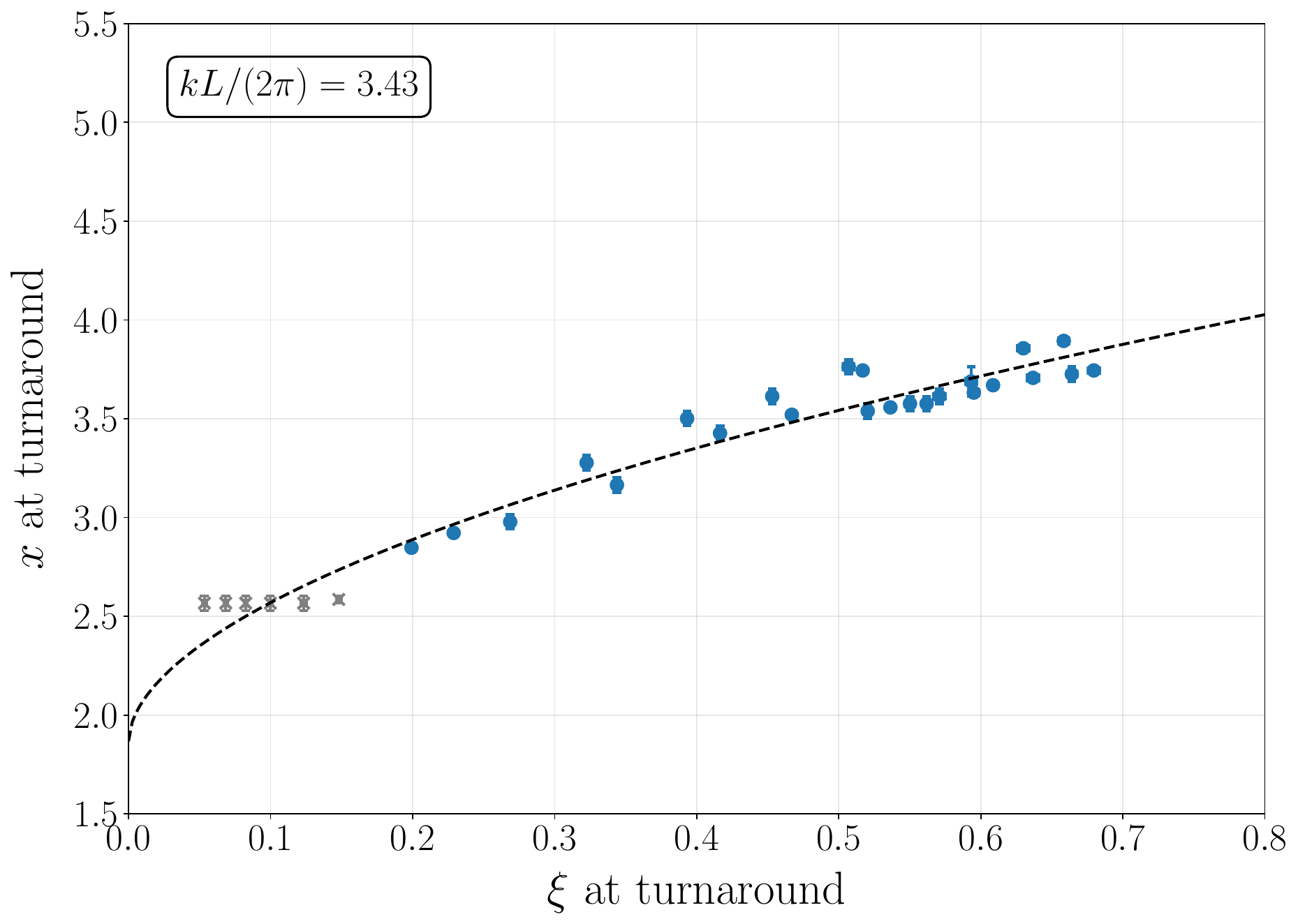}}
\end{array}$
\caption{The relation between the string density parameter $\xi$ and the value of $x=k\tau$ at turnaround,
obtained from the data of the evolution of $\mathcal{F}$ of one Fourier mode specified by $kL/(2\pi) = 2.41$ (top panel) and $kL/(2\pi) = 3.43$ (bottom panel) 
for various different values of the initial strings density.
The results are obtained from simulations of physical strings with $2048^3$ lattice sites.
Dashed lines represent the fits to a function given by Eq.~\eqref{x_ta_model}.
The data points marked by gray crosses are not used for the fit, since in those cases the first maximum of $\mathcal{F}$ appears at around the initial time of the simulation.}
\label{fig:ext_x0_vs_xi}
\end{figure}

The relation between the string density $\xi$ and the value of $x$ at turnaround for 30 sets of simulations with different initial string densities is summarised in Fig.~\ref{fig:ext_x0_vs_xi}.
Note that the values of $\xi$ shown in that figure do not represent the initial string densities (evaluated at $\ell=3$) but the values of $\xi$ evaluated at the time of the turnaround.
The figure clearly shows the trend that the value of $x$ at turnaround increases with $\xi$.
The error bars represent the statistical uncertainties, which become larger for large $\xi$,
since in that case the mode starts to oscillate at late times of the simulation, where there are not many Hubble patches in the simulation box and the value of $\mathcal{F}$ becomes more uncertain.

Now let us model the relation between $\xi$ and $x$ at turnaround  in terms of the following function,
\begin{align}
x_{\rm t.a.} = 1.87 + b_1\xi^{b_0}, \label{x_ta_model}
\end{align}
where the value of the constant term is chosen such that $x_{\rm t.a.}$ approaches to the value inferred from
the misalignment oscillation of the free axion field in the limit $\xi\to 0$.
We fit the data of $x$ and $\xi$ at turnaround to the above model by utilising the ODR algorithm,
where statistical errors of both $x$ and $\xi$ are taken into account as weights for the evaluation of the squared distance.
When we use the data of $kL/(2\pi)=3.43$, we omit six data points from the lowest string density (gray cross markers in the bottom panel of Fig.~\ref{fig:ext_x0_vs_xi}),
since in those cases the first maximum in $\mathcal{F}$ appears at around the initial time of the simulation, which could bias the location of the turnaround.
With this procedure, the best-fit parameters are found to be
\begin{equation}
    \begin{aligned}
&b_0 =0.53(6),\ b_1 =2.3(1)\  &&\text{for}\ \ kL/(2\pi)=2.41,\\
&b_0 =0.54(4),\ b_1 =2.44(7)\  &&\text{for}\ \ kL/(2\pi)=3.43. \label{x_ta_parameters}
\end{aligned}
\end{equation}
The function~\eqref{x_ta_model} with the above best-fit parameters is plotted in Fig.~\ref{fig:ext_x0_vs_xi} as dashed lines.
Remarkably, the value of the exponent $b_0$ is consistent with the naive expectation $x_{\rm t.a.}\propto \sqrt{\xi}$.
Combined with the expectation that $\xi$ increases linearly with $\ell$, the result implies that the location of the IR cutoff 
can be as large as $x_0 \sim x_{\rm t.a.} \approx 10\text{--}15$ at $\ln(m_r/H) = 70$.
We will use this extrapolation for the estimation of the axion relic abundance in the next subsection.

\subsection{Axion relic abundance}
\label{sec:axion_abundance}

For the calculation of the axion abundance, the evolution of the axion field has to be treated with caution if the emission spectrum is completely dominated by IR ($q\gg1$).
In Ref.~\cite{Gorghetto:2020qws}, it was argued that for such an IR dominated spectrum the axion field becomes highly non-linear, which could lead to a transient regime of axion number non-conserving processes
around the epoch of the QCD phase transition and result in an alleviation of the enhancement of the axion abundance due to the large value of $\xi\ln(m_r/H)$.

To check the relevance of the possible non-linear effect, it is instructive to calculate $\sqrt{\langle \theta^2 \rangle}$, the average square amplitude of the axion field given by Eq.~\eqref{amp_analytical_2}.
Here we compute it as
\begin{align}
\langle\theta^2\rangle= 
\int^{\ell}_{\ell_{\rm sim}}\frac{d\ell'}{2}\frac{e^{\ell'}}{e^{\ell}}\frac{\Gamma_a'}{f_a^2H'^3}\langle x^{-2}\rangle(y') + \frac{e^{\ell_{\rm sim}}}{e^{\ell}}
\langle\theta^2\rangle_{\ell_{\rm sim}},
\label{amp_calculation}
\end{align}
where $\langle\theta^2\rangle_{\ell_{\rm sim}}$
is obtained by directly integrating $(1/k^2)\partial\rho_a/\partial k$ [see Eq.~\eqref{amp_analytical_1}] from the simulation result at $\ell_{\rm sim} = 7$.
With this initial value, we perform the integration from $\ell_{\rm sim}$ to $\ell$.
For $\Gamma_a$ in the integrand, we use $\Gamma_{\rm th}$ given by Eq.~\eqref{Gamma_th} with parameters given by Eqs.~\eqref{fL_fit_result} and~\eqref{eta_fit_result}. 
The approximation $\Gamma_a \approx \Gamma_{\rm th}$ is guaranteed if we assume $\Gamma_a \gg \Gamma_r$ at large $\ell$ (see Fig.~\ref{fig:disc_Gamma_physical_comparison}).
For $\langle x^{-2}\rangle(y)$ in the integrand, we use 
a simple power law approximation for $\mathcal{F}(x)$ given by Eq.~\eqref{F_simple_power_law}.
Here we adopt $x_0 = x_{\rm t.a.}$ with $x_{\rm t.a.}$ given by Eq.~\eqref{x_ta_model} and coefficients given by Eq.~\eqref{x_ta_parameters}.
Furthermore, we extrapolate the string density parameter $\xi$ by numerically solving Eq.~\eqref{eq:attractive_xi_evol} for two models of $\xi_c$ [Eqs.~\eqref{eq:xi_c_fit_lin} and~\eqref{eq:xi_c_fit_sat}],
and estimate the uncertainty associated with the modeling of $\xi_c$.

\begin{figure}[htbp]
\includegraphics[width=0.48\textwidth]{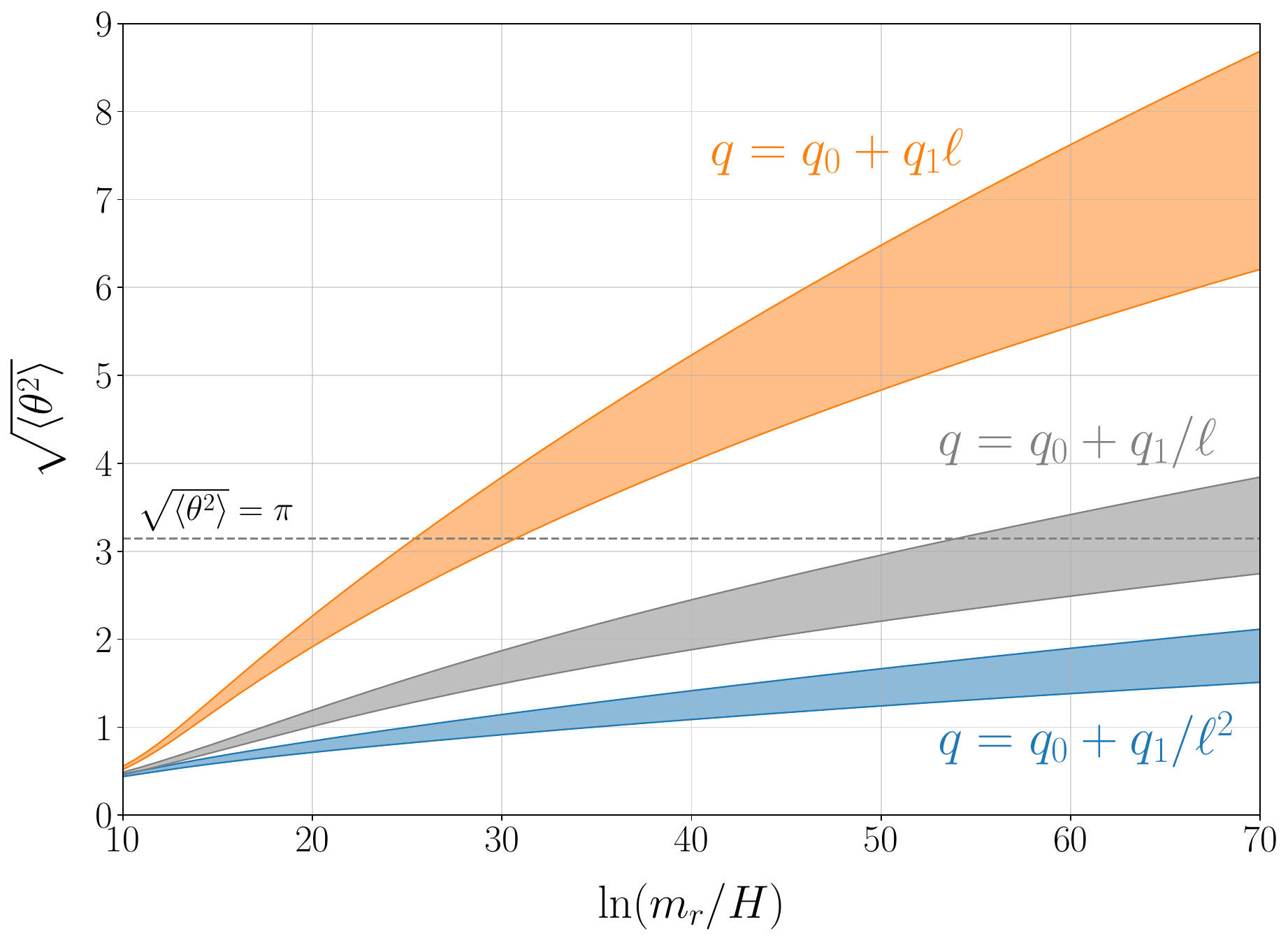}
\caption{The $\ln(m_r/H)$ dependence of the average square amplitude of the axion field computed by using Eq.~\eqref{amp_calculation}
for three different scenarios of the extrapolation of $q$. 
The shaded regions correspond to the uncertainty given by the maximum and minimum values among one sigma variations of the fit-parameters in
$q$, $\xi_c$, or $x_0$, for each model of $\xi_c$ [i.e. $\xi_c^{\rm lin}$ given by Eq.~\eqref{eq:xi_c_fit_lin} and $\xi_c^{\rm sat}$ given by Eq.~\eqref{eq:xi_c_fit_sat}].}
\label{fig:ext_amplitude}
\end{figure}

Figure~\ref{fig:ext_amplitude} shows the estimates of $\sqrt{\langle \theta^2 \rangle}$ for different assumptions on $q$.
Here we consider three models that give good fits to the simulation results (Model A, C, and D in Table~\ref{tab:fits_q_results}).
We see that the system indeed becomes highly non-linear ($\sqrt{\langle \theta^2 \rangle} \gg \pi$) at large $\ell$ for $q_{\rm model} = q_0+q_1\ell$, 
while it stays linear for $q_{\rm model} = q_0+q_1/\ell^2$.
For $q_{\rm model} = q_0+q_1/\ell$, the average square amplitude takes a value around the boundary between the linear and non-linear regime ($\sqrt{\langle \theta^2 \rangle} \sim \pi$) at $\ell=70$.

If the system remains in the linear regime, the axion number at around the QCD phase transition can be estimated by using Eq.~\eqref{na_analytical_2}.
In that case we can compute it in a similar manner as the average square amplitude,
\begin{align}
\frac{n_a}{f_a^2H} &= \int^{\ell}_{\ell_{\rm sim}}\frac{d\ell'}{2}\frac{e^{\ell'/2}}{e^{\ell/2}}\frac{\Gamma_a'}{f_a^2H'^3}\langle x^{-1}\rangle(y') \nonumber\\
&\quad + \frac{e^{\ell_{\rm sim}/2}}{e^{\ell/2}}\left.\frac{n_a}{f_a^2H}\right|_{\ell_{\rm sim}},
\label{na_calculation}
\end{align}
where $n_a/(f_a^2H)|_{\ell_{\rm sim}}$ is obtained by directly integrating $(1/k)\partial\rho_a/\partial k$ [see Eq.~\eqref{na_analytical_1}] from the simulation result at $\ell_{\rm sim} = 7$.
For $\langle x^{-1}\rangle(y)$ in the integrand, we use a simple power law approximation for $\mathcal{F}(x)$ given by Eq.~\eqref{F_simple_power_law}.
Other setups are the same as the calculation of the average square amplitude described below Eq.~\eqref{amp_calculation}.
After obtaining the axion number in this way, we can straightforwardly obtain the relic axion abundance at the present time by applying the appropriate red-shift factor.

The above estimate of the axion abundance could be modified if the system becomes non-linear. 
In that case, the axion number is given by~\cite{Gorghetto:2020qws}
\begin{align}
n_a(t_{\ell}) = c_nc_V z m_a(T_1)f_a^2\left(\frac{R(t_1)}{R(t_{\ell})}\right)^3, \label{abundance_nonlinear}
\end{align}
where
\begin{align}
m_a(T)^2 = \frac{\chi(T)}{f_a^2},
\end{align}
is the axion mass at high temperatures defined by the topological susceptibility $\chi(T)$, and in the relevant temperature range,
we adopt a power law approximation parameterised by an exponent $n_{\rm QCD}$,
\begin{align}
\chi(T) \propto T^{-n_{\rm QCD}}. \label{chi_T_dependence}
\end{align}
Recent lattice QCD simulations have found a value around $n_{\rm QCD} \simeq 8$~\cite{Borsanyi:2016ksw}. 
The quantity $z$ in Eq.~\eqref{abundance_nonlinear} is defined as
\begin{align}
z \equiv \left(\frac{m(T_{\ell})}{H_1}\right)^{1+\frac{6}{n_{\rm QCD}}}, \label{z_definition}
\end{align}
which represents the duration of the non-linear transient.
$t_1$ and $t_{\ell}$ correspond to the time defined by $H_1 = m_a(T_1)$ and that at the end of the non-linear transient, respectively.
$c_n$ and $c_V$ are coefficients that can be determined numerically.

From the requirement that the non-linear transient is terminated when the gradient energy of the IR modes becomes comparable to the potential energy from the axion mass, 
the value of $z$ can be fixed and given in terms of infinitely nested logarithms~\cite{Gorghetto:2020qws},
\begin{widetext}
\begin{align}
z = \left[\frac{4\pi f_L\xi_1\ell_1}{c_V}\left[1-\frac{2}{n_{\rm QCD} + 4}\right]\ln\left(\frac{4\pi f_L\xi_1\ell_1}{c_V}\left[1-\frac{2}{n_{\rm QCD} + 4}\right]\left[\frac{c_m}{x_0}\right]^{2\left(1+\frac{2}{n_{\rm QCD}+2}\right)}\ln(\dots)\right)\right]^{\frac{1}{2}\left(1 + \frac{2}{n_{\rm QCD} + 4}\right)}, \label{z_formula}
\end{align}
\end{widetext}
where $\xi_1$ and $\ell_1$ are values of $\xi$ and $\ell$ evaluated at $H=H_1$.
$c_m$ is another $\mathcal{O}(1)$ coefficient that can also be determined from the numerical simulations. 
In Ref.~\cite{Gorghetto:2020qws}, it was shown that the above analytical estimates show a good agreement
with the numerical results for the following values,
\begin{align}
c_n = 1.35,\quad c_V = 0.13,\quad c_m = 2.08, \label{parameters_nonlinear}
\end{align}
which we adopt in our analysis. 

Some comments are in order.
First, in Eq.~\eqref{z_formula} we have applied the replacement $\xi_1 \to f_L\xi_1$ to Eq.~(34) of Ref.~\cite{Gorghetto:2020qws},
taking into account that the emission rate~\eqref{Gamma_th} is accompanied by the overall factor of $f_L$.
Second, the analytical expressions [Eqs.~\eqref{abundance_nonlinear},~\eqref{z_definition}, and \eqref{z_formula}] and parameters [Eq.~\eqref{parameters_nonlinear}]
are obtained based on the assumption of $q \gg 1$.\footnote{In Ref.~\cite{Gorghetto:2020qws}, $q = 5$ was used for simulations of the axion field evolution at the non-linear regime.}
In that limit, the dependence on $q$ disappears in Eq.~\eqref{z_formula}, while the result substantially depends on $q$ when it is not so large (see Fig. 26 of Ref.~\cite{Gorghetto:2020qws}).
Hence, those analytical formulae cannot be used reliably in the case where $q$ is not much larger than one (e.g. $q\lesssim 2$).
Finally, though Eqs.~\eqref{abundance_nonlinear} and \eqref{z_formula} were derived based on the assumption that $x_0$ is constant,
they can be used in a good approximation even if we include the possible (slow) evolution of $x_0 \approx x_{\rm t.a.}$ [cf. Eq.~\eqref{x_ta_model}].
Since the dependence on $x_0$ is only logarithmic, the change in this quantity just gives rise to subdominant effects, which are negligible for $\ln(m_r/H_1) \gg 1$.

After the non-linear transient ($t > t_{\ell}$), the comoving axion number density is conserved, and the axion abundance at the present time can be estimated by using Eq.~\eqref{abundance_nonlinear}.
Note that the axion number density at the present time $t_0$ is given by
\begin{align}
n_a(t_0) &= \left(\frac{R_{\ell}}{R_0}\right)^3 n_a(t_{\ell}) \nonumber\\
&= \left(\frac{R_{\ell}}{R_0}\right)^3c_nc_V z m_a(T_1)f_a^2 \left(\frac{R_1}{R_{\ell}}\right)^3 \nonumber\\
&= c_nc_V z H_1f_a^2 \left(\frac{R_1}{R_0}\right)^3.
\end{align}
Hence, we can think of $c_nc_V z$ as an enhancement factor compared to the naive estimate $H_1f_a^2 \left(R_1/R_0\right)^3$.

In order to quantify the effect of strings on the estimation of the dark matter abundance,
let us define the production efficiency from strings compared to the usual misalignment estimate,
\begin{align}
K \equiv \frac{n_a^{\rm str}(t_0)}{n_a^{\rm mis}(t_0)}, \label{K_definition}
\end{align}
where $n_a^{\rm mis}(t_0) = c_{\rm mis}H_1f_a^2 \left(R_1/R_0\right)^3$ with $c_{\rm mis} = 2.31$
is the fiducial axion number representing a typical result from angle-average misalignment estimation.\footnote{The coefficient is motivated by the averaging of the initial misalignment angle $\theta_i$,
$c_{\rm mis} = \langle\theta_i^2\rangle/2$ with $\sqrt{\langle\theta_i^2\rangle} \approx 2.15$, which was obtained numerically by including anharmonicities in the QCD potential~\cite{GrillidiCortona:2015jxo,Borsanyi:2016ksw}.}
In terms of $K$, the axion CDM abundance can be estimated as
\begin{align}
\Omega_a h^2 = K\Omega_a^{\rm mis}h^2,
\end{align}
where
\begin{align}
\Omega_a^{\rm mis}h^2 &= \frac{m_a n_a^{\rm mis}(t_0)}{\rho_{\rm crit}/h^2} \nonumber\\
 &= c_{\rm mis}\sqrt{\frac{20}{\pi}}g_{*s}(T_0)\frac{\sqrt{g_{*\rho}(T_1)}}{g_{*s}(T_1)}\Omega_{\gamma}h^2\frac{m_af_a^2}{T_1T_0M_P}, \label{Omega_mis_formula}
\end{align}
$g_{*\rho}(T)$ and $g_{*s}(T)$ are the effective degrees of freedom for the energy density and entropy density, respectively,
$\rho_{\rm crit}$ is the critical density of the Universe today, $g_{*s}(T_0) \simeq 3.931$ are the effective degrees of freedom for the entropy density today~\cite{Saikawa:2018rcs}, 
$\Omega_{\gamma}h^2 \simeq 2.473 \times 10^{-5}$ is the photon density parameter, $T_0 \simeq 2.349 \times 10^{-4}\,\mathrm{eV}$ is the photon temperature today~\cite{Fixsen:2009ug}, 
and $M_P \simeq 1.221\times 10^{19}\,\mathrm{GeV}$ is the Planck mass.

\begin{figure}[htbp]
\includegraphics[width=0.48\textwidth]{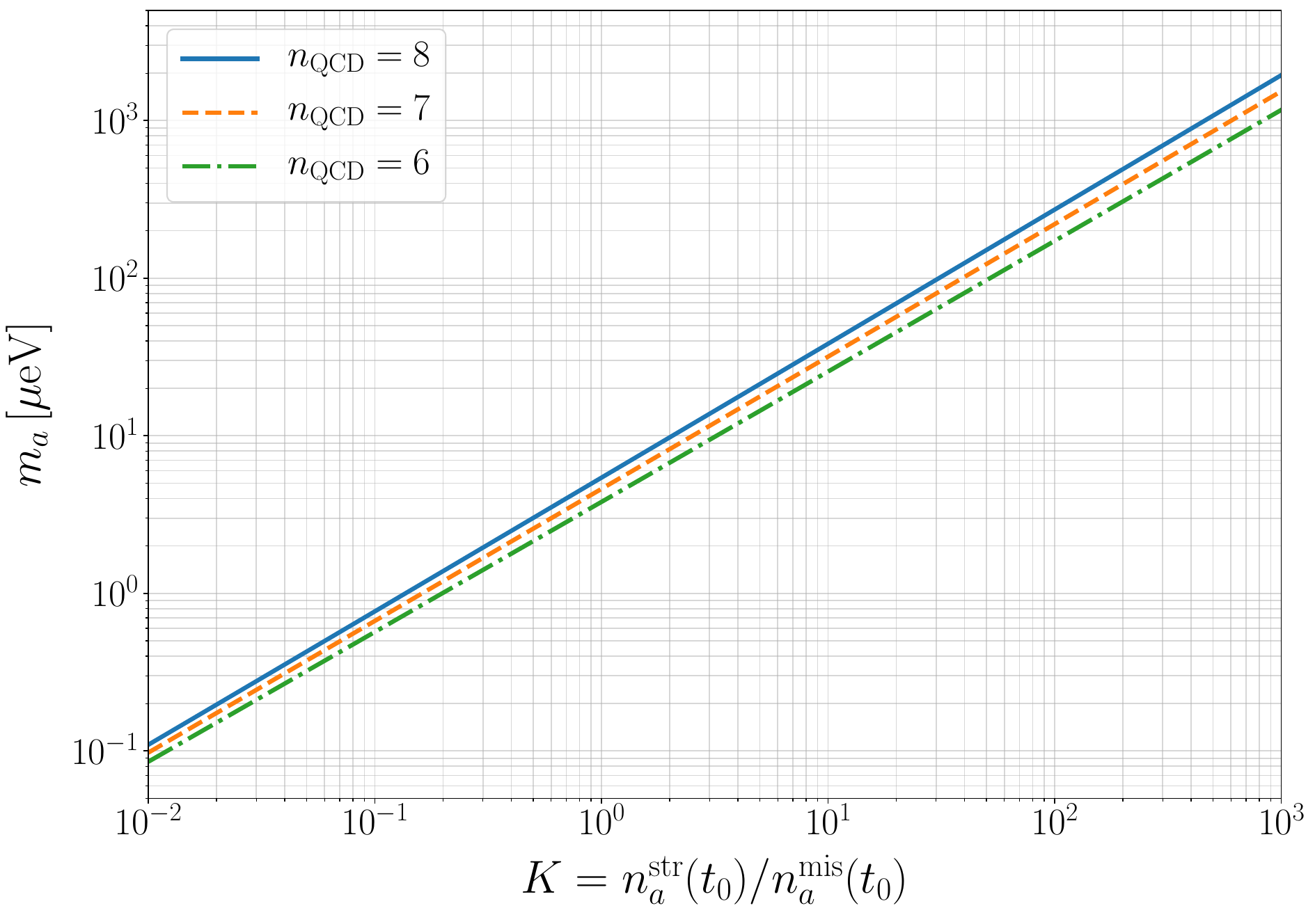}
\caption{The relation between the axion production efficiency defined by Eq.~\eqref{K_definition} and the axion mass explaining the observed CDM abundance.
Different lines correspond to different values of the exponent $n_{\rm QCD}$ characterising the temperature dependence of the topological susceptibility [Eq.~\eqref{chi_T_dependence}].}
\label{fig:ext_K_vs_ma}
\end{figure}

For a given value of $K$, one can determine the axion mass $m_a$ satisfying the condition $\Omega_ah^2 = \Omega_{\rm CDM}h^2 = 0.12$~\cite{Planck:2018vyg},
as shown in Fig.~\ref{fig:ext_K_vs_ma}.
To produce this relation between $K$ and $m_a$, we used tabulated data of $g_{*\rho}(T)$ and $g_{*s}(T)$ obtained in Ref.~\cite{Saikawa:2018rcs}.
As shown in the figure, the result slightly depends on the exponent $n_{\rm QCD}$ characterising the temperature dependence of the axion mass.
A smaller value of $n_{\rm QCD}$ makes the axion massive at earlier time, shifting $T_1$ to a higher value. This results in a suppression $\Omega_a^{\rm mis}h^2 \propto 1/T_1$ [see Eq.~\eqref{Omega_mis_formula}],
and requires a higher value of $f_a$ to explain the observed dark matter abundance. Therefore, a smaller mass is predicted for a smaller $n_{\rm QCD}$.
We also note that the uncertainty due to the equation of state $g_{*\rho}(T)$ and $g_{*s}(T)$ quantified in Ref.~\cite{Saikawa:2018rcs} is negligibly small compared to the effect of $n_{\rm QCD}$,
since a small variation in the Hubble parameter due to $g_{*\rho}(T)$ does not have a big impact on the determination of $T_1$ for a fast temperature dependence of $m_a(T)$.

Now we can compute the axion relic abundance relative to the fiducial estimate,
\begin{align}
K = \begin{cases}
{\displaystyle \frac{1}{c_{\rm mis}}\frac{n_a(t_1)}{f_a^2H_1}} & \text{(linear)},\\[2.5ex]
{\displaystyle \frac{c_nc_V z}{c_{\rm mis}}} & \text{(non-linear)},
\end{cases}
\label{K_how_to_estimate}
\end{align}
where Eq.~\eqref{na_calculation} can be used for the linear case, while Eq.~\eqref{z_formula} should be used for the non-linear case.
Once we obtain $K$ for a given value of $\ell$, we can translate it to the prediction for the dark matter mass by using the $K$-$m_a$ relation shown in Fig.~\ref{fig:ext_K_vs_ma}.

In order to perform the extrapolation to large $\ell$, we need to fix the assumption on how the values of $q$ and $\xi$ change with $\ell$.
For the value of $q$, among three different possibilities considered in Fig.~\ref{fig:ext_amplitude}, the linear estimate can be used for the case with $q=q_0+q_1/\ell^2$,
while the non-linear estimate should be used for that with $q=q_0+q_1\ell$.
Hereafter we estimate the axion relic abundance and the dark matter mass prediction for these two cases, and call the scenario with $q=q_0+q_1/\ell^2$
the \emph{plateau extrapolation} and that with $q=q_0+q_1\ell$ the \emph{linear extrapolation}.
It might also be possible to consider yet another type of extrapolation by using the model with $q=q_0+q_1/\ell$,
but in that case both the formulae shown in Eq.~\eqref{K_how_to_estimate} would not hold to a good approximation,
since the system lies in the boundary between linear and non-linear regime at $\ell \sim 70$ (see Fig.~\ref{fig:ext_amplitude}), which requires some additional numerical study to obtain quantitative results.
For this technical reason, in the following we do not perform the extrapolation based on the model with $q=q_0+q_1/\ell$.
Nevertheless, it is certainly true that the result of the extrapolation with $q=q_0+q_1/\ell$ would lie between the value predicted in the case with $q=q_0+q_1/\ell^2$ and that with $q=q_0+q_1\ell$.

For the value of $\xi$, we use the extrapolation based on two models of $\xi_c$ given by Eqs.~\eqref{eq:xi_c_fit_lin} and~\eqref{eq:xi_c_fit_sat}.
Here we adopt two models of $\xi_c$ for each of the two possible extrapolations of $q$, 
so that we consider four different possibilities for the values of $q$ and $\xi$ in total,
and quantify their impact on the estimation of the axion abundance and the dark matter mass prediction.

\begin{figure*}[htbp]
$\begin{array}{cc}
\subfigure{
\includegraphics[width=0.48\textwidth]{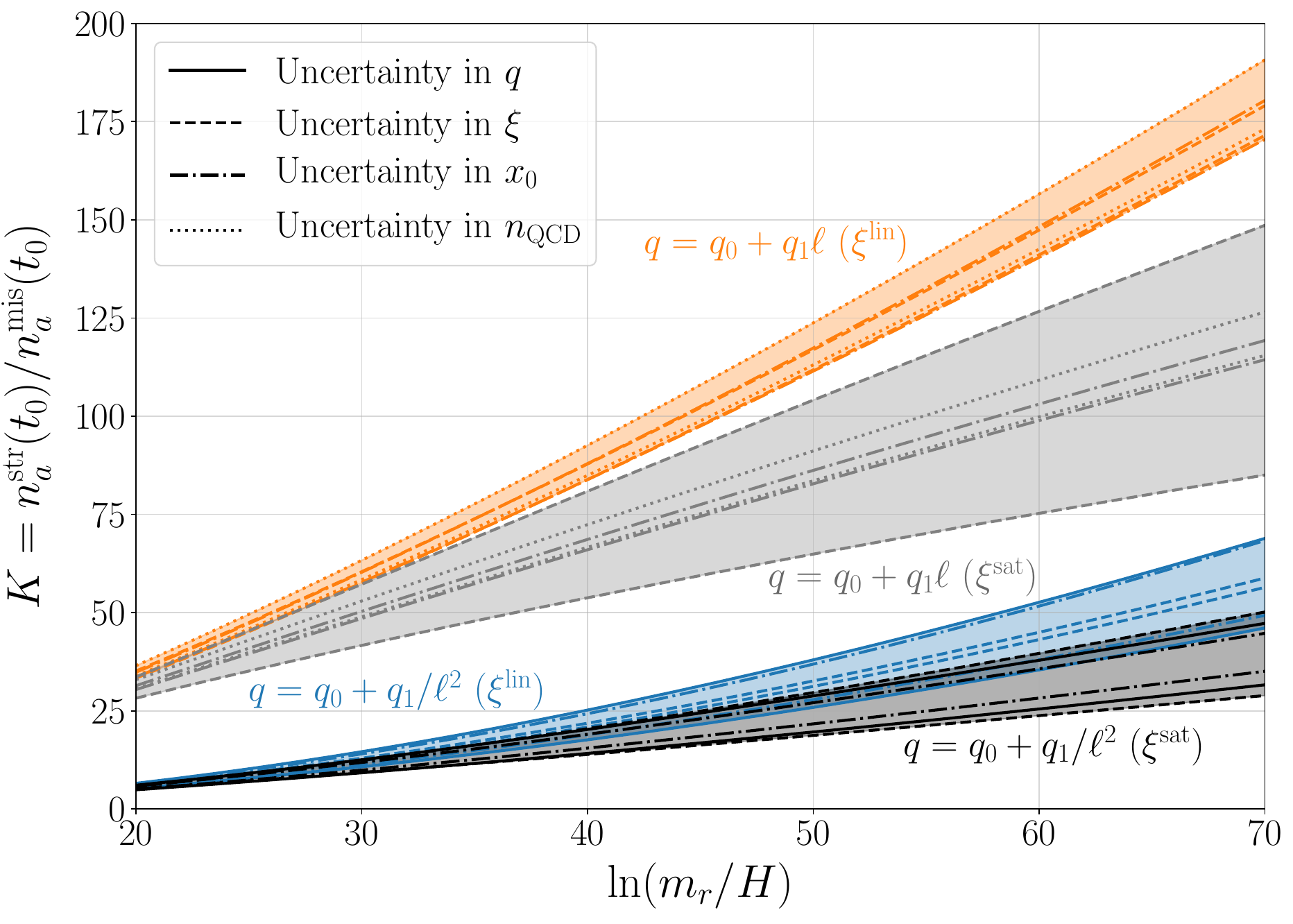}}
\hspace{1mm}
\subfigure{
\includegraphics[width=0.48\textwidth]{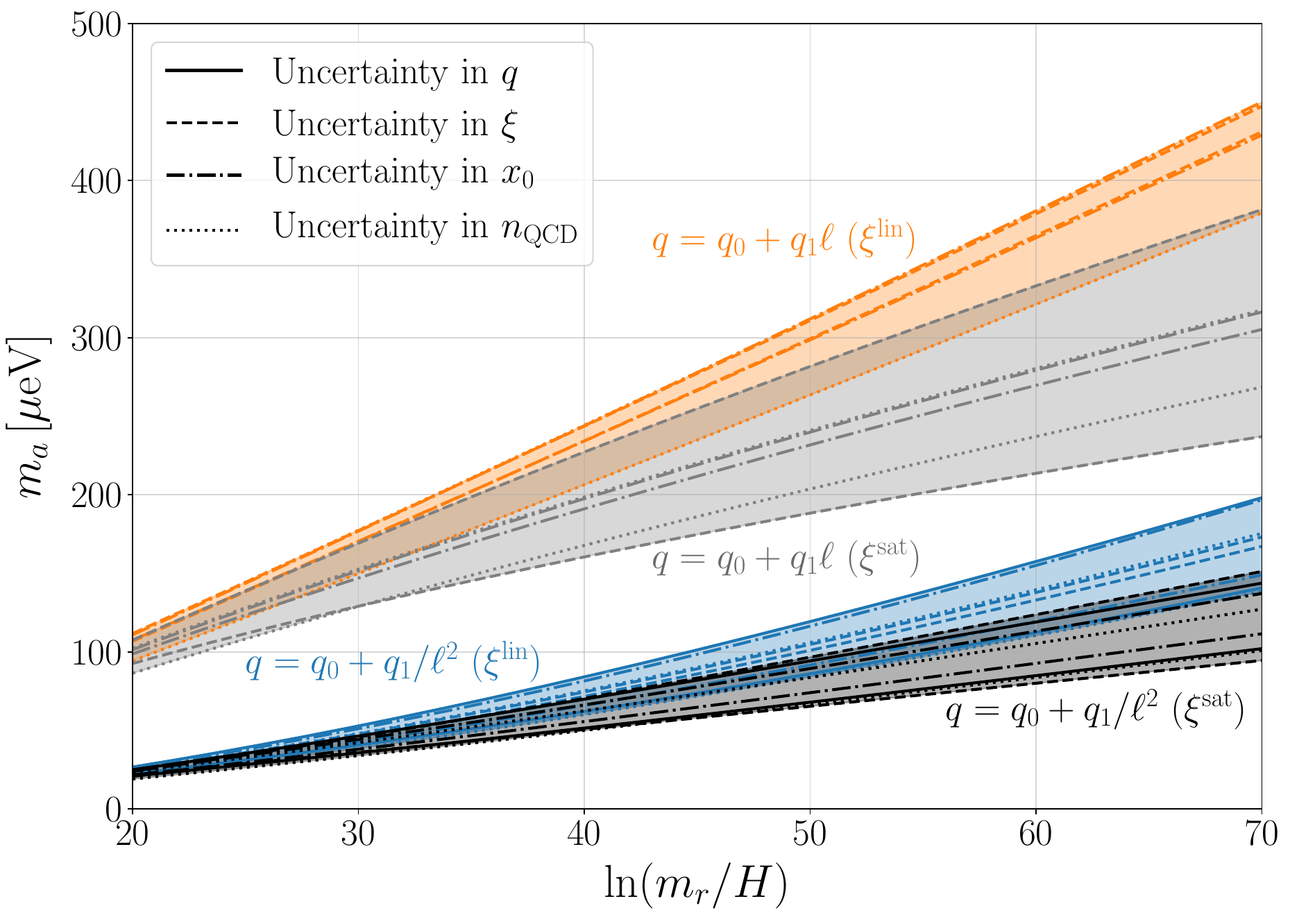}}
\end{array}$
\caption{The axion production efficiency $K$ (left) and corresponding dark matter mass (right) obtained by extrapolating numerical results to larger values of $\ln(m_r/H)$.
Different colours represent four different assumptions on the extrapolation of $q$ and $\xi$:
$q=q_0+q_1/\ell^2$ with $\xi_c=\xi_c^{\rm lin}$ (blue), $q=q_0+q_1\ell$ with $\xi_c=\xi_c^{\rm lin}$ (orange), 
$q=q_0+q_1/\ell^2$ with $\xi_c=\xi_c^{\rm sat}$ (black), and $q=q_0+q_1\ell$ with $\xi_c=\xi_c^{\rm sat}$ (gray). 
For each type of extrapolation, uncertainties due to the change in the parameters of $q$, $\xi$, $x_0$, and $n_{\rm QCD}$ (see text for details) are shown by the region enclosed by
solid, dashed, dash-dotted, and dotted lines, respectively.}
\label{fig:ext_ma}
\end{figure*}

In Fig.~\ref{fig:ext_ma}, we show the results of the calculation of $K$ (left panel) and the corresponding axion dark matter mass prediction (right panel) for different possibilities on the extrapolation of $q$ and $\xi$.
In these plots, we also show the uncertainties due to the change in the values of parameters used for the calculation.
The uncertainty due to $q$ is derived from the fit parameters $q_0$ and $q_1$ (see Table~\ref{tab:fits_q_results}).
The uncertainty due to $\xi$ is derived from the fit parameters of $\xi_c$ shown in Eqs.~\eqref{eq:xi_c_fit_lin} and~\eqref{eq:xi_c_fit_sat}.
The uncertainty due to $x_0$ is derived from the fit parameters $b_0$ and $b_1$ shown in Eq.~\eqref{x_ta_parameters}, where we use the results obtained from data with $kL/(2\pi)=2.41$ as the fiducial values.
As for the uncertainty due to $n_{\rm QCD}$, we change the value of $n_{\rm QCD}$ from 7 to 8.16 (estimate based on the dilute instanton gas approximation~\cite{Borsanyi:2016ksw}),
where $n_{\rm QCD}=8$ is used as the fiducial value.\footnote{In our analysis, the dependence on $n_{\rm QCD}$ appears only through the determination of $T_1$ in Eq.~\eqref{Omega_mis_formula} for the linear case.
It should be noted that there can be additional effect of $n_{\rm QCD}$ through the modification of the axion field evolution around the epoch of the QCD phase transition, which we have not included in our analysis.
Simulations with toy models performed in Ref.~\cite{Chaumet:2021gaz} suggest that the axion production efficiency is significantly higher than the misalignment estimate for $n_{\rm QCD} = 0$ compared to $n_{\rm QCD} = 7$.
A similar trend was also found in the simulations performed in Ref.~\cite{OHare:2021zrq}, showing that the $n_{\rm QCD}=0$ case predicts 25\% higher dark matter abundance than the misalignment estimate.
Such an enhancement at small values of $n_{\rm QCD}$ could compensate the reduction due to the change in $T_1$ and modify the error estimation for the linear case.
On the other hand, in the non-linear case the effect may not be so significant as the linear case, since the gradient energy dominates over the potential energy during the non-linear transient, 
and the effect of the QCD potential on the axion field dynamics is expected to be subdominant in that regime.}
Note that the error in $n_{\rm QCD}$ is irrelevant for the calculation of $K$ in the plateau extrapolation, and the corresponding lines (blue or black dotted lines) are not shown in the left panel of Fig.~\ref{fig:ext_ma}.
Furthermore, the error in $q$ (would be orange or gray solid lines) is irrelevant for the linear extrapolation, 
since we have used the formula given by Eq.~\eqref{z_formula}, which is derived under the assumption of $q\gg 1$ and does not depend on the detailed values of $q$.
It turns out that, for the plateau extrapolation the largest uncertainty comes from the error in the value of $q$ (for $\xi_c=\xi_c^{\rm lin}$) or $\xi$ (for $\xi_c=\xi_c^{\rm sat}$).
On the other hand, for the linear extrapolation with $\xi_c=\xi_c^{\rm lin}$ the dominant source of the uncertainty is the value of $n_{\rm QCD}$, which affects the timescale of the dynamics of the system,\footnote{Since 
Eq.~\eqref{z_formula} implies that $K \propto (\xi_1\ell_1)^{\frac{1}{2}+\frac{1}{n_{\rm QCD}+4}}$, the production efficiency becomes larger for smaller values of $n_{\rm QCD}$.
We can see this trend in the left panel of Fig.~\ref{fig:ext_ma}, where the upper dotted lines correspond to the case with $n_{\rm QCD} = 7$.
However, this enhancement is overwhelmed by the shift in $T_1$, and the resulting dark matter mass becomes smaller for $n_{\rm QCD} = 7$, as shown by the lower orange or gray dotted line in the right panel of Fig.~\ref{fig:ext_ma}.}
though the error in $\xi$ dominates if instead we adopt $\xi_c=\xi_c^{\rm sat}$.

\begin{table*}[ht]
\caption{Axion dark matter mass predicted at $\ln(m_r/H)=70$ and its error budget.
For each extrapolation of $q$ (plateau extrapolation with $q=q_0+q_1/\ell^2$ and linear extrapolation with $q=q_0+q_1\ell$), 
we consider two models of $\xi_c$.}\label{tab:mass_predictions}
\begin{tabular*}{0.7\textwidth}{@{\extracolsep{\fill}}lcccc}
\hline \hline
\multirow{2}{*}{Parameter} & \multicolumn{2}{l}{$m_a\,[\mu\mathrm{eV}]$ (Plateau extrapolation)} & \multicolumn{2}{l}{$m_a\,[\mu\mathrm{eV}]$ (Linear extrapolation) } \\
 & $\xi_c^{\rm lin}$ & $\xi_c^{\rm sat}$ & $\xi_c^{\rm lin}$ & $\xi_c^{\rm sat}$\\
\hline
$q$ & 141--198 & 102--144 & 439 & 311\\
$\xi$ & 167--173 & 95--151 & 431--447 & 237--381\\
$x_0$ & 149--197 & 111--137 & 429--450 & 305--316\\
$f_L$ & 168--172 & 122--125 & 435--443 & 308--313\\
$n_{\rm QCD}$ & 138--175 & 101--127 & 379--449 & 268--317\\
\hline
\end{tabular*}
\end{table*}

Finally, in Table~\ref{tab:mass_predictions} we summarise the values of the predicted axion dark matter mass at $\ln(m_r/H) = 70$ and its uncertainty due to the change in the parameters within the range described above
for four assumptions on the extrapolation of $q$ and $\xi$.
In addition to the errors due to $q$, $\xi$, $x_0$, and $n_{\rm QCD}$, here we also show the effect of the change in $f_L$ within the range shown in Eq.~\eqref{fL_fit_result},
though it is subdominant compared to other uncertainties.
Taking the maximum and minimum values from Table~\ref{tab:mass_predictions} (including the difference between two models of $\xi_c$), 
we end up with the following estimate for the axion dark matter mass for the plateau extrapolation of $q$:
\begin{align}
m_a = 95\text{--}198\,\mu\mathrm{eV}, \label{mass_prediction_plateau}
\end{align}
which corresponds to the value of the axion decay constant, $f_a \approx (2.9\text{--}6.0)\times 10^{10}\,\mathrm{GeV}$.
On the other hand, if we take the extrapolation along the linear increase model of $q$, the predicted mass reads
\begin{align}
m_a = 237\text{--}450\,\mu\mathrm{eV}, \label{mass_prediction_linear}
\end{align}
which corresponds to $f_a \approx (1.3\text{--}2.4)\times 10^{10}\,\mathrm{GeV}$.

\section{Conclusions}
\label{sec:conclusions}

In this paper, we have performed large scale simulations of global string networks and analysed the spectrum of axions radiated from them.
Our main findings are summarised as follows.
\begin{enumerate}
\item We have shown that the attractor behaviour of the system can be described by 
a simple differential equation~\eqref{eq:attractive_xi_evol} with an appropriate choice of the coefficient function $C(x)$.
This interpretation is complemented by looking at how the system radiates saxions.
It is also found that the spectral index $q$ of the instantaneous axion emission spectrum can be over/under estimated when the string density is lower/higher than the attractor.
\item The instantaneous emission spectrum is largely contaminated by the oscillations in the axion energy density spectrum.
These oscillations are expected on physical grounds due to
the misalignment oscillation of the Fourier mode of the axion field after the corresponding mode enters the horizon, 
or the production of high momentum modes due to the parametric resonance effect
induced by the oscillating radial field when the momentum of the corresponding mode becomes comparable to $m_r/2$.
The estimates of $q$ can be biased unless such oscillations are properly handled.
\item We have observed that the spectra of axions and saxions are distorted by discretisation effects, which bias $q$ toward larger values. The effects blow up rapidly at large $\ell=\ln(m_r/H)$.
We have performed the global fits to extract those effects and characterised the evolution of $q$ in the continuum limit in terms of some simple functions of $\ell$.
It turns out that several possibilities on the $\ell$ dependence of $q$ can fit the data, including the scenario where $q$ approaches $q \approx 1$ as $1/\ell^2$ at large $\ell$ (plateau model),\footnote{We note that 
there are several heuristic arguments that show a preference for $q\lesssim 1$ in the literature.
The scenario with $q=1$ was indeed claimed a few decades ago in Refs.~\cite{Harari:1987ht,Hagmann:1990mj,Hagmann:2000ja}.
More recently, Ref.~\cite{Dine:2020pds} provided intuitive explanation on why we should expect $q\le 1$.
The authors of Ref.~\cite{Buschmann:2021sdq} also suggested an interpretation to have $q=1$ by focusing on the distribution of string loops.
Our result that the plateau model shows a good fit to the data could be compatible with these arguments, 
but we cannot exclude the other possibility that $q$ becomes greater than 1 at large $\ell$ in the current analysis.} 
and the scenario where $q$ increases linearly with $\ell$ (linear increase model).
\item We have found that the turnaround in the evolution of $\mathcal{F}$ around the horizon crossing depends on the string density parameter $\xi$, which is well described by the function given by Eq.~\eqref{x_ta_model}
with parameters determined as Eq.~\eqref{x_ta_parameters}. This dependence on $\xi$ remarkably agrees with our expectation on the behaviour of the IR peak of the spectrum, $x_0\propto \sqrt{\xi}$.
\end{enumerate}

By combining the numerical results, we have performed the extrapolation to large $\ell$ and estimated the prediction for the axion dark matter mass.
The prediction basically differs according to the assumption on how $q$ depends on $\ell$: For the plateau model it is given by Eq.~\eqref{mass_prediction_plateau},
while for the linear increase model it is given by Eq.~\eqref{mass_prediction_linear}.
Since both models fit the data almost equally well, we expect that yet other models which lead to a value of $q$ in-between two models could also be allowed.\footnote{For instance, 
we found that the model $q = q_0 + q_1/\ell$ also gives a good fit to the data (see Table~\ref{tab:fits_q_results}), 
and the extrapolation with that model should lead to a value which lies between the range shown in Eq.~\eqref{mass_range_conclusion}, as discussed in Sec.~\ref{sec:axion_abundance}.}
Therefore, we assess the uncertainty in the dark matter mass prediction at the difference between the results of two different ways of extrapolation:
\begin{align}
95\,\mu\mathrm{eV} \lesssim m_a \lesssim 450\,\mu\mathrm{eV}. \label{mass_range_conclusion}
\end{align}
The lower part of the mass range obtained by the extrapolation with $q \approx 1$ will be probed by future haloscope experiments,
including MADMAX~\cite{MADMAX:2019pub}, ALPHA~\cite{ALPHA:2022rxj}, and ORGAN~\cite{McAllister:2017lkb}.\footnote{A part of this mass range is actually probed in recent experiments~\cite{Quiskamp:2023ehr}.}
On the other hand, accessing the higher end of the mass range $m_a \sim 450\,\mu\mathrm{eV}$ obtained by the extrapolation with $q>1$ would remain a challenge even for future experiments.

Note that the axion dark matter mass~\eqref{mass_range_conclusion} obtained by extrapolating our simulation results to large $\ell$ does not agree with the value $m_a=26.2\pm3.4\,\mu\mathrm{eV}$ obtained in Ref.~\cite{Klaer:2017ond},
which made use of a different method to simulate high-tension strings \emph{directly} by introducing extra field degrees of freedom~\cite{Klaer:2017qhr}.
Understanding the origin of such a discrepancy should be a crucial step to improve the robustness of our indirect method to predict the dark matter mass,
or to find any unaccounted systematics (if exists) in the extrapolation.
In this regard, it would be interesting to calculate the axion emission spectrum precisely in the simulation framework of Refs.~\cite{Klaer:2017ond,Klaer:2017qhr}
to make a close comparison between two methods.\footnote{See Ref.~\cite{Pierobon:2023ozb} for the comparison of direct and indirect simulation methods on the prediction for the formation of axion miniclusters.}

The main source of the uncertainty in the axion dark matter mass shown in Eq.~\eqref{mass_range_conclusion}
is the different scenario on the value of $q$ at large $\ell$. In order to reduce the uncertainty,
it is thus important to see if $q$ actually continues to increase or saturates to $q\approx 1$ when we further increase the value of $\ell$ with suppressing the discretisation effects.
To this end, a dedicated study by using some advanced numerical scheme that allows for even larger dynamical ranges [such as the AMR framework~\cite{2020arXiv200912009Z,Andrade:2021rbd}] is warranted.
We leave it as the subject of our future work.

\begin{acknowledgments}
Part of this research was supported by the Munich Institute for Astro- and Particle Physics (MIAPP) which is funded by DFG
under Germany's Excellence Strategy -- EXC-2094 -- 390783311.
The work of K.S. is supported by Leading Initiative for Excellent Young Researchers (LEADER), the Ministry of Education, Culture, Sports, Science, and Technology (MEXT), Japan. 
The work of K.S. is also supported by JSPS KAKENHI Grant Number JP24K07015.
This article is based upon work from COST Action COSMIC WISPers CA21106, 
supported by COST (European Cooperation in Science and Technology).
The work of J.R., A.V. and M.K. is supported by Grants PGC2022-126078NB-C21 funded by MCIN/AEI/ 10.13039/501100011033 and “ERDF A way of making Europe” and Grant  DGA-FSE grant 2020-E21-17R Aragon Government and the European
Union - NextGenerationEU Recovery and Resilience Program on `Astrofísica y Física de Altas Energías' CEFCA-CAPA-ITAINNOVA. Additionally, the work of M.K. is supported by the Government of Aragón, Spain, with a PhD fellowship as specified in ORDEN CUS/702/2022.  
A.V. is further supported by AEI (Spain) under Grant No. RYC2020-030244-I / AEI / 10.13039/501100011033.
Computations were performed on the HPC system Raven and Cobra at the Max Planck Computing and Data Facility.
This work was partly achieved through the use of SQUID at the Cybermedia Center, Osaka University.
\end{acknowledgments}

\appendix 
\numberwithin{equation}{section}

\section{Code and simulation setups}
\label{app:code}

In this work, we utilise the {\tt Jaxions} code,\footnote{\url{https://github.com/veintemillas/jaxions}} 
which is a massively parallel code to study the evolution of the PQ field in a FRW background,
originally developed in Ref.~\cite{Vaquero:2018tib}. This appendix is devoted to the description of some details of the code and simulation setups.

For the study of the axion production from strings, one has to solve the following classical equation of motion derived from the Lagrangian~\eqref{Lagrangian_PQ},
\begin{align}
\ddot{\phi} + 3H\dot{\phi} - \frac{1}{R^2}\nabla^2\phi +\lambda\phi\left(|\phi|^2 - f_a^2\right) = 0.
\end{align}
In the numerical study, it is convenient to define the rescaled variables as
\begin{align}
\phi \to \frac{\phi}{f_a}\frac{R}{R_1}, \quad x^i \to R_1H_1x^i, \quad \tau \to R_1H_1\tau,
\end{align}
where $x^i$ are comoving spatial coordinates, $\tau$ is the conformal time, 
$H_1$ is the Hubble parameter characterising a reference timescale of the system [see Eq.~\eqref{H1_estimate}], and $R_1$ is the scale factor at that time.
If we assume the radiation dominated Universe ($R\propto \tau$) and use
the rescaled variables defined above, we can simplify the equation of motion as
\begin{align}
\phi_{\tau\tau} - \nabla^2\phi + \bar{\lambda}\phi(|\phi|^2 - \tau^2) = 0,
\label{EOM_code}
\end{align}
which is used in the numerical scheme implemented in the {\tt Jaxions} code.
Note that the equation of motion can be characterised by only one free parameter,
\begin{align}
\bar{\lambda} \equiv \frac{\lambda f_a^2}{H_1^2} = \frac{m_r^2}{2H_1^2}.
\label{lambda_definition}
\end{align}

A natural choice for the typical timescale of the system is the time at which the axion mass starts to be relevant due to the QCD effects, $H_1 = m_a(T_1)$.
This amounts to~\cite{Vaquero:2018tib}
\begin{align}
H_1 \simeq 3.45\times 10^{-3}\,\mu\mathrm{eV}\,\left(\frac{m_a}{50\,\mu\mathrm{eV}}\right)^{0.338}, \label{H1_estimate}
\end{align}
where the exponent is found numerically by using the lattice QCD output of Ref.~\cite{Borsanyi:2016ksw}.
Substituting Eq.~\eqref{H1_estimate} into Eq.~\eqref{lambda_definition}, we find that the realistic value of the $\bar{\lambda}$-parameter should be
\begin{align}
\bar{\lambda} \simeq 1.1\times 10^{57}\,\lambda\left(\frac{m_a}{50\,\mu\mathrm{eV}}\right)^{-2.676}. 
\end{align}
It is not possible to perform simulations for such a large value of $\bar{\lambda}$, since the ratio of the lattice spacing, which is at most of order the string core radius $\sim m_r^{-1}$,
to the size of the simulation box, which is at least of the order of the Hubble radius $\sim H_1^{-1}$, is limited by the computing resources.
This fact requires an adequate treatment to extrapolate the numerical results to extremely large values of $\bar{\lambda}$.

The equation of motion~\eqref{EOM_code} is solved numerically by defining the PQ field on a static cubic lattice with periodic boundary conditions.
The simulation box is given by a constant comoving volume with size $L^3$, which means that the physical size of the lattice spacing $a \equiv R(t)L/N$
increases proportionally to the scale factor $R(t)$, where $N$ is the number of lattice sites per dimension.
Since the width of the string core $\sim m_r^{-1} \propto (\sqrt{\lambda}f_a)^{-1}$ is constant, the ratio of the physical lattice spacing to the string core width, $m_ra$,
increases with time, which leads to serious discretisation effects as described in Sec.~\ref{sec:disc_effect}. 
The effects may be alleviated by artificially replacing $\bar{\lambda}$ by $(R/R_1)^{-2}\bar{\lambda}$ such that $m_ra$ remains constant (the PRS method).
In this work, we perform both types of simulations: the physical type, where $\bar{\lambda}$ is constant throughout the simulation,
and the PRS type, where $\bar{\lambda}$ decreases as $\propto R(t)^{-2}$.

The Laplacian is reproduced by building the stencil involving $N_g$ neighbouring points around a given grid point labeled by indices ${\bm i} = (i_x,i_y,i_z)$,
\begin{align}
(\nabla^2 \phi)_{{\bm i}} = \frac{1}{\delta^2}\sum_{u=x,y,z}\sum_{n=1}^{N_g}C_n(\phi_{{\bm i}+n{\bm n}_u} + \phi_{{\bm i}-n{\bm n}_u} - 2\phi_{{\bm i}}),
\end{align}
where $\delta = L/N$ and ${\bm n}_u$ $(u=x,y,z)$ are unit vectors representing displacements of one lattice spacing in $u$-direction.
In {\tt Jaxions}, the above discretisation scheme is implemented up to $N_g = 4$, whose coefficients are summarised in Table~\ref{C_Laplacian}.
The effect of different choices of $N_g$ is discussed in Sec.~\ref{sec:lap}.

\begin{table}
\centering
\caption{Coefficients for Laplacian stencils}\label{C_Laplacian}
\begin{tabular}{c c c c c}
\hline\hline
$N_g$ & $C_1$ & $C_2$ & $C_3$ & $C_4$ \\
\hline
1 & 1 & 0 & 0 & 0 \\
2 & $16/12$ & $-1/12$ & $0$ & $0$ \\
3 & $3/2$ & $-3/20$ & $1/90$ & $0$ \\
4 & $8/5$ & $-1/5$ & $8/135$ & $-1/560$ \\
\hline
\end{tabular}
\end{table}

The time evolution of the PQ field $\phi$ is computed by converting the second-order differential equations~\eqref{EOM_code}
into a couple of first-order equations for $\phi$ and $\phi_{\tau}$.
For the time propagation we adopt the fourth-order McLachlan-Atela (Runge-Kutta-Nystr\"om) method~\cite{R_I_McLachlan_1992},
which is a highly accurate, explicit method optimized for the system described by separable Hamiltonian with quadratic kinetic energy. 
For the size of the time step, we use $Rd\tau = 1.5/\omega_{\rm max}$, where $\omega_{\rm max} = \sqrt{m_r^2 + k_*^2/R^2}$ with $k_*^2 = 12/\delta^2$
is the characteristic frequency of the fastest mode.\footnote{Strictly speaking, $k_*^2=12/\delta^2$ corresponds to the maximum momentum for the Laplacian with $N_g=1$.
For general $N_g$, the maximum momentum takes a larger value given by $k_*^2 = 12\sum_{n=1}^{N_g}C_n/\delta^2$.
Nevertheless, we use the same criterion ($k_*^2=12/\delta^2$) to determine $d\tau$ even for the case with larger $N_g$ in the code.}
This amounts to $d\tau = 1.5\delta/\sqrt{(m_ra)^2+12} < 0.433\delta$ for any value of $m_ra>0$ and satisfies the Courant condition.
For the update of the field variables at each time step, we use advanced vector extensions (AVX512),
which can process multiple sets of data simultaneously, in addition to the standard MPI/OpenMP parallelization.
Furthermore, the loops for the field update are broken up into sub-blocks, whose size is adjusted to the cache size of the processor.
Since the cache size differs according to the hardware, the program executes a function that tunes the block size to find 
an optimal loop tiling at the beginning of the simulation.
With these functionalities in {\tt Jaxions}, we can achieve a substantial speed-up in the calculation of the field evolution.

Parameters used in the simulations performed in this work 
and the number of simulations executed for each choice of the parameters
are summarised in Table~\ref{tab:simu_parameters}.
For each set of parameters, simulations are performed up to a conformal time $\tau_f$ shown in the fourth column of the table (in terms of the ratio $\tau_f/L$).
For physical-type simulations, we performed four sets of $4096^3$ simulations to study discretisation effects due to the Laplacian, and
four sets of $3072^3$ simulations with different values of $\bar{\lambda}$ to study discretisation effects due to the resolution of the string core width,
in addition to the largest $11264^3$ simulations, which reach $\ln(m_r/H) = 9.08$ at the final time. 
Furthermore, we also performed 30 sets of $2048^3$ simulations with different initial string densities to study the attractor behaviour of the system.
Other relatively small scale simulations (from $1024^3$ to $3072^3$) are performed for the studies of the initial conditions (Appendix~\ref{app:initial_conditons}) and the finite volume effects (Appendix~\ref{app:syst_finite_V}).
Except for the simulations with $N=2048$ and $\bar{\lambda}=6400$, all physical-type simulations begin with the initial string density corresponding to the attractor, 
$\xi_{\ell=3}= 0.3$ (see Sec.~\ref{sec:attractor} for identification of the attractor).

\begin{table*}[ht]
\begin{minipage}{\textwidth}
\caption{Parameters used in the simulations and the number of simulations executed for each choice of them.
The list of the number of figures in which the results are used is also shown in the last column.}
\label{tab:simu_parameters}
 \begin{tabular*}{\textwidth}{@{\extracolsep{\fill}}lllllllll}
\hline\hline 
 Type\footnote{For all physical-type simulations (except for $N=1024$, see footnote~\ref{fn:1024}) we set the initial time $\tau_i$ such that the simulation starts from $\ln(m_r/H) = 3$, 
 and for all PRS-type simulations we set $\tau_i$ such that the simulation starts from $\ln(m_r/H) =0$.} 
 & Grid size  & Laplacian & Final time & $\ln(m_r/H)$  & Parameter & Number of & List of figures \\
         & ($N^3$)    &               &  ($\tau_f/L$) & at $\tau_f$ & & simulations &\\
\hline
 Physical & $11264^3$ & 4-neighbours & 0.625 & 9.08 & $\bar{\lambda}=195799$ & 20 & \ref{fig:attr_xi_physical},\ref{fig:attr_xi_PRS_vs_physical},\ref{fig:osc_mode_evol_lowk},\ref{fig:osc_mode_evol_highk},\ref{fig:osc_F_diff_vs_fit},\ref{fig:disc_spectrum_physical},\ref{fig:disc_F_physical},\ref{fig:disc_Gamma_physical},\ref{fig:disc_mra_q},\ref{fig:disc_Gamma_tot},\ref{fig:disc_Gamma_physical_comparison},\ref{fig:disc_q_physical_continuum},\ref{fig:syst_cutoffs_physical},\ref{fig:comparison}\\
 Physical & $4096^3$ & 1-neighbour & 0.625 & 8.07 & $\bar{\lambda}=25890.8$ & 30 & \ref{fig:disc_lap_spectrum},\ref{fig:disc_lap_q}\\
 Physical & $4096^3$ & 2-neighbours & 0.625 & 8.07 & $\bar{\lambda}=25890.8$ & 30 & \ref{fig:disc_lap_spectrum},\ref{fig:disc_lap_q}\\
 Physical & $4096^3$ & 3-neighbours & 0.625 & 8.07 & $\bar{\lambda}=25890.8$ & 30 & \ref{fig:disc_lap_spectrum},\ref{fig:disc_lap_q}\\
 Physical & $4096^3$ & 4-neighbours & 0.625 & 8.07 & $\bar{\lambda}=25890.8$ & 30 & \ref{fig:osc_q_cUV},\ref{fig:disc_lap_spectrum},\ref{fig:disc_lap_q},\ref{fig:mask_spectrum},\ref{fig:mask_q}\\
 Physical & $3072^3$ & 4-neighbours & 0.5 & 7.34 & $\bar{\lambda}=14563.6$ & 30 & \ref{fig:attr_xi_PRS_vs_physical},\ref{fig:disc_Gamma_physical},\ref{fig:disc_mra_q},\ref{fig:disc_Gamma_tot},\ref{fig:disc_q_physical_continuum},\ref{fig:mask_EA_ES}\\
 Physical & $3072^3$ & 4-neighbours & 0.5 & 7.74 & $\bar{\lambda}=32768$ & 30 & \ref{fig:attr_xi_PRS_vs_physical},\ref{fig:disc_Gamma_physical},\ref{fig:disc_mra_q},\ref{fig:disc_Gamma_tot},\ref{fig:disc_q_physical_continuum}\\
 Physical & $3072^3$ & 4-neighbours & 0.5 & 8.08 & $\bar{\lambda}=64225.3$ & 30 & \ref{fig:attr_xi_PRS_vs_physical},\ref{fig:disc_Gamma_physical},\ref{fig:disc_mra_q},\ref{fig:disc_Gamma_tot},\ref{fig:disc_q_physical_continuum}\\
 Physical & $3072^3$ & 4-neighbours & 0.5 & 8.37 & $\bar{\lambda}=114178$ & 30 & \ref{fig:attr_xi_PRS_vs_physical},\ref{fig:disc_Gamma_physical},\ref{fig:disc_Gamma_tot}\\
 Physical & $2048^3$ & 4-neighbours & 0.55 & 7.12 & $\bar{\lambda}=6400$ & 30$\times$30\footnote{For simulations of physical strings with $N=2048$, we chose 30 different initial string densities
 ranging from $\xi_{\ell=3}\simeq 0.022$ to 0.828 as shown in Fig.~\ref{fig:attr_xi}. For each value of the initial string density, we executed 30 simulations with randomly generated initial field configurations.}
  & \ref{fig:attr_xi},\ref{fig:attr_ES_invariant},\ref{fig:attr_ES_rmass},\ref{fig:attr_q},\ref{fig:attr_xi_physical},\ref{fig:ext_F_evolution},\ref{fig:ext_x0_vs_xi},\ref{fig:evol_EA}\\
 Physical & $1024^3$ & 4-neighbours & 0.5 & 6.23 & $\bar{\lambda}=1600$ & 30$\times$4\footnote{\label{fn:1024}Simulations of physical strings with $N=1024$ are performed for the check of initial conditions. 
 For these simulations we used 4 different ways to generate initial conditions, including a case where the simulation starts from $\ln(m_r/H) = 1$. See Appendix~\ref{app:initial_conditons} for details. 
 For each type of the initial condition, we executed 30 simulations with randomly generated initial field configurations.} & \ref{fig:init_xi},\ref{fig:init_mode_evol},\ref{fig:init_avrho},\ref{fig:init_spectrum}\\
 Physical & $3072^3$ & 4-neighbours & 0.458367 & 7.5 & $\bar{\lambda}=28571.2$ & 30 & \ref{fig:syst_finiteV}\\
 Physical & $2560^3$ & 4-neighbours & 0.550042 & 7.5 & $\bar{\lambda}=13778.5$ & 30 & \ref{fig:syst_finiteV}\\
 Physical & $2048^3$ & 4-neighbours & 0.687552 & 7.5 & $\bar{\lambda}=5643.68$ & 30 & \ref{fig:syst_finiteV}\\
 Physical & $1536^3$ & 4-neighbours & 0.916735 & 7.5 & $\bar{\lambda}=1785.69$ & 30 & \ref{fig:syst_finiteV}\\
 Physical & $1024^3$ & 4-neighbours & 1.3751 & 7.5 & $\bar{\lambda}=352.73$ & 30 & \ref{fig:syst_finiteV}\\
  PRS & $8192^3$ & 4-neighbours & 0.55 & 6.80 & $m_ra = 0.2$ & 20 & \ref{fig:attr_xi_PRS_vs_physical},\ref{fig:disc_Gamma_PRS},\ref{fig:disc_mra_q},\ref{fig:disc_Gamma_tot} \\
  PRS & $8192^3$ & 4-neighbours & 0.55 & 7.21 & $m_ra = 0.3$ & 20 & \ref{fig:attr_xi_PRS_vs_physical},\ref{fig:disc_mra_spectrum_PRS},\ref{fig:disc_Gamma_PRS},\ref{fig:disc_mra_q},\ref{fig:disc_Gamma_tot} \\
  PRS & $8192^3$ & 4-neighbours & 0.55 & 7.72 & $m_ra = 0.5$ & 20 & \ref{fig:attr_xi_PRS_vs_physical},\ref{fig:disc_mra_spectrum_PRS},\ref{fig:disc_Gamma_PRS},\ref{fig:disc_mra_q},\ref{fig:disc_Gamma_tot} \\
  PRS & $8192^3$ & 4-neighbours & 0.55 & 8.06 & $m_ra = 0.7$ & 20 & \ref{fig:attr_xi_PRS_vs_physical},\ref{fig:disc_mra_spectrum_PRS},\ref{fig:disc_Gamma_PRS},\ref{fig:disc_mra_q},\ref{fig:disc_Gamma_tot} \\
  PRS & $8192^3$ & 4-neighbours & 0.55 & 8.41 & $m_ra = 1.0$ & 20 & \ref{fig:attr_xi_PRS_vs_physical},\ref{fig:disc_mra_spectrum_PRS},\ref{fig:disc_Gamma_PRS},\ref{fig:disc_mra_q},\ref{fig:disc_Gamma_tot},\ref{fig:syst_mrRL},\ref{fig:syst_cutoffs_PRS} \\
  PRS & $8192^3$ & 4-neighbours & 0.55 & 8.82 & $m_ra = 1.5$ & 20 & \ref{fig:attr_xi_PRS_vs_physical},\ref{fig:disc_mra_spectrum_PRS},\ref{fig:disc_Gamma_PRS},\ref{fig:disc_mra_q},\ref{fig:disc_Gamma_tot} \\
  PRS & $4096^3$ & 4-neighbours & 0.55 & 7.72 & $m_ra = 1.0$ & 30 & \ref{fig:syst_mrRL} \\
  PRS & $2048^3$ & 4-neighbours & 0.55 & 7.03 & $m_ra = 1.0$ & 30 & \ref{fig:syst_mrRL} \\
  PRS & $1024^3$ & 4-neighbours & 0.55 & 6.33 & $m_ra = 1.0$ & 30 & \ref{fig:syst_mrRL} \\
  PRS & $2048^3$ & 4-neighbours & 0.5 & 6.93 & $m_ra = 1.0$ & 1 & \ref{fig:mask_correction} \\
 \hline
\end{tabular*}
\end{minipage}
\end{table*}

\begin{figure}[htbp]
\includegraphics[width=0.48\textwidth]{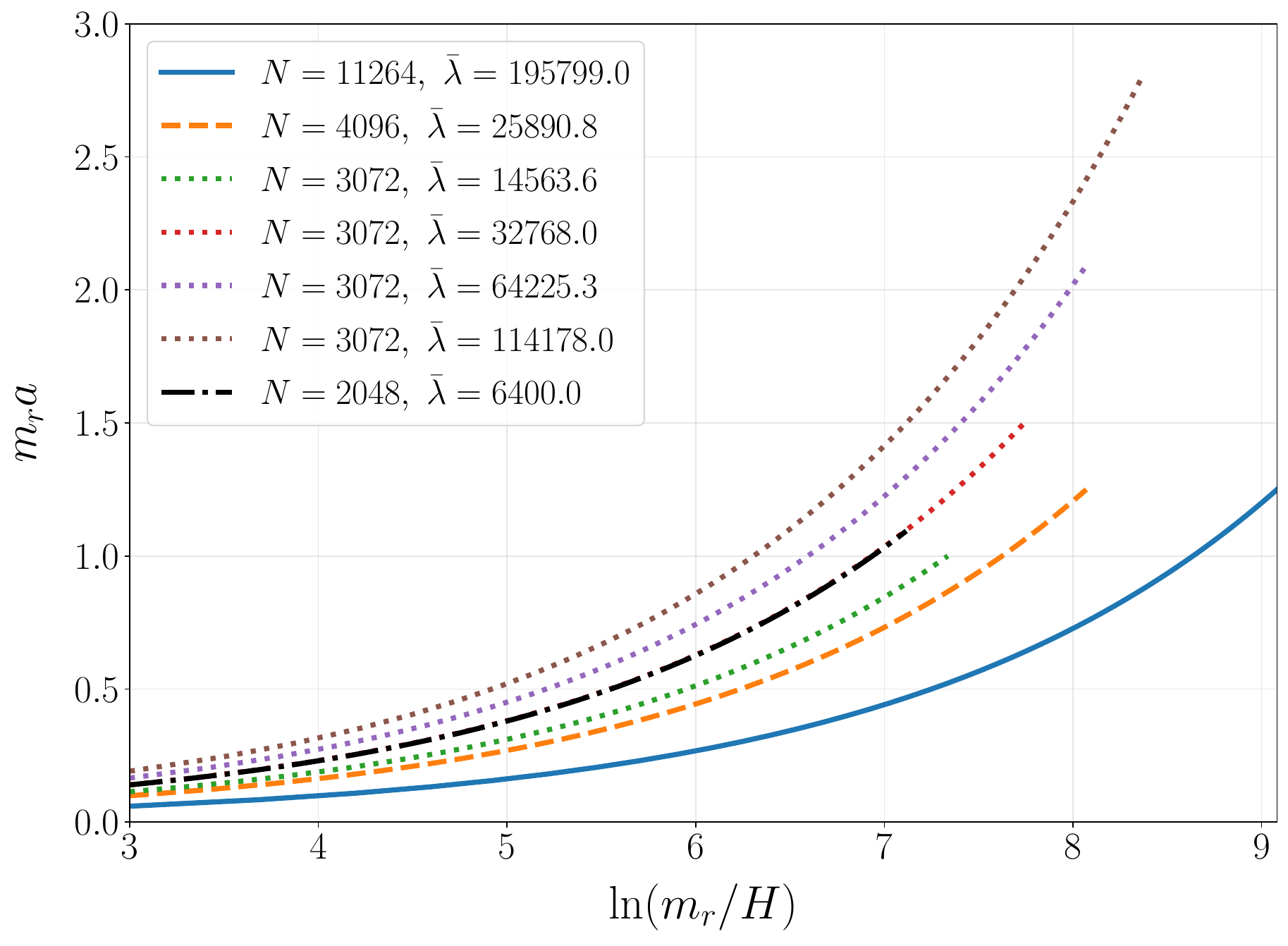}
\caption{Value of $m_ra$ as a function of $\ln(m_r/H)$ for different sets of parameters used in the simulations of physical strings.}
\label{fig:code_mra}
\end{figure}

As mentioned above, the quantity $m_ra$, which can be regarded as a measure of the string core resolution, 
increases with time for physical-type simulations.
In terms of the simulation parameters, it is given (in rescaled units) by, 
\begin{align}
m_ra = \frac{(2\bar{\lambda})^{1/2}L}{N}\tau = \frac{(2\bar{\lambda})^{1/4}L}{N}e^{\ell/2}. \label{mra_vs_parameters}
\end{align}
Figure~\ref{fig:code_mra} shows the evolution of $m_ra$ for parameters of the physical-type simulations used in the main analysis.
As explicitly shown in Eq.~\eqref{mra_vs_parameters}, the dependence on $\ell=\ln(m_r/H)$ is exponential, 
and the resolution gets worse rapidly at large $\ell$.
Obviously, the best case is the simulation with $N=11264$, though the value of $m_ra$ reaches 1.25 at the end of the simulation.

For PRS-type simulations the value of $m_ra$ remains constant.
Hence for such cases we use $m_ra$ as an input parameter instead of $\bar{\lambda}$.
As shown in Table~\ref{tab:simu_parameters}, we performed PRS-type simulations with $8192^3$ lattice sites for 6 different choices of $m_ra$.
In addition, we performed three sets of simulations ($1024^3$, $2048^3$, and $4096^3$ lattice sites) with $m_ra=1$ for the study of the finite volume effects (Appendix~\ref{app:syst_finite_V})
and one $2048^3$ simulation with $m_ra=1$ for the check of the systematics associated with the masking procedure (Appendix~\ref{app:correction matrix}).
For smaller values of $m_ra$ the resolution of the string core improves,  but the maximum reachable value of $\ln(m_r/H)$ decreases.

For PRS-type simulations, the initial conditions are produced as random fluctuations obeying Gaussian distribution in the Fourier space
such that the string density parameter at the initial time ($\ell=0$) becomes $\xi_{\ell=0} \simeq 0.011$, a value found by extrapolating 
the curve of ``attractor" in Ref.~\cite{Gorghetto:2018myk} into $\ell = 0$.
Since the convergence to the attractor in the PRS case is faster than the physical case,
we expect that this choice for the initial condition is enough for the study of the attractor in the PRS-type simulations.
For physical-type simulations, the initial condition should be chosen much more carefully, and we will discuss this issue in details in Appendix~\ref{app:initial_conditons}.

\section{Initial conditions}
\label{app:initial_conditons}

The choice of the initial condition is a tricky part of the simulation, since it is impossible to produce strings in a physically correct way.
Ideally, one should simulate the process of the PQ phase transition to produce strings, 
but assuming that it happens thermally at the temperature $T_{\rm PQ} \sim f_a$ (corresponding to the Hubble parameter $H_{\rm PQ}$) in the radiation dominated epoch, 
a huge scale separation $\ln(m_r/H_{\rm PQ}) \sim 19$ (for $m_r \sim f_a \sim 10^{10}\,\mathrm{GeV}$) is required.
It is still possible to mimic such a thermal initial condition with artificially small $\ln(m_r/H_{\rm PQ})$ 
[as done in some earlier works~\cite{Buschmann:2021sdq,Hiramatsu:2010yu,Kawasaki:2018bzv,Buschmann:2019icd,Yamaguchi:1998gx,Hiramatsu:2012gg,Kawasaki:2014sqa}],
but it is not straightforward to confirm if the unphysical nature of ``thermal" fluctuations can settle down within the limited dynamical range of the simulation.
In this work, we instead follow a perspective initiated by Ref.~\cite{Gorghetto:2018myk}, 
which is to start the simulation from the condition closest to the attractor rather than including the dynamics at the PQ phase transition,
since the latter may be irrelevant at the epoch of the QCD phase transition in which we are mostly interested.

In Sec.~\ref{sec:attractor}, we have identified the initial string density that can be regarded as the attractor. 
However, just tuning the initial string density is not enough to get an accurate result.
The problem is associated with the systematics induced by the oscillation in the spectrum discussed in Sec.~\ref{sec:oscillations}.
There we have seen that the oscillation in the spectrum is produced through physical processes, 
the misalignment oscillation after the horizon crossing and the parametric resonance effect after the saxion mass crossing.
On the other hand, these oscillations can also be produced by some unphysical effects, as we will see below.
One has to choose the initial condition that minimises such unphysical oscillations as much as possible.

To produce an appropriate initial condition, it may be convenient to {\it prepropagate} the field configuration before the beginning of the (physical) simulation.
Namely, we let the fields evolve with some modified (unphysical) equation of motion at the early stage of the simulation, and switch to the physical evolution at some point.
During the prepropagation phase, we modify the equation of motion such that the PQ field quickly settles down into the minimum of the potential, suppressing unphysical field fluctuations.
One possible choice for the prepropagation is to use the PRS-type evolution: At the early times we make a replacement $\bar{\lambda} \to (R_i/R)^2\bar{\lambda}$,
such that the value of $\bar{\lambda}$ is matched to the input value at the beginning of the physical evolution ($R=R_i$).
In addition, we can also prepropagate by only allowing for an evolution in the radial direction $\rho$ of the PQ field $\phi=\rho e^{\theta}$, 
while the angular direction $\theta$ is kept frozen (hereafter we call it the $\rho$-only prepropagation).\footnote{The equation of motion for the $\rho$-only evolution can be implemented
by projecting the equation of motion of $\phi$ into the radial direction.
This can be done by translating the complex space of $\phi=\rho e^{i\theta}$ into cylindrical coordinates $(\rho,\theta)$, and artificially setting the components in the $\theta$ direction to zero.}
In this way, we expect to reduce the fluctuations in the radial direction in the prepropagation phase.

To see how different choices of the initial condition affect the simulation results, 
we performed simulations of physical strings with $1024^3$ lattice sites and the parameter $\bar{\lambda}=1600$.
In this study, we use 4 different ways to produce the initial condition:
the prepropagation with the PRS-type evolution, $\rho$-only prepropagation, no prepropagation with the simulation starting from $\ell=1$,
and no prepropagation with the simulation starting from $\ell=3$.
For the case with the PRS-type prepropagation, we let the prepropagation start from $\ell=2$ and the physical evolution from $\ell=3$.
For the case with the $\rho$-only prepropagation, we let the prepropagation start from $\ell=1$ and the physical evolution from $\ell=3$.
In all cases, the field configurations at the very beginning of the simulation (including the prepropagation phase) are produced as random fluctuations 
obeying a Gaussian distribution in Fourier space, with tuning the cutoff in the distribution such that the string density parameter at $\ell=3$ becomes $\xi_{\ell = 3} \simeq 0.3$.
For each choice of the initial condition, we executed 30 simulations with randomly generated initial field configurations and averaged them.

\begin{figure}[htbp]
$\begin{array}{c}
\subfigure{
\includegraphics[width=0.48\textwidth]{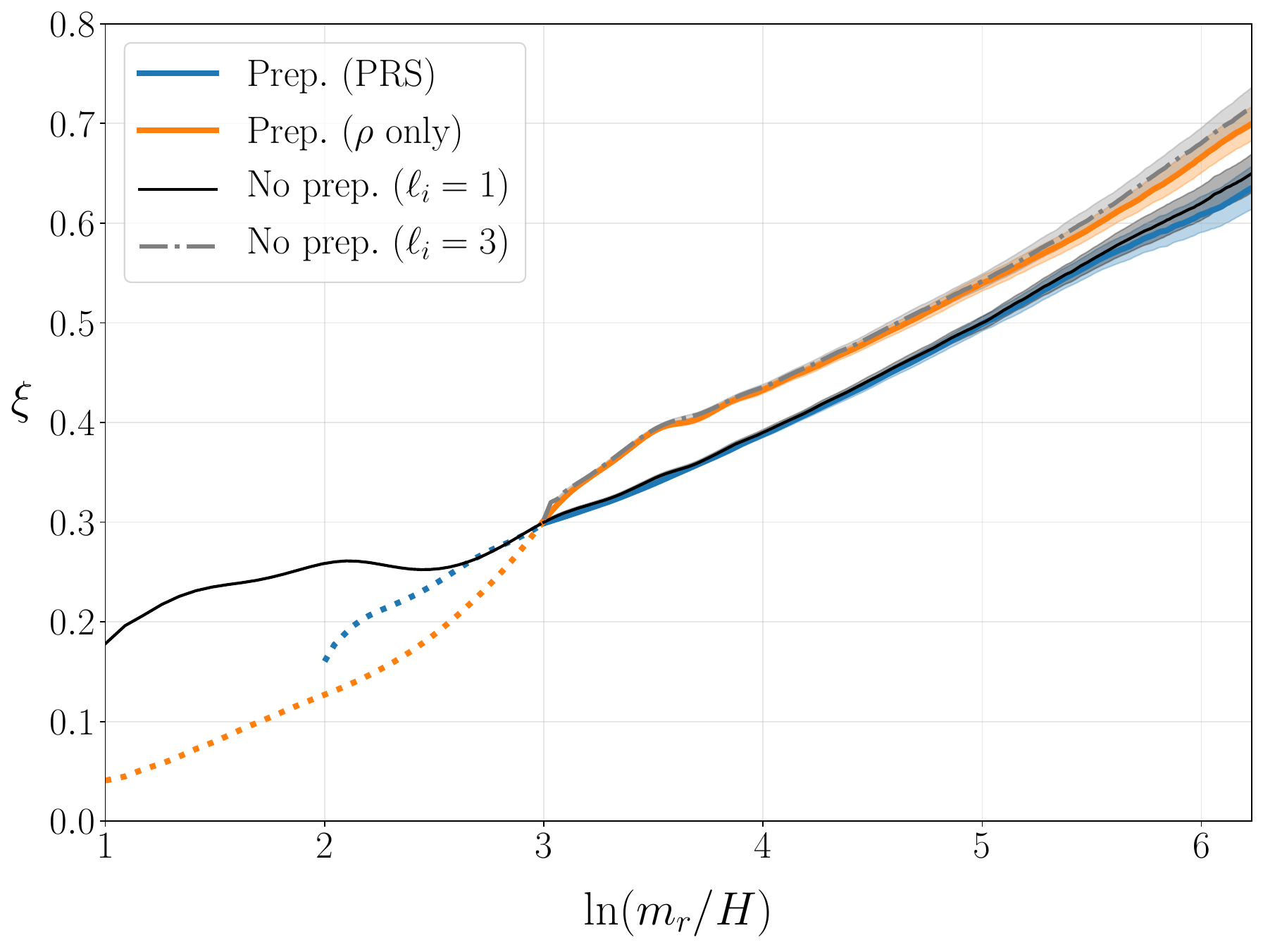}}
\\
\subfigure{
\includegraphics[width=0.48\textwidth]{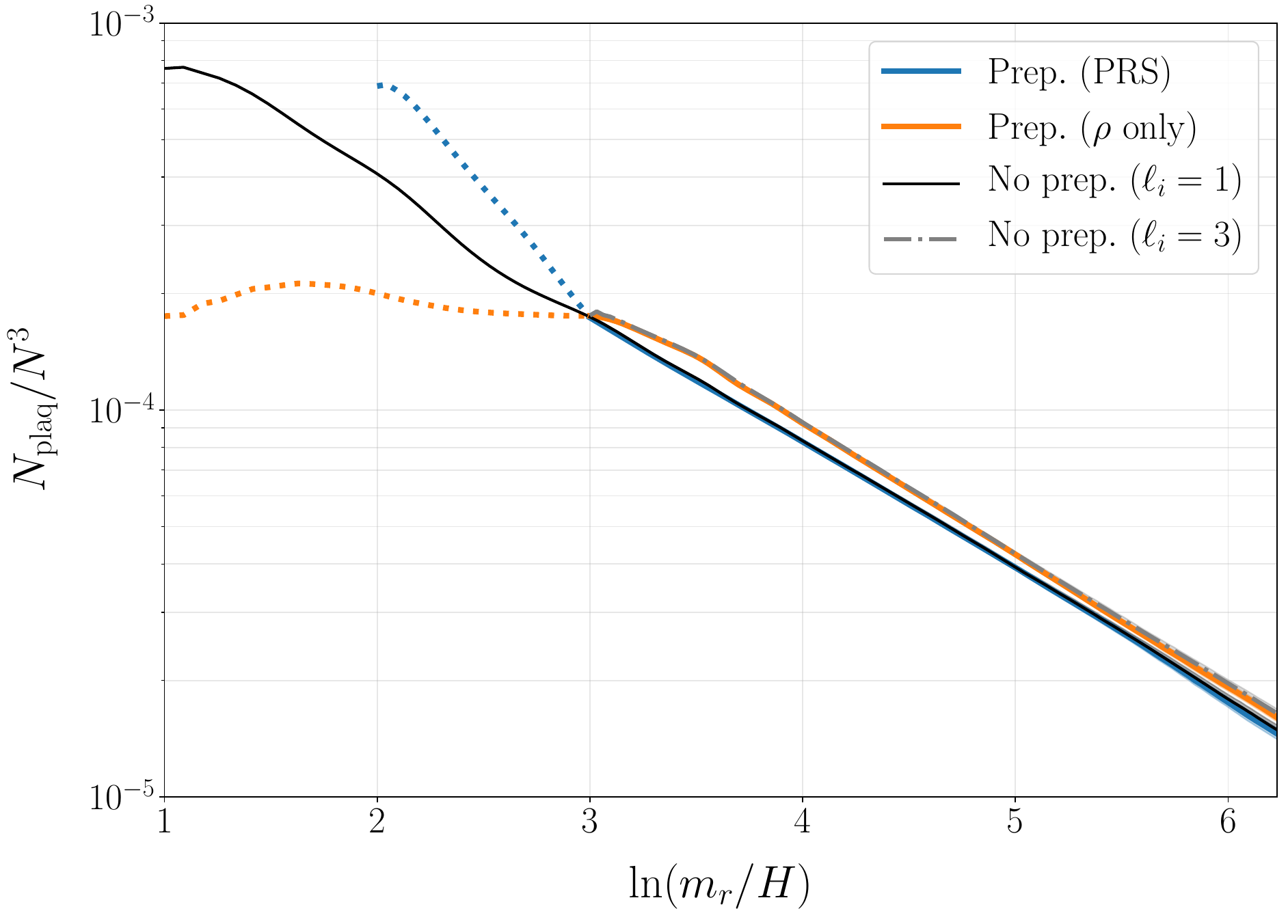}}
\end{array}$
\caption{Evolution of the string density parameter $\xi$ (top panel) and the ratio of the number $N_{\rm plaq}$ of 
plaquettes pierced by strings to the total number of lattice sites $N^3 = 1024^3$ in the simulation box (bottom panel)
for different choices of the initial conditions.
Blue and orange thick lines correspond to the cases with the PRS-type prepropagation and $\rho$-only prepropagation, respectively.
The thin black line and gray dot-dashed line correspond to the cases without prepropagation starting from $\ell=1$ and $\ell=3$, respectively.
The coloured bands represent statistical uncertainties,
and the evolutions during the prepropagation phase are represented by dotted lines.}
\label{fig:init_xi}
\end{figure}

The top panel of Fig.~\ref{fig:init_xi} shows the evolution of the string density parameter $\xi$ for different choices of the initial condition.
We see that $\xi$ becomes larger for the cases with the $\rho$-only prepropagation and without prepropagation starting from $\ell=3$
compared to those with the PRS-type prepropagation and without prepropagation starting from $\ell=1$.
This difference can be easily understood by looking at the evolution of the number $N_{\rm plaq}$ of 
plaquettes pierced by strings,\footnote{We use the statistical method to calculate $\xi$~\cite{Fleury:2015aca},
where the number of plaquettes pierced by strings is related to $\xi$ as
\begin{align}
\xi = \frac{\delta}{6}N_{\rm plaq}\frac{\tau^2}{L^3}. \label{xi_from_Nplaq} 
\end{align} 
We confirmed that the output of this algorithm reproduces the result of the method of directly computing the string length by connecting points pierced 
by a string in a cube, which was used in Refs.~\cite{Yamaguchi:2002zv,Yamaguchi:2002sh,Hiramatsu:2010yu,Hiramatsu:2012gg,Kawasaki:2014sqa}.
} 
which is shown in the bottom panel of Fig.~\ref{fig:init_xi}.
For the case with the $\rho$-only prepropagation, the angular mode is frozen, and hence $N_{\rm plaq}$ remains almost constant during the prepropagation phase.
Since the physical evolution starts from such a frozen configuration at $\ell=3$, it takes some time for the system to reduce $N_{\rm plaq}$ and reach the attractor.
Such a delay in the annihilation of strings results in higher string densities after the prepropagation phase.
It is not surprising that the simulations starting from $\ell=3$ without prepropagation show a similar behaviour, 
since in that case we just put strings in the simulation box with the appropriate number of $N_{\rm plaq}$ 
but do not allow them to have a velocity to annihilate quickly enough to follow the attractor.
On the other hand, for the case with the PRS-type prepropagation, strings acquire an appropriate velocity to annihilate during the prepropagation phase,
and exhibit a smooth transition to the physical evolution.

\begin{figure}[htbp]
\includegraphics[width=0.48\textwidth]{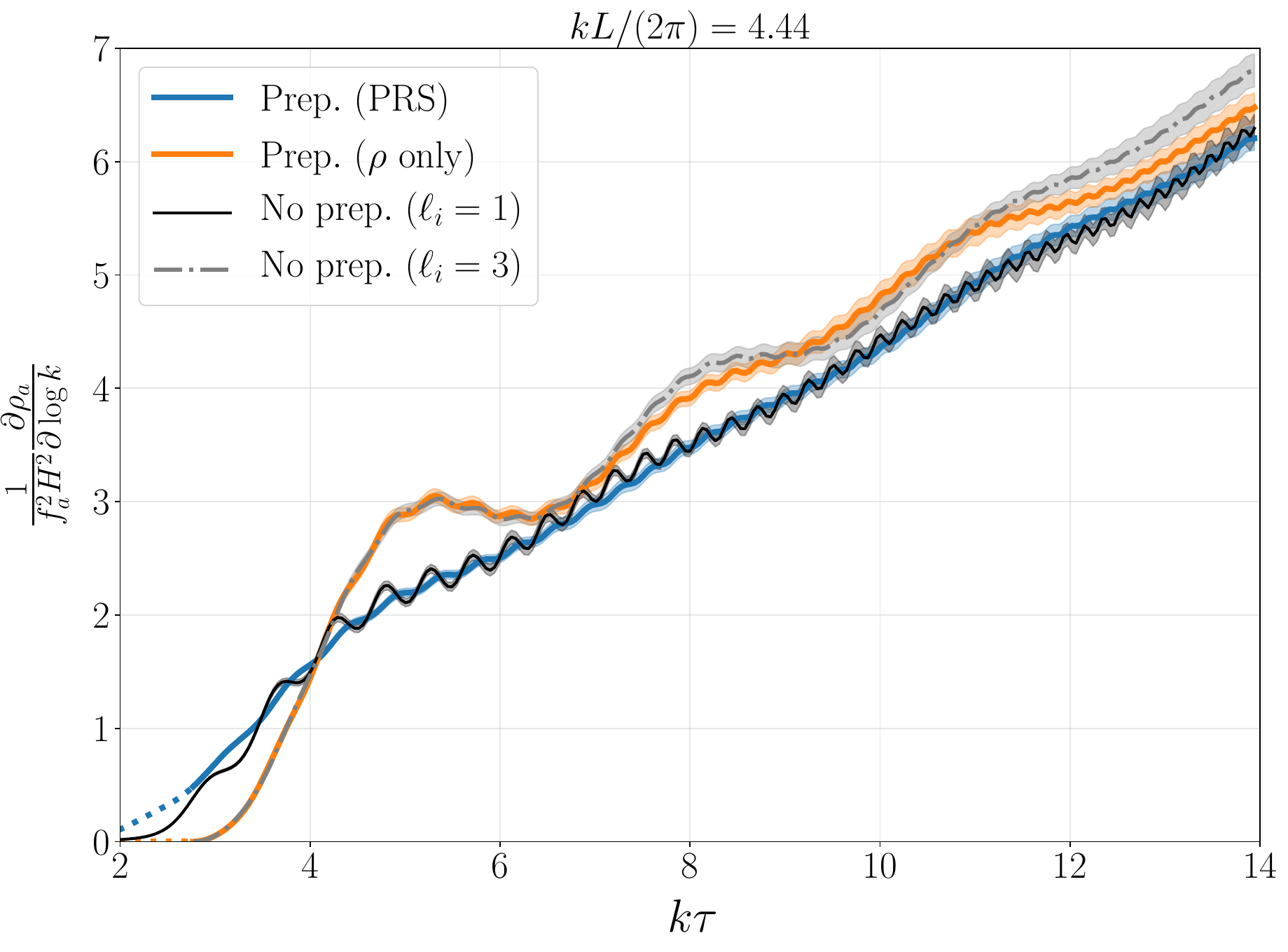}
\caption{Evolution of the energy density of one Fourier mode of the axion field specified by $kL/(2\pi)=4.44$
for different choices of the initial conditions.}
\label{fig:init_mode_evol}
\end{figure}

The unphysical nature of the initial conditions is most noticeable in the time evolution of one Fourier mode of the axion field, which is shown in Fig.~\ref{fig:init_mode_evol}.
From the figure we see that the results of the simulations with the $\rho$-only prepropagation and those without prepropagation starting from $\ell=3$ 
exhibit low frequency oscillations with relatively large amplitudes.
These features can be associated with the misalignment oscillations of the free axion field after the horizon crossing, 
as their frequency roughly corresponds to $2k$ in conformal time.
In these simulations, the field configurations are frozen at the initial time and gradually start to move at $\ell \gtrsim 3$.
During such a process, the angular component of the PQ field tries to realign itself, 
acquiring large kinetic and gradient energies that enhance the amplitude of the misalignment oscillation.
The appearance of such low frequency oscillations can be regarded as unphysical effect, since they originate from the initial conditions where the field configuration is artificially frozen.

\begin{figure}[htbp]
\includegraphics[width=0.48\textwidth]{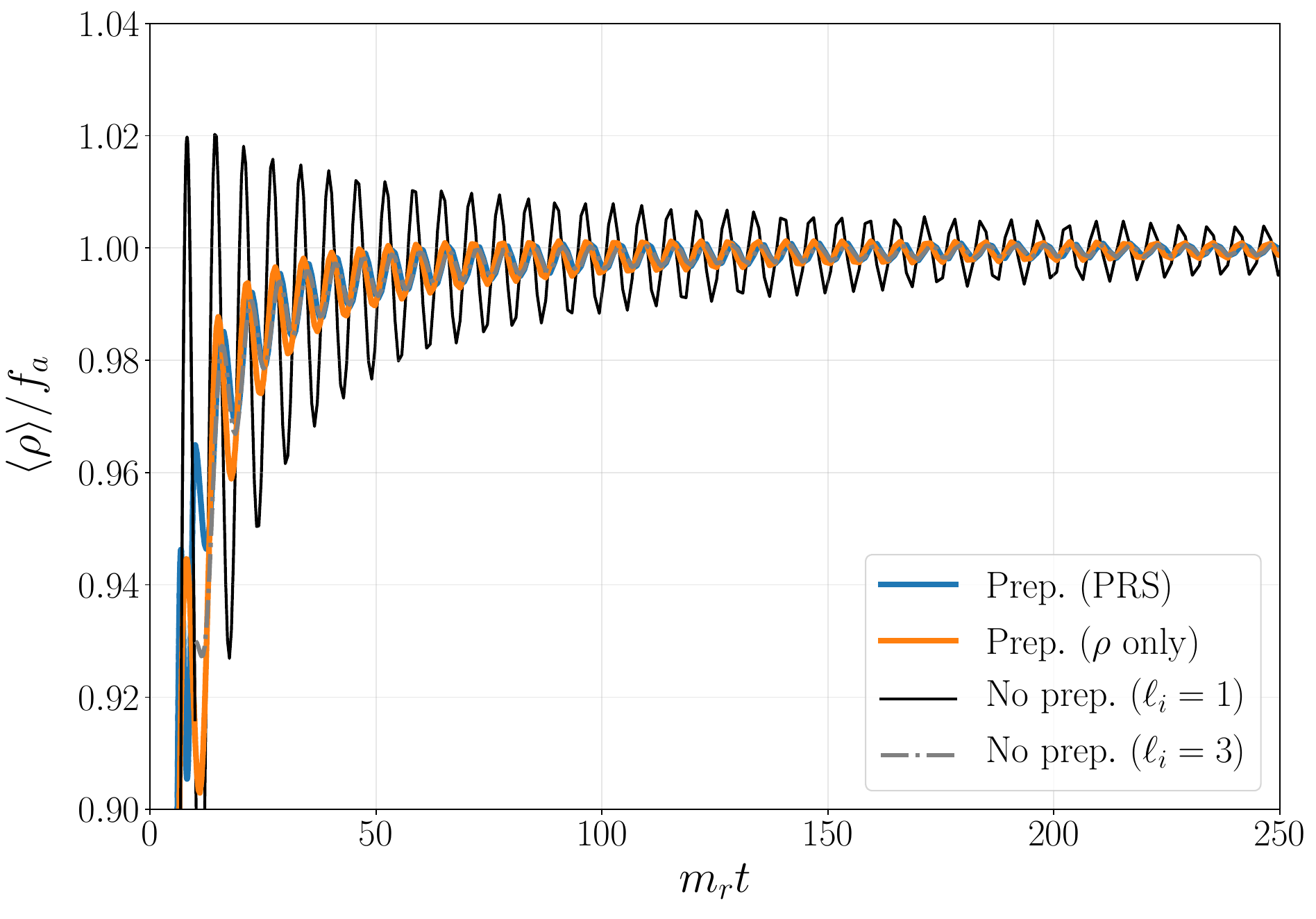}
\caption{Evolution of the spatially averaged value of $\rho=|\phi|$ 
for different choices of the initial conditions. Note that the $x$-axis is given by $m_rt$.}
\label{fig:init_avrho}
\end{figure}

In addition to the aforementioned low frequency oscillations, there are high frequency oscillations in all cases plotted in Fig.~\ref{fig:init_mode_evol}.
We find that these oscillatory features have a constant frequency $m_r$ in cosmic time, as one period of oscillation looks comparable to the
change in $m_rt$ by $2\pi$ when plotted as a function of $m_rt$ rather than $k\tau$.
Such high frequency oscillations can be attributed to the oscillation in the background PQ field around the minimum of the potential.
During the simulation, the background PQ field may have a residual oscillation with frequency $m_r$.
This background oscillation shakes the effective amplitude of the ``axion decay constant" $|\phi|$, 
and hence induces oscillations in the overall amplitude of the axion energy density.
In fact, we have confirmed that the phase of the high frequency oscillations in the axion mode evolution shown in Fig.~\ref{fig:init_mode_evol} 
is aligned with that of the oscillation in the spatial average of the absolute value of the PQ field around $|\phi|=f_a$, which is shown in Fig.~\ref{fig:init_avrho}.

Looking at Figs.~\ref{fig:init_mode_evol} and~\ref{fig:init_avrho}, we see that the amplitude of the high frequency oscillation is particularly large in the case
where simulations start from $\ell=1$ without prepropagation. 
This can be understood as follows.
In the case of no prepropagation with $\ell_i=1$, $\xi$ takes a larger value at early times and relaxes into the attractor until $\ell \lesssim 3$ (see top panel of Fig.~\ref{fig:init_xi}).
The annihilation of such dense string networks leads to larger fluctuations in the radial component of the PQ field,
and hence gives rise to a larger amplitude for the residual oscillation of its averaged value.

\begin{figure}[htbp]
\includegraphics[width=0.48\textwidth]{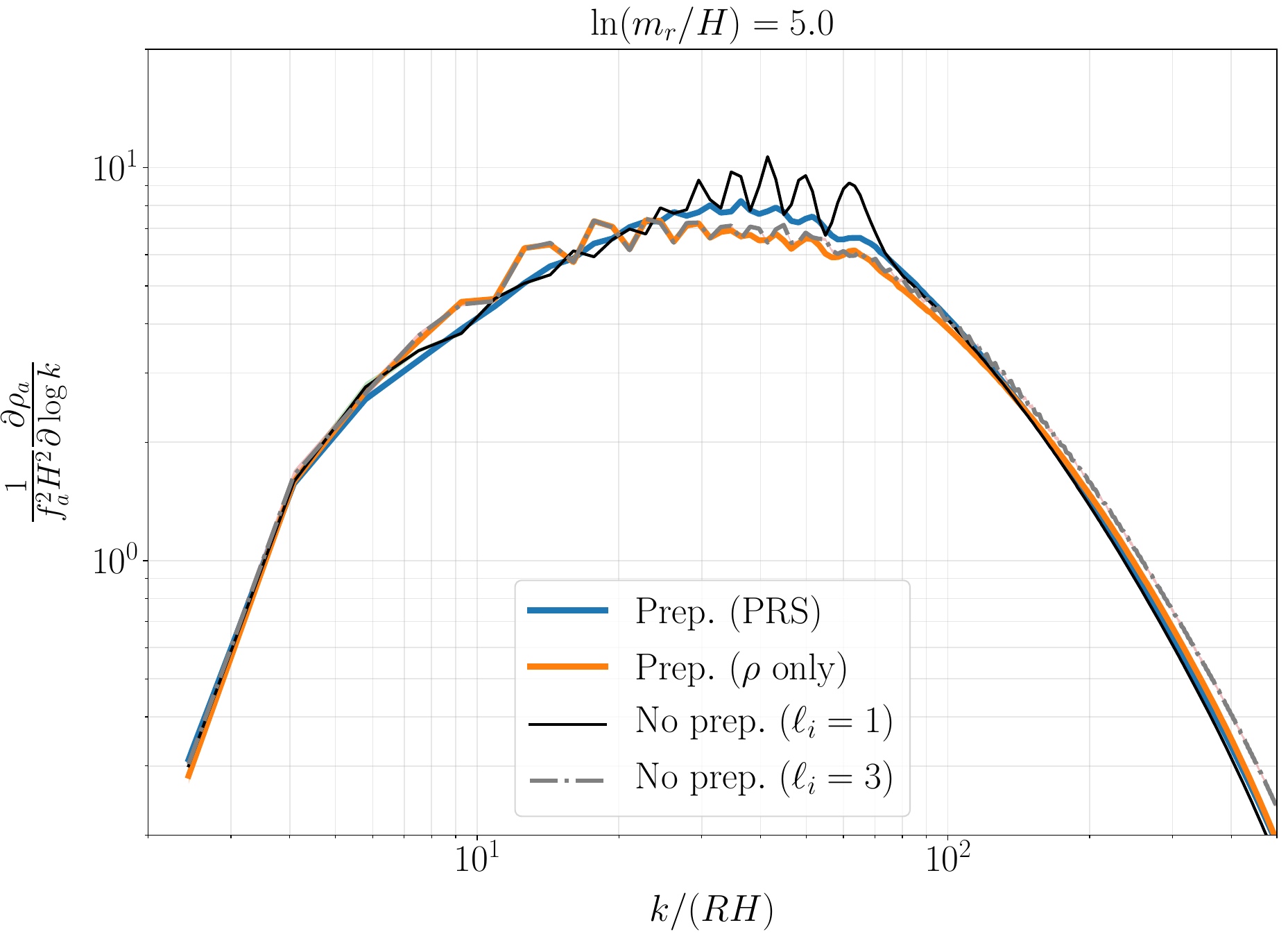}
\caption{Energy density spectrum of axions measured at $\ln(m_r/H)=5$ for different choices of the initial conditions.}
\label{fig:init_spectrum}
\end{figure}

Finally, we compare the energy density spectrum of axions among different choices of the initial conditions in Fig.~\ref{fig:init_spectrum}.
We see that the spectra obtained from the simulations with the $\rho$-only prepropagation and those starting from $\ell=3$ without prepropagation
exhibit relatively large oscillatory features at lower momenta. These can be regarded as artificial features caused by the process of the realignment of frozen initial configurations.
We can also see that the spectrum is weighted towards lower momenta in those cases.
On the other hand, such oscillations at lower momenta are less pronounced than in the other two cases.
In the case where simulations start from $\ell=1$ without prepropagation, there are large oscillatory features at higher momenta.
They can be interpreted as the consequence of the parametric resonance effect induced by the saxion field oscillating with a larger amplitude
due to the imperfect initial condition of over-dense strings.

In summary, we have found that there are two kinds of unphysical effects that can be caused by improper choices of initial conditions.
One is the enhancement in the amplitude of the $2k$-oscillation of the axion field, 
which occurs when the field configurations are frozen (or strings do not have appropriate velocities) at the initial time.
The other is the oscillation with the frequency $m_r$, which can be associated with a large amplitude oscillation of the saxion field induced by an over-dense initial condition.
These artificial features appear to be minimised when we produce the initial condition with the PRS-type prepropagation,
and we adopt it (PRS-type prepropagation from $\ell=2$) for all physical-type simulations performed in this work, except for those presented in this appendix.

\section{Masking method}
\label{app:masking}

In the numerical analysis, we calculate the spectra of axions and saxions by using the Fourier transform of 
the time derivative of the corresponding field components.  Typically, we use twice their kinetic energy densities, $\rho_a =  2\rho_{a,\mathrm{kin}} = \langle \dot{a}^2\rangle$ and $\rho_r = 2\rho_{r,\mathrm{kin}} = \langle \dot{r}^2\rangle$, because the gradients of the axion field are strongly contaminated from the intrinsic gradients surrounding the strings. Moreover, it is much more economical than evaluating gradients. 
In terms of their kinetic energy densities, the spectra of axions and saxions are given by
\begin{align}
\frac{\partial\rho_a}{\partial\log k} &= \frac{k^3}{2\pi^2 L^3}\int\frac{d\Omega_k}{4\pi}|\tilde{\dot{a}}({\bm k})|^2, \label{axion_energy_spectrum}\\
\frac{\partial\rho_r}{\partial\log k} &= \frac{k^3}{2\pi^2 L^3}\int\frac{d\Omega_k}{4\pi}|\tilde{\dot{r}}({\bm k})|^2, \label{saxion_energy_spectrum} 
\end{align}
where $\tilde{\dot{a}}({\bm k})$ and $\tilde{\dot{r}}({\bm k})$ are the Fourier transform of $\dot{a}({\bm x})$ and $\dot{r}({\bm x})$, respectively,
$L^3$ is the comoving volume of the simulation box, and $\int d\Omega_k$ is the integration over the solid angle in the Fourier space.

These kinetic energies can be overestimated around the region close to the string core, and
it was argued that the inclusion of the axion field around such a region can contaminate the spectrum of radiated axions~\cite{Yamaguchi:1998gx,Hiramatsu:2010yu}.
In this appendix, we revisit this issue and study how different methods to remove the data around the string core affect the calculation of
energy densities and spectra of radiated fields.

\subsection{Mask field}
\label{app:mask_field}

To mitigate the contamination from the string core, we may replace the time derivative of the axion or saxion field
with the masked field,
\begin{align}
\dot{X}^{\rm mask}({\bm x}) = M({\bm x})\dot{X}({\bm x}), \label{masked_field}
\end{align}
where $X({\bm x}) = a({\bm x})$, $r({\bm x})$ denotes either the axion field or the saxion field,
and the mask field $M({\bm x})$ vanishes around the string core and becomes unity outside.
Different choice of $M({\bm x})$ would affect the shape of the resulting spectrum of the field.
Here we describe some possible choices of the mask field implemented in our simulation code.

A simple choice is to use the fact that the value of the radial field $|\phi|$ approaches to zero inside the string core:
\begin{align}
M({\bm x}) = \left(\frac{|\phi({\bm x})|}{f_a}\right)^k, \label{mask_phi_k}
\end{align}
where $k$ is a positive number. This function is inspired by Ref.~\cite{Gorghetto:2018myk}, in which $k=1$ was used.
One could choose other values of $k$, which affects the profile of the masked region.
A larger value of $k$ would effectively increase the radius of the masked region and make the mask more abrupt.

Another possible choice of $M({\bm x})$ is the top-hat mask, which is obtained as follows.
We select all four corners of a plaquette pierced by a string.
These points are closer to the string core center than the lattice spacing $\delta = L/N$. We create an anti-mask field,
$W_0({\bm x})$, where all the points of a plaquette are valued as $W_0 = 1$, and the rest as $W_0 = 0$. This
would be a pretty sharp top-hat distribution corresponding to a masking length $\sim \delta$ (in the comoving coordinate). 
We now apply a Gaussian filter to this distribution, with a smoothing length $\tilde{l}$,
\begin{align}
W_1({\bm x}) = \int\frac{d^3p}{(2\pi)^3}e^{-i{\bm p} \cdot {\bm x}}e^{-|{\bm p}|^2\tilde{l}^2/2}\int d^3 x'W_0({\bm x'})e^{i{\bm p} \cdot {\bm x'}}.
\end{align}
Applied to $W_0({\bm x}) = \delta^3\cdot\delta^{(3)}({\bm x}-{\bm x}_0)$ 
for a point ${\bm x}_0$ pierced by a string in a continuum limit, it gives,
\begin{align}
W^{\rm point}_1({\bm x}) &= \delta^3\int\frac{d^3p}{(2\pi)^3}e^{-|{\bm p}|^2\tilde{l}^2/2}e^{i{\bm p}\cdot({\bm x}_0-{\bm x})} \nonumber\\
&= \left(\frac{\delta}{\sqrt{2\pi}\tilde{l}}\right)^3e^{-\frac{|{\bm x}_0 - {\bm x}|^2}{2\tilde{l}^2}},
\label{W_1_evaluated}
\end{align}
effectively distributing the value of $W_0({\bm x} \sim {\bm x}_0) = 1$ in a region of radius $\sim \tilde{l}$.

A collection of plaquette vertices along a string forms a sort of one dimensional distribution of points,
that will be smoothed to a cylinder of radius $\sim\tilde{l}$ when applying the same filter. 
One can compute the effective value of $W_1$ at the desired distance as a function of $\tilde{l}$.

Assuming $\delta\ll\tilde{l}$, we can use the continuum description, and we only have to integrate~\eqref{W_1_evaluated}
with ${\bm x}_0$ along a line with a point number per unit length $p_L$ (or points per lattice spacing $\delta$, $p_{\delta}$). 
As a function of the comoving distance to the center of the string core $r_c$, we have
\begin{align}
W_1^{\rm string} &\simeq p_L\int dL\left(\frac{\delta}{\sqrt{2\pi}\tilde{l}}\right)^3e^{-\frac{L^2+r_c^2}{2\tilde{l}^2}} \nonumber\\
&=p_{\delta}\left(\frac{\delta}{\sqrt{2\pi}\tilde{l}}\right)^2e^{-\frac{r_c^2}{2\tilde{l}^2}}.
\end{align}
Therefore, choosing a filter with $\tilde{l} = r_c$, the points satisfying
\begin{align}
W_1({\bm x}) > p_{\delta}\left(\frac{a}{\sqrt{2\pi}r_{\rm mask}}\right)^2e^{-\frac{1}{2}}
\end{align}
are closer than $r_{\rm mask}$ to the string core, where $r_{\rm mask} = Rr_c$ is the physical radius.

In the discrete case, rather than computing it, we can calibrate it by direct calculation.
We build the mask by choosing the points where $W_1$ has values above a critical value,
which was calibrated to be:
\begin{align}
(W_1)_c = p_c\left(\frac{1}{\sqrt{2\pi}l_c}\right)^2e^{-\frac{n^2}{2l_c^2}},\quad
l_c = cn/2,
\end{align}
with $p_c = 2.5$ and $c = 1.25$,
where $n$ is a distance from the string core (in lattice units) which we want to mask.
The top-hat mask is thus defined as
\begin{align}
M({\bm x}) =
\begin{cases}
0\ (\mathrm{inside}) & \mathrm{for}\quad W_1 > (W_1)_c, \\
1\ (\mathrm{outside}) & \mathrm{for}\quad W_1 < (W_1)_c. \label{mask_top_hat}
\end{cases}
\end{align}
The algorithm to build the top-hat mask based on the above procedure is implemented in {\tt Jaxions},
which enables us to mask the fields for arbitrary values of $r_{\rm mask}$.

\subsection{Energy density and spectrum}
\label{app:mask_energy_spectrum}

From the data of the spatial distribution of the PQ scalar $\phi({\bm x})$ and $\dot{\phi}({\bm x})$ in the simulation box, we can compute the energy densities of axions and saxions.
Substituting the decomposition~\eqref{field_decomposition} into the kinetic energy density of the PQ field, we have
$\frac{1}{2}|\dot{\phi}|^2 = \frac{1}{2}\dot{r}^2 + \frac{1}{2}|\phi|^2\dot{\theta}^2$,
and hence it is natural to define the kinetic energy densities of radial and angular fields as
\begin{align}
\rho_{r,\mathrm{kin}} = \frac{1}{2}\dot{r}^2,\quad \rho_{a,\mathrm{kin}} = \frac{1}{2}|\phi|^2\dot{\theta}^2 \label{kinetic_energy_density_r_a}
\end{align}
After obtaining the top-hat mask, we can compute the masked energy density as an average over all masked points,
\begin{align}
\rho^{\rm mask}_i = \frac{2}{N^3 - N_{\rm mask}}\left(\sum_{{\bm x}}\rho_{i,\mathrm{kin}}({\bm x}) - \sum_{\text{inside}}\rho_{i,\mathrm{kin}}({\bm x})\right),
\end{align}
for $i=r,a$, where $N_{\rm mask}$ is the number of masked points, $\sum_{\text{inside}}$ denotes the sum over the points inside the masked region (i.e. points on which $M({\bm x})=0$), 
and the overall factor $2$ represents the fact that we estimate the energy density based on twice of the kinetic energy density 
in order to avoid including the long range gradient energy, which should be regarded as a part of the energy density of strings.

\begin{figure}[htbp]
$\begin{array}{c}
\subfigure{
\includegraphics[width=0.48\textwidth]{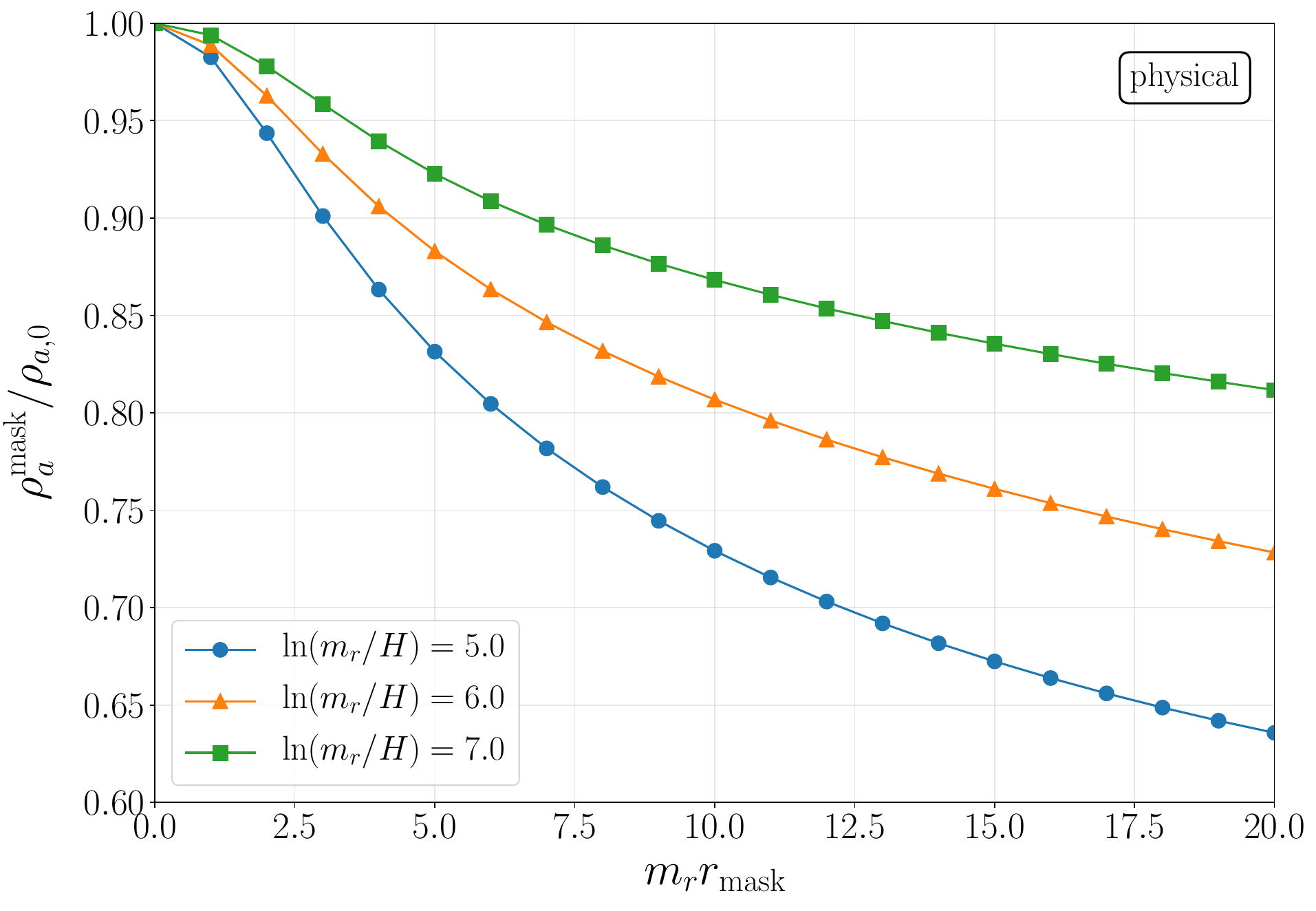}}
\\
\subfigure{
\includegraphics[width=0.48\textwidth]{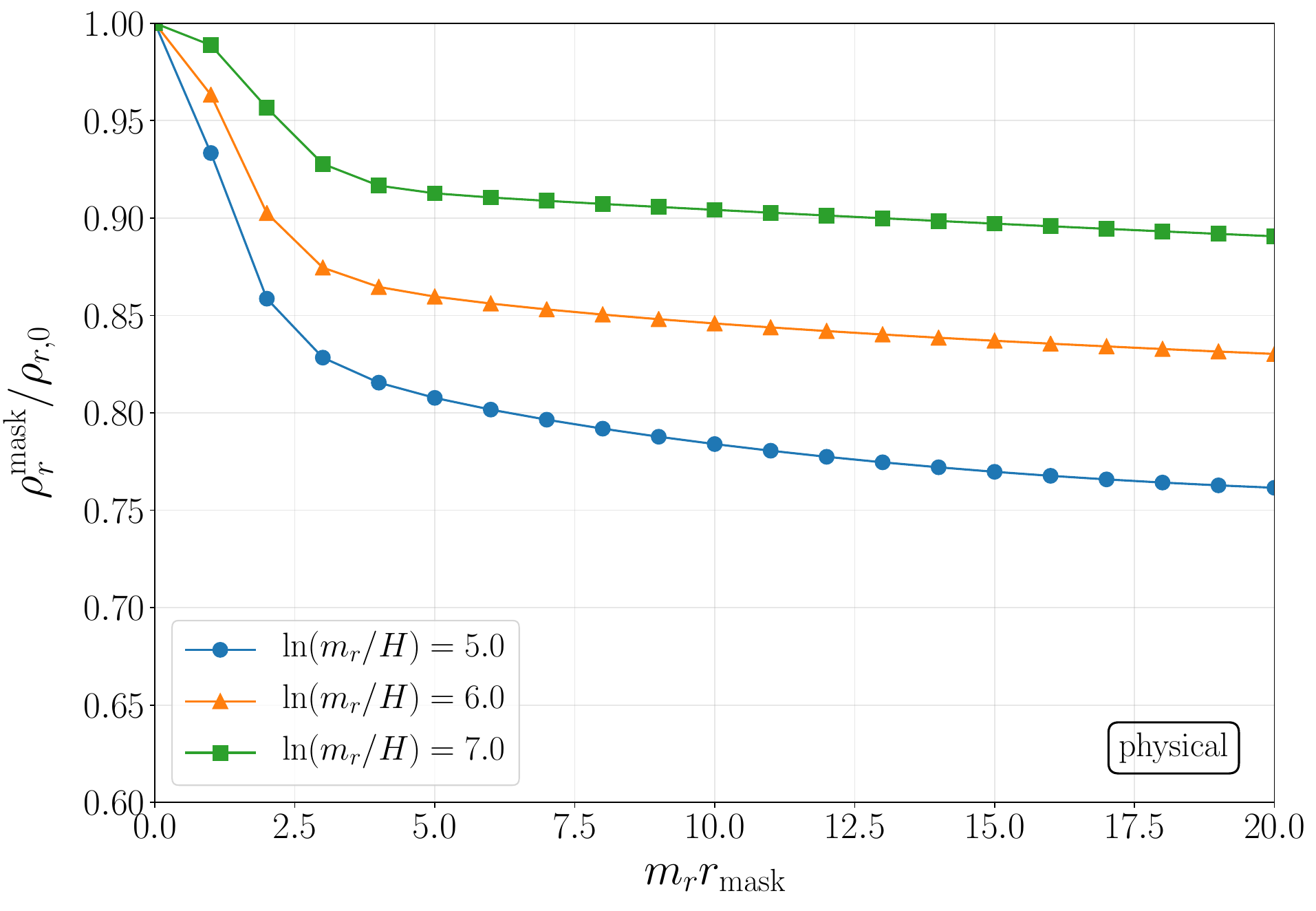}}
\end{array}$
\caption{The ratio of the masked energy density to that without the mask for axions (top panel) and saxions (bottom panel)
plotted as functions of the radius $r_{\rm mask}$ of the masked region (in the unit of $m_r$).
The different colours and markers correspond to the different values of $\ln(m_r/H)$.
The results are obtained from simulations of physical strings with $3072^3$ lattice sites and the parameter $\bar{\lambda}=14563.6$.
Each marker represents the average of 30 simulations, where the statistical uncertainty is negligibly small.}
\label{fig:mask_EA_ES}
\end{figure}

Figure~\ref{fig:mask_EA_ES} shows the ratio of the masked energy density $\rho^{\rm mask}_i$ to that without the mask $\rho_{i,0}$ 
for different choices of the radius $r_{\rm mask}$ of the top-hat mask, obtained from simulations of physical strings.
Here we computed energy densities with 20 different choices of the mask radius, ranging from $r_{\rm mask}=m_r^{-1}$ to $r_{\rm mask}=20m_r^{-1}$.
We see that the overall suppression due to the mask becomes weaker for larger values of $\ln(m_r/H)$ both for the axion and saxion energy densities.
This is simply because the fraction of the masked points (or the number of grid points pierced by strings) in the simulation box gets smaller at late times.
From the figure we can also see that there is a noticeable difference between the energy density of axions and that of saxions.
For axions, the ratio $\rho_a^{\rm mask}/\rho_{a,0}$ continues to decrease with increasing $r_{\rm mask}$, while
for saxions the suppression becomes milder for $r_{\rm mask}\gtrsim 5m_r^{-1}$.
This fact implies that there is a rather clear distinction between the kinetic energy of the moving string core and that of radiations for the radial field,
while it is hard to find such a distinction for the angular field.

\begin{figure*}[htbp]
\includegraphics[width=0.85\textwidth]{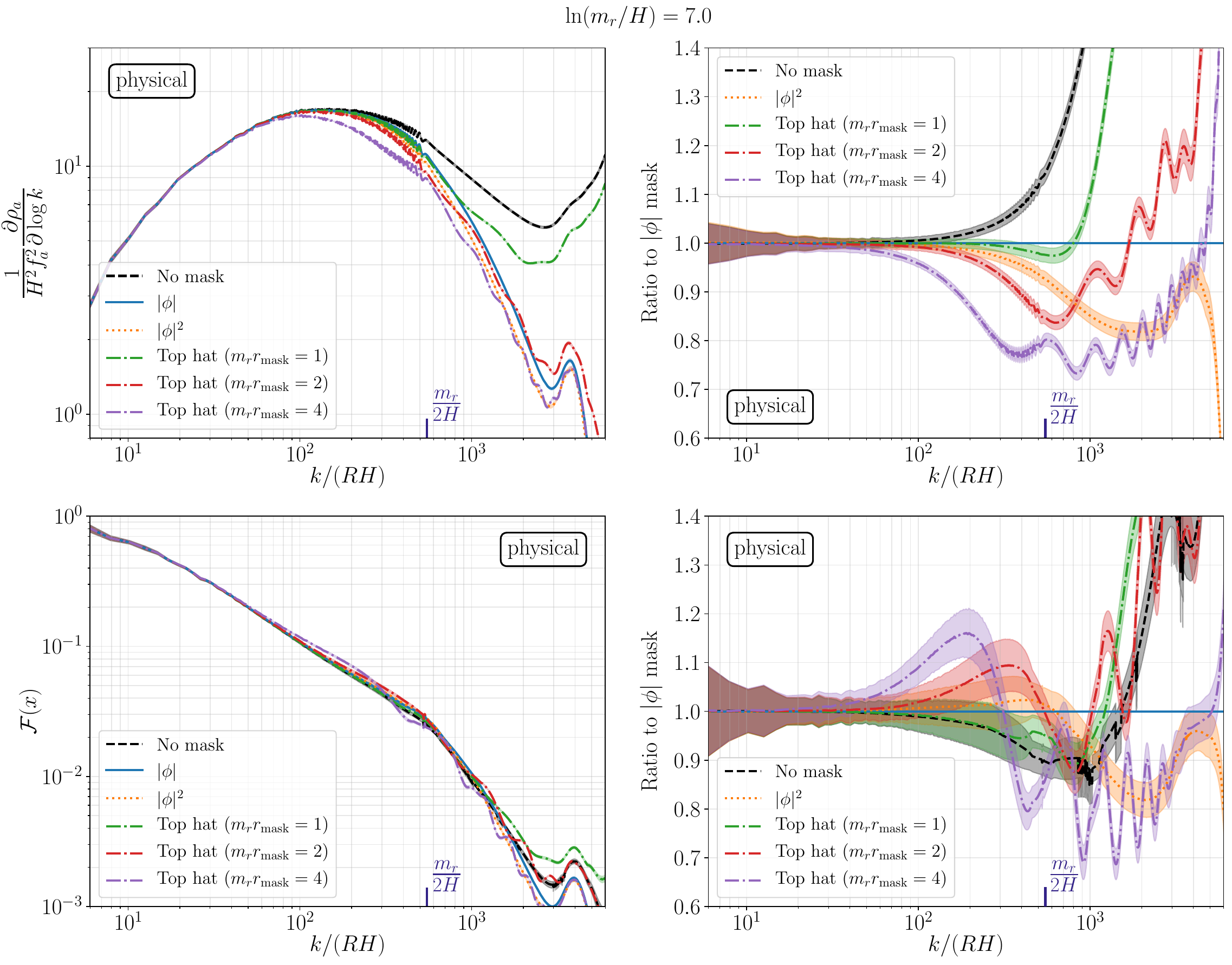}
\caption{The axion energy density spectrum (top left) and instantaneous emission spectrum (bottom left) at $\ln(m_r/H)=7$
obtained from simulations of physical strings with $4096^3$ lattice sites for different choices of the mask method.
The results without the mask [$M({\bm x}) = 1$ for all ${\bm x}$] are shown by black dashed lines,
those with the $|\phi|^k$ mask [Eq.~\eqref{mask_phi_k}] for $k=1$ and $k=2$ are shown by blue solid lines and orange dotted lines, respectively,
and those with the top-hat mask [Eq.~\eqref{mask_top_hat}] for three different choices of $r_{\rm mask}$ are shown by coloured dash-dotted lines.
The ratio of the energy density spectrum (instantaneous emission spectrum) for each choice of the mask to that for the $|\phi|$ mask is also shown in the top right (bottom right) panel.
The coloured bands represent statistical uncertainties, and the momentum corresponding to $k/R = m_r/2$ is marked with dark blue ticks.}
\label{fig:mask_spectrum}
\end{figure*}

We can also compute the masked energy spectrum as
\begin{align}
\frac{\partial\rho^{\rm mask}}{\partial\log k} &= \frac{k^3}{2\pi^2 L^3}\int\frac{d\Omega_k}{4\pi}|\tilde{\dot{X}}^{\rm mask}({\bm k})|^2,
\end{align}
where $\tilde{\dot{X}}^{\rm mask}({\bm k})$ is the Fourier transform of the masked field~\eqref{masked_field}.
Figure~\ref{fig:mask_spectrum} shows a comparison between different choices of the mask field.
We see that the difference is most pronounced at higher momenta,
and the energy density spectrum (top panel of Fig.~\ref{fig:mask_spectrum}) is more suppressed for larger $r_{\rm mask}$ of the top-hat mask or larger $k$ of the $|\phi|^k$ mask.
For the top-hat mask with $r_{\rm mask}=4m_r^{-1}$, there is a substantial suppression even at lower momenta, $k/R \lesssim m_r/2$.
Such a suppression leads to an underestimate of the axion energy density, and is compatible with the trend observed in the top panel of Fig.~\ref{fig:mask_EA_ES}.
We also observe that the amplitude of the instantaneous emission spectrum $\mathcal{F}(x)$ increases at intermediate momenta for larger values of $r_{\rm rmask}$.
This can be understood by noting that the effect of the mask gets weaker at late times as shown in Fig.~\ref{fig:mask_EA_ES},
which can result in a relative increase in the energy density for certain modes.
Such an artificial increase due to the mask is more pronounced for larger values of $r_{\rm rmask}$, leading to higher values of $\mathcal{F}(x)$.

\begin{figure}[htbp]
\includegraphics[width=0.48\textwidth]{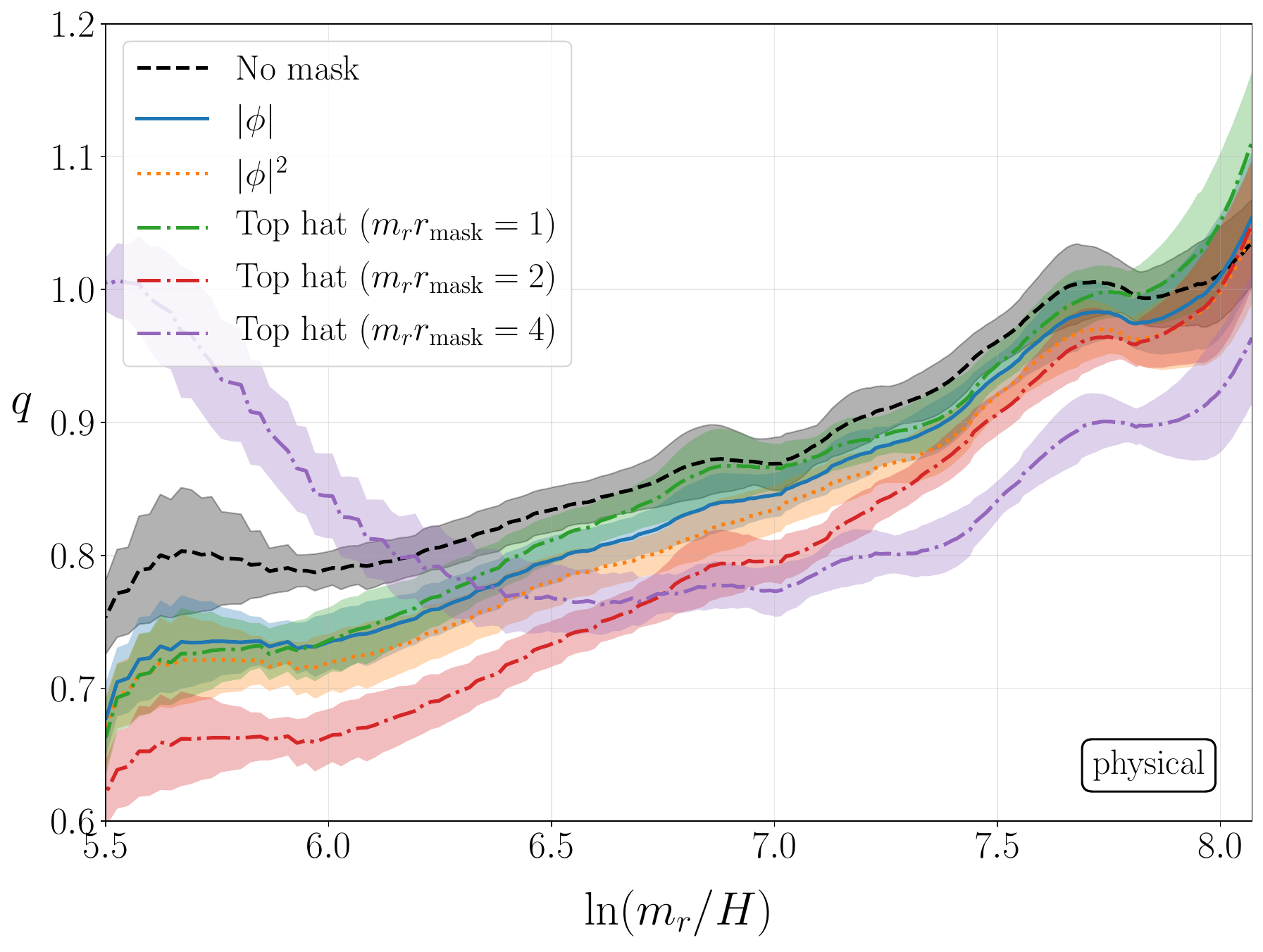}
\caption{Evolution of the spectral index $q$ of the instantaneous emission spectrum for different choices of the mask method,
obtained from simulations of physical strings with $4096^3$ lattice sites. 
The coloured bands represent the error induced by changing the parameter $\sigma_{\rm filter}$ for the filtering procedure to calculate $\mathcal{F}$ [Eq.~\eqref{filter_for_F}] in addition to statistical uncertainties.}
\label{fig:mask_q}
\end{figure}

The impact of the different choices of the masking method on the estimation of the spectral index $q$ is shown in Fig.~\ref{fig:mask_q}.
We see that $q$ takes larger values for results without the mask, and smaller values when we increase the radius of the masked region.
This trend is consistent with the above observation that the amplitude of $\mathcal{F}(x)$ becomes larger at intermediate momenta when we increase the value of $r_{\rm mask}$.
We can also see that the difference in $q$ becomes less pronounced at larger values of $\ln(m_r/H)$. 
This implies that the effect of the masking of the spectrum is relevant only at higher momenta, whose relative importance becomes diminished 
as the IR and UV parts of the spectrum are sufficiently separated at large $\ln(m_r/H)$.

Since the different choice of $r_{\rm mask}$ is likely to modify the axion spectrum at intermediate momenta,
the applicability of the top-hat mask to the estimation of the axion spectrum seems obscure.
In the main analysis of this paper, we use the $|\phi|$ mask [Eq.~\eqref{mask_phi_k} with $k=1$] as our fiducial choice for the calculation of the energy density and spectrum of axions, 
since it is equivalent to the $|\phi|^2$ factor in Eq.~\eqref{kinetic_energy_density_r_a}, being the natural normalization factor for the energy density of the angular field.
On the other hand, for the case of saxions, it is clear from the bottom panel of Fig.~\ref{fig:mask_EA_ES} that most of the contribution of the string core can be screened for $r_{\rm mask}\gtrsim 5m_r^{-1}$,
and we use the top-hat mask with $r_{\rm mask}=5m_r^{-1}$ for the calculation of the energy density and spectrum of saxions in our main analysis.

\subsection{Comment on the correction matrix method}
\label{app:correction matrix}

The spectrum built from the masked field $\dot{X}^{\rm mask}({\bm x})$ may still be imperfect as it contains
``defects" at the masked points. It was argued that they can be corrected 
by using the Pseudo Power Spectrum Estimator (PPSE)~\cite{Hivon:2001jp,Hiramatsu:2010yu}.
Here we investigate its effect on the spectrum and show that the correction does not play an important role in the large scale simulations.

For a spectrum of a masked field $P^{\rm mask}_X(k) \equiv \int\frac{d\Omega_k}{4\pi}|\tilde{\dot{X}}^{\rm mask}({\bm k})|^2$, its PPSE is given by~\cite{Hivon:2001jp}
\begin{align}
P_X(k) = \int\frac{dk'k'^2}{2\pi^2}\mathcal{M}^{-1}(k,k')\langle P^{\rm mask}_X(k')\rangle, \label{PPSE_formula}
\end{align}
where $\langle\dots\rangle$ denotes an ensemble average over many realisations, 
$\mathcal{M}^{-1}(k,k')$ satisfies $\int\frac{dk'k'^2}{2\pi^2}\mathcal{M}^{-1}(k,k')\mathcal{M}(k',k'') = (2\pi^2/k^2)\delta(k-k'')$,
and the matrix $\mathcal{M}(k,k')$ can be built from the Fourier transform of the mask field $\tilde{M}({\bm k})$,
\begin{align}
&\mathcal{M}(k,k') = \frac{1}{L^3}\int\frac{dk''k''^2}{2\pi^2}\int\frac{d\Omega_{k''}}{4\pi}\langle|\tilde{M}({\bm k''})|^2\rangle J(k,k',k''), \\
&J(k,k',k'') = \int\frac{d\Omega_k}{4\pi}\int\frac{d\Omega_{k'}}{4\pi}(2\pi)^3\delta^{(3)}({\bm k}+{\bm k'}+{\bm k''}). \label{J_function}
\end{align}
In practice, we may calculate $\mathcal{M}(k,k')$ without taking the ensemble average of $|\tilde{M}({\bm k''})|^2$, 
and expect that Eq.~\eqref{PPSE_formula} holds as an approximation.

An important feature of the correction matrix $\mathcal{M}^{-1}$ is that it compensates the suppression of the masked spectrum at lower $k$.
Let us consider the $k\to 0$ limit of the masked field $\tilde{\dot{X}}^{\rm mask}({\bm k})$.
Introducing the new anti-mask field $W_2({\bm x}) \equiv 1 - M({\bm x})$, which becomes unity inside the masked region
and vanishes outside it, we have $\tilde{\dot{X}}^{\rm mask}({\bm k}\to {\bm 0}) = \int d^3x(1-W_2({\bm x}))\dot{X}({\bm x})$.
Since $W_2({\bm x})$ is relevant only inside the masked region, the amplitude of the lower modes is of order
$\tilde{\dot{X}}^{\rm mask}({\bm k}\to {\bm 0}) \sim (1-N_{\rm mask}/N^3)\tilde{\dot{X}}({\bm k}\to {\bm 0})$.
Hence we expect that the lower modes are suppressed compared to the bare ones by a factor given by the masked volume,
\begin{align}
P_X^{\rm mask}(k\to 0) \sim \left(1 - \frac{2N_{\rm mask}}{N^3}\right)P_X(k\to 0).
\nonumber
\end{align}
Since Eq.~\eqref{J_function} gives $J(k,k',k'') = \pi^2/(kk'k'')$ for $|k'-k''|\le k\le k'+k''$, which implies that the contributions from higher $k$ are suppressed by
some inverse power of $k$, the dominant contribution to the $k\to 0$ mode is
$\mathcal{M}(0,0) \sim L^{-6}\langle|M({\bm k''})|^2\rangle_{k''\to 0}J(0,0,0) \sim L^{-3}(1-2N_{\rm mask}/N^3)$.
Therefore, by evaluating the corrected spectrum as Eq.~\eqref{PPSE_formula},
we expect a cancellation of the suppression effect on the masked spectrum.

\begin{figure}[htbp]
$\begin{array}{c}
\subfigure{
\includegraphics[width=0.48\textwidth]{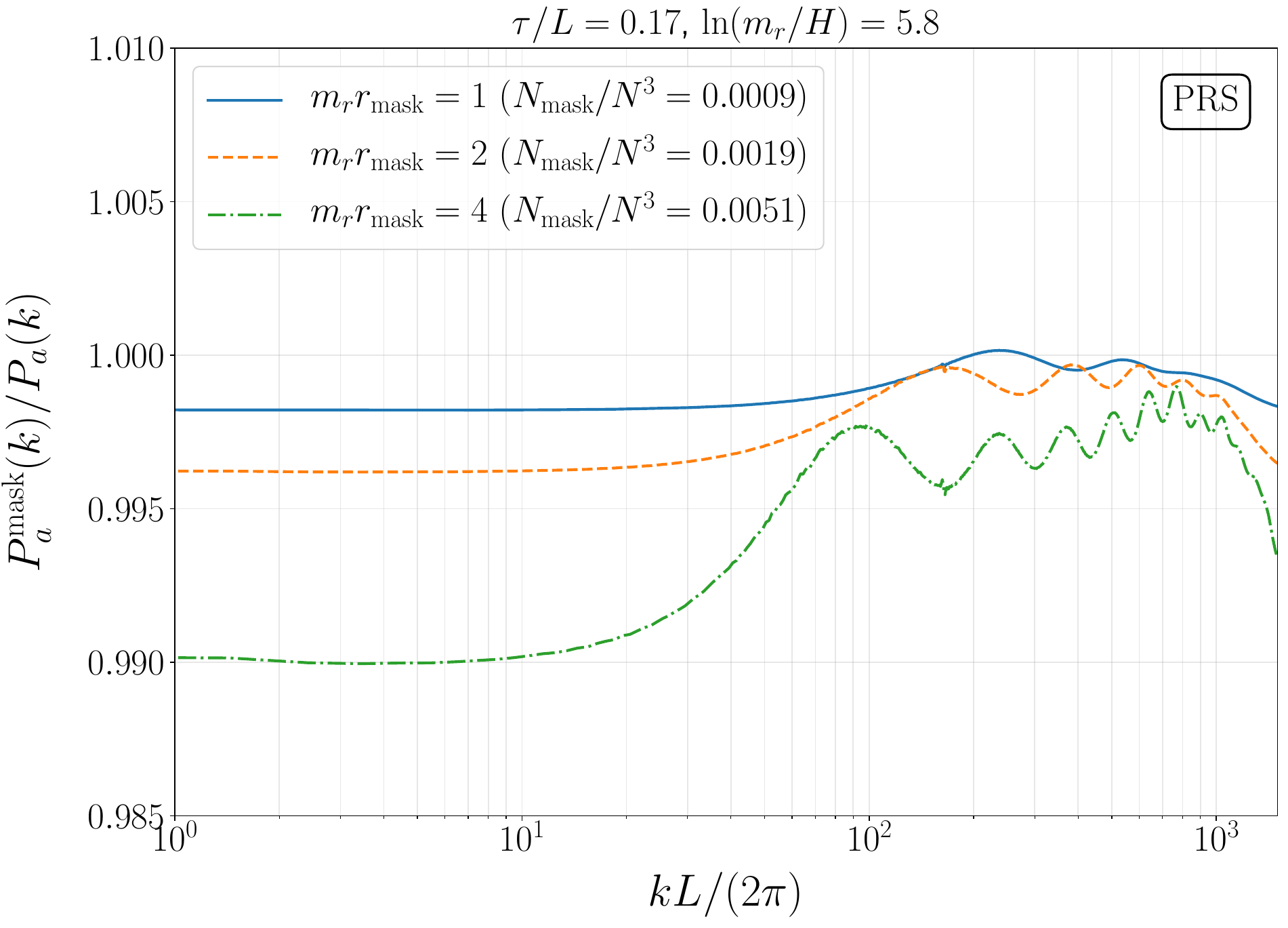}}
\\
\subfigure{
\includegraphics[width=0.48\textwidth]{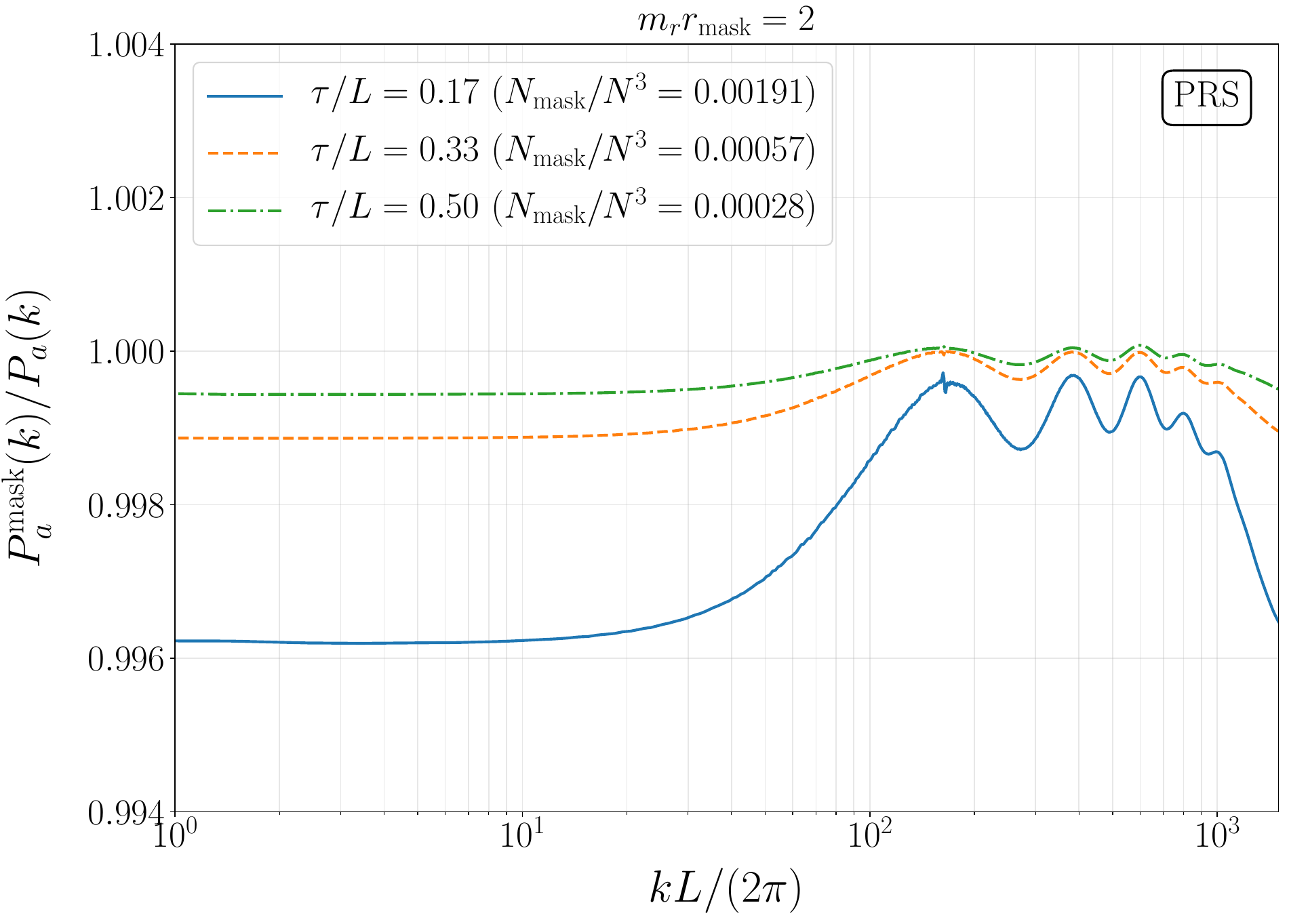}}
\end{array}$
\caption{Ratio of the axion spectrum built with the top-hat mask to that corrected by the matrix via Eq.~\eqref{PPSE_formula}.
Top panel shows a comparison among different choices of $r_{\rm mask}$ at the same conformal time $\tau/L=0.17$,
and bottom panel shows a comparison among spectra measured at different times for $r_{\rm mask}=2m_r^{-1}$.
For each line, the ratio of the number of the masked points $N_{\rm mask}$ to the total number of lattice sites $N^3$ in the simulation box is also shown in the legend. 
The results are obtained from simulations of PRS strings with $2048^3$ lattice sites and $m_ra=1$.}
\label{fig:mask_correction}
\end{figure}

Figure~\ref{fig:mask_correction} shows a comparison between the axion spectrum built from the top-hat mask and that corrected by the matrix $\mathcal{M}^{-1}(k,k')$.
As expected, there is a suppression in the uncorrected spectrum at low $k$, and the amount of the suppression scales according to $N_{\rm mask}/N^3$.
If we take a larger value of $r_{\rm mask}$, the value of $N_{\rm mask}$ increases, and the spectrum becomes more suppressed.
However, the difference between the corrected spectrum and the uncorrected spectrum decreases with time,
since the number of plaquettes pierced by strings, which determines the number of masked points $N_{\rm mask}$, becomes smaller at late times.

From the above arguments, we expect that the ``imperfectness'' of the masked spectrum $P_X^{\rm mask}(k)$ is of order $2N_{\rm mask}/N^3$.
Since we are interested in measuring the axion spectrum at late times of the simulation where the string length becomes comparable to the size of the simulation box,
a reasonable estimate is that $N_{\rm mask} \sim \mathcal{O}(N)$, and hence $2N_{\rm mask}/N^3 \sim \mathcal{O}(N^{-2})$ in the relevant time period.
Therefore, we expect that this effect is negligibly small in the simulations with large $N$. For instance, the bottom panel of Fig.~\ref{fig:mask_correction} tells us that
the correction is already as small as $\mathcal{O}(0.01$\textendash$0.1)\%$ for $N=2048$, which can become even smaller for larger values of $N$,
though the computation of the correction matrix becomes much more expensive.
From this reason, we do not apply the correction matrix and simply use the masked spectrum $P_X^{\rm mask}(k)$ in our main analysis.

\section{Calculation of $\mathcal{F}(x,y)$, $\Gamma$, and $q$}
\label{app:calc_F_and_q}

In this appendix, we describe technical details on the calculations of the instantaneous emission spectrum $\mathcal{F}(x,y)$,
energy density emission rate $\Gamma$, and spectral index $q$.

\subsection{Instantaneous emission spectrum}
\label{app:method_F}

We compute $\mathcal{F}(x,y)$ by analysing the time evolution of the energy density spectrum for each Fourier component of the axion field specified by a fixed value of the comoving momentum $k$.
Let us introduce the following notation to describe such a time evolution of one Fourier mode,
\begin{align}
\mathcal{E}(x) \equiv \frac{1}{f_a^2H^2}\frac{\partial\rho_a}{\partial\log k}, \label{E_definition}
\end{align}
where $x=k\tau$ should be regarded as a time variable for given $k$.\footnote{Note that here we use $x$ as a time variable.
This is different from the parametrization of $\mathcal{F}(x,y)$ defined in Eq.~\eqref{F_definition},
where $y =m_r/H$ represents the time evolution, and $x=k/(RH) = k\tau$ represents the momentum dependence for given $\tau$. 
Nevertheless, in the absence of strings, the energy density spectrum could actually be described in terms of a single variable function,
where $x$ represents both the time and momentum dependencies (see Appendix~\ref{app:evolution_analytical}).}
Writing the derivative with respect to the conformal time as $\partial/\partial\tau = k\partial/\partial x$
and using $H^2 \propto R^{-4}$ for the radiation dominated background, the definition of $\mathcal{F}$ [Eq.~\eqref{F_definition}] takes a simple form,
\begin{align}
\mathcal{F} = \frac{\partial\mathcal{E}}{\partial x}. \label{F_dEdx}
\end{align}

As shown in Sec.~\ref{sec:oscillations}, just computing the derivative in the right-hand side of Eq.~\eqref{F_dEdx} via a finite difference method 
gives rise to a problem because of the existence of oscillations in the spectrum.
In order to suppress the contamination from the oscillations, we compute the time derivative in the following way.
First, we fit the analytical function given by Eq.~\eqref{mode_fit_function} to the data of the time evolution $\mathcal{E}(x)$ for each $k$, 
and subtract it from the data to obtain the residue,
\begin{align}
\mathcal{R}(x) \equiv \mathcal{E}_{\rm data}(x) - \mathcal{E}_{\rm fit}(x).
\end{align}
Furthermore, we subtract a linear function from the residue,
\begin{align}
\overline{\mathcal{R}}(x) = \mathcal{R}(x) - (\alpha x + \beta),
\end{align}
where $\alpha$ and $\beta$ are determined such that $\overline{\mathcal{R}}(x_{\rm ini}) = \overline{\mathcal{R}}(x_{\rm fin}) = 0$,
and $x_{\rm ini}$ and $x_{\rm fin}$ represent the initial time and final time of the data used for the fit, respectively.
In the main analysis, we choose $x_{\rm ini}$ as a time corresponding to $\ln(m_r/H)=4$, and $x_{\rm fin}$ as the final time of the simulation.
The residue $\mathcal{R}(x)$ [or $\overline{\mathcal{R}}(x)$] may contain oscillating components, whose dominant angular frequency is given by $\sim 2k$ in the conformal time, 
as shown in the bottom panel of Fig.~\ref{fig:osc_mode_evol_lowk}.
To remove those oscillations, we apply the type-I discrete sine transform (DST) to the data of $\overline{\mathcal{R}}(x)$,\footnote{
We arrange $\overline{\mathcal{R}}(x)$ such that  $\overline{\mathcal{R}}(x_{\rm ini}) = \overline{\mathcal{R}}(x_{\rm fin}) = 0$ 
because otherwise some unnecessary high frequency components can arise due to the discontinuity at the boundary.
With this preparation, the type-I DST, which assumes the odd extension on both sides of the boundary,
reproduces a non-vanishing value of $d\overline{\mathcal{R}}/dx$ at the boundary
that can correct any bias caused by $d\mathcal{E}_{\rm fit}/dx$ in Eq.~\eqref{F_method_filtered}.} 
and multiply it by a Gaussian filter,
\begin{align}
\exp\left(-\frac{(f/f_{\rm osc})^2}{2\sigma_{\rm filter}^2}\right), \label{filter_for_F}
\end{align}
where $f$ is the frequency of DST components, $f_{\rm osc} = k/\pi$ the frequency of $2k$-oscillations, and $\sigma_{\rm filter}$ the parameter to be adjusted appropriately.
After that, we perform the inverse DST to obtain the filtered residue $\overline{\mathcal{R}}_{\rm filtered}(x)$, and calculate $\mathcal{F}$ as
\begin{align}
\mathcal{F} = \frac{d\mathcal{E}_{\rm fit}}{dx} + \alpha + \frac{d\overline{\mathcal{R}}_{\rm filtered}}{dx}, \label{F_method_filtered}
\end{align}
where $d\mathcal{E}_{\rm fit}/dx$ can be computed analytically by differentiating Eq.~\eqref{mode_fit_function}.
Expecting that $\overline{\mathcal{R}}_{\rm filtered}(x)$ becomes smooth enough, we compute $d\overline{\mathcal{R}}_{\rm filtered}/dx$ just by using the finite difference method.

In order to deal with the $2k$-oscillations properly, it is important to measure the spectrum with a rate faster than the oscillation frequencies.
In the simulations performed in this work, the axion energy density spectrum is measured 300 times from $\ln(m_r/H)=4$ to the final time in a linear interval in the conformal time
for the simulations with $11264^3$ lattice sites and 250 times for others.\footnote{The measurements are executed in a linear interval in $\tau$, since we need to resolve the oscillations whose frequencies 
are constant in the conformal time. This means that more measurements are done at larger $\ln(m_r/H)$.}
The maximum value of the comoving momentum that can be resolved by these measurements is given by $k_{\rm res}  = (\pi/2)f_s = (\pi N_{\rm meas}/2)/(\tau_{\rm fin}-\tau_{\rm ini})$, 
where $f_s = N_{\rm meas}/(\tau_{\rm fin}-\tau_{\rm ini})$ is the sampling frequency, and $N_{\rm meas} = 300$ or $250$ is the number of measurements.
In Table~\ref{tab:k_res}, we summarise the value of $k_{\rm res}$, its ratio to the Hubble parameter, and that to the saxion mass at the final time of the simulations.
We see that the oscillations for the modes with $kL/(2\pi) \lesssim 100$ are well resolved with those measurements, 
and hence the $2k$-oscillations can be properly filtered through the above procedure for these lower momentum modes.

\begin{table*}
\caption{The maximum value of the comoving momentum $k_{\rm res}$ of the Fourier component of the axion field whose oscillations can be resolved by $N_{\rm meas}$ measurements 
for each choice of the simulation parameters used in the main analysis.
The ratio of $k_{\rm res}/R$ to the Hubble parameter and that to the saxion mass at the final time of the simulations are also shown.}
\label{tab:k_res}
 \begin{tabular*}{0.8\textwidth}{@{\extracolsep{\fill}}llllllll}
\hline\hline 
 Type & Grid size & Final time & Parameter & $N_{\rm meas}$ & $k_{\rm res}L/(2\pi)$ & $k_{\rm res}/(RH)$ & $k_{\rm res}/(Rm_r)$ \\
         & ($N^3$)  & ($\tau_f/L$) &             &                           &                                  & at $\tau_f$            & at $\tau_f$\\
\hline
 Physical & $11264^3$ & 0.625 & $\bar{\lambda}=195799$ & 300 & 130.260 & 511.531 & 0.0581285 \\
 Physical & $4096^3$ & 0.625 & $\bar{\lambda}=25890.8$ & 250 & 115.025 & 451.701 & 0.141156  \\
 Physical & $3072^3$ & 0.5 & $\bar{\lambda}=14563.6$ & 250 & 154.042 & 483.939 & 0.315064 \\
 Physical & $3072^3$ & 0.5 & $\bar{\lambda}=32768$ & 250 & 147.743 & 464.150 & 0.201454 \\
 Physical & $3072^3$ & 0.5 & $\bar{\lambda}=64225.3$ & 250 & 143.695 & 451.431 & 0.139953 \\
 Physical & $3072^3$ & 0.5 & $\bar{\lambda}=114178$ & 250 & 140.872 & 442.563 & 0.102903 \\
 Physical & $2048^3$ & 0.55 & $\bar{\lambda}=6400$ & 250 & 143.936 & 497.408 & 0.403720 \\
  PRS & $8192^3$ & 0.55 & $m_ra = 0.2$ & 250 & 120.966 & 418.027 & 0.463897 \\
  PRS & $8192^3$ & 0.55 & $m_ra = 0.3$ & 250 & 118.420 & 409.229 & 0.302756 \\
  PRS & $8192^3$ & 0.55 & $m_ra = 0.5$ & 250 & 116.459 & 402.453 & 0.178646 \\
  PRS & $8192^3$ & 0.55 & $m_ra = 0.7$ & 250 & 115.638 & 399.617 & 0.126705 \\
  PRS & $8192^3$ & 0.55 & $m_ra = 1.0$ & 250 & 115.030 & 397.516 & 0.0882271 \\
  PRS & $8192^3$ & 0.55 & $m_ra = 1.5$ & 250 & 114.562 & 395.897 & 0.0585786 \\
 \hline
\end{tabular*}
\end{table*}

Table~\ref{tab:k_res} also shows that $k_{\rm res}/R$ is smaller than $m_r/2$ at the final time of the simulations in all cases.
Hence, in such simulations the oscillations at the UV part of the spectrum are not adequately resolved by the measurements.
This is not an issue for the PRS-type simulations, since the amplitude of the oscillations induced by the horizon crossing are expected to be smaller at higher momenta (see Sec.~\ref{sec:oscillations}),
and those induced by the parametric resonance effect are relevant only at the comoving momentum given by $k = m_rR/2$.
On the other hand, for the physical-type simulations, the $2k$-oscillations can be produced by the parametric resonance effect over a broad range of momenta satisfying $k \lesssim m_rR/2$.
In order to monitor such oscillations at higher momenta, one needs to increase the number of measurements at the cost of more time complexity, which is practically unfeasible.

We note that it is still possible to grasp the oscillations at higher momenta even if $k > k_{\rm res}$.
It is known that when the actual oscillation frequency $f$ of the data exceeds the Nyquist frequency $f_s/2 = (N_{\rm meas}/2)/(\tau_{\rm fin}-\tau_{\rm ini})$,
it shows up as a spurious low frequency oscillation (called alias), whose frequency is given by $f_{\rm al} = |f - nf_s|$, where $n$ is an integer that gives $f_{\rm al} < f_s/2$.
We actually confirmed that the frequencies of the oscillations for higher momentum modes in our data agree with those predicted by $f_{\rm al}$.
Based on this observation, we replace $f_{\rm osc}$ in Eq.~\eqref{filter_for_F} with $\max(|f_{\rm osc}-nf_s|,f_{\rm min})$ for the modes with $k > k_{\rm res}$
when calculating the instantaneous emission spectrum for physical-type simulations.
Here we have introduced a minimum frequency $f_{\rm min}$, since otherwise some artificial features arise in $\mathcal{F}(x)$ for the modes whose aliasing frequencies $|f_{\rm osc}-nf_s|$ become close to zero.
In the main analysis, we take $f_{\rm min}L=60$ to reduce artificial features at the corresponding momenta.

Finally, let us comment on the choice of the parameter $\sigma_{\rm filter}$ in Eq.~\eqref{filter_for_F}.
Obviously, taking a too large value of $\sigma_{\rm filter}$ does not help to remove the oscillations, 
while taking a too small value of $\sigma_{\rm filter}$ leads to $\overline{\mathcal{R}}_{\rm filtered}(x)\to 0$, which can bias the result.\footnote{In the absence of the filtered residue,
$\mathcal{F}$ is strongly biased by $d\mathcal{E}_{\rm fit}/dx$ in Eq.~\eqref{F_method_filtered}.
For instance, we observed that the result of $q$ takes different values when we use some functions different from Eq.~\eqref{mode_fit_function},
if we do not include the $d\overline{\mathcal{R}}_{\rm filtered}/dx$ term.} 
We found that the results are not too biased and at the same time the fluctuations in $q$ are adequately suppressed if we take $\sigma_{\rm filter} = 0.02\text{--}0.05$.
Changing $\sigma_{\rm filter}$ in this range slightly modifies the final results on $q$, and we add this variation in $q$ to the statistical uncertainty when we refer to the errors in $q$ in the main analysis.

\subsection{Energy density emission rate}
\label{app:method_Gamma}

We compute the energy density emission rate in a similar way to the instantaneous emission spectrum.
From Eq.~\eqref{Gamma_definition}, we have
\begin{align}
\frac{\Gamma_a}{f_a^2H^3} = \tau\frac{d\overline{\rho_a}}{d\tau},\quad
\frac{\Gamma_r}{f_a^2H^3} = \tau^{5-\langle z\rangle}\frac{d}{d\tau}\left(\tau^{\langle z\rangle-4}\overline{\rho_r}\right), \label{Gamma_definition_dimensionless}
\end{align}
where $\overline{\rho_a} = \rho_a/(f_a^2H^2)$, $\overline{\rho_r} = \rho_r/(f_a^2H^2)$, and
\begin{align}
\langle z\rangle = 
\begin{cases}
{\displaystyle \frac{1}{\rho_r}\int dk z[k]\frac{\partial\rho_r}{\partial k}} & (\text{physical}), \\
4 & (\text{PRS}).
\end{cases}
\label{saxion_mean_z}
\end{align}
Here the mean redshift exponent $\langle z\rangle$ takes account of the non-trivial redshift of massive modes, 
and is different according to whether $m_r$ is constant (physical) or $m_r \propto R^{-1}$ (PRS) during simulations. 
For the physical case, $z[k]$ is given by
\begin{align}
z[k] = 3 - \frac{d\log\omega(k)}{d\log R} = 3 + \frac{\left(\frac{k}{m_rR}\right)^2}{1+\left(\frac{k}{m_rR}\right)^2},
\label{saxion_zk}
\end{align}
where $\omega(k) = \sqrt{m_r^2 + k^2/R^2}$.

To compute the time derivatives on the right-hand side of Eq.~\eqref{Gamma_definition_dimensionless},
we fit the following function to the data of $\ln\overline{\rho_a}$ and $\ln(\tau^{\langle z\rangle-4}\overline{\rho_r})$,
\begin{align}
\mathcal{G}_{\rm fit}(\ln\tau) = a_0 + a_1\ln\tau + a_2(\ln\tau)^2 + a_3(\ln\tau)^3,
\end{align}
where $a_i$ ($i=0,1,2,3$) are constants to be determined by the fit.
After that, we calculate the residue $\mathcal{R} = \overline{\rho}_{\rm data} - \overline{\rho}_{\rm fit}$, where $\overline{\rho}=\overline{\rho_a}$ for axions
and $\overline{\rho}=\tau^{\langle z\rangle-4}\overline{\rho_r}$ for saxions.
Then we subtract the linear function, $\overline{\mathcal{R}}= \mathcal{R} -(\alpha\tau+\beta)$, where $\alpha$ and $\beta$ are fixed such that
$\overline{\mathcal{R}}(\tau_{\rm ini})=\overline{\mathcal{R}}(\tau_{\rm fin})=0$, and $\tau_{\rm ini}$ and $\tau_{\rm fin}$ correspond to the conformal time
at $\ln(m_r/H)=4$ and the final time of the simulation, respectively. After obtaining $\overline{\mathcal{R}}$, we apply the type-I DST with a Gaussian filter to it.
Here the filter is chosen as 
\begin{align}
\exp\left(-\frac{(f/f_s)^2}{2\sigma^2_{\Gamma}}\right),
\end{align}
where $f_s=N_{\rm meas}/(\tau_{\rm fin}-\tau_{\rm ini})$, and $\sigma_{\Gamma}$ is the parameter to be adjusted.
Now the energy density emission rate is calculated as
\begin{align}
\frac{\Gamma}{f_a^2H^3} = \tau^{\epsilon}\left[\frac{\exp(\mathcal{G}_{\rm fit})}{\tau}\frac{d\mathcal{G}_{\rm fit}}{d\ln\tau} + \alpha + \frac{d\overline{\mathcal{R}}_{\rm filtered}}{d\tau}\right],
\end{align}
where $\epsilon=1$ for axions, and $\epsilon = 5-\langle z\rangle$ for saxions.
In the main analysis, we use $\sigma_{\Gamma}=0.01$, which is sufficient to remove rapid fluctuations on the data.

\subsection{Spectral index}
\label{app:method_q}

Assuming that the instantaneous emission spectrum is actually given by a power law, 
\begin{align}
\mathcal{F}(x) \propto \frac{1}{x^q},
\end{align}
between the comoving momenta
related to the Hubble radius $k \sim HR$, and the size of the string core $k \sim m_rR$,
we would like to determine the value of $q$ from simulations and explore how it depends on $\ell = \ln(m_r/H)$.
To this end, we attempt to fit the power law to the numerical results in the following manner.

The instantaneous emission spectrum computed from the output of the simulations is given in discrete bins $\mathcal{F}_i$, where the index $i$ increases linearly with $k$.
Since we are interested in the behaviour at intermediate momenta, we first select the data points from a range defined by
\begin{align}
c_{\rm IR}HR < k < c_{\rm UV} m_r R,
\label{interval_for_fit}
\end{align}
where the coefficients $c_{\rm IR}$ and $c_{\rm UV}$ should be adjusted appropriately (see Sec.~\ref{sec:oscillations} and Appendix~\ref{app:syst_cutoffs}).
The selected data is rebinned such that a new set of bins are
spaced homogeneously in $\ln k$, which is more convenient when we fit a power law.
If instead we perform the fits in linear-spaced bins, results would be biased by  data at higher $k$,
since they are more abundant in a logarithmic interval.

The rebinned data is simply obtained by averaging over data within an interval $\delta l$ defined by log-spacing,
\begin{align}
\mathcal{F}_n = \frac{1}{N_n}\sum_{\ln k_i \in (l_n,l_n+\delta l)}\mathcal{F}_i \quad \text{with} \quad n = 1,2,\dots, N_{\rm bin},
\end{align}
where $N_n$ is the number of linear bins in the log-bin, and
$N_{\rm bin}$ is the total number of log-bins.
Throughout the analysis in this work, we fix the number of the rebinned data as $N_{\rm bin} = 30$.

From the rebinned data, we define the $\chi^2$ function in terms of logarithmic variables,
\begin{align}
\chi^2 = \sum_n\frac{(L_n - M_n)^2}{\sigma^2},
\end{align}
where
\begin{align}
L_n = \ln \mathcal{F}_n, \quad M_n = m - l_n q, \quad \text{and} \quad l_n = \ln x.
\end{align}
The model $M_n$ corresponds to $\mathcal{F} = e^m/x^q$ described by two parameters, $m$ and $q$.
We define $\sigma^2$ from the residuals of the different bins at the best fit model,
\begin{align}
\sigma^2 \equiv \frac{1}{N_{\rm bin}}\sum_n(L_n - M_n(m_{\rm min},q_{\rm min}))^2,
\end{align}
where $m_{\rm min}$ and $q_{\rm min}$ are the best fit values that minimise $\sum_n(L_n-M_n)^2$.
It is possible to estimate a confidence interval $[q_{\rm min}-\sigma_q,q_{\rm min}+\sigma_q]$ 
as an interval of $q$ satisfying $\Delta\chi^2 \equiv \chi^2(\overline{m}_q,q) - \chi^2_{\rm min} = 1$,
where we have marginalized over the parameter $m$ by using the value $\overline{m}_q$ that minimises $\chi^2$ for given $q$.
However, we find that $\sigma_q$ is typically smaller than the statistical uncertainty from the simulations with different random initial field configurations.

\section{Other systematics}
\label{app:other_systematics}

This appendix is devoted to discussions on a couple of effects that are not thoroughly elaborated on in the main text:
One is the finite volume effect, and the other is the effect of the momentum range for the fit of the instantaneous emission spectrum.

\subsection{Finite volume}
\label{app:syst_finite_V}

\begin{figure*}[htbp]
\includegraphics[width=0.85\textwidth]{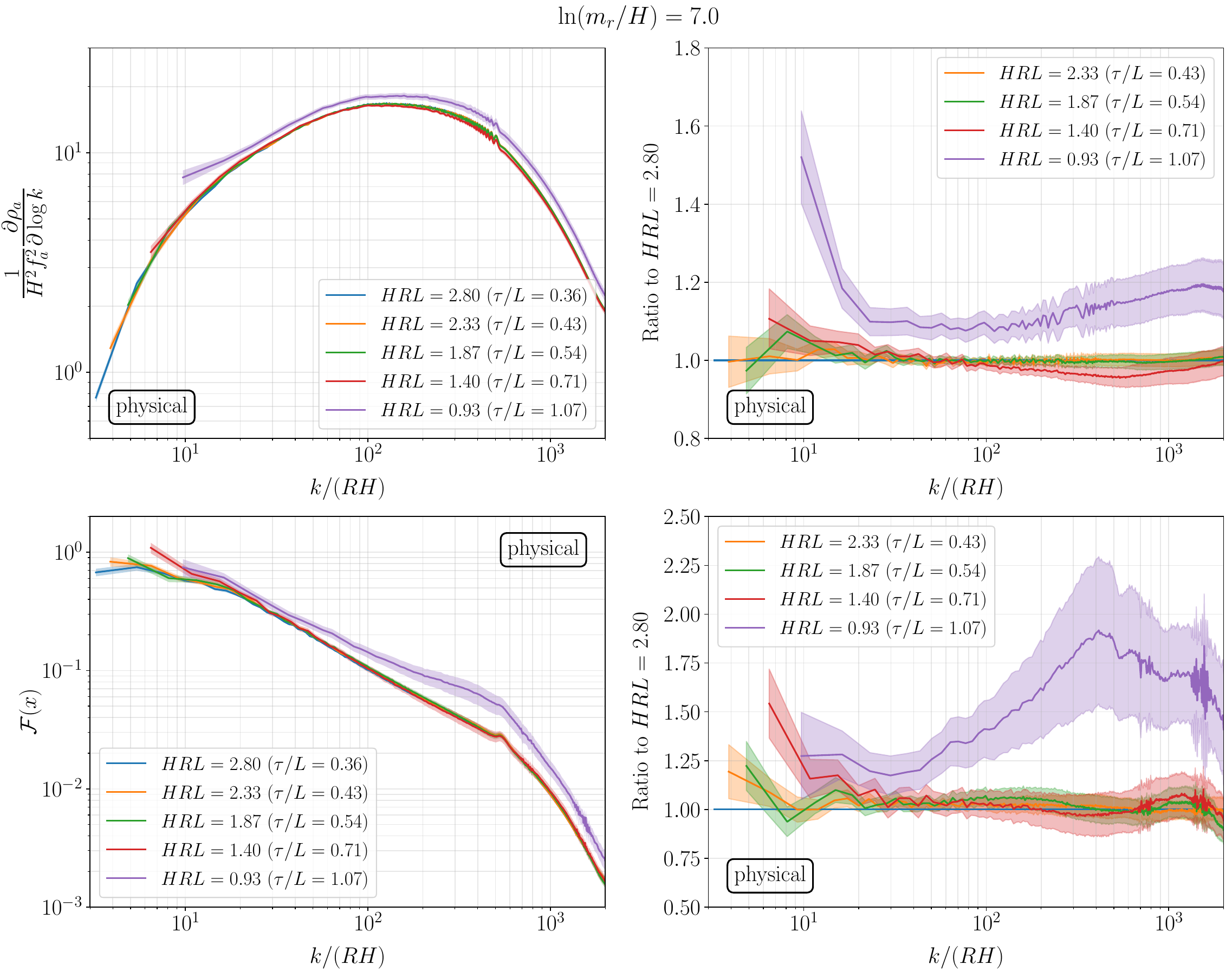}
\caption{The axion energy density spectrum (top left) and instantaneous emission spectrum (bottom left) at $\ln(m_r/H)=7$
obtained from simulations of physical strings with different values of the ratio $HRL$ (the physical box size to the Hubble radius).
These simulations are performed on different number of lattice sites, and for each of them
we chose the value of the coupling parameter $\bar{\lambda}$ such that $m_ra$ takes the same value ($m_ra=1$) at $\ln(m_r/H)=7$ (see Table~\ref{tab:simu_parameters}).
The ratio of the energy density spectrum (instantaneous emission spectrum) for each value of $HRL$ to that for $HRL=2.80$ is also shown in the top right (bottom right) panel.
The coloured bands represent statistical uncertainties.}
\label{fig:syst_finiteV}
\end{figure*}

To check the finite volume effect, we performed additional simulations of physical strings with $1024^3$, $1536^3$, $2048^3$, $2560^3$, and $3072^3$ lattice sites, 
such that they have the same value of $m_ra=1.0$ but different values of the ratio of the physical box size $RL$ to the Hubble radius $H^{-1}$ at $\ln(m_r/H)=7$.
We show the comparison of the results of those simulations in Fig.~\ref{fig:syst_finiteV}.
Although the spectrum can be largely distorted for $HRL \lesssim 1$, we can see that the results are convergent for $HRL \gtrsim 1.4$ (or $\tau/L \lesssim 0.7$).
Since we terminate all simulations used in the main analysis at $\tau/L \le 0.625$ (see Table~\ref{tab:simu_parameters}), 
we expect that the finite volume effects on the spectrum is not an issue in this work.

\begin{figure}[htbp]
\includegraphics[width=0.48\textwidth]{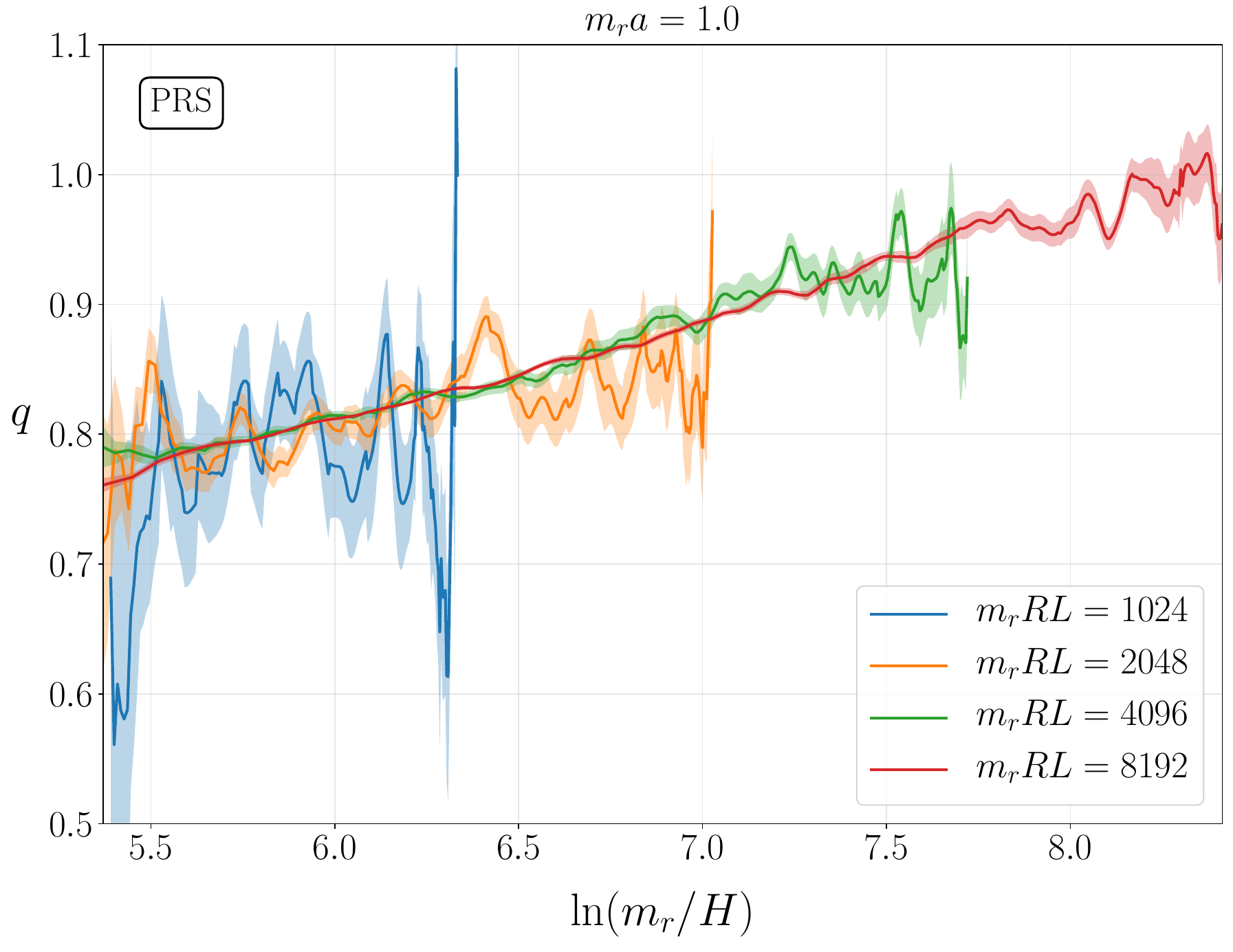}
\caption{Evolution of the spectral index $q$ of the instantaneous emission spectrum for different values of $m_rRL$ in PRS-type simulations.
Coloured lines correspond to the results of four sets of simulations with the same value of $m_ra=1$ but different numbers of lattice sites, for which $m_rRL$ becomes equal to $N$.
The coloured bands represent statistical uncertainties.
In these plots, we used $\sigma_{\rm filter}=0.2$ (a larger value than the one used in the main analysis) for the filtering procedure to calculate $\mathcal{F}$ [Eq.~\eqref{filter_for_F}]
to enlarge the fluctuations for illustrative purposes.}
\label{fig:syst_mrRL}
\end{figure}

Even if the effect of the finite volume does not significantly alter the shape of the spectrum, it can have some impact on the estimation of $q$.
To demonstrate it, in Fig.~\ref{fig:syst_mrRL} we show the evolution of $q$ obtained from simulations of PRS strings 
with the same value of $m_ra=1$ but different numbers ($N^3$) of lattice sites.
When we fix the value of $m_ra$, changing the value of $N$ corresponds to changing the ratio of the box size $RL$ to the string core width $m_r^{-1}$, since $m_rRL=(m_ra)N$.
From Fig.~\ref{fig:syst_mrRL} we see that the results with smaller $m_rRL$ exhibit more fluctuations in $q$. This can be understood as follows.
For simulations with smaller $m_rRL$ (or smaller $N$), the number of momentum bins in the interval $c_{\rm IR}HR < k < c_{\rm UV}m_rR$ used for the fit to determine $q$
becomes smaller, which implies that less data points are available at the IR part of the spectrum.
As a result, oscillations in the IR part of the spectrum have more impact on the result of the fit, which leads to larger fluctuations in $q$.

Note that the limit of small $m_rRL$ can also be realised by taking $m_ra$ to small values with $N$ fixed,
though in that case the discretisation effects due to the different value of $m_ra$ are not disentangled.
This is exactly what we plotted in the top panel of Fig.~\ref{fig:disc_mra_q}.
There we can see that $q$ exhibits larger oscillations for smaller values of $m_ra$.
These features can again be attributed to the lack of the momentum bins at the IR part of the spectrum.

The above observations may pose a challenge to the precise estimation of $q$:
We need to make $m_ra$ smaller to avoid discretisation effects, but at the same time keep $m_rRL$ sufficiently large
in order to have enough data at the IR part of the spectrum to avoid large fluctuations in $q$.
In the main analysis, these fluctuations are alleviated by having large values of $m_rRL$ at the cost of discretisation effects.

\subsection{IR and UV cutoffs}
\label{app:syst_cutoffs}

\begin{figure}[htbp]
$\begin{array}{c}
\subfigure{
\includegraphics[width=0.48\textwidth]{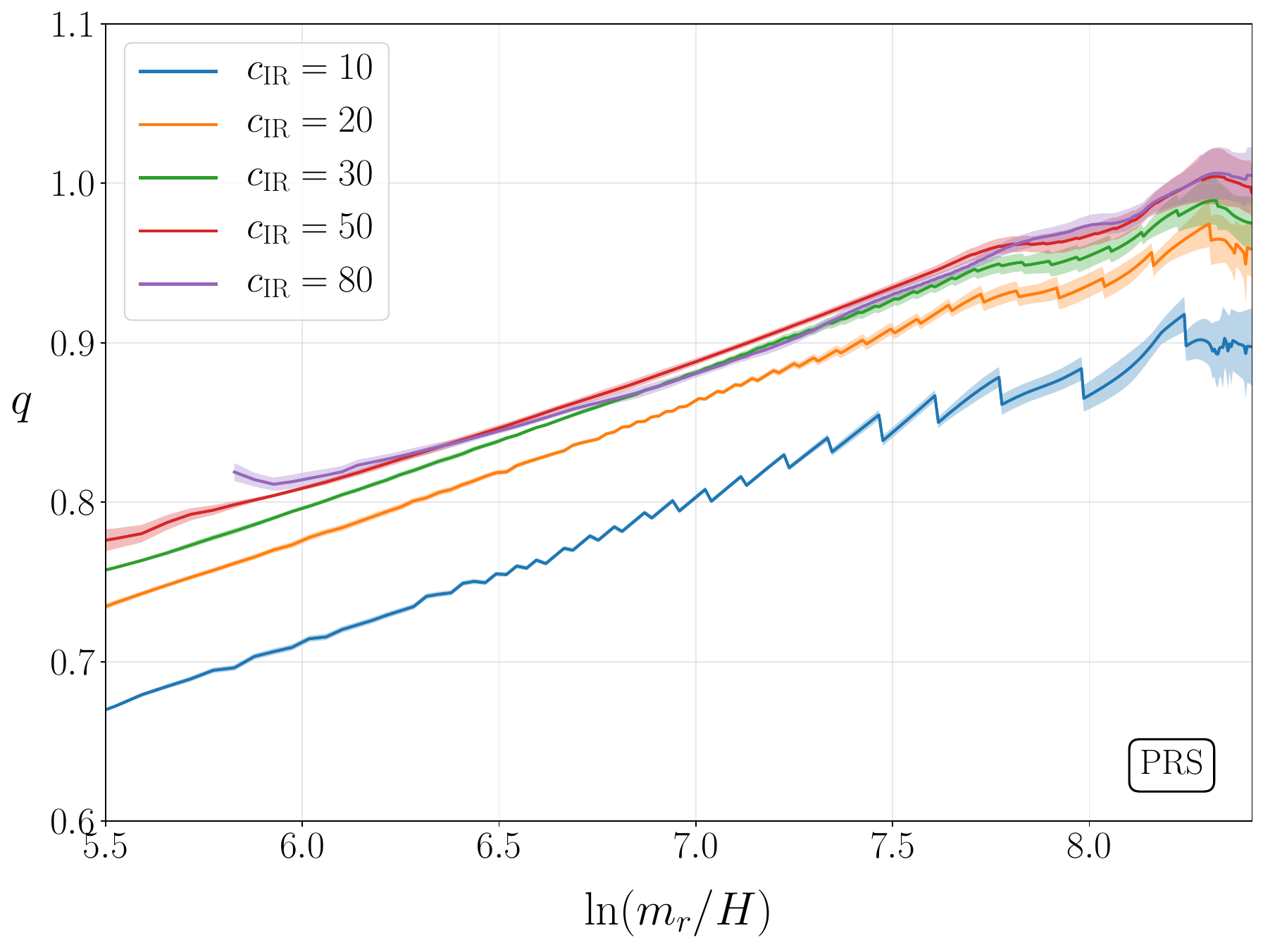}}
\\
\subfigure{
\includegraphics[width=0.48\textwidth]{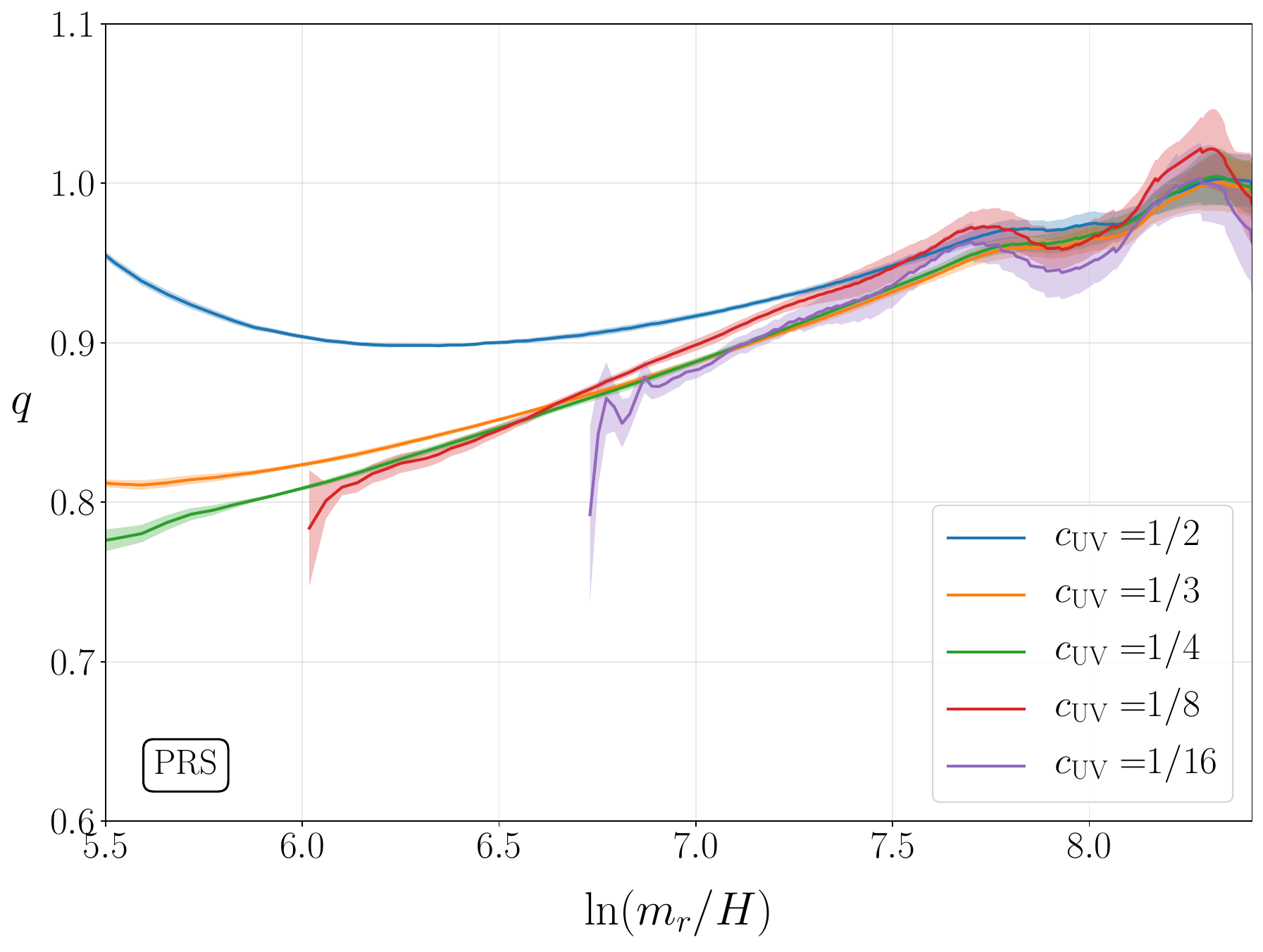}}
\end{array}$
\caption{Evolution of the spectral index $q$ of the instantaneous emission spectrum for different choices of the IR cutoff $c_{\rm IR}$ (top panel) and the UV cutoff $c_{\rm UV}$ (bottom panel)
of the momentum interval [Eq.~\eqref{interval_for_fit}] used for the fit.
In the top panel, we change the value of $c_{\rm IR}$ with the fixed UV cutoff $c_{\rm UV}=1/4$. In the bottom panel, we instead change the value of $c_{\rm UV}$ with the fixed IR cutoff $c_{\rm UV}=50$.
The results are obtained from simulations of PRS strings with $8192^3$ lattice sites and $m_ra=1$. 
The coloured bands represent the error induced by changing the parameter $\sigma_{\rm filter}$ for the filtering procedure to calculate $\mathcal{F}$ [Eq.~\eqref{filter_for_F}] in addition to statistical uncertainties.}
\label{fig:syst_cutoffs_PRS}
\end{figure}

The result of the power law fit depends on the choice of the momentum interval shown in Eq.~\eqref{interval_for_fit}.
In Fig.~\ref{fig:syst_cutoffs_PRS}, we show how the results of $q$ for simulations of PRS strings
change when we use different values of $c_{\rm IR}$ and $c_{\rm UV}$.
From the top panel of Fig.~\ref{fig:syst_cutoffs_PRS}, we see that for smaller $c_{\rm IR}$
the fit results are more affected by the feature around the IR peak, being biased toward smaller values of $q$.
This trend is diminished for $c_{\rm IR} \gtrsim 30$, as the lower end of the interval is located sufficiently far from the IR peak.
On the other hand, taking even larger values (e.g. $c_{\rm IR} = 80$) appear to introduce another bias at smaller $\ln(m_r/H)$, 
since in that case the momentum interval becomes too short to have enough data points to extract the global feature of the spectrum.

We see that the curve of $q$ becomes jagged for smaller values of $c_{\rm IR}$.
This feature is merely due to the structure of the binning. 
Note that the lower end of the interval $c_{\rm IR}HR$ decreases with time. 
This implies that more IR bins are included in the interval at later times.
Since IR bins are sparse in log-spaced bins, 
the structure of the log-bin changes suddenly when a bin with wavenumber $k$
crosses the threshold value $c_{\rm IR}HR$ and enters into the interval.
At that point the fit is biased by the lowest bin, which results in the sudden decrease of the value of $q$.
This feature is more pronounced for smaller $c_{\rm IR}$, as the data interval contains the feature around the IR peak,
where the spectrum becomes less steep.

By comparing the results for different values of the higher end of the interval $c_{\rm UV}$, we also find that 
the values of $q$ can be overestimated when we choose a value of $c_{\rm UV}$
that is not too far from the UV cutoff, as shown in the bottom panel of Fig.~\ref{fig:syst_cutoffs_PRS}.
This can be avoided if we choose a sufficiently small value of $c_{\rm UV}$, but for such a small $c_{\rm UV}$
we again observe a bias due to the short interval of the data points.
As discussed in Sec.~\ref{sec:oscillations}, for smaller $c_{\rm UV}$ the impact of the oscillations in the IR part of the spectrum becomes more pronounced,
which leads to larger oscillatory features in the evolution of $q$.

\begin{figure}[htbp]
$\begin{array}{c}
\subfigure{
\includegraphics[width=0.48\textwidth]{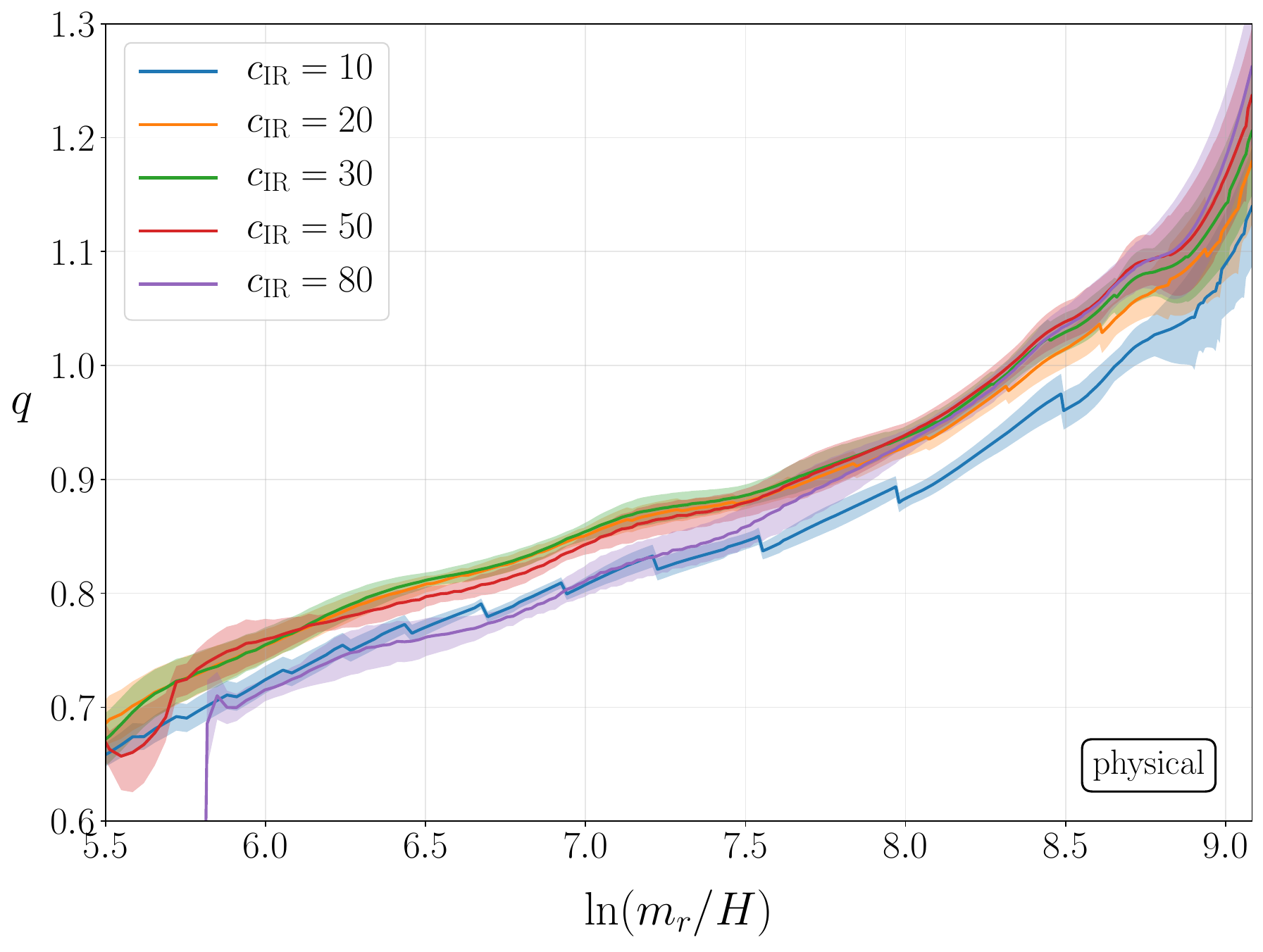}}
\\
\subfigure{
\includegraphics[width=0.48\textwidth]{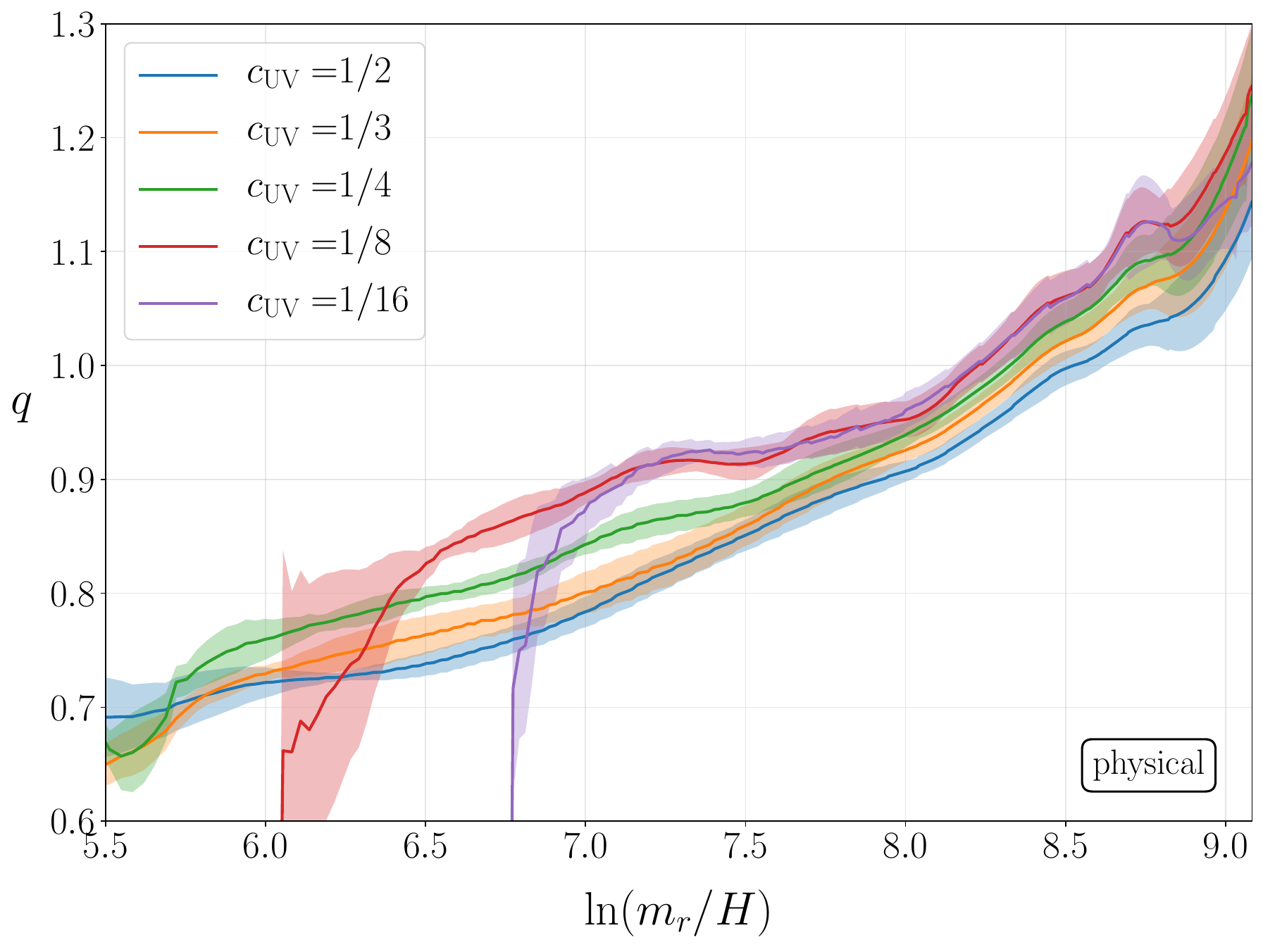}}
\end{array}$
\caption{The same figure as Fig.~\ref{fig:syst_cutoffs_PRS} but the dependences on the IR and UV cutoffs are plotted for physical strings,
obtained from simulations with $11264^3$ lattice sites.}
\label{fig:syst_cutoffs_physical}
\end{figure}

The biases due to the choice of IR and UV cutoffs are also observed in the results of physical-type simulations, as shown in Fig.~\ref{fig:syst_cutoffs_physical}.
We see that the values of $q$ are underestimated for $c_{\rm IR}$ as small as $c_{\rm IR}=10$, while the results are compatible within uncertainties for $20\lesssim c_{\rm IR}\lesssim 50$.
For $c_{\rm IR}$ as large as $c_{\rm IR}=80$, the result is again distorted particularly at smaller $\ln(m_r/H)$ due to the short momentum interval.
The trend that $q$ takes smaller values at smaller $\ln(m_r/H)$ for $c_{\rm IR}=80$ can be attributed to the feature at the UV part of the instantaneous emission spectrum
for physical strings, which is described below.

From the bottom panel of Fig.~\ref{fig:syst_cutoffs_physical}, we see that $q$ behaves somewhat differently than the case of PRS strings when we change the UV cutoff $c_{\rm UV}$.
For the physical case $q$ takes smaller values when $c_{\rm UV}$ becomes larger, while for the PRS case it takes larger values.
This difference can be associated with the feature at the UV part of the instantaneous emission spectrum:
In the PRS case, we observed that $\mathcal{F}(x)$ just becomes steeper when we increase $x$ across the momentum corresponding to $k/R=m_r/2$, which leads to larger values of $q$ for larger $c_{\rm UV}$.
On the other hand, in the physical case $\mathcal{F}(x)$ appears to have a plateau-like feature at around $k/R=m_r/2$, and $q$ tends to take smaller values when the data contains such a feature in the UV.
The occurrence of such a feature could be attributed to the fact that some amount of axions are produced at $k \lesssim m_rR/2$ due to the parametric resonance effect after the saxion mass crossing.

In order to avoid the biases due to the features in the IR and UV part of the spectrum, we need to take a sufficiently larger value of $c_{\rm IR}$ and smaller value of $c_{\rm UV}$, respectively,
but taking too small (large) value of $c_{\rm UV}$ ($c_{\rm IR}$) can again lead to large contaminations from oscillations due to the short interval available for the fit.
In the main analysis in this paper, we take $c_{\rm IR} = 50$ and $c_{\rm UV} = 1/4$ both for the PRS and physical cases as the fiducial choice.

\section{Evolution of the free axion field around the horizon crossing}
\label{app:evolution_analytical}

In the following, we perform the analytical calculation on the evolution of the axion field in the absence of string networks,
aiming at estimating the location of the turnaround used in Sec.~\ref{sec:IRpeak}.

We start from the equation of motion for the free massless axion field in a FRW background,
\begin{align}
\ddot{\theta} + 3H\dot{\theta}-\frac{1}{R^2}\nabla^2\theta = 0.
\end{align}
Introducing a new field variable $\psi\equiv \tau\theta$ and replacing the time derivative with the derivative with respect to the conformal time $\tau$,
the above equation reduces to
\begin{align}
\psi_{\tau\tau} - \nabla^2\psi = 0.
\end{align}
The solution in Fourier space reads $\widetilde \psi({\bm k}) \propto \sin(k\tau)$ or $\cos(k\tau)$, and for the Fourier component of the $\theta$ field,
\begin{align}
\widetilde\theta({\bm k}) = A_{\bm k}\frac{\sin(k\tau)}{k\tau} + B_{\bm k}\frac{\cos(k\tau)}{k\tau},
\end{align}
where $k = |{\bm k}|$, and $A_{\bm k}$ and $B_{\bm k}$ are constant coefficients.
It turns out that the $B_{\bm k}$ term must be negligibly small, $B_{\bm k} \lesssim T_{\rm QCD}/T_{\rm PQ} \ll 1$,
where $T_{\rm QCD}$ and $T_{\rm PQ}$ are the critical temperature of the QCD phase transition and that of the PQ phase transition, respectively,
in order to guarantee a finite field amplitude at the epoch of the PQ phase transition~\cite{Kolb:1993hw}.

We assume that the spatial distribution of $\theta$ is given by
a superposition of plane waves with random amplitudes $A_{\bm k}$:
\begin{align}
\theta({\bm x},\tau) = \int\frac{d^3k}{(2\pi)^3}\frac{\sin(k\tau)}{k\tau}A_{\bm k}e^{-i{\bm k}\cdot{\bm x}}. \label{theta_evolution_analytical}
\end{align}
Furthermore, we assume a scale invariant spectrum for the mode amplitude~\cite{Dine:2020pds},
\begin{align}
\langle A_{\bm k}A^*_{\bm k'}\rangle = \frac{\mathcal{C}}{k^3}(2\pi)^3\delta^{(3)}({\bm k} - {\bm k'}), \label{Ak_spectrum}
\end{align} 
where $\langle\dots\rangle$ represents an average over the random distribution of $A_{\bm k}$, and $\mathcal{C}$ is a dimensionless constant.
Using Eqs.~\eqref{theta_evolution_analytical} and~\eqref{Ak_spectrum}, we can compute the contributions to the axion energy density from kinetic and gradient terms:
\begin{align}
\rho_{a,\mathrm{kin}} &= \frac{1}{2}f_a^2\langle\dot{\theta}^2\rangle = \frac{\mathcal{C}}{4\pi^2}\frac{f_a^2}{R^2}\int kdk\left(\frac{\cos k\tau}{k\tau}-\frac{\sin k\tau}{(k\tau)^2}\right)^2,\\
\rho_{a,\mathrm{grad}} &= \frac{1}{2}f_a^2\frac{1}{R^2}\langle|\nabla\theta|^2\rangle = \frac{\mathcal{C}}{4\pi^2}\frac{f_a^2}{R^2}\int kdk\left(\frac{\sin k\tau}{k\tau}\right)^2.
\end{align}
Note that the kinetic energy $\rho_{a,\mathrm{kin}}$ vanishes in the super-horizon limit $k\tau \ll 1$, while the gradient energy acquires a non-vanishing contribution,
\begin{align}
(\rho_{a,\mathrm{grad}})_{\rm sup} \lesssim \frac{\mathcal{C}}{4\pi^2}\frac{f_a^2}{R^2}\int^{\tau^{-1}}_0 k dk = \frac{\mathcal{C}}{8\pi^2}f_a^2H^2,
\end{align}
where in the first inequality we have used the fact that only the modes with $k \ll \tau^{-1}$ can contribute to the integral, $\int k dk \lesssim \int^{\tau^{-1}}_0kdk$.
This result agrees with a naive estimate $\rho_a \sim f_a^2\langle|\nabla \theta|^2\rangle \sim f_a^2 H^2$, which stems from an expectation that
the misalignment angle $\theta$ varies randomly on the scale corresponding to the Hubble radius.

We can also compute the contributions of subhorizon modes ($k\tau\gg 1$),
\begin{align}
(\rho_{a,\mathrm{kin}})_{\rm sub} &= \frac{\mathcal{C}}{4\pi^2}\frac{f_a^2}{R^2}\int_{k\gg \tau^{-1}}kdk\frac{\cos^2(k\tau)}{(k\tau)^2},\nonumber\\
(\rho_{a,\mathrm{grad}})_{\rm sub} &= \frac{\mathcal{C}}{4\pi^2}\frac{f_a^2}{R^2}\int_{k\gg \tau^{-1}}kdk\frac{\sin^2(k\tau)}{(k\tau)^2},
\end{align}
from which the total energy density reads 
\begin{align}
(\rho_a)_{\rm sub} &= (\rho_{a,\mathrm{kin}})_{\rm sub} + (\rho_{a,\mathrm{grad}})_{\rm sub} \nonumber\\
&= \frac{\mathcal{C}}{4\pi^2}f_a^2H^2\int\frac{dk}{k} \nonumber\\
&\approx \frac{\mathcal{C}}{4\pi^2}f_a^2H^2\ln\left(\frac{\Lambda}{H}\right).
\end{align}
This quantity is logarithmically divergent, and we have introduced the UV cutoff $\Lambda$.
A natural choice for the cutoff scale would be the saxion mass, $\Lambda \approx m_r$.
Then, this result implies that the energy density of misalignment axions $\rho_a \sim f_a^2H^2\ln(m_r/H)$
can be comparable to the typical energy density of strings in the scaling regime~\cite{Dine:2020pds}.

Since we calculate the axion spectrum based on the kinetic energy in the numerical study,
let us focus on the (differential) spectrum for the kinetic term.
It is straightforward to obtain
\begin{align}
\mathcal{E}(x) = \frac{2}{(f_aH)^2}\frac{d\rho_{a,\mathrm{kin}}}{d\log k} = \frac{\mathcal{C}}{2\pi^2}x^2[j_1(x)]^2, \label{E_free}
\end{align}
where we use the same notation as Eq.~\eqref{E_definition}, and $j_n(x)$ are spherical Bessel function of the first kind.
Note that this quantity depends only on $x = k\tau$. Namely, its time evolution is the same for all $k$, when viewed as a function of $k\tau$.
This observation can explain the numerical result shown in the top panel of Fig.~\ref{fig:osc_mode_evol_lowk} that the evolution is almost the same among different modes when they are outside the horizon.

\begin{figure}[htbp]
\includegraphics[width=0.48\textwidth]{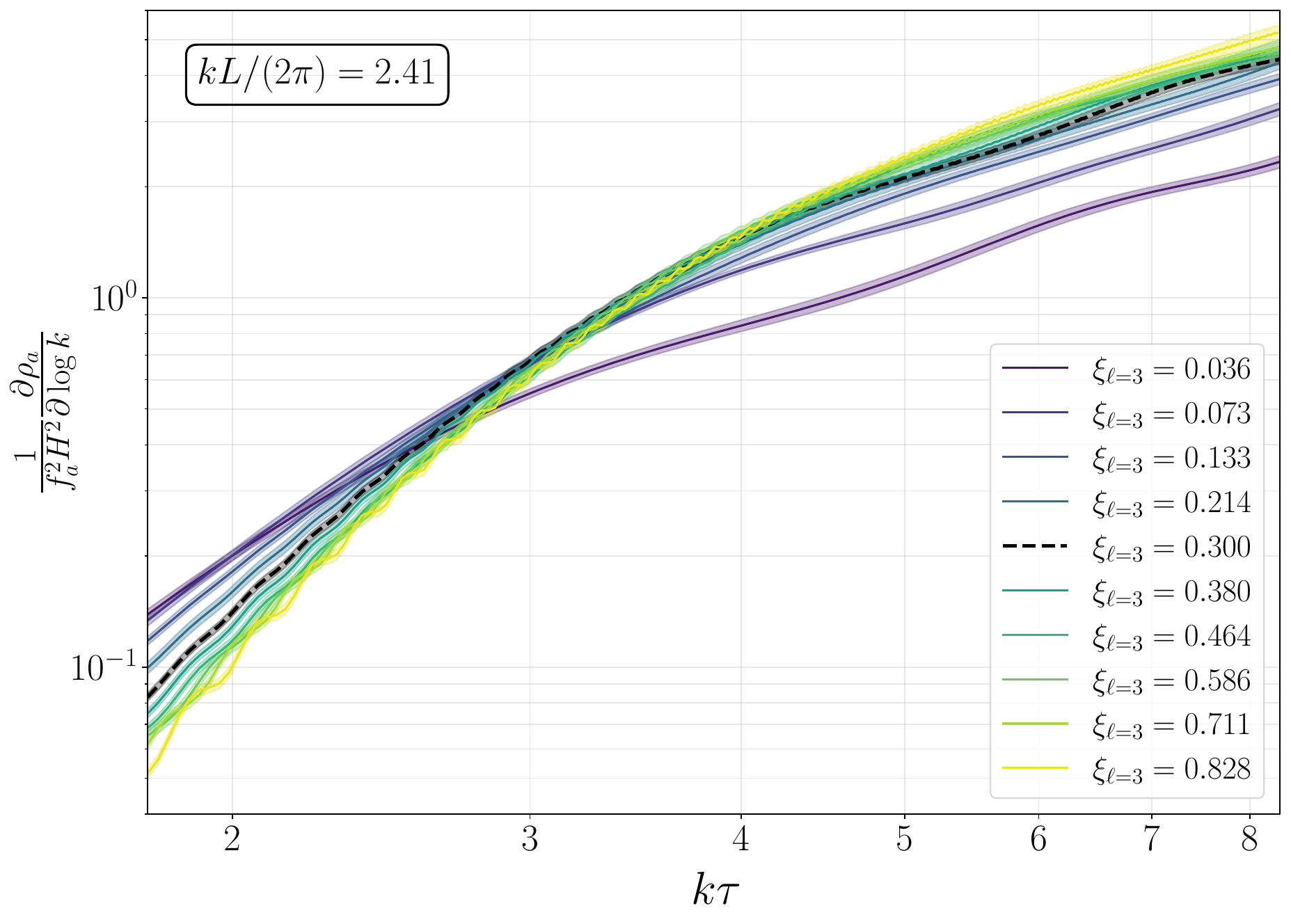}
\caption{Time evolution of the energy density of one Fourier mode of the axion field specified by $kL/(2\pi) = 2.41$
for various different values of the initial strings density $\xi_{\ell=3}$.
The coloured bands represent statistical uncertainties, and the black dashed line corresponds to the attractor with $\xi_{\ell=3} = 0.3$. 
The results are obtained from simulations of physical strings with $2048^3$ lattice sites.}
\label{fig:evol_EA}
\end{figure}

We also note that the asymptotic behaviour in the super-horizon limit reads
\begin{align}
\frac{2}{(f_aH)^2}\frac{d\rho_{a,\mathrm{kin}}}{d\log k} \xrightarrow{k\tau \to 0} \frac{\mathcal{C}}{2\pi^2}\frac{1}{9}(k\tau)^4.
\end{align}
This implies that the IR part of the spectrum (or its evolution before the horizon crossing)
scales as $\propto (k\tau)^4$ in the absence of strings.
For a comparison, in Fig.~\ref{fig:evol_EA} we show the evolution of the mode with the fixed comoving momentum $kL/(2\pi)=2.41$ obtained from simulations with various different initial string densities.
In those simulations the slope of $\mathcal{E} \propto (k\tau)^p$ at small $k\tau$ takes a value between $4\lesssim p \lesssim 6$ according to the value of the string density $\xi$,
indicating that the IR falloff of the spectrum becomes steeper in the presence of strings.\footnote{We may describe this effect by introducing a phase $\alpha_{\rm k}$ in the evolution of the mode
[i.e. replacing $\sin(k\tau )$ with $\sin(k\tau + \alpha_{\bm k})$ in Eq.~\eqref{theta_evolution_analytical}] and allowing a slow time variation in
$A_{\bm k}$ and $\alpha_{\bm k}$, though we do not pay much attention to their precise time dependence here.}
Furthermore, there is a trend that the slope approaches to the free field value $p\to 4$ with decreasing $\xi$.
We observed a similar trend when we plot the spectra as a function of $k$ rather than $\tau$. 

\begin{figure}[htbp]
\includegraphics[width=0.48\textwidth]{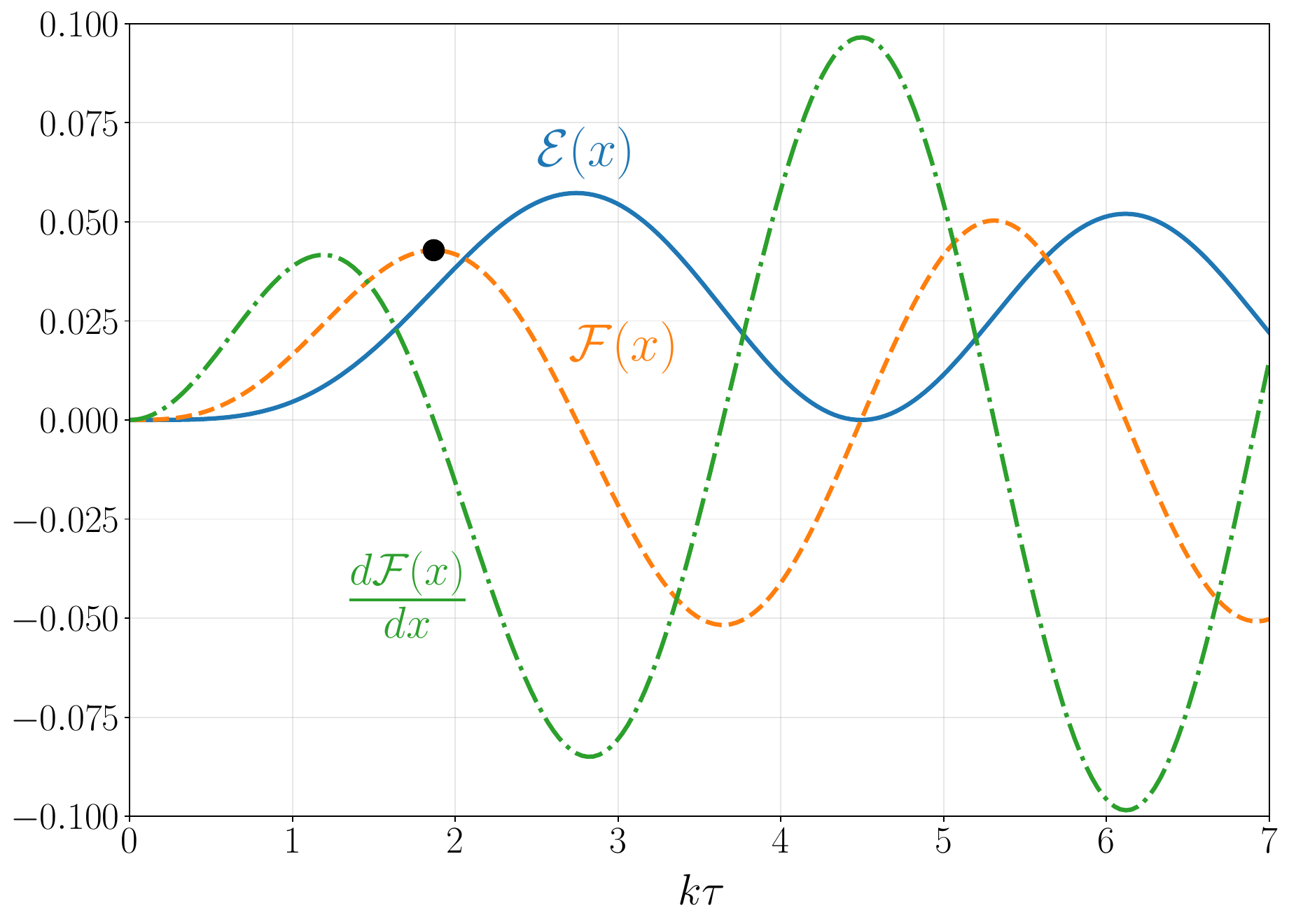}
\caption{Evolution of $\mathcal{E}(x)$ (blue solid line), $\mathcal{F}(x)$ (orange dashed line), and $\partial\mathcal{F}(x)/\partial x$ (green dash-dotted line) as functions of $x= k\tau$.
In these plots, we take $\mathcal{C}=1$ for simplicity. The location of the first maximum of $\mathcal{F}$ (first node of $\partial\mathcal{F}/\partial x$) is shown as a black dot.}
\label{fig:evol_analytical}
\end{figure}

The fact that $\mathcal{E}(x)$ exhibits a non-trivial time dependence implies that there is a contribution to the instantaneous emission spectrum even in the absence of strings:
\begin{align}
\mathcal{F}(x) = \frac{\partial\mathcal{E}(x)}{\partial x} = \frac{\mathcal{C}}{\pi^2}xj_1(x)[xj_0(x)-j_1(x)], \label{F_evolutuon_analytical}
\end{align}
where the relation $\mathcal{F}=\partial\mathcal{E}/\partial x$ follows from the definition of $\mathcal{F}$ [see Eq.~\eqref{F_dEdx}].
We can also obtain its derivative with respect to $x$,
\begin{align}
\frac{\partial\mathcal{F}(x)}{\partial x}
= \frac{\mathcal{C}}{\pi^2}x\left(x[j_0(x)]^2-j_1(x)[j_0(x)+xj_1(x)-j_2(x)]\right). \label{dFdx_analytical}
\end{align}
Figure~\ref{fig:evol_analytical} shows the evolution of $\mathcal{E}$, $\mathcal{F}$, and $\partial\mathcal{F}/\partial x$ given by Eqs.~\eqref{E_free},~\eqref{F_evolutuon_analytical}, and~\eqref{dFdx_analytical}.
The location of the first turnaround of $\mathcal{F}$ ($\partial\mathcal{F}/\partial x = 0$) is found to be $x = 1.86765$, which is shown as a black dot in Fig.~\ref{fig:evol_analytical}.

\section{Comparisons}
\label{app:comparisons}

\begin{figure*}[htbp]
$\begin{array}{cc}
\subfigure{
\includegraphics[width=0.48\textwidth]{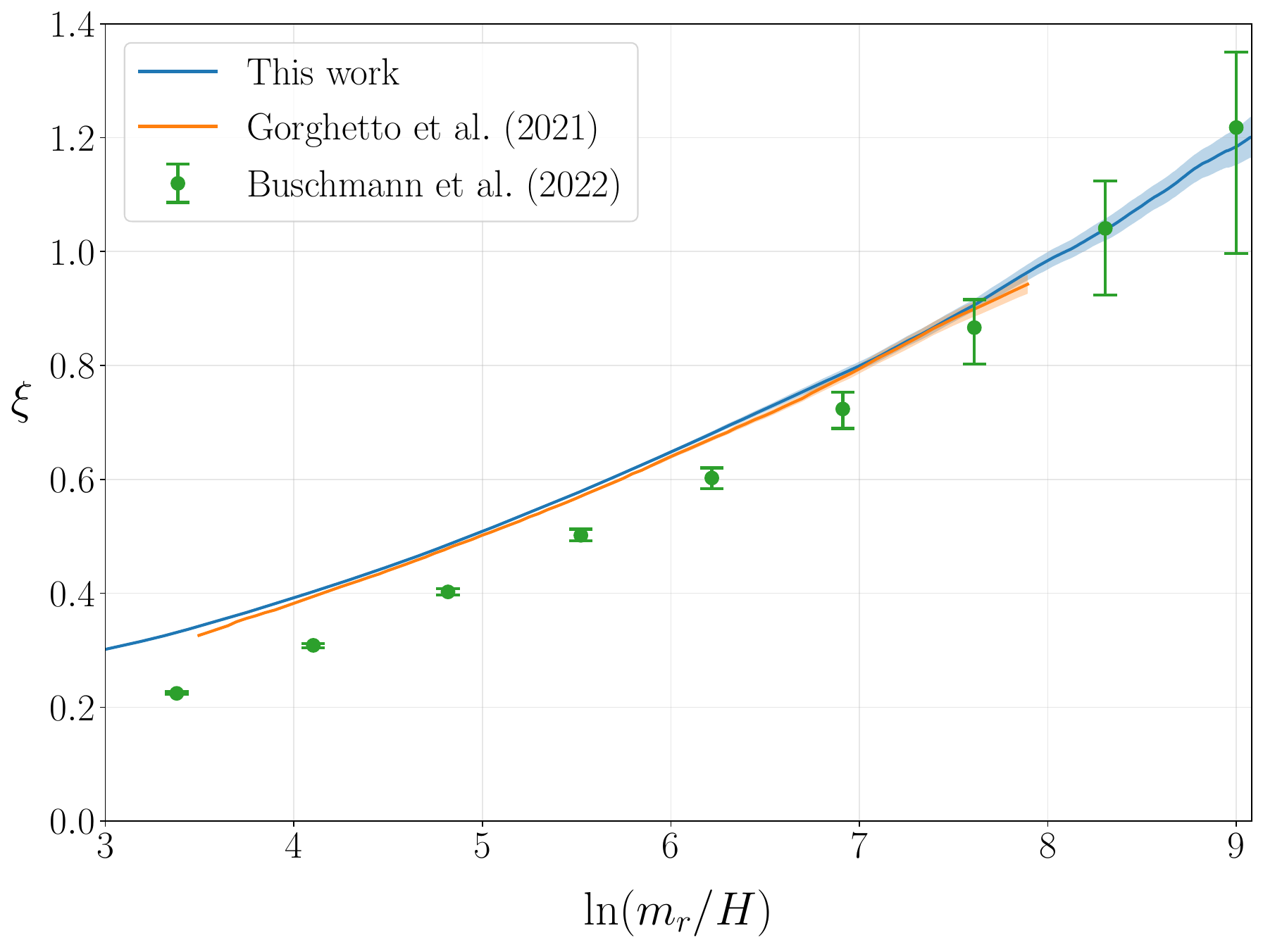}}
\hspace{1mm}
\subfigure{
\includegraphics[width=0.48\textwidth]{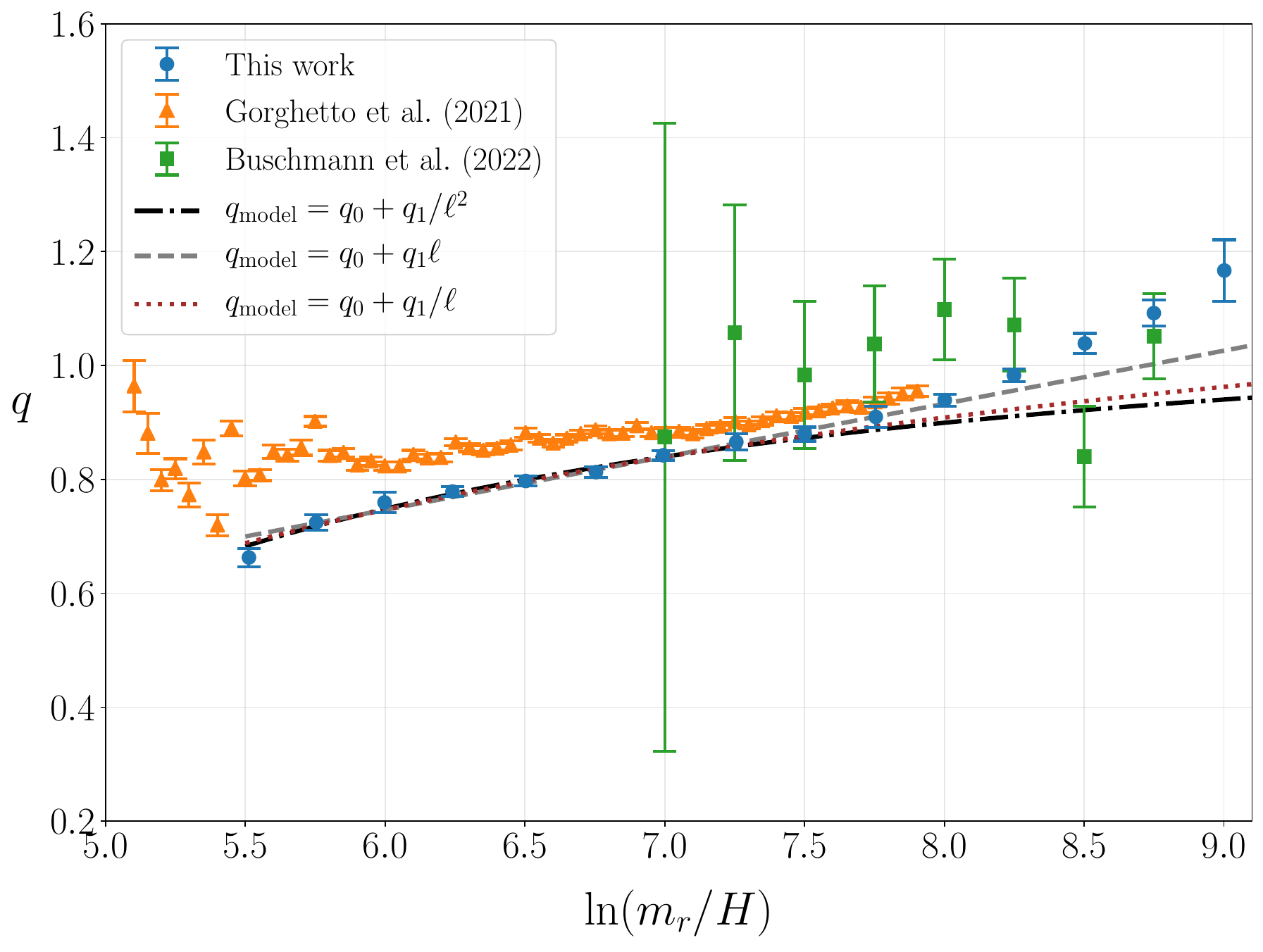}}
\end{array}$
\caption{Comparisons of the main results of this work with those of Gorghetto {\it et al}. (2021)~\cite{Gorghetto:2020qws} and Buschmann {\it et al}. (2022)~\cite{Buschmann:2021sdq}.
Left panel: Comparison of the evolution of the string density parameter $\xi$.
The data is digitised from Fig.~1 of Ref.~\cite{Gorghetto:2020qws} and Supplementary Fig.~13 of Ref.~\cite{Buschmann:2021sdq}.
For the results of this work and Ref.~\cite{Gorghetto:2020qws}, the coloured bands represent statistical uncertainties. 
The error bars for the result of Ref.~\cite{Buschmann:2021sdq} do not represent the statistical uncertainties but an educated guess involving the analytical modeling and fit.
Right panel: Comparison of the evolution of the spectral index $q$ of the axion emission spectrum.
The data is digitised from the right panel of Fig.~16 of Ref.~\cite{Gorghetto:2020qws} and Supplementary Fig.~10 of Ref.~\cite{Buschmann:2021sdq}.
For the results of this work, the error bars include systematics induced by changing the parameter $\sigma_{\rm filter}$ for the filtering procedure to calculate $\mathcal{F}$ [Eq.~\eqref{filter_for_F}] in addition to statistical uncertainties.
For the results of Ref.~\cite{Gorghetto:2020qws}, the error bars represent statistical uncertainties. 
The error bars for the result of Ref.~\cite{Buschmann:2021sdq} represent the error of the fit, rather than statistical uncertainties.
Three possible models for the continuum extrapolation of the results of this work obtained in Sec.~\ref{sec:continuum_ext_q} 
are also shown by the dash-dotted line ($q_{\rm model}=q_0+q_1/\ell^2$), dashed line ($q_{\rm model}=q_0+q_1\ell$), and dotted line ($q_{\rm model}=q_0+q_1/\ell$).}
\label{fig:comparison}
\end{figure*}

Our results on $\xi$ and $q$ obtained from simulations with $11264^3$ lattice sites
are compared with those from recent simulations performed by the authors of Refs.~\cite{Gorghetto:2020qws,Buschmann:2021sdq} in Fig.~\ref{fig:comparison}.
Here we summarise differences in the simulation/analysis methods and discuss possible sources of discrepancies.

The authors of Ref.~\cite{Gorghetto:2020qws} performed simulations up to $4500^3$ lattice sites.
Among multiple sets of simulations with different choices of parameters, they used the results from simulations with $m_ra=1$ at the final time in the main analysis, and we replot them in Fig.~\ref{fig:comparison}.
From the left panel of that figure, we see that the string density parameter at the attractor found in Sec.~\ref{sec:attractor} is very compatible with their result.
On the other hand, the values of $q$ obtained from our analysis become systematically smaller than those obtained in Ref.~\cite{Gorghetto:2020qws}.
One possible origin of this discrepancy is attributed to the fact that we use the 4-neighbour discretisation scheme of the Laplacian, while they used the 2-neighbour scheme.
In Sec.~\ref{sec:lap}, we have shown that a lower level of discretisation of the Laplacian can distort the spectrum and lead to larger values of $q$.
We have also shown that the 2-neighbour Laplacian may not be enough to suppress that effect.

Other than the effect of the Laplacian, there are a few effects that can cause the difference.
The fact that the string core resolution $m_ra = 1$ at $\ln(m_r/H)\simeq 7.9$ in Ref.~\cite{Gorghetto:2020qws} is lower than ours [$m_ra\simeq 0.69$ at $\ln(m_r/H)=7.9$]
can also result in larger values of $q$ and explain the discrepancy at large $\ln(m_r/H)$.
Furthermore, the values of $\xi$ in Ref.~\cite{Gorghetto:2020qws} remains slightly smaller than our results (see left panel of Fig.~\ref{fig:comparison}) 
throughout the simulated range of $\ln(m_r/H)$, and such smaller values of $\xi$ could lead to larger values of $q$ (cf. Fig.~\ref{fig:attr_q}), though the effect would be small.
The discrepancy in $q$ appears to be more pronounced at smaller $\ln(m_r/H)$, but it is not straightforward to offer a simple explanation for that,
since at smaller $\ln(m_r/H)$ the range of the momentum used for the fit of the instantaneous emission spectrum becomes very short,
which may lead to potentially larger systematic uncertainties.
In particular, they used the finite difference method to compute the instantaneous emission spectrum, which could 
be more sensitive to the effect of the oscillations in the spectrum and result in larger fluctuations in $q$ as illustrated in Sec.~\ref{sec:oscillations}.
We also note that their fiducial choice of the IR cutoff $c_{\rm IR} = 30$ for the fit deviates from the critical value $x_{\crit} \simeq 25.1$,
which could potentially amplify the effect of oscillations in the IR part of the spectrum (see Sec.~\ref{sec:oscillations}).

The authors of Ref.~\cite{Buschmann:2021sdq} carried out a simulation based on the AMR technique rather than the conventional method with the static lattice.
The simulation was performed with a uniform grid of $2048^3$ cells and up to five levels of refinement around the string core, which would effectively amount to static grid simulation with $65536^3$ lattice sites.
It should be noted that only one large scale simulation was performed in Ref.~\cite{Buschmann:2021sdq}, and it is not possible to assign the statistical uncertainties to the results.
The error bars for $\xi$ shown in the left panel of Fig.~\ref{fig:comparison} were determined by treating them as a nuisance parameter in the fit of 
a model of $\xi$ given by a simple function of $\ell$
to the data.
The error bars for $q$ shown in the right panel of Fig.~\ref{fig:comparison} was obtained from the second partial derivatives of the Gaussian likelihood used for the fit of 
a power law model to the data of the instantaneous emission spectrum.

There are several factors that could give rise to a different assessment of $q$ in Ref.~\cite{Buschmann:2021sdq}.
First, the simulation performed in Ref.~\cite{Buschmann:2021sdq} corresponds to under-dense strings compared to ours as shown in the left panel of Fig.~\ref{fig:comparison},
and such an under-dense case is likely to give a higher value of $q$, as shown in Fig.~\ref{fig:attr_q}.
Second, in Ref.~\cite{Buschmann:2021sdq} the 1-neighbour finite difference was used for the discretisation of the Laplacian, while we use the Laplacian with 4-neighbours.
As shown in Fig.~\ref{fig:disc_lap_q}, the value of $q$ computed in the 1-neighbour case becomes substantially larger than other cases.
Finally, in Ref.~\cite{Buschmann:2021sdq} the instantaneous emission spectrum was computed based on the finite difference method with the interval $\Delta \ell=0.25$,
and a smaller value $c_{\rm UV}=1/16$ than ours ($c_{\rm UV}=1/4$) was used for the UV cutoff of the momentum range used in the fit to determine $q$.
As discussed in Sec.~\ref{sec:oscillations}, due to the existence of the $2k$-oscillations in the spectrum 
the values of $q$ could be biased at smaller values of $c_{\rm UV}$
when the instantaneous emission spectrum is computed based on the finite difference method, 
and the effect could be amplified by statistical fluctuations.
The fact that they used a thermal initial condition could also lead to some bias, since there might be a high frequency oscillation of the PQ field 
that can also lead to contaminations in the spectrum, as discussed in Appendix~\ref{app:initial_conditons}.
Moreover, the momentum range available for the fit becomes even shorter at smaller $\ln(m_r/H)$, which together with the existence of the $2k$-oscillations makes
the fit more uncertain at smaller $\ln(m_r/H)$. This can actually be seen in the right panel of Fig.~\ref{fig:comparison}, 
where the error bars of $q$ in the results of Ref.~\cite{Buschmann:2021sdq} become particularly large at smaller $\ln(m_r/H)$.
The trend of the logarithmic increase of $q$ observed in Ref.~\cite{Gorghetto:2020qws} and this work could become less apparent in Ref.~\cite{Buschmann:2021sdq} because of these effects.

Compared to the results of Ref.~\cite{Buschmann:2021sdq}, our results give rise to larger values of $q$ at later times of the simulation [$\ln(m_r/H)\gtrsim 8.5$].
This is attributed to the fact that the resolution of the string core becomes worse at late times in our simulations [$m_ra=1.25$ at the final time corresponding to $\ln(m_r/H)=9.08$], 
and $q$ can be biased towards larger values due to the discretisation effect discussed in Sec.~\ref{sec:msa}.
This effect should not be a problem in the simulation performed in Ref.~\cite{Buschmann:2021sdq},
since the resolution of the string core is kept as good as $(m_ra)^{-1} > 3$ throughout the simulation by adding 5 refinement levels appropriately.

After the first version of this paper was submitted, the work~\cite{Kim:2024wku} by yet another group appeared.
In Ref.~\cite{Kim:2024wku}, simulations were performed on a static lattice with $4096^3$ sites. 
Parameters were chosen such that $m_ra=1$ and $\tau/L = 4^{-1/3} \simeq 0.63$ at the end of the simulations, which correspond to $\ln(m_r/H) \lesssim 7.86$.
Two different methods to produce initial conditions (a PRS-type prepropagation and a thermal initial condition) were applied, 
and it was confirmed that both cases lead to similar results except that the results from the thermal initial condition give rise to more fluctuations in the axion radiation spectrum and 
take more time to be relaxed to the regime close to the attractor.
The initial string density was set to $\xi \simeq 0.2$ at $\ln(m_r/H)=2$, which makes the evolution of $\xi$ quite similar to Ref.~\cite{Buschmann:2021sdq}
and hence corresponds to an under-dense network compared to the attractor [$\xi\simeq 0.3$ at $\ln(m_r/H)=3$] found in Sec.~\ref{sec:attractor}.
The instantaneous emission spectrum was computed by using the finite difference method with $\Delta\ell = 0.25$ as a default choice, 
but it was observed that higher values of $\Delta \ell$ make the spectrum smoother.
This trend is consistent with what we found in Fig.~\ref{fig:osc_F_diff_vs_fit}, and we provided some analytical understanding of this feature [see Eq.~\eqref{F_finite_diff_analytical} and the corresponding discussion].
We note that the value of $c_{\rm IR}=15$ used to measure $q$ in the main analysis in Ref.~\cite{Kim:2024wku} is smaller than the critical value $x_{\crit} \simeq 25.1$ for $\Delta\ell = 0.25$, 
and hence the effect of the oscillations in the IR part of the spectrum might not be minimised.

We find that the values of $q$ obtained in Ref.~\cite{Kim:2024wku} are more or less consistent with our results at small $\ell$, showing the trend of the logarithmic increase.
On the other hand, its values at higher $\ln(m_r/H)$ appear to be slightly larger than our results and those of Ref.~\cite{Gorghetto:2020qws}.
This disagreement could be attributed to the smaller grid size, which leads to higher values of $m_ra$ and could pronounce the discretisation effects.
The fact that the value of $\xi$ in the simulations in Ref.~\cite{Kim:2024wku} remains smaller than the attractor
could also make the value of $q$ slightly larger.
Furthermore, although it is not clear how the Laplacian was computed in the simulations in Ref.~\cite{Kim:2024wku}, 
the discretisation of the Laplacian could be an issue, as we discussed in Sec.~\ref{sec:lap}.

\bibliography{main}
\bibliographystyle{utphys}

\end{document}